\pdfoutput=1

\documentclass[11pt,twoside,a4paper,cmspaper,final,collab]{cms-tdr}
\def\svnVersion{753d5d2-D}\def\svnDate{2020/05/30}\def\cmsCernNoTag{CERN-EP-2020-037}\def\cmsCernDate{\today}\def\cmsMessage{"Published in the Journal of Instrumentation as \href{http://dx.doi.org/10.1088/1748-0221/15/06/P06005}{\doi{10.1088/1748-0221/15/06/P06005}.}"}

\begin{document}\cmsNoteHeader{JME-18-002}

\hyphenation{had-ron-i-za-tion}
\hyphenation{cal-or-i-me-ter}
\hyphenation{de-vices}
\RCS$HeadURL$
\RCS$Id$

\newcommand{\met}{\ptmiss}
\newcommand{\ptrel}{\ensuremath{p_{\text{T,rel}}}\xspace}
\newcommand{\esig}{\ensuremath{\epsilon_{\mathrm{S}}}\xspace}
\newcommand{\ebkg}{\ensuremath{\epsilon_{\mathrm{B}}}\xspace}
\newcommand{\mjet}{\ensuremath{m_{\text{jet}}}\xspace}
\newcommand{\tauthreetwo}{\ensuremath{\tau_{32}}\xspace}
\newcommand{\tautwoone}{\ensuremath{\tau_{21}}\xspace}
\newcommand{\sdmass}{m\ensuremath{{}_{\mathrm{SD}}}\xspace}
\newcommand{\nvtx}{\ensuremath{N_{\mathrm{PV}}}\xspace}
\newcommand{\puppi}{\ensuremath{\text{PUPPI}}\xspace}
\newcommand{\pp}{\ensuremath{\Pp\Pp}\xspace}
\newcommand{\photonsample}{\ensuremath{\Pgg\text{+jets}}\xspace}
\newcommand{\msd}{\ensuremath{m_{\mathrm{SD}}}\xspace}
\newcommand{\tauthreetwosd}{\ensuremath{\tau_{32}^{\mathrm{SD}}}\xspace}
\newcommand{\msdtop}{\ensuremath{m_\mathrm{SD}+{\tau_{32}}}\xspace}
\newcommand{\msdtopbtag}{\ensuremath{m_\mathrm{SD}+{\tau_{32}+\PQb}}\xspace}
\newcommand{\msdv}{\ensuremath{m_\text{SD}+{\tau_{21}}}\xspace}
\newcommand{\ecftop}{\ensuremath{N_{3}\text{-}\mathrm{BDT}\,(\mathrm{CA}15)}\xspace}
\newcommand{\ecfv}{\ensuremath{m_\mathrm{SD}+N_{2}}\xspace}
\newcommand{\ecfvddt}{\ensuremath{m_\mathrm{SD}+N_{2}^{\mathrm{DDT}}}\xspace}
\newlength\cmsTabSkip\setlength\cmsTabSkip{1ex}

\cmsNoteHeader{JME-18-002}
\title{Identification of heavy, energetic, hadronically decaying particles using  machine-learning techniques}

\author{The CMS Collaboration}

\date{\today}

\abstract{
Machine-learning (ML) techniques are explored to
identify and classify hadronic decays of highly Lorentz-boosted
\PW{}/\PZ{}/Higgs bosons and top quarks. 
Techniques without ML have also been evaluated and are
included for comparison.
The identification performances of a variety of algorithms are characterized in
simulated events and directly compared with data. The
algorithms are validated using proton-proton collision data
 at $\sqrt{s} =  13\TeV$, corresponding to an integrated luminosity of 35.9\fbinv.
Systematic uncertainties are assessed by comparing the results obtained using
 simulation and collision data.
 The new techniques studied in this paper
provide significant performance improvements over non-ML techniques,
reducing the background rate by up to an order of magnitude at the same
signal efficiency.
}

\hypersetup{%
pdfauthor={CMS Collaboration},%
pdftitle={Machine learning-based identification of highly Lorentz-boosted hadronically decaying particles at the CMS experiment},%
pdfsubject={CMS},%
pdfkeywords={CMS, physics, jet substructure, machine learning, jet tagging}}

\maketitle

\section{Introduction}
\label{sec:introduction}

At the CERN LHC~\cite{LHC}, an efficient classification 
 of hadronic decays of heavy standard-model (SM) particles (objects) that are
reconstructed within a single jet would provide a significant improvement in
 the sensitivity of searches for physics beyond the SM (BSM)
 and in measurements of SM parameters.
The understanding of jet substructure in  highly Lorentz-boosted \PW{}/\PZ{}/\PH bosons
(where \PH is the Higgs boson)
and top (\PQt) quark
jets has advanced dramatically in recent years, both
experimentally~\cite{Asquith:2018igt} and
theoretically~\cite{Larkoski:2017jix}.
 For a particle with a Lorentz boost of $\gamma$,
the angular separation between its decay products scales as
$\theta \sim 2/\gamma$ in radians.
A knowledge of the radiation patterns of these jets and their substructure is an important topic in theoretical and experimental research.

In this paper, we present studies using the CMS detector~\cite{JINST} at the LHC 
to evaluate and compare the
performances of a variety of algorithms (``taggers'') designed to distinguish
hadronically decaying massive SM particles with large Lorentz boosts,
namely \PW{}/\PZ{}/\PH bosons and \PQt~quarks, from other jets originating
from lighter quarks (\PQu{}/\PQd{}/\PQs{}/\PQc{}/\PQb)
or gluons (\Pg). We refer to
such jets as ``boosted \PW{}/\PZ{}/\PH{}/\PQt~jets,'' or ``\PW{}/\PZ{}/\PH{}/\PQt-tagged jets''.
The machine-learning (ML) algorithms  include the energy correlation functions tagger (ECF),
the boosted event shape tagger (BEST), the ImageTop tagger, and the DeepAK8
tagger.  Algorithms without ML techniques have also been evaluated and are
included for comparison. An alternative approach for jet clustering and identification, named the ``heavy object with variable R (HOTVR)'', where the heavy object is a \PW{}/\PZ{}/\PH boson or \PQt quark,
 is also studied.

The theoretical and experimental understanding of jet substructure
has gained significant precision in recent years.
The CMS Collaboration has made many relevant measurements of jet
substructure, including measurements of the cross
section of highly Lorentz-boosted
\PQt quarks~\cite{Khachatryan:2016gxp}, jet mass in
\ttbar~\cite{Sirunyan:2017yar},
dijet~\cite{Chatrchyan:2013vbb,Sirunyan:2018xdh},
samples enriched in light-flavors~\cite{Chatrchyan:2013vbb}, and
substructure observables in jets of different light-quark
flavors~\cite{Sirunyan:2018asm} in resolved \ttbar events.
Similar measurements by the ATLAS Collaboration are found in
Refs.~\cite{ATLAS:2012am,Aaboud:2017qwh,Aad:2014haa,Aaboud:2018eqg,Aaboud:2019aii}.
Overall, the systematic effects of jet substructure are
well understood and, after correcting for detector effects, the results are
generally consistent with theoretical expectations as expressed in simulations.
Residual differences between data and simulation can be adjusted using scale factors.

ML-based approaches can be tailored to suit the needs of
individual analyses. Some analyses require as pure a sample
as possible, with optimized signal efficiency for a fixed background rejection.
Others require well-behaved background estimates as a function
of kinematic variables. A characteristic example 
is the use of jet mass sidebands
for the background estimation. In this case, removing dependencies on the jet mass is collectively
referred to as ``mass decorrelation'', as described in
Ref.~\cite{Dolen:2016kst}. This paper provides tools derived from 
 a strong program of previous
study~\cite{CMS-PAS-JME-13-007,CMS-PAS-JME-15-002,CMS-PAS-JME-16-003,Khachatryan:2014vla,Sirunyan:2017ezt}
for both the jet-mass-decorrelated and nominal scenarios.

The paper is organized as follows.
A brief description of the CMS detector is presented in
Section~\ref{sec:cmsdetector}. The Monte Carlo (MC) simulated events
used for the results are discussed in Section~\ref{sec:samples},
and details of the CMS event reconstruction and the event
selections used for the studies are summarized in
Sections~\ref{sec:evtreco} and~\ref{sec:evtselection},
respectively. Section~\ref{sec:hrtalgorithms} presents an overview of
the methods currently used in CMS for heavy-resonance (i.e., \PW{}/\PZ{}/\PH bosons and \PQt quarks) identification,
and describes a set of novel algorithms that utilize ML methods and
observables for this task. Our discussion of the CMS methods
 builds on the work documented
in Refs.~\cite{CMS-PAS-JME-13-007,CMS-PAS-JME-15-002,CMS-PAS-JME-16-003,Khachatryan:2014vla,Sirunyan:2017ezt}.
Section~\ref{sec:performanceinmc} details the analyses performed
to understand the complementarity
between the algorithms using simulated events. The performance of
the algorithms is validated in data samples collected in
proton-proton (\pp) collisions at $\sqrt{s} =  13\TeV$ by the CMS
experiment at the LHC in 2016, and corresponding to an integrated
luminosity of 35.9\fbinv. The results, along with the effect of
systematic uncertainties in their measurement, are presented in
Section~\ref{sec:performanceindata}, followed by a discussion of the results
and a summary in Section~\ref{sec:summary}.

\section{The CMS detector}
\label{sec:cmsdetector}
The central feature of the CMS apparatus is a superconducting solenoid
of 6\unit{m} internal diameter, providing a magnetic field of
3.8\unit{T}. Within the solenoid volume are a silicon
pixel and strip tracker, a lead tungstate crystal electromagnetic
calorimeter (ECAL), and a brass and scintillator hadron calorimeter
(HCAL), each composed of a barrel and two endcap sections. Forward
calorimeters extend the pseudorapidity ($\eta$) coverage provided
by the barrel and endcap detectors~\cite{JINST}. Muons are
measured in gas-ionization chambers embedded in the steel flux-return
yoke outside the solenoid.

In the barrel section of the ECAL, an energy resolution of about 1\%
is achieved for unconverted or late-converting photons in the tens of
GeV energy range. The remaining barrel photons have a resolution of
about 1.3\% up to $\abs{\eta} = 1$, rising to about
2.5\% at $\abs{\eta} = 1.4$. In the endcaps, the resolution of unconverted
or late-converting photons is about 2.5\%, while the remaining endcap
photons have a resolution between 3 and
4\%~\cite{CMS:EGM-14-001}.

In the region $\abs{ \eta } < 1.74$, the HCAL cells have widths of
0.087 in $\eta$ and 0.087 radians in azimuth ($\phi$). In the
$\eta$-$\phi$ plane, and for $\abs{\eta} < 1.48$, the HCAL cells map
onto $5{\times}5$ ECAL crystals arrays to form calorimeter towers
projecting radially outwards from close to the nominal interaction
point. At larger values of $\abs{ \eta }$, the size of the towers
increases and the matching ECAL arrays contain fewer crystals.

Muons are measured in the $\eta$ range $\abs{\eta} < 2.4$,
with detection planes made using three technologies: drift tubes,
cathode strip chambers, and resistive-plate chambers. Matching muons
to tracks measured in the silicon tracker results in a relative
transverse momentum ($\pt$) resolution for muons with $20 <\pt < 100\GeV$ of
1.3--2.0\% in the barrel and better than 6\% in the endcaps. The \pt
resolution in the barrel is better than 10\% for muons with \pt up to
1\TeV~\cite{Chatrchyan:2012xi}.

The silicon tracker measures charged particles within the
pseudorapidity range $\abs{\eta} < 2.5$. It consists of 1440 silicon
pixel and 15\,148 silicon strip detector modules. Isolated particles
of $\pt = 100\GeV$ emitted at $\abs{\eta} < 1.4$ have track resolutions
of 2.8\% in \pt and 10 (30)\mum in the transverse (longitudinal)
impact parameter \cite{TRK-11-001}.

Events of interest are selected using a two-tiered trigger
system~\cite{Khachatryan:2016bia}. The first level (L1), composed of
custom hardware processors, uses information from the calorimeters and
muon detectors to select events at a rate of around 100\unit{kHz}. 
The second level, known as
the high-level trigger (HLT), consists of a farm of processors running
a version of the full event reconstruction software optimized for fast
processing, and reduces the event rate to around 1\unit{kHz} before
data storage.

A more detailed description of the CMS detector, together with the
 definition of the coordinate system used and the relevant kinematic
variables, is given in Ref.~\cite{JINST}.

\section{Simulated event samples}
\label{sec:samples}

Simulated pp collision events are generated at $\sqrt{s}=13\TeV$
using various generators described below. They are used for the design
and the performance studies of the heavy-resonance identification
algorithms to compare with data and to estimate systematic uncertainties.
 The
signal samples, enriched in one or more \PW{}/\PZ{}/\PH{}/-tagged
jets, are obtained from the simulation of BSM processes. The \PQt and \PW jet
signal samples are obtained from heavy spin-1 \PZpr resonances decaying
to either a pair of \PQt quarks (\ttbar) or a pair of \PW bosons,
respectively. These resonances are narrow, having intrinsic widths
equal to 1\% of the resonance mass. The \PZ- and \PH-tagged jet
samples are obtained from decays of spin-2 Kaluza--Klein graviton
resonances in the Randall--Sundrum model~\cite{Randall:1999ee,Randall:1999vf}
 to a pair
of \PZ or \PH bosons, following the narrow-width
assumption.
The \PZpr and graviton samples are simulated at leading order (LO)
with \MGvATNLO 2.2.2~\cite{Alwall:2014hca} interfaced with
\PYTHIA 8.212~\cite{Sjostrand:2007gs, Skands:2014pea} with
the CUETP8M1 underlying event tune~\cite{Khachatryan:2015pea}
for the fragmentation and hadronization description.
Signal events are generated over a wide range of \pt 
for different \PZpr and graviton mass values.
The background
sample is represented by jets produced via the strong interaction of
quantum chromodynamics (QCD), referred to as ``QCD multijet''
processes. The QCD multijet events are generated using \PYTHIA\
in exclusive $\hat{\pt}$ bins using the
NNPDF2.3 LO~\cite{Ball:2012cx} parton distribution function (PDF)
set.

A variety of MC simulations are needed for the study of the performance of the
tagging algorithms in data. The \ttbar process is generated with the
next-to-leading-order (NLO) generator \POWHEG v2.0
\cite{Nason:2004rx,Frixione:2007nw,Alioli:2010xd} interfaced with
\PYTHIA for the fragmentation and hadronization description.
 Simulated events originating from
\PW{}+jets, \PZ{}+jets, and \photonsample, are generated using \MGvATNLO
at LO accuracy using the
NNPDF3.0 LO~\cite{Ball:2012cx} PDF set. The \PW{}\PZ{}, \PZ{}\PZ{}, \ttbar{}\PW{},
 and \ttbar{}\photonsample processes are generated using
\MGvATNLO at NLO accuracy, the single \PQt quark process
in the \PQt\PW{} channel and the \PW{}\PW{} process are
generated at NLO accuracy with \POWHEG, all using the NNPDF3.0 NLO PDF
set. In all of these cases, parton showering and
hadronization are simulated in \PYTHIA. Double
counting of partons generated using \PYTHIA and \MGvATNLO
is eliminated using the MLM~\cite{Alwall:2007fs} and
FxFx~\cite{Frederix:2012ps} matching schemes for the LO and NLO
samples, respectively.

The systematic uncertainties associated with the performance of the
taggers are evaluated using simulated events produced with alternative
generation settings. For the \ttbar~process, an additional sample
is generated using \POWHEG interfaced
with \HERWIGpp v2.7.1 \cite{Bahr:2008pv,Bellm:2013hwb}
with the UE-EE-5C underlying event tune~\cite{Seymour:2013qka}
to assess systematic uncertainties related to
the modeling of the parton showering and hadronization.
Additional QCD multijet samples are generated at LO accuracy using
\MGvATNLO, interfaced with \PYTHIA to test the
modeling of the hard scattering in background events, or generated solely
with \HERWIGpp 
with the CUETHppS1 underlying event tune~\cite{Khachatryan:2015pea}
to provide an alternative model of the background jets.

The most precise cross section calculations available are used to normalize the
SM simulated samples. In most cases, this is next-to-NLO
 accuracy in the inclusive cross section. Finally, the \pt~spectrum of top quarks
in \ttbar~events is reweighted (referred to as ``top quark 
\pt~reweighting'') to account for effects due to missing higher-order
corrections in MC simulation, according to the results presented in
Ref.~\cite{PhysRevD.95.092001}. The simulation of the QCD multijet and $\gamma+$jets processes
is based on LO calculations.
To account for missing higher-order corrections, the simulated QCD
multijet events and the $\gamma+$jets events are reweighted such that
the \pt distribution of the leading jet in simulation matches that in
data. Before extracting the weights, contributions from other processes
 are subtracted from data using the predicted cross sections in both cases.

A full \GEANTfour-based model~\cite{Agostinelli:2002hh} is used to
simulate the response of the CMS detector to SM background
samples. Event reconstruction is performed in the same manner for MC
simulation as for collision data. A nominal distribution of multiple pp
collisions in the same or neighboring bunch crossings (referred to as
``pileup'') is used to overlay the simulated events. The events are
then weighted to match the pileup profile observed in the
data.  For the data used in this paper, there were an average of 23
interactions per bunch crossing.

\section{Event reconstruction and physics objects}
\label{sec:evtreco}

Events are reconstructed using the CMS particle-flow (PF)
algorithm~\cite{CMS-PRF-14-001}, which aims to reconstruct and identify
each individual particle in the event with an optimized combination of information
from the various elements of the detector. Particles are identified as
charged or neutral hadrons, photons, electrons, or muons, and
 cannot be classified into multiple categories.
 The PF candidates are then used to build higher-level objects,
such as jets. Events are required to have at
least one reconstructed vertex. The physics objects are those returned by a
jet-finding algorithm~\cite{Cacciari:2008gp,Cacciari:2011ma} applied
to the tracks associated with the vertex, and the associated missing transverse momentum
\ptvecmiss, taken as the negative vector sum of the \pt of those jets.
In the case of multiple overlapping events with
multiple reconstructed vertices, the vertex with the largest value of
summed physics object $\pt^2$ is defined to be the primary \pp\
interaction vertex (PV).

Photons are reconstructed from energy depositions in the ECAL using
identification algorithms that use a collection of variables
related to the spatial distribution of shower energy in the
supercluster (a group of $5{\times}5$ ECAL crystals), the photon isolation, and
the fraction of the energy deposited in the HCAL behind the
supercluster relative to the energy observed in the
supercluster~\cite{CMS:EGM-14-001,Khachatryan:2015hwa}. The
requirements imposed on these variables ensure an efficiency of 80\%
in selecting prompt photons. Photon candidates are required to be
reconstructed with $\pt>200\GeV$ and $\abs{\eta} < 2.5$. Simulation-to-data
correction factors are used to correct photon identification performance in MC.

Electrons are reconstructed by combining information from the inner
tracker with energy depositions in the ECAL~\cite{Khachatryan:2015hwa}. 
Muons are reconstructed by combining
tracks in the inner tracker and in the muon system~\cite{Chatrchyan:2012xi}. 
Tracks associated with electrons or muons
are required to originate from the PV, and a set of quality criteria
is imposed to assure efficient identification~\cite{Khachatryan:2015hwa,Chatrchyan:2012xi}. 
To suppress
misidentification of charged hadrons as leptons, we require electrons
and muons to be isolated from jet activity within a \pt-dependent cone
  in the $\eta$-$\phi$ plane, $\Delta R = \sqrt{\smash[b]{(\Delta\eta)^2+(\Delta\phi)^2}}$, 
where $\phi$ is the azimuthal angle in radians. The relative
isolation, $I_\text{rel}$, is defined as the $\pt$ sum
of the PF candidates  within the cone divided by the lepton
\pt. Neither charged PF candidates not originating from the PV, nor 
those  identified as electrons or muons, are included in the
sum.

The isolation sum $I_\text{rel}$ is corrected for contributions of
neutral particles originating from pileup interactions using an
area-based estimate~\cite{CMS-PAS-JME-14-001} of pileup energy
deposition in the cone. The requirements imposed on the electron and
muon candidates lead to an average identification efficiency of~70 and 95\%,
respectively. In addition, the electron and muon candidates are
required to have $\pt>40\GeV$ and be within the tracker acceptance of
$\abs{\eta} < 2.5$. The electron and muon identification performance in simulation
is corrected to match the performance in data.

The primary jet collection in this paper, referred to as ``AK8 jets'', 
is produced by clustering PF
candidates using the anti-\kt algorithm \cite{Cacciari:2008gp} with
a distance parameter of $R=0.8$ with the \FASTJET3.1 software
package~\cite{Cacciari:2008gp,Cacciari:2011ma}. 

 A collection of jets produced
using the Cambridge--Aachen
(CA)~\cite{Dokshitzer:1997in, Wobisch:1998wt} clustering algorithm
with $R=1.5$, referred to as ``CA15 jets'', is also used in this paper. 
In both jet collections,
the ``PileUp Per Particle Identification (PUPPI)''~\cite{Bertolini:2014bba} method
is used to mitigate the effect of pileup on jet observables. This method
makes use of local shape information around each particle in
the event, the event pileup properties, and tracking information. This PUPPI algorithm
operates at the PF candidate level, before any jet clustering is
performed. A local variable $\alpha$ is computed for each PF candidate,
which contrasts the
collinear structure of QCD with the low-$\pt$ diffuse radiation
arising from pileup interactions. This $\alpha$ variable is used to
calculate a weight correlated with the probability that an individual PF
candidate originates from a pileup collision. These per PF candidate
weights are used to rescale the four-momenta of each PF candidate to
correct for pileup. The resulting PF candidate list is used as an input
to the clustering algorithm. A detailed description of the PUPPI implementation in CMS
can be found in Ref.~\cite{puppicms}. No additional pileup corrections are
applied to jets clustered from these weighted inputs. Corrections are
applied to the jet energy scale to compensate for nonuniform detector
response \cite{Khachatryan:2016kdb}. Jets are required to have
$\pt>200\GeV$ and $\abs{\eta}<2.4$.

A collection of jets, reconstructed with the anti-\kt algorithm and a smaller 
distance parameter $R = 0.4$, referred to as ``AK4 jets'', are used to define 
the event samples for the validation of the algorithms.
To reduce
the effect of pileup collisions, charged PF candidates identified as
originating from pileup vertices are removed before the jet
clustering, based on the method known as ``charged-hadron
subtraction''~\cite{Khachatryan:2016kdb}. An event-by-event
correction based on jet area~\cite{Khachatryan:2016kdb} is applied to
the jet four-momenta to remove the remaining neutral energy from pileup
vertices. As with the AK8 and CA15 jets described above, additional
corrections to the jet energy scale are applied to compensate for
nonuniform detector response. The AK4 jets are required to have
$\pt>30\GeV$ and be contained within the tracker volume of
$\abs{\eta}<2.4$.

Jets originating from the hadronization of bottom (\PQb) quarks are
identified, or ``tagged'', using the combined secondary vertex
(CSVv2) \PQb tagging algorithm \cite{Sirunyan:2017ezt}. The working point, 
i.e., a selection on the algorithm's discriminant providing 
a well defined signal (e.g., \PQb quarks) and background 
(e.g., light quarks) efficiency, used
provides an efficiency for the \PQb tagging of jets originating from \PQb\
quarks that varies from 60 to 75\%, depending on \pt, whereas the
misidentification rate for light quarks or gluons is $\sim$1\%, and
$\sim$15\% for charm quarks.

\begin{figure}
\centering
\includegraphics[width=0.47\textwidth]{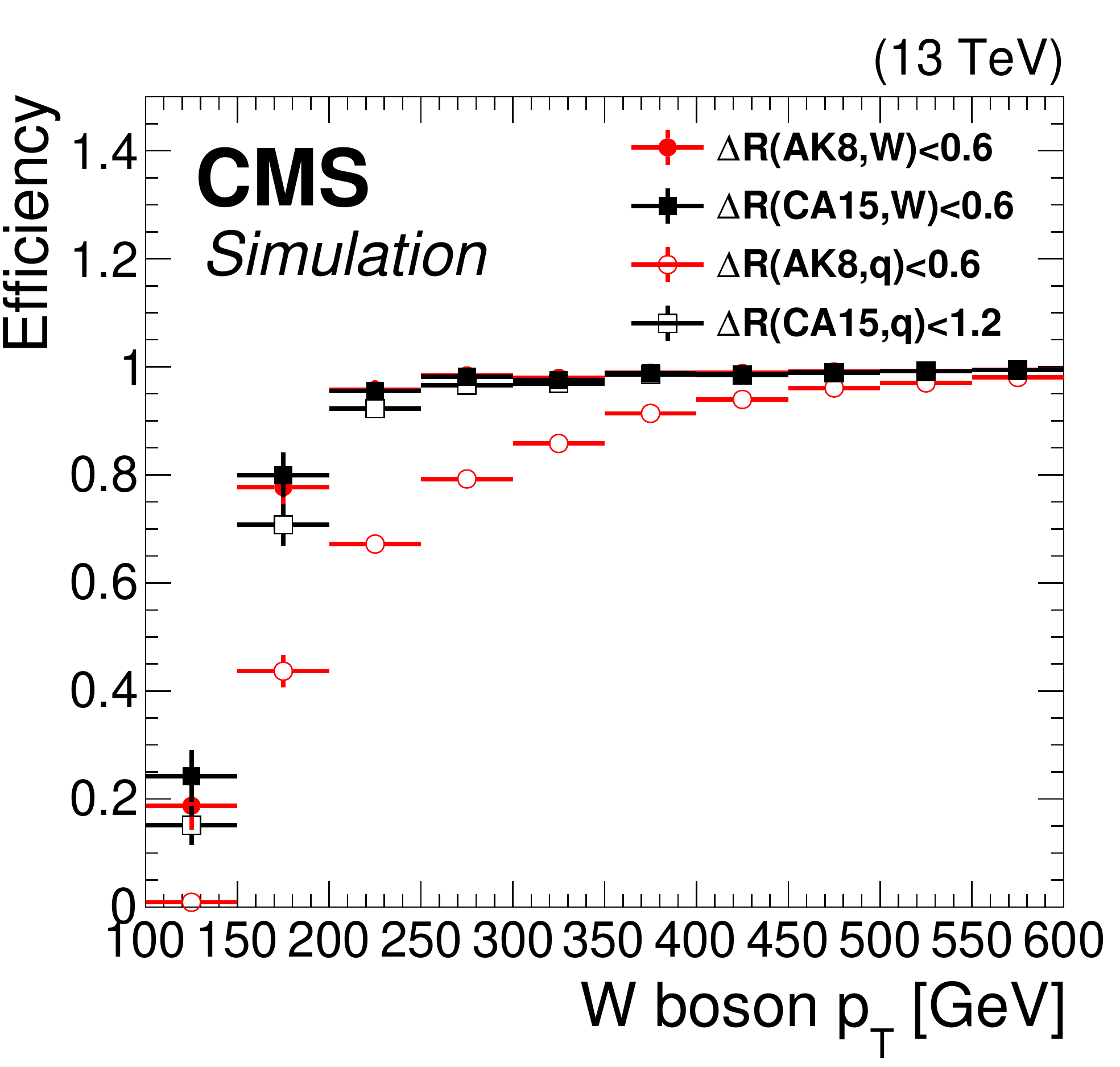}
\includegraphics[width=0.47\textwidth]{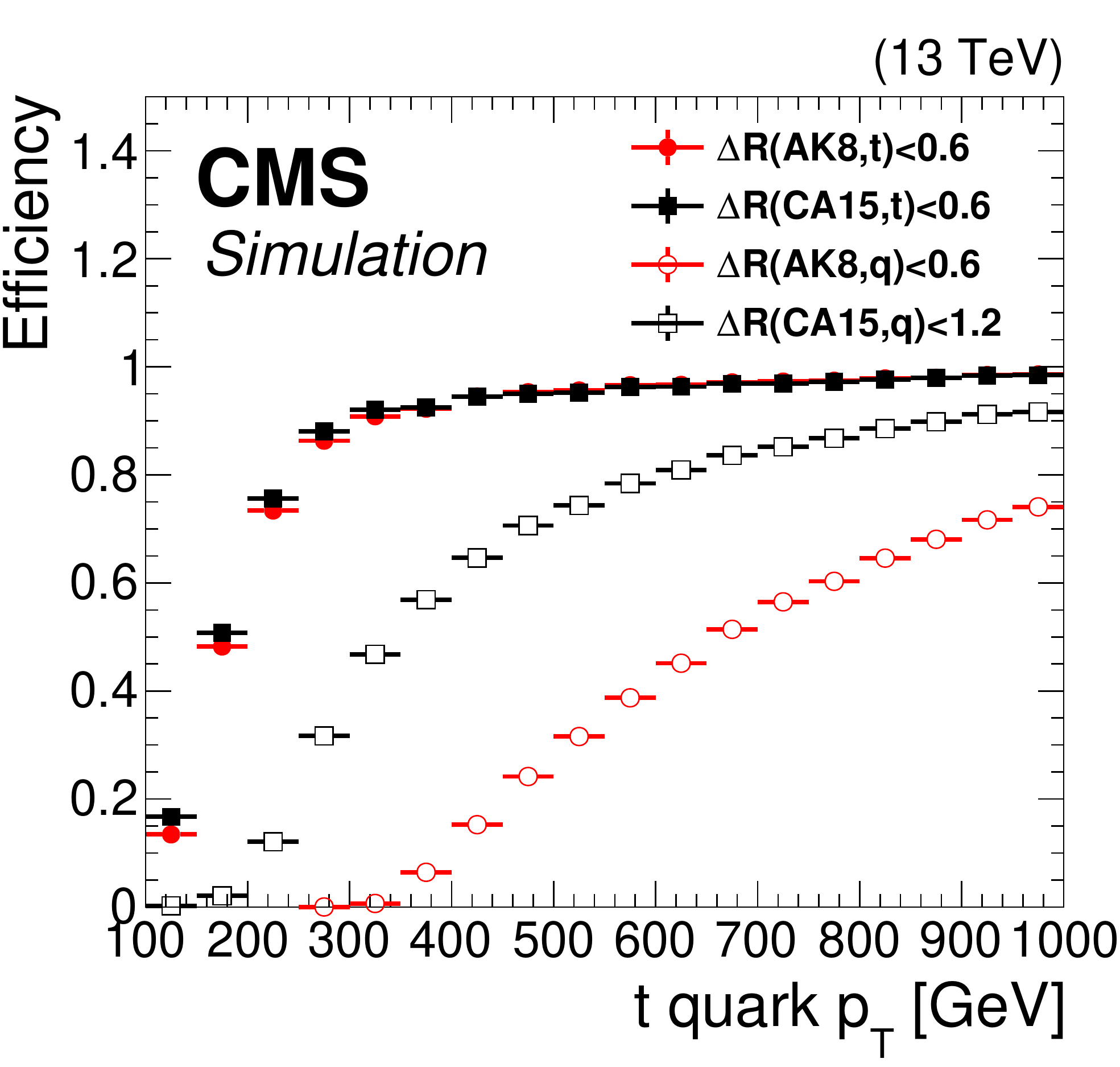}
\caption{\label{fig:truthmatch_dr_effvspt} Matching efficiency
  as a function of the \pt~of the generated particle, for hadronically
  decaying \PW{} bosons (left) and \PQt quarks (right). This
  efficiency is defined as the fraction of the generated particles (\PQt\
  quarks or \PW  bosons) that are within $\Delta R<0.6$  with an
  AK8 or CA15 jet with $\pt>200\GeV$ and $\abs{\eta}<2.4$. Superimposed
  is the merging efficiency as a function of the generated particle \pt when all
  decay products are within $\Delta R (\text{AK8}, \Pq_{i}) < 0.6$
  ($\Delta R (\text{CA15}, \Pq_{i}) < 1.2$) with an AK8 (CA15) jet. }
\end{figure}

{\tolerance=1200
For the studies presented in this paper, the simulated signal jets (AK8 or CA15
jets) are identified as
\PW{}/\PZ{}/\PH/\PQt-tagged jets when the $\Delta R$ between the reconstructed jet and
the closest generated particle (\PW{}/\PZ{}/\PH boson or \PQt quark) before the
decay, denoted as $\Delta R (\text{jet}, \text{generated particle})$, is
less than 0.6. This definition allows for a
consistent comparison of the performance of the algorithms using
collections of jets clustered with different $R$. The choice of the 0.6 value
approximately corresponds to the minima of the $\Delta R$ distribution
between jets and the closest generated particle based on studies reported
in Ref.~\cite{CMS-PAS-JME-15-002}.
The fraction of AK8
jets with $\Delta R (\text{AK8}, \text{generated particle})<0.6$ as a
function of the \pt of the generated particle for jets initiated from
the decay of a \PW boson (left) or \PQt quark (right) is shown in
Fig.~\ref{fig:truthmatch_dr_effvspt}. This ``matching'' efficiency of
\PW bosons (\PQt quarks) reaches a plateau of nearly 100\% for 
$\pt\gtrsim 200$ (400)\GeV.
The corresponding efficiency curve for CA15 jets is superimposed on the plots,
and shows consistent efficiency with AK8 jets. A similar efficiency is obtained when a
relaxed selection of $\Delta R (\text{CA15}, \text{generated particle})<1.2$ is applied.
This justifies the use of the same $\Delta R (\text{jet}, \text{generated particle})$
reconstruction criteria for both jet collections.
\par}

Additional criteria are applied to simulated jets for the evaluation of the
performance in data and for the calibration of the algorithms. The
partonic decay products (b, $\Pq_1$, $\Pq_2$ for \PQt quarks, or
$\Pq_1$, $\Pq_2$ for \PW, \PZ or \PH bosons) are required to be
fully contained in the AK8 (CA15) jet, satisfying
$\Delta R(\text{AK8}, \Pq_{i})<0.6$ ($\Delta R (\text{CA15}, \Pq_{i})<1.2$). These
 requirements were derived from the studies in Ref.~\cite{CMS-PAS-JME-15-002}.
The ``merging'' probability as a
function of the \pt of the generated particle (i.e., the efficiency
for the decay products of the \PQt quark or \PW boson to be fully contained
in a single jet based on the above requirements) is also shown in
Fig.~\ref{fig:truthmatch_dr_effvspt}. For \PW bosons (\PQt
quarks) with $\pt  \gtrsim 200$\,(650)\GeV, at least 50\% of the AK8
jets fully contain the \PW (\PQt)  decay products. In the case of
CA15 jets, similar efficiency is achieved for \PW bosons (\PQt\
quarks) with $\pt \gtrsim 150$\,(350)\GeV.

In the case of background jets, partons (\PQu, \PQd, \PQs, \PQc, \PQb, and
gluon) from the hard scattering are required to be contained in the
jet cone for the jet to be classified as such.

Finally, the \ptvecmiss is defined as the negative of the vectorial
sum of the $\ptvec$ of all PF candidates in the
event~\cite{Sirunyan:2019kia}. Its magnitude is denoted as
\ptmiss. The jet energy scale corrections applied to the jets are
propagated to \ptvecmiss.

\section{Event selection}
\label{sec:evtselection}
Several samples are used to validate the performance of the
tagging algorithms in data. A single-$\mu$~signal sample is
used to calibrate the \PQt quark and \PW boson identification
performance in a sample enriched in hadronically decaying \PQt
quarks, as explained below. A dijet sample, dominated by light-flavor quarks and
gluons, enables the study of the identification probability of
background jets (misidentification rate) in a wide range of \pt. The
misidentification rate depends on the flavor of the parton that initiated
the jet. Thus, in addition to the dijet sample, the
single-$\gamma$  background sample is further used. The dijet
and single-$\gamma$ samples differ in the light-flavor quark and gluon
fractions. The former has a larger fraction of gluon jets than the
latter.

Systematic effects are quantified using these samples 
to determine uncertainties in measurements corrected for the detector effects.

\subsection{The single-\texorpdfstring{$\mu$}{mu} signal sample}
\label{sec:evtselection_tt1l}
The single-$\mu$~signal sample was recorded using a
single-muon trigger that selects events online based on the muon $\pt$.
Candidate events are required to have exactly one muon with
$\pt>55\GeV$, satisfying the identification criteria defined in
Section~\ref{sec:evtreco}, except for the requirement related to the
isolation of leptons $I_\text{rel}$. In high-\pt leptonic decays of the
\PQt quarks, the lepton from the \PW boson decay often overlaps with the \PQb jet from
the \PQt quark decay, leading to large values of
$I_\text{rel}$, causing the event to be rejected. 
Therefore, a custom isolation criterion is applied by
requiring a minimal distance between the muon and the nearest AK4 jet,
$\Delta R(\mu, \text{AK4})>0.4$, or the perpendicular component of the
muon $\pt$ with respect to the nearest AK4 jet, $\ptrel > 25\GeV$.
This has been extensively used in measurements~\cite{Khachatryan:2016gxp} and
searches~\cite{Chatrchyan:2013lca,Khachatryan:2015sma,Sirunyan:2017uhk,Sirunyan:2018ryr}
involving high momentum \PQt quarks in the single-$\mu$ sample.

The
AK4 jets used in this selection are clustered from PF candidates after
removing muons with $\pt>55\GeV$. The custom isolation requirement
results in an up to 40\% increase in the statistical power of the
sample. To suppress the contribution from QCD multijet processes we
require $\met>50\GeV$. To enhance the sample purity in $\ttbar$
events, we require the presence of two or more AK4 jets, at least one
of which is reconstructed as a \PQb jet. In addition, to probe high momentum
topologies, we require the \ptvec~of the leptonically decaying \PW bosons,
defined as $\ptvec(\PW)=\ptvec(\mu)+\ptvecmiss$, and the scalar \pt~sum
of the AK4 jets, denoted as $\HT$, to be greater than 250\GeV. 
The \PQt{}/\PW candidate is the highest \pt~AK8 or CA15 jet in
the event with $\pt>200\GeV$, satisfying the criteria discussed in
Section~\ref{sec:evtreco}. To further improve the purity, we require
the azimuthal angle $\Delta\phi$ between the AK8 or CA15 jet and the
muon to be greater than 2 radians. The purity of the sample in
semileptonic \ttbar events is $\sim$70\%; other contributions arise
from QCD multijet ($\sim$15\%) and \PW{}+jets ($\sim$10\%) processes.

\subsection{The dijet background sample}
\label{sec:evtselection_qcd}
The dijet background sample was recorded with a trigger that
uses $\HT$. Events with $\HT>1000\GeV$ are selected to ensure
100\% trigger efficiency. Events are required to have at least one AK8
or CA15 jet meeting the requirements presented in
Section~\ref{sec:evtreco}, and the absence of electrons or muons,
leading to a sample dominated by jets from the QCD multijet
process, which are backgrounds to the algorithms presented here.

\subsection{The single-\texorpdfstring{$\gamma$}{gamma} background sample}
\label{sec:evtselection_pho}
The single-$\gamma$ background sample was collected using an 
isolated-single-photon trigger. Events with a photon with $\pt>200\GeV$ are
selected to ensure 100\% trigger efficiency. The photon is further
required to satisfy the criteria presented in
Section~\ref{sec:evtreco}. In addition to the photon, the
single-$\gamma$ sample is required to have at least one AK8 or CA15
jet and no electrons or muons. The sample consists of $\sim$80\%
\photonsample events, but only $\sim$15\% QCD multijet
events.

\section{Overview of the algorithms}
\label{sec:hrtalgorithms}
This section presents recently developed ML-based CMS heavy-object tagging
methods. However, to understand the historical developments and their
limitations, we first present tagging algorithms that do not rely
on selections involving
ML-based methods, but instead rely on selections based on a set of jet
substructure observables (``cutoff-based'' approaches). To
better explore the complementarity between the jet substructure
variables, alternative tagging algorithms were developed using
multivariate methods. Lastly, to exploit the full potential of the CMS
detector and event reconstruction, methods based on Deep Neural
Networks (DNNs) are explored using either high level inputs (e.g., jet
substructure observables), or lower level inputs, such as PF
candidates and secondary vertices. Finally, dedicated versions of the
algorithms are developed that are only loosely correlated with the jet
mass. A detailed discussion of each algorithm is presented in
this Section and a summary of all \PQt quark, \PW, \PZ or \PH
boson identification algorithms is given in
Table~\ref{tab:list_of_algos}.

\begin{table}[!ht]
\centering
\topcaption{\label{tab:list_of_algos} Summary of the CMS algorithms
  for the identification of hadronically decaying \PQt~quarks and \PW,
  \PZ and \PH bosons. See text for explanation of the algorithm names.
The column ``Subsection'' indicates the subsection
where the algorithm is described, and the column ``jet \pt [\GeVns{}]'' indicates the jet \pt
threshold to be used in each algorithm. The $^{*}$ in DeepAK8 and DeepAK8-MD algorithms
indicates the ability of these algorithm to also identify
the decay modes of each particle. }

\resizebox*{\textwidth}{!}{
\begin{tabular}{lcccccc}  \hline
  Algorithm & Subsection & jet \pt [\GeVns{}] & \PQt~quark & \PW boson & \PZ boson & \PH boson\\ \hline
\msdtop & 6.1 & 400 & \checkmark & & & \\
\msdtopbtag & 6.1 & 400 &  \checkmark & & & \\
\msdv & 6.1 & 200 & \checkmark & \checkmark & &  \\
HOTVR & 6.2 & 200 & \checkmark & & &  \\
\ecftop & 6.3 & 200 & \checkmark & & &  \\
\ecfv & 6.3 & 200 & & \checkmark & \checkmark & \checkmark  \\
BEST & 6.5 & 500 & \checkmark & \checkmark & \checkmark & \checkmark  \\
ImageTop & 6.6 & 600 & \checkmark &  &  &   \\
DeepAK8$ ^{(*)}$ & 6.7 & 200 & \checkmark & \checkmark & \checkmark & \checkmark \\[\cmsTabSkip]
\multicolumn{6}{c}{Jet mass decorrelated algorithms}\\
\ecfvddt & 6.3 & 200 &  & \checkmark & \checkmark & \checkmark \\
double-\PQb & 6.4 & 300 & & & \checkmark & \checkmark \\
ImageTop-MD & 6.6 & 600 & \checkmark &  &  &   \\
DeepAK8-MD$ ^{(*)}$ & 6.7 & 200 & \checkmark & \checkmark & \checkmark & \checkmark  \\ \hline
\end{tabular}
}
\end{table}

\subsection{Substructure variable based algorithms}
\label{sec:subsec_softdrop}
Historically, the high momentum \PQt quark and \PW{}/\PZ{}/\PH boson tagging methods
used by the CMS Collaboration are
based on a combination of selection criteria on the jet mass and
the energy distribution inside the jet~\cite{CMS-PAS-JME-13-007,CMS-PAS-JME-15-002,CMS-PAS-JME-16-003,Khachatryan:2014vla,Sirunyan:2017ezt}.

The jet mass is one of the most powerful observables to discriminate
\PQt quark and \PW{}/\PZ{}/\PH boson jets from background jets (i.e., jets stemming
from the hadronization of light-flavor quarks or gluons).
The QCD radiation will cause a radiative shower of quarks and gluons, which will be
collimated within a jet. The probability for a gluon to be radiated
from a propagating quark or gluon is inversely proportional to the
angle and energy of the radiated gluon. Hence, the radiated gluon will tend to appear
close to the direction of the original quark or gluon. These radiated
gluons tend to be soft, resulting in a characteristic ``Sudakov peak''
structure. This is explained in detail in
Ref.~\cite{Sirunyan:2018xdh}. Contributions from initial-state
radiation, the underlying event, and pileup also contribute strongly
to the jet mass, especially at larger values of $R$. As such, jet
mass from QCD radiation scales as the product of the jet \pt and $R$.

Several methods have been developed to remove soft or uncorrelated
radiation from jets, a procedure generally called ``grooming''.
These methods strongly reduce
the Sudakov peak structure in the jet mass distribution. The removal of 
the soft and uncorrelated radiation results in a much weaker
dependence of the jet mass on its $\pt$.

The \PQt quark and \PW{}/\PZ{}/\PH bosons have an intrinsic mass,
and the jet substructure tends to be dominated by electroweak
splittings~\cite{Chen:2016wkt} at larger angles than QCD. This can be exploited to separate
such jets from jets arising from heavy SM particles.

The grooming method used most often in CMS is the ``modified mass drop
tagger'' algorithm (mMDT)~\cite{Dasgupta:2013ihk}, which is a special case of
the ``soft drop'' (SD) method~\cite{Larkoski:2014wba}. This algorithm
systematically removes the soft and collinear radiation from the jet in
a manner that can be theoretically
calculated~\cite{Frye:2016aiz,Marzani:2017mva} (comparisons
to data are found in Ref.~\cite{Sirunyan:2018xdh}).

The first step
in the SD algorithm is the reclustering of the jet
constituents with the CA algorithm, and then the identification of two
``subjets'' within the main jet by reversing the CA clustering history. The jet is
considered as the final jet if the two subjets meet the SD condition:
\begin{equation}
\label{eq:sdcondition}
\frac{ \min (p_{\text{T}1},p_{\text{T}2})}{p_{\text{T}1}+p_{\text{T}2}} > z_{\text{cut}} \left( \frac{\Delta R_{12}}{R_0} \right)^\beta,
\end{equation}
where $R_{0}$ is the distance parameter used in jet clustering algorithm,
 $p_{\text{T}1}$ ($p_{\text{T}2}$) is the \pt of the leading (subleading)
subjet and $\Delta R_{12}$ is their angular separation. The parameters
$z_{\text{cut}}$ and $\beta$ define what the algorithm considers
``soft'' and ``collinear,'' respectively. The values used in CMS are
$z_{\text{cut}}=0.1$ and $\beta=0$ (making this identical to the
mMDT algorithm, although for notation we still denote this as
SD). If the SD condition is not met, the subleading subjet is removed and
the same procedure is followed until Eq.~(\ref{eq:sdcondition}) is
satisfied or no further declustering can be performed.

\begin{figure}[htb!]
  \centering
  \includegraphics[width=0.45\textwidth]{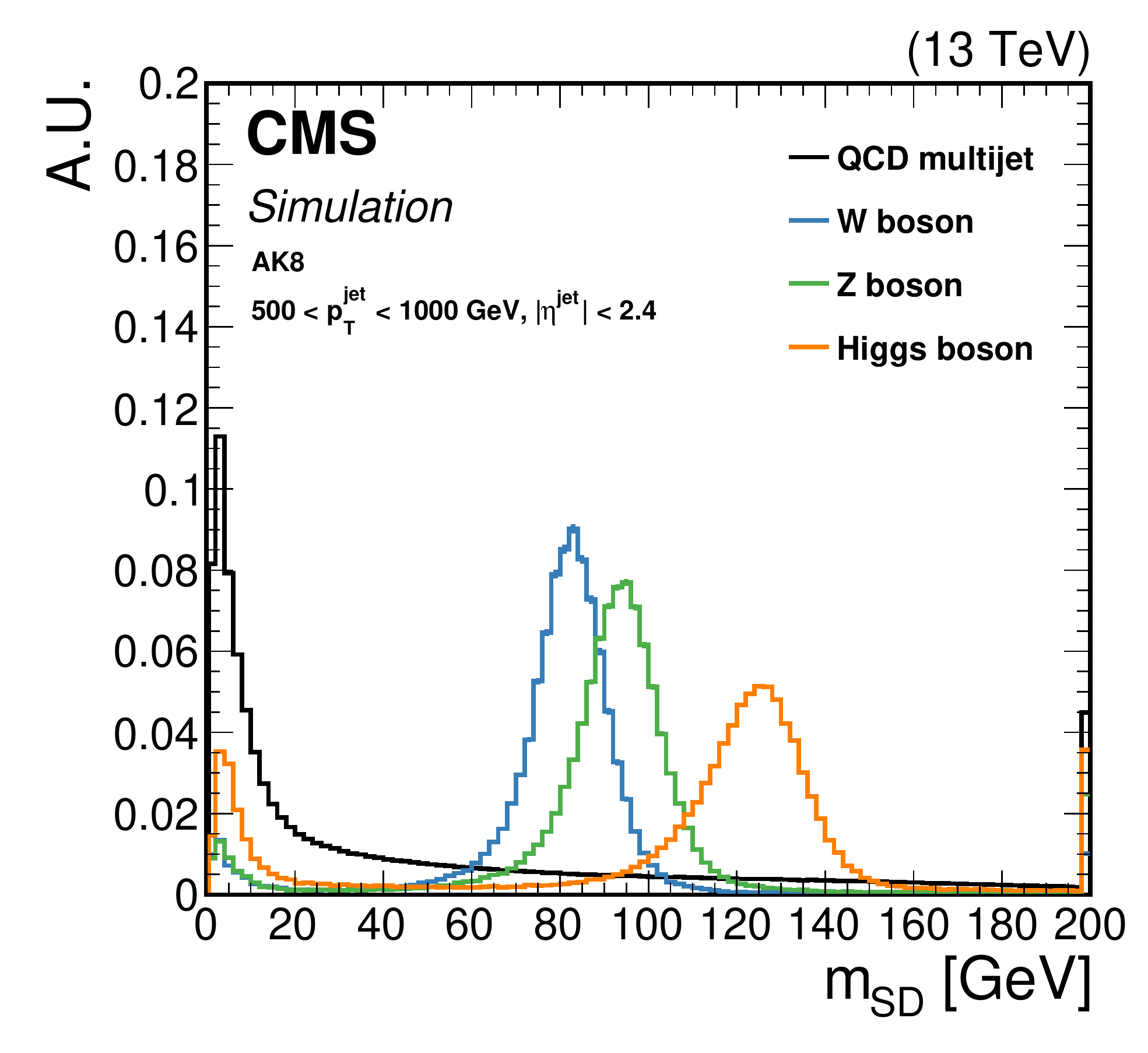}
  \includegraphics[width=0.47\textwidth]{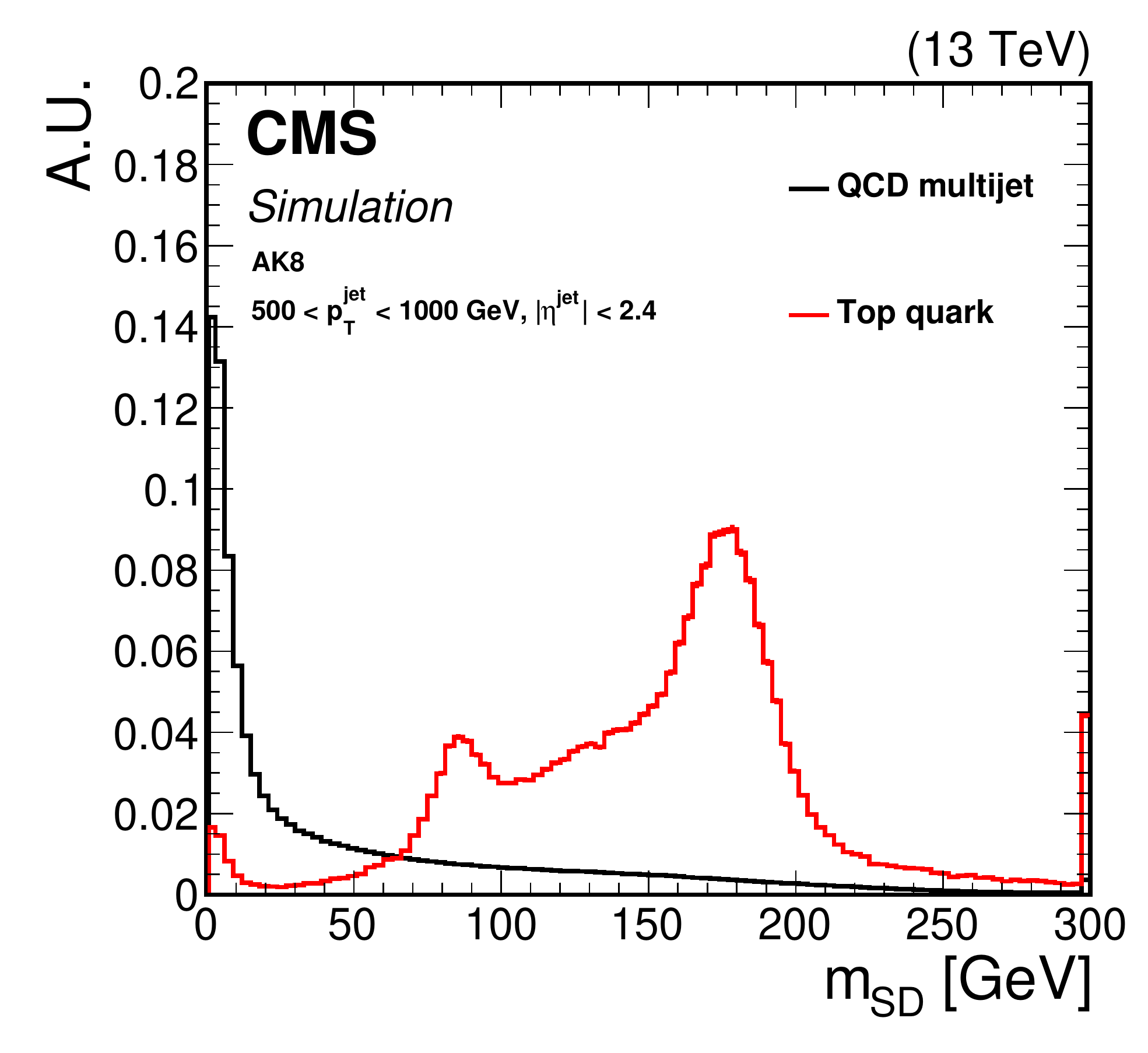}
  \caption{\label{fig:baselinetag_msd}Comparison of the \msd shape in
    signal and background AK8 jets in simulation. The fiducial selection on the jets is displayed
    on the plots. Signal jets are defined as jets arising from
    hadronic decays of \PW{}/\PZ{}/\PH bosons (left) or \PQt
    quarks (right), whereas background jets are obtained from the QCD
    multijet sample.}
\end{figure}

The two subjets returned by the SD algorithm are used to calculate the
jet mass. Figure~\ref{fig:baselinetag_msd} shows the
distribution of the AK8 jet mass after applying the SD algorithm
(\msd) in simulated signal and background jets.
The jet mass has been measured in data in previous papers by CMS for
\PQt-tagged~\cite{Sirunyan:2017yar} and QCD
jets~\cite{Chatrchyan:2013vbb,Sirunyan:2018xdh}.

The \msd in
background jets peaks close to zero because of the suppression of the
Sudakov peak~\cite{Dasgupta:2013ihk}, whereas the \msd\
for signal jets
peaks around the mass of the heavy SM particle (\PQt quark, or
\PW{}/\PZ{}/\PH bosons). In Fig.~\ref{fig:baselinetag_msd} (right),
the peak around 80\GeV is from jets that contain just the
two quarks from the \PW decay and not all three quarks from the \PQt decay.
 Similar
conclusions also hold for CA15 jets.  Based on these observations, we
define three regions in \msd. The ``\PW/\PZ mass region'' with
$65<\msd<105\GeV$, the ``\PH mass region'' with $90<\msd<140\GeV$,
and the ``\PQt mass region'' with $105<\msd<210\GeV$. These definitions
will be used throughout this paper unless stated
otherwise.

\begin{figure}[htb!]
  \centering
  \includegraphics[width=0.45\textwidth]{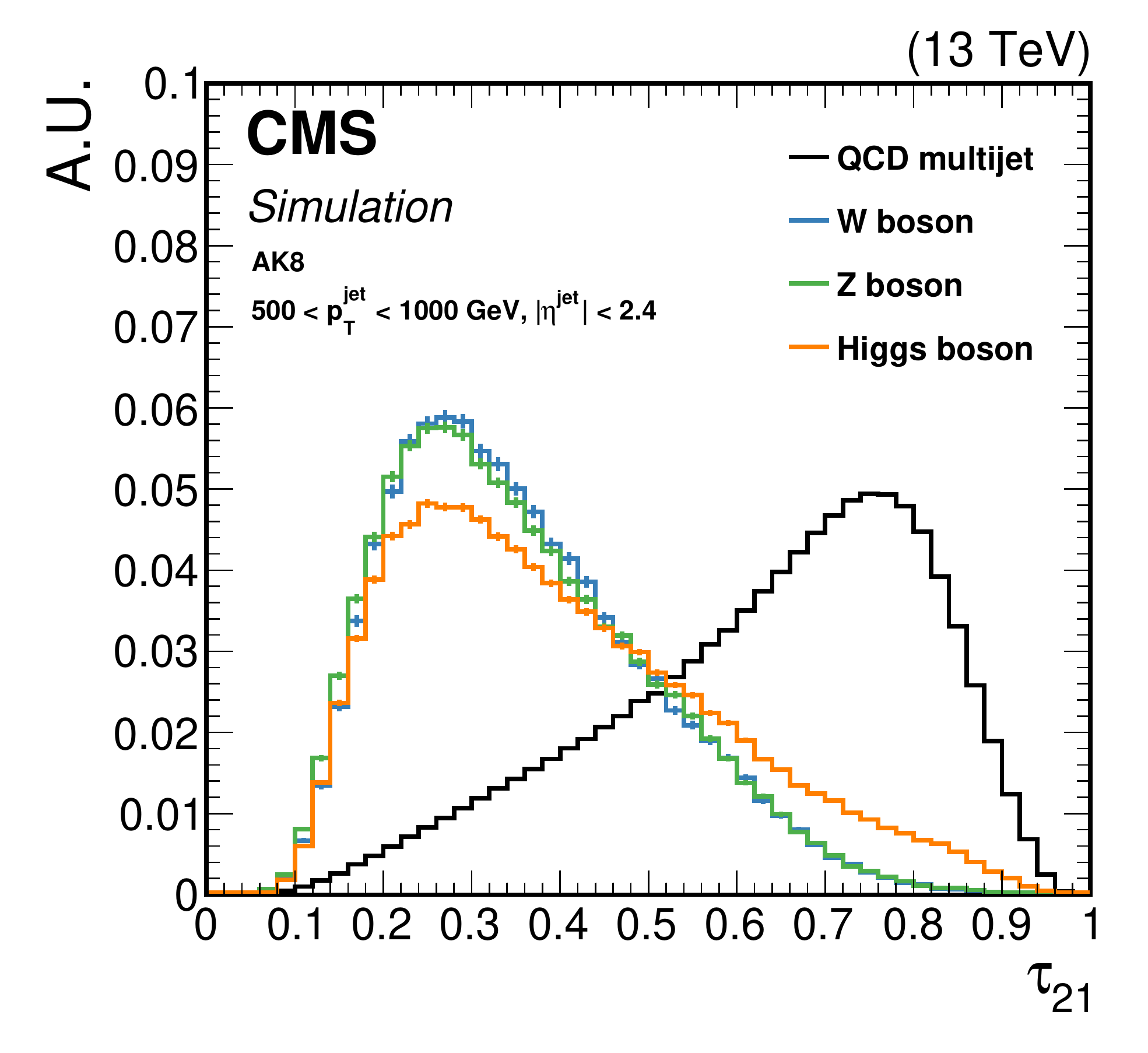}
  \includegraphics[width=0.45\textwidth]{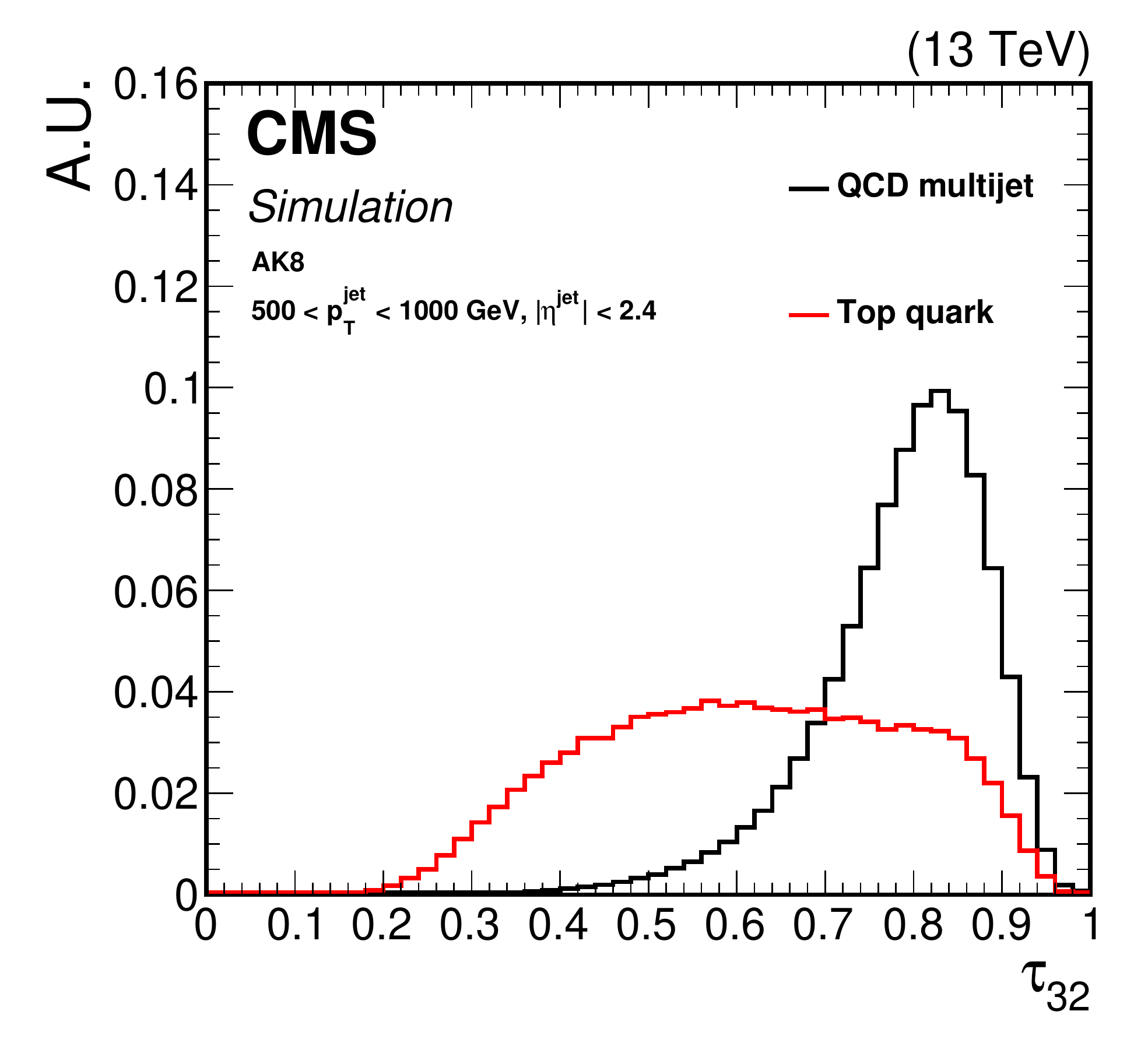}
  \caption{\label{fig:baselinetag_tau}Comparison of the \tautwoone
    (left) and \tauthreetwo (right) shape in signal and background
    AK8 jets.
    The fiducial selection on the jets is displayed in the plots. As
    signal jets we consider jets stemming from hadronic decays of \PW,
    \PZ, or \PH bosons (left), or \PQt quarks (right), whereas
    background jets are obtained from the QCD multijet sample.}
\end{figure}

An additional handle to separate signal from background events is to
exploit the energy distribution inside the jet. Jets resulting from
the hadronic decays of a heavy particle to $N$ separate quarks or gluons
are expected to have $N$ subjets. For two-body decays like \PW{}/\PZ{}/\PH, there are two
subjets, while for \PQt quarks, there are three. In contrast, jets
arising from the hadronization of light quarks or gluons are expected
to only have one or two  (in the case of gluon splitting) subjets. The $N$-subjettiness variables
\cite{Thaler:2010tr,Thaler:2011gf},
\begin{equation}
\label{eq:njettiness}
\tau_N = \frac{1}{d_0} \displaystyle\sum_i p_{\mathrm{T},i}\min\left[\Delta R_{1,i}, \Delta R_{2,i}, \dots, \Delta R_{N,i}\right],
\end{equation}
provide a measure of the number of subjets that
can be found inside the
jet.  The index $i$ refers to the jet constituents, while the $\Delta R$
terms represent the spatial distance between a given jet
constituent and the subjets.  The quantity $d_0$ is a
normalization constant.   The centers of hard radiation are found by
applying  the exclusive \kt
algorithm~\cite{Catani:1993hr,Ellis:1993tq} on the jet constituents
before the use of any grooming techniques. The values of the
$\tau_N$ variables are typically  small if the jet is compatible with
having $N$ or more subjets. However, a more
discriminating observable is the ratio of different $\tau_N$
variables. For this purpose, the ratio $\tau_3/\tau_2\equiv$\tauthreetwo is used for \PQt\
quark identification, whereas the ratio \tautwoone is used for \PW{}/\PZ{}/\PH
boson identification. The distribution \tautwoone and \tauthreetwo\
for signal and background AK8 jets is shown in
Fig.~\ref{fig:baselinetag_tau}.
Measured values of these distributions at CMS can also be found for
light-flavor jets in Ref.~\cite{Sirunyan:2018asm}.
Typical operating regions for \tautwoone (\tauthreetwo) are 0.35--0.65 
(0.44--0.89), which correspond to a misidentification rate after the
\msd selection of 0.1--10\% for both.

The baseline \PW and \PZ boson (collectively referred to as \PV boson) tagging algorithm, based on
selections on \msd and \tautwoone, will be labelled
 as ``\msdv'' in this paper. The \PV tagging with this method is used frequently in
current analyses (e.g., in Refs.~\cite{Sirunyan:2019jbg,Sirunyan:2017wto,Sirunyan:2018omb,Sirunyan:2019sza}) starting at approximately 200\GeV in \pt.

For \PQt quark tagging we studied a tagger based on \msd and \tauthreetwo, which will
be referred to as ``\msdtop''. An additional improvement in the performance of the \PQt quark
identification is achieved by applying the CSVv2 \PQb tagging
algorithm discussed in Section~\ref{sec:evtreco} on the subjets
returned by the SD algorithm. In the studies presented in this paper
we require at least one of the two subjets to pass the loose working
point of the CSVv2 algorithm, corresponding to the \PQb quark jet
identification efficiency $\sim$85\%, with a misidentification rate for
light-flavor quarks and gluon jets of $\sim$10\%, and $\sim$60\% for the \PQc
quark jets. This version of the baseline \PQt quark tagging algorithm
is referred to as ``\msdtopbtag''.  Top-quark tagging with this method is used extensively in physics analyses
(e.g., in Refs.~\cite{Sirunyan:2018ryr,Sirunyan:2018fki,Sirunyan:2018rfo,Sirunyan:2017ukk})
tagging high-\pt \PQt quarks, which start to merge into the AK8 cone at $\pt\sim 350\GeV$ and are 50\% efficient at around 600\GeV.
For applications below this mass range, analyses can profit from the larger (or variable) $R$ clustering algorithms discussed in the
following sections.

\subsection{Heavy object tagger with variable \texorpdfstring{$R$}{R}}

The heavy object tagger with variable $R$ (HOTVR)~\cite{Lapsien2016}
is a new cutoff-based algorithm for the identification of
jets originating from hadronic decays of boosted heavy objects.
 It introduces a
new jet clustering technique with a variable $R$ and removal of soft
contributions during the clustering. The clustering is similar to
other standard sequential clustering algorithms such as the CA algorithm,
where particles are sequentially added. However, instead of a fixed
$R$, HOTVR uses a \pt-dependent $R$ ($R_{\text{HOTVR}}$), defined as:
\begin{linenomath}
\begin{equation}
\label{eq:hotvr}
 R_{\text{HOTVR}} =
 \begin{cases}
   R_\text{min}, \text{ for }  \rho / \pt < R_{\text{min}} \\
   R_\text{max}, \text{ for }  \rho / \pt > R_{\text{max}} \\
   \rho / \pt, \text{ elsewhere} \\
 \end{cases}.
\end{equation}
\end{linenomath}
The value of $\rho$ is chosen to correspond to a typical energy scale of the event ($\mathcal{O}(100)\GeV$). In the
case of $\rho \to 0$, the algorithm is identical to the CA
algorithm for $R=R_{\mathrm{min}}$, whereas for $\rho \to\infty$ it
is identical to the CA algorithm for
$R=R_{\text{max}}$. Higher values of $\rho$ result in larger jet
sizes. The parameters $R_{\text{min}}$ and $R_{\text{max}}$ are
introduced for robustness of the algorithm with respect to
detector effects.

Inspired by Ref.~\cite{Lapsien2016}, at each clustering step, the invariant
mass $m_{ij}$ between two subjets $i$ and $j$
 is calculated. If $m_{ij}$ is greater than a
threshold, $\mu$, the following condition is verified:
\begin{equation}
\label{eq:hotvr_massdrop}
\theta m_{ij} > \max(m_i, m_j) ,
\end{equation}
where $m_{i}$ and $m_{j}$ are the masses of the two subjets, and
$\theta$ is a parameter that determines the strength of the condition
and ranges from 0 to 1. If the condition in
Eq.~(\ref{eq:hotvr_massdrop}) is not fulfilled, the subjet with the
lower mass is discarded; otherwise depending on the relative \pt\
difference of the subjets they are either combined into a single
subjet or the softer one is discarded. The algorithm continues until
no other subjet is found. The detailed description of the HOTVR
algorithm is presented
in Ref.~\cite{Lapsien2016}. Table~\ref{tab:HOTVR_params} lists the values
of HOTVR parameters used in CMS. In the CMS implementation, HOTVR
jets are clustered using \puppi corrected PF candidates.
\begin{table}[!ht]
\centering
\topcaption{\label{tab:HOTVR_params} Summary of the HOTVR
  parameters used in CMS. The $\pt{}_{\text{sub}}$ is the minimum \pt threshold of
  each subjet.}
\begin{tabular}{cccccc}  \hline
$R_{\text{min}}$ & $R_{\text{max}}$ & $\rho$ [\GeVns{}] & $\mu$  [\GeVns{}] & $\pt{}_{\text{sub}}$  [\GeVns{}] & $\theta$  \\ \hline
0.1 & 1.5 & 600 & 30 & 30 & 0.7 \\ \hline
\end{tabular}
\end{table}

\begin{figure}[htbp]
  \centering
  \includegraphics[width=0.32\textwidth]{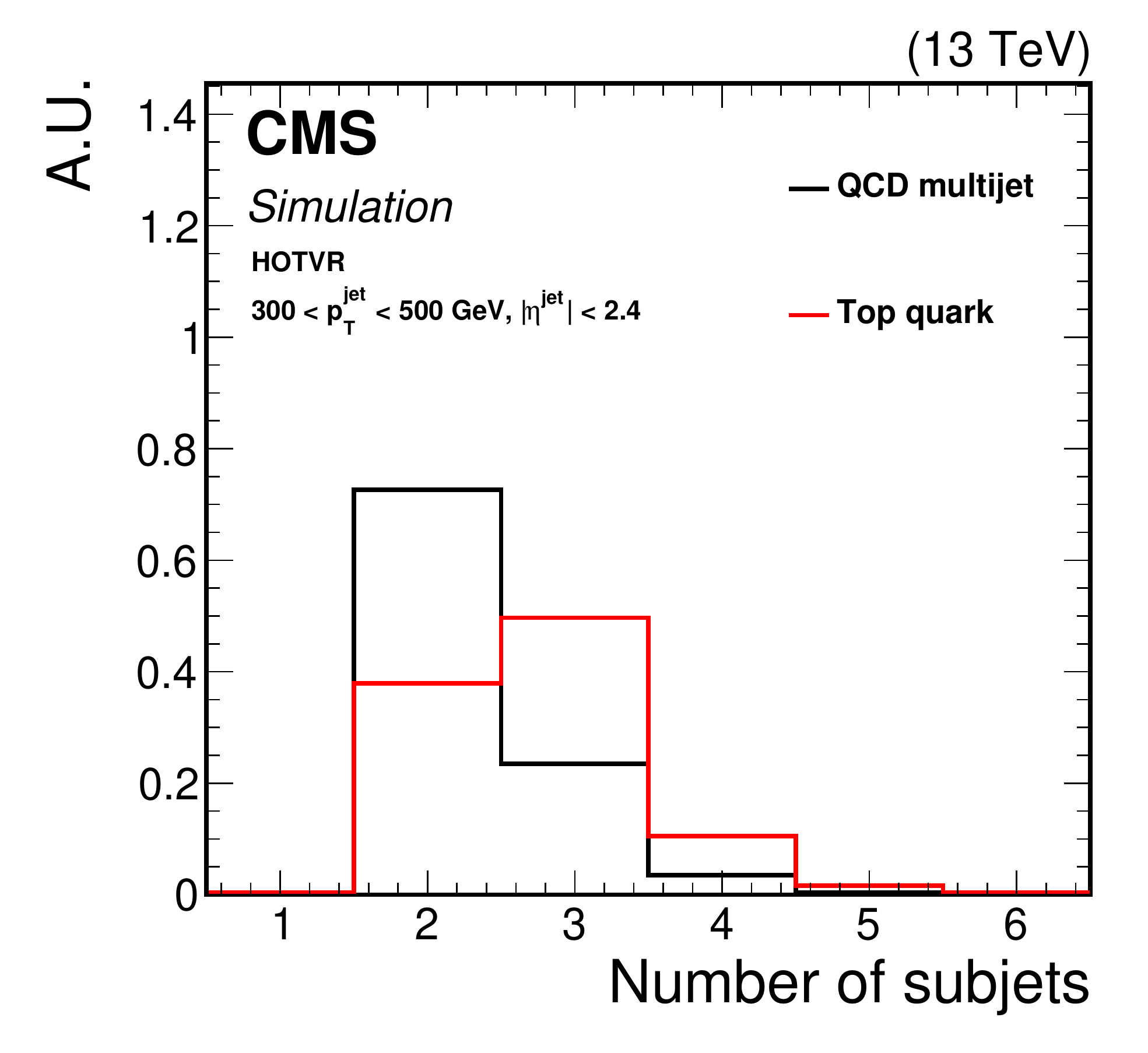}
  \includegraphics[width=0.32\textwidth]{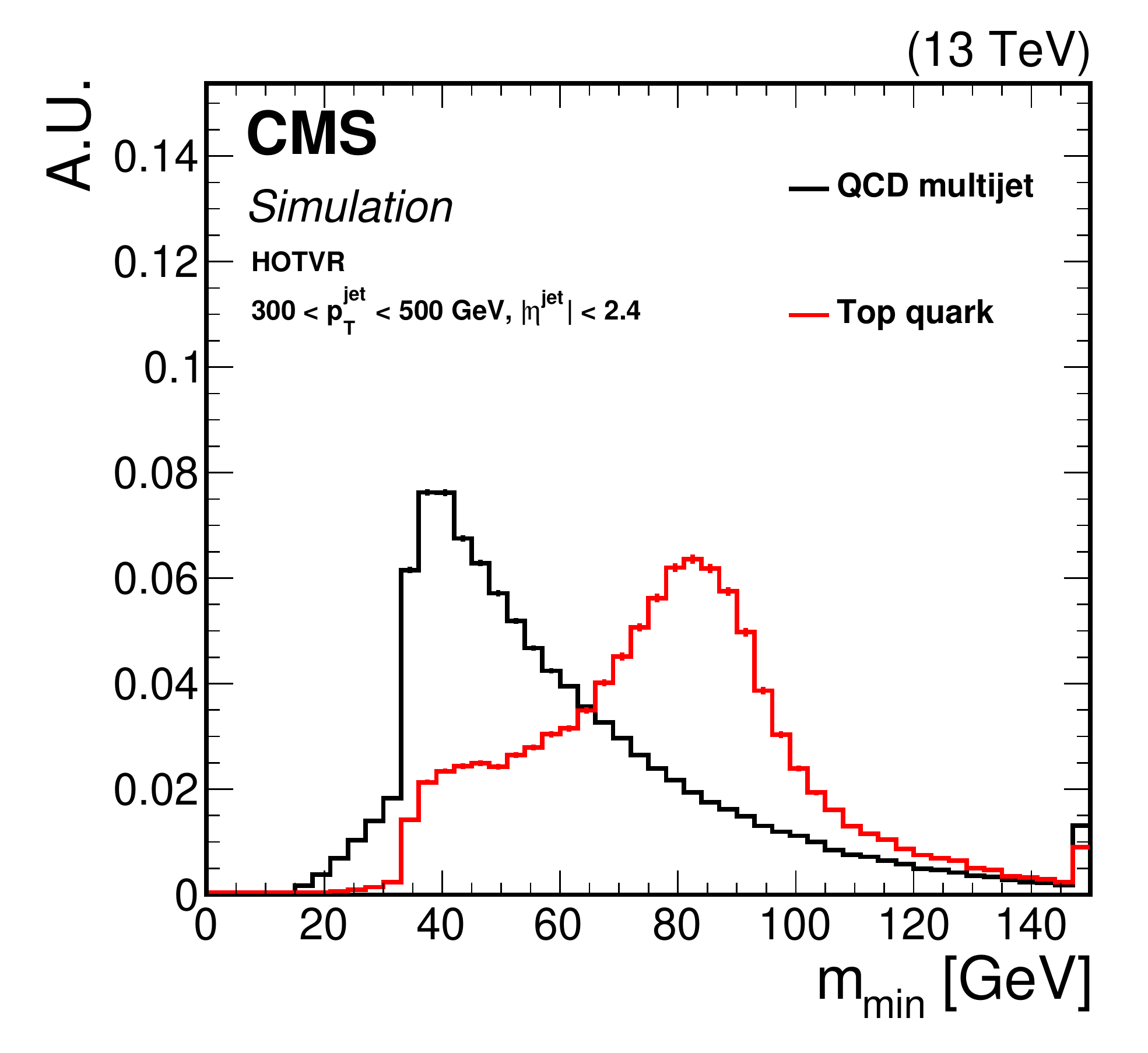}
  \includegraphics[width=0.32\textwidth]{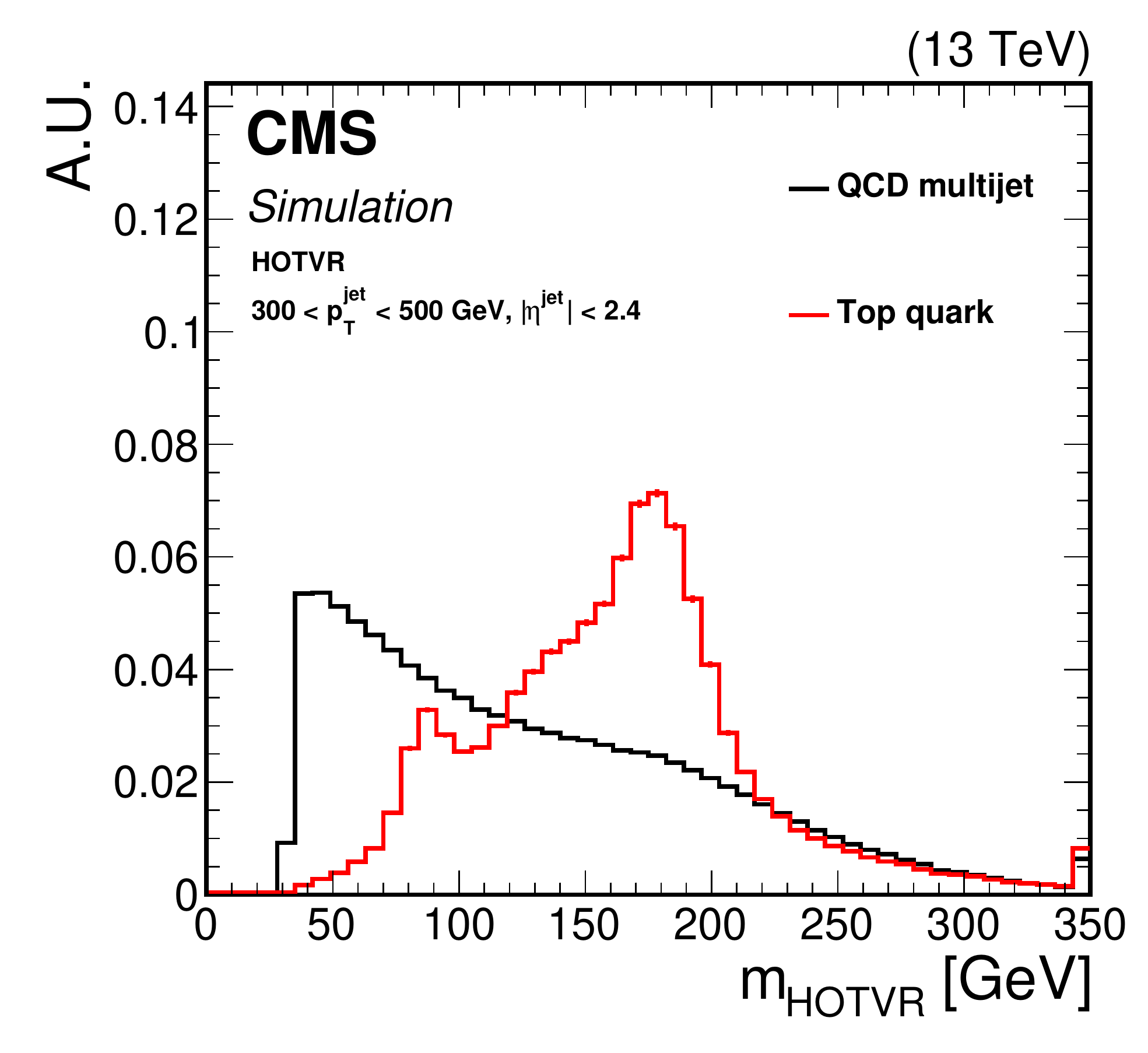}

  \includegraphics[width=0.32\textwidth]{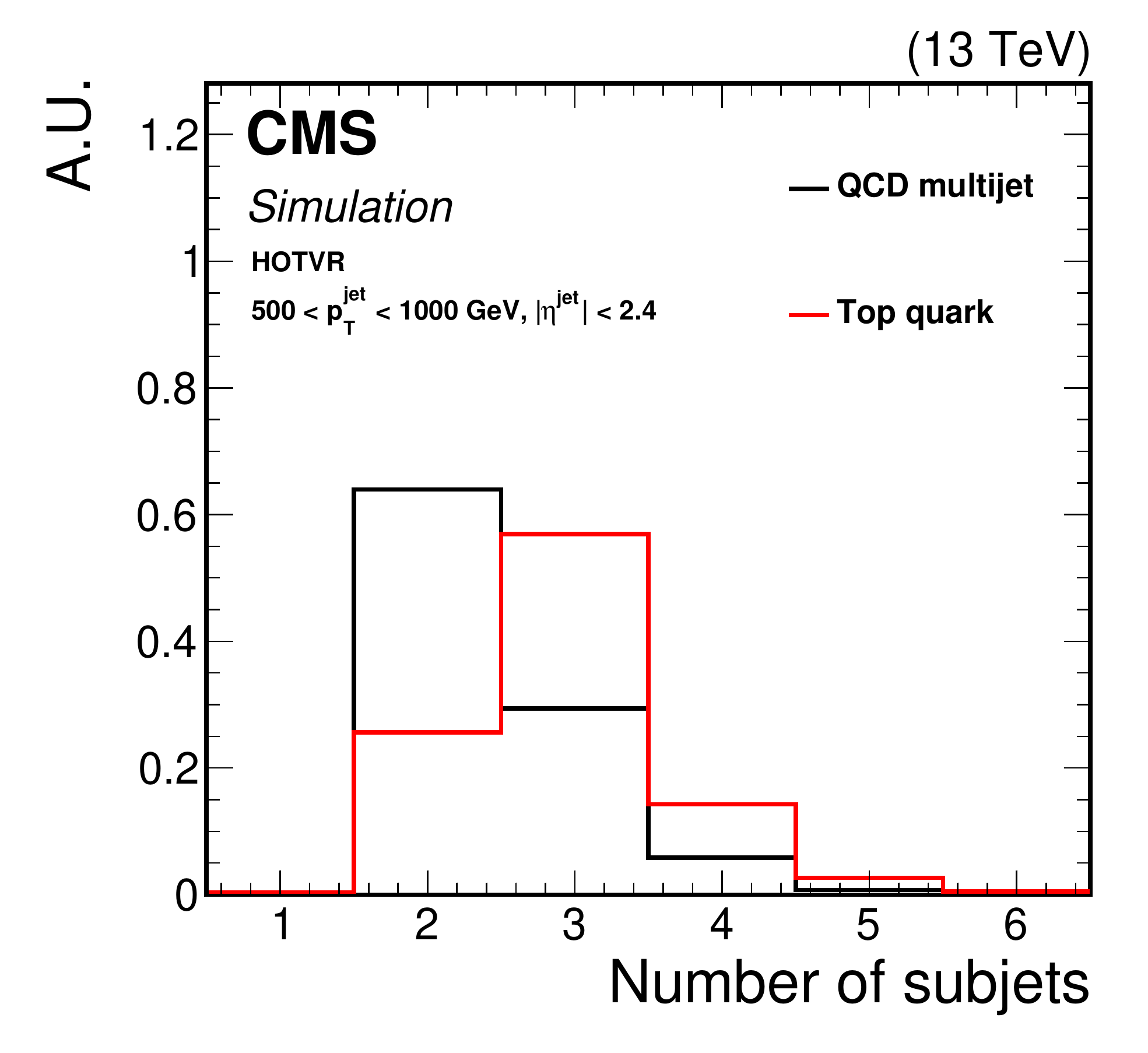}
  \includegraphics[width=0.32\textwidth]{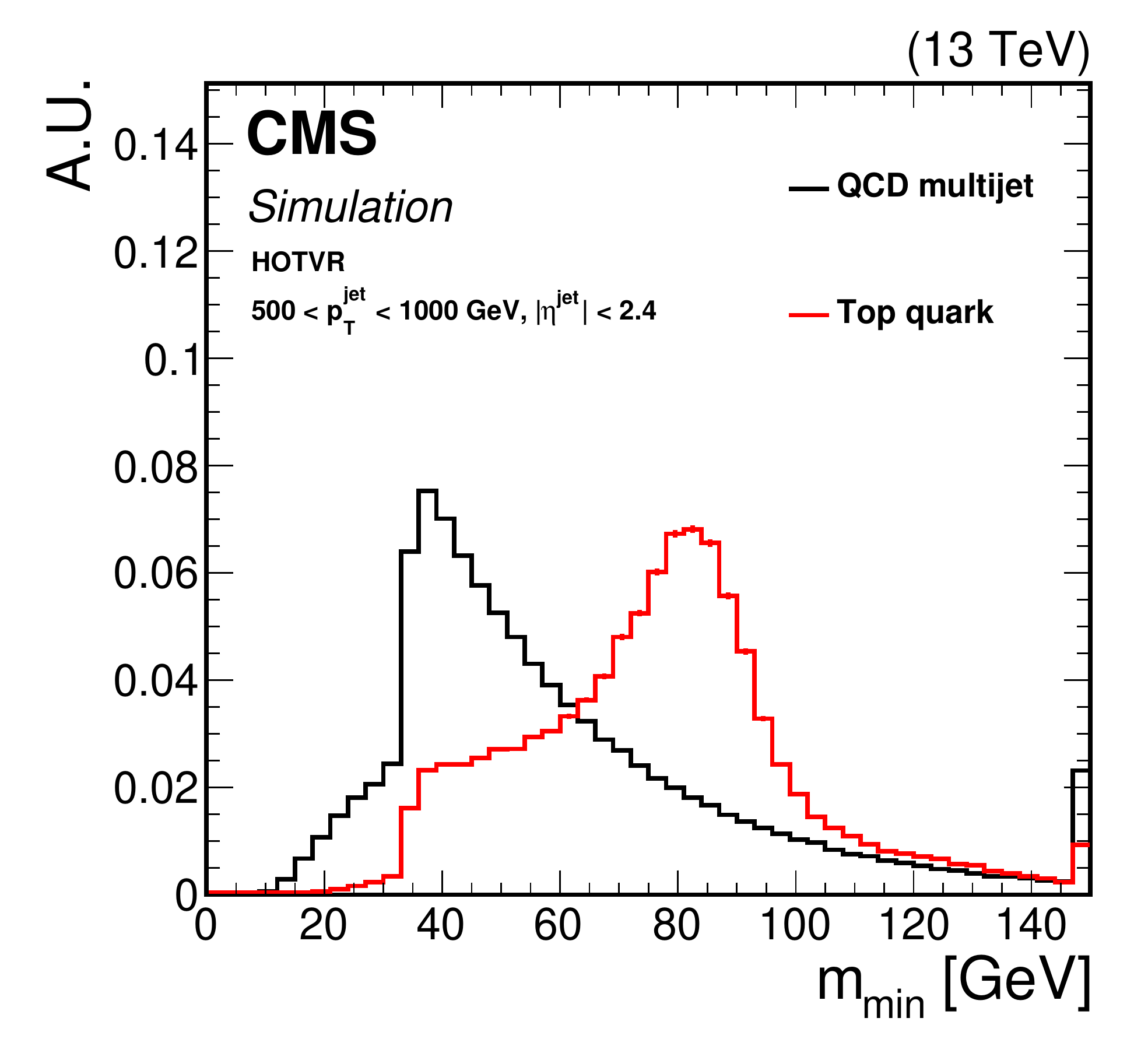}
  \includegraphics[width=0.32\textwidth]{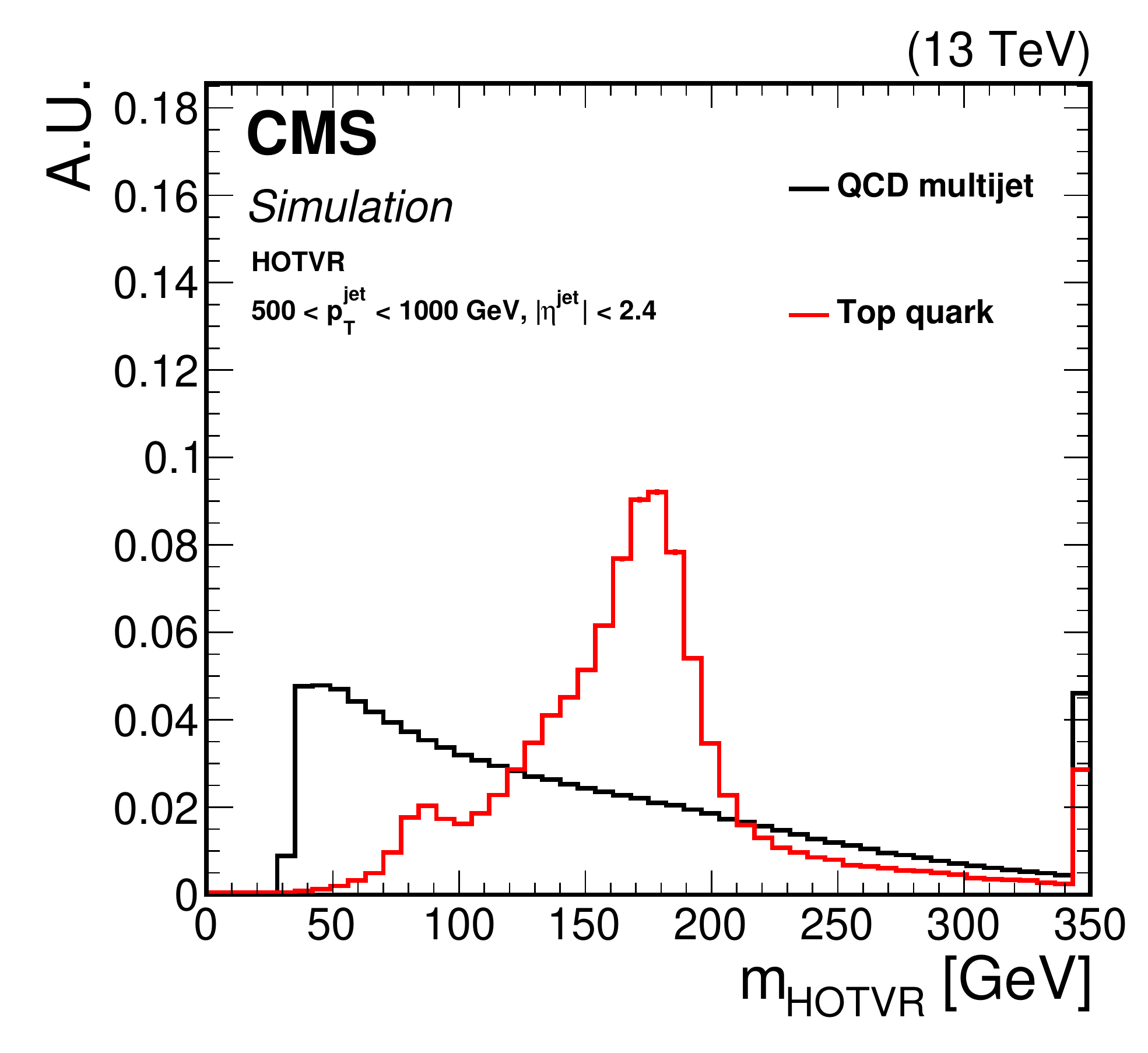}
  \caption{Shape comparison of the main variables of the HOTVR
    algorithm for signal and background jets, in two different regions
    of the jet \pt as displayed in the plots.}
  \label{fig:hotvr_mc}
\end{figure}

The HOTVR clustering algorithm is currently being explored in CMS for
\PQt quark identification. The jets returned by HOTVR (i.e., ``HOTVR
jets'') are required to have mass consistent with $m_{\PQt}$, namely
$140<m_{\text{HOTVR}}<220\GeV$, and at least three subjets,
$N_{\text{sub,~HOTVR}} \geq 3$, the minimum pairwise mass of which
should be $m_{\text{disub,~min}}>50\GeV$. In addition, the \pt of the
hardest subjet must be less than 80\% of the HOTVR jet \pt.
Lastly, to further improve the discrimination,
$\tauthreetwo<0.56$ is required. The shape comparison of the main
variables of the HOTVR algorithm for signal and background, for
different parton \pt ranges, is shown in Fig.~\ref{fig:hotvr_mc}.

\subsection{Energy correlation functions}
A new set of $N$-prong identification algorithms,
the generalized energy correlation functions
(ECFs)~\cite{Moult2016}, are now used by the CMS Collaboration. The ECFs explore the
energy distribution inside a jet by aiming to quantify the number of
centers of hard radiation using an axis-free approach, differing from
the axis-dependent definition used by $N$-subjettiness, which reduces
the dependence of the observable on the jet \pt. This allows
the exploration of complementary information between the two
techniques.

For a jet containing $N_{\text{C}}$ particles, an ECF is defined as:
\begin{equation}
  _q e_N^\beta = \sum _{1 \leq i_1<i_2<\cdots<i_{N} \leq N_{\text{C}}}  \left[\prod_{1\leq k \leq N} \frac{\pt^{i_{k}}}{\pt^{J}}\right]  \prod_{m=1}^{q} \min_{i_j<i_k\in\{i_1,i_2,\cdots,i_{N}\}}^{(m)} \left\{ \Delta R_{i_j,i_k}^\beta\right\},
\end{equation}
where $1 \leq i_1<i_2<\cdots<i_{N} \leq N_{\text{C}}$ range over the jet constituents. The
symbols $\pt^{i_{k}}$ and $\pt^{J}$ are the \pt of the constituent
$i_k$ and the \pt of the jet, respectively. The notation $\text{min}^{(m)}$
refers to the $m$th smallest element, and $\Delta R_{i_j,i_{k}}$ 
is the angular distance between constituents $i_j$ and
$i_k$. The parameters $N$ and $q$ must be positive integers, and the exponent
$\beta$ must be positive as well. For a concrete example, we calculate the ECF
corresponding to $q=2, N=3, \beta=1$. This ECF tests the compatibility
of a jet with three centers of hard radiation, but  only considering
the two smallest angles ($q=2$):
\begin{equation}
  _2e_3^1 =  \sum _{1\leq a<b<c \leq M} \frac{\pt^{a}\pt^{b}\pt^{c}}{(\pt^{J})^{3}}  \min\{\Delta R_{ab}\Delta R_{ac},\Delta R_{ab}\Delta R_{bc},\Delta R_{bc}\Delta R_{ac}\}.
\end{equation}
Moreover, there is the possibility to select subsets of
the jet that contain large energy fractions and pairwise opening
angles only if the size of the subset is less than or equal to the
number of the centers of radiation in the jet. In general, a jet with
$N$ centers of radiation has $e_N \gg e_M$, for $M>N$.

\subsubsection{The ECFs for 3-prong decay identification}
The ratios of type $(N=4)/(N=3)$ can identify the
hadronic 3-body decays, such as those of \PQt quarks.
Reference~\cite{Moult2016} proposes
to use the specific ratio $N_3$ for this purpose:
\begin{equation}
  N_3^{(\beta)} = \frac{_2 e_{4}^{\beta}} {(_1 e_{3}^{\beta})^{2}}.
\end{equation}
Since a jet contains $N_C\sim \mathcal{O}(\pt/\mathrm{GeV})$
constituents, and the sum has $\binom{N_C}{N}$ terms, it is
prohibitively expensive to compute $e(N=4)$ on high-\pt\
jets. For example, about 10--15\% of CA15 jets with $\pt\sim500\GeV$
have more than 100 particles.
However, we find that these functions are dominated by the hardest
particles, and therefore limiting to the 100 hardest particles makes
the calculation tractable without significant performance degradation.

\begin{figure}[htbp]
  \centering
  \includegraphics[width=0.45\textwidth]{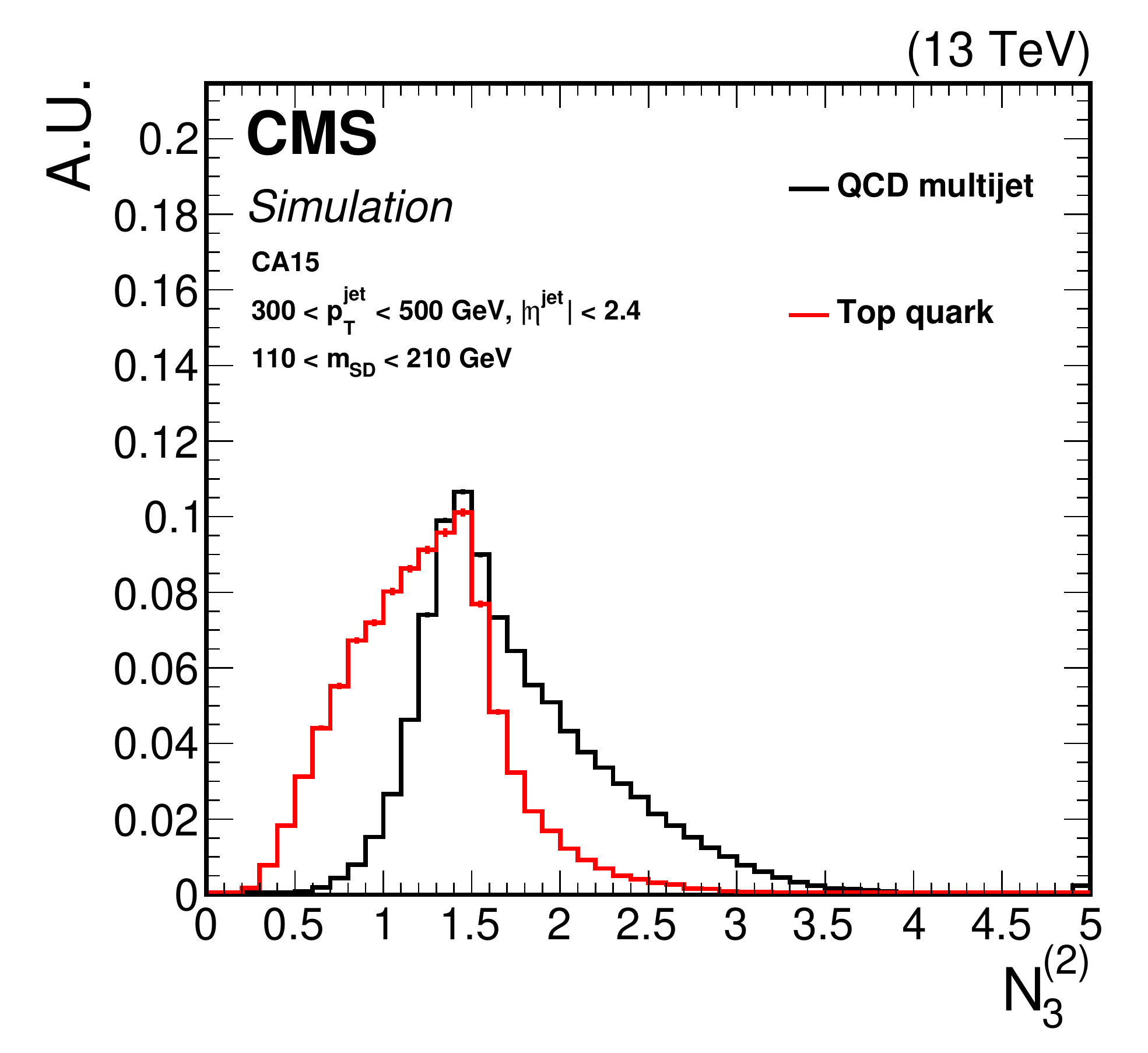}
  \includegraphics[width=0.45\textwidth]{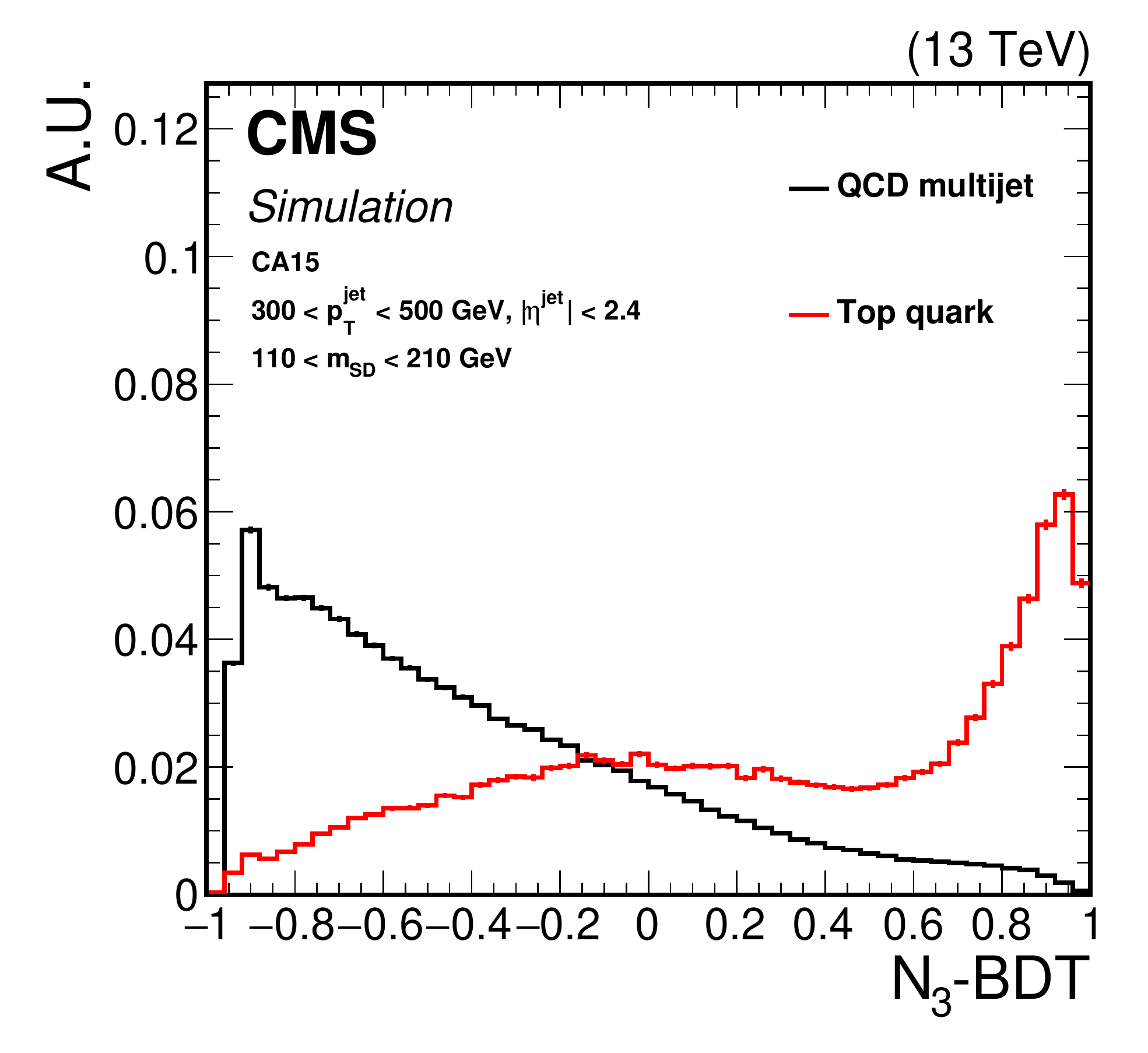}
  \caption{Comparison of the distribution of $N_3^{(2)}$
    (left) and the \ecftop discriminant (right)
    in \PQt quarks jets (signal) and jets from QCD
    multijet processes (background). }
  \label{fig:n3tau32}
\end{figure}

In our reconstruction, the ECF ratios are calculated for jets
after the SD grooming is applied, which improves the stability of
ECF as a function of jet mass and \pt. An example of the ECF ratios is shown
in the left plot of Fig.~\ref{fig:n3tau32} for simulated \PQt quark and QCD jets.
The ECF ratios are measured in data in
Ref.~\cite{Sirunyan:2018asm} showing reasonable agreement with the expectation from simulation.
 While $N_3$ is designed to have comparable
performance with \tauthreetwo, its dependence on \pt is reduced.

Therefore, a set of ECFs is chosen based on the improvement in the
performance of the \PQt tagging algorithm, while in parallel
maintaining small dependence on jet \pt. Despite the fact
that the terms of the ECFs are dimensionless, the angular component of
ECF function is modified according to the boost of the
parent particle. Hence, scale invariant ECF ratios are constructed by only
considering those ratios that satisfy:
\begin{equation}
  \frac{_ae_N^{\alpha}}{(_b e_M^\beta)^x}\text{, where } M\leq N\text{ and }x=\frac{a\alpha}{b\beta}.
\end{equation}
Only ratios that are not highly correlated among themselves are considered
for the \PQt quark tagging algorithm, and ECF ratios that are not
well described by simulation are discarded. The following 11 ECF
ratios are
finally selected:
\begin{linenomath}
\begin{equation}\begin{aligned}
     \frac{_1e_2^{(2)}}{\Bigl({}_1e_2^{(1)}\Bigr)^2},
    \frac{_1e_3^{(4)}}{{}_2e_3^{(2)}},
    \frac{_3e_3^{(1)}}{\Bigl({}_1e_3^{(4)}\Bigr)^{3/4}},
    &\frac{_3e_3^{(1)}}{\Bigl({}_2e_3^{(2)}\Bigr)^{3/4}},
    \frac{_3e_3^{(2)}}{\Bigl({}_3e_3^{(4)}\Bigr)^{1/2}},
    \\
     \frac{_1e_4^{(4)}}{\Bigl({}_1e_3^{(2)}\Bigr)^{2}},
    \frac{_1e_4^{(2)}}{\Bigl({}_1e_3^{(1)}\Bigr)^2},
    \frac{_2e_4^{(1/2)}}{\Bigl({}_1e_3^{(1/2)}\Bigr)^{2}},
    &\frac{_2e_4^{(1)}}{\Bigl({}_1e_3^{(1)}\Bigr)^{2}},
     \frac{_2e_4^{(1)}}{\Bigl({}_2e_3^{(1/2)}\Bigr)^{2}},
      \frac{_2e_4^{(2)}}{\Bigl({}_1e_3^{(2)}\Bigr)^{2}}.
     \label{eq:goodecfs}
\end{aligned}\end{equation}
\end{linenomath}
In addition to the ECFs, two jet substructure observables are employed
to further distinguish \PQt quark jets from light quarks or
gluons. The first observable is \tauthreetwo  calculated for CA15 jets,
after applying the SD grooming, defined as \tauthreetwosd\,
and the second is the $f_{\text{rec}}$ variable of the HEPTopTagger
algorithm~\cite{Plehn:2009rk,Plehn:2010st,Kasieczka:2015jma}, which
quantifies the difference between the reconstructed \PW boson and
\PQt quark masses and their expected values, and is defined as:
\begin{linenomath}
\begin{equation}
 f_\text{rec} = \min_{i,j} \left|\frac{m_{ij}/m_{123}}{m_\text{W}/m_\text{t}} - 1\right|,
\end{equation}
\end{linenomath}
where $i,j$ range over the three chosen subjets, $m_{ij}$ is the mass
of subjets $i$ and $j$, and $m_{123}$ is the mass of all three
subjets.

The ECF-based \PQt quark tagger, referred to as ``\ecftop'', is
based on a boosted decision tree (BDT)~\cite{tmva} with
the 11 ECF ratios, the \tauthreetwosd, and the $f_{\text{rec}}$ as
inputs. The \ecftop algorithm was trained using jets with 
$110 < \msd <210\GeV$. To avoid possible bias in the identification performance due
to differences in the \pt spectrum of the signal (\PQt quarks) and
background (light quarks or gluons) jets, their contributions are
reweighted such that they have a flat distribution in jet \pt.

Figure~\ref{fig:n3tau32} (right) shows a comparison of the \ecftop discriminant
distribution between signal and background jets.
The final \ecftop algorithm also requires at least one of the two
subjets returned by the SD method to be identified as a \PQb jet by
the CSVv2 algorithm using the loose working point.  The ECF BDT tagger is used for \PQt quark
jet identification in the context of dark matter production in association with a single
\PQt quark in the $\pt>250\GeV$ range \cite{Sirunyan:2018gka}.

\subsubsection{The ECFs for  2-prong decay identification}
\label{sec:ecfv}
The use of ECFs is also explored for the identification of 2-prong decays, such as
hadronic decays of \PW{}/\PZ{}/\PH bosons. In this case, the signal jets have a
stronger 2-point correlation than a 3-point correlation and the
discriminant variable $N_{2}^{1}$ can be used to separate jets
originating from \PW{}/\PZ{}/\PH bosons. The $N_{2}$ variable is
constructed via the ratio
\begin{equation}
N^{1}_{2} \equiv N_{2}^{1} = \frac{_{2}e_{3}^{1}}{(_{1}e_{2}^{1})^2},
\end{equation}
and shows similar performance to $N$-subjettiness ratio \tautwoone,
with the advantage that it is more stable as a function of the jet
mass and \pt. This method is referred to as ``\ecfv''.

\begin{figure}[htbp]
  \centering
  \includegraphics[width=0.45\textwidth]{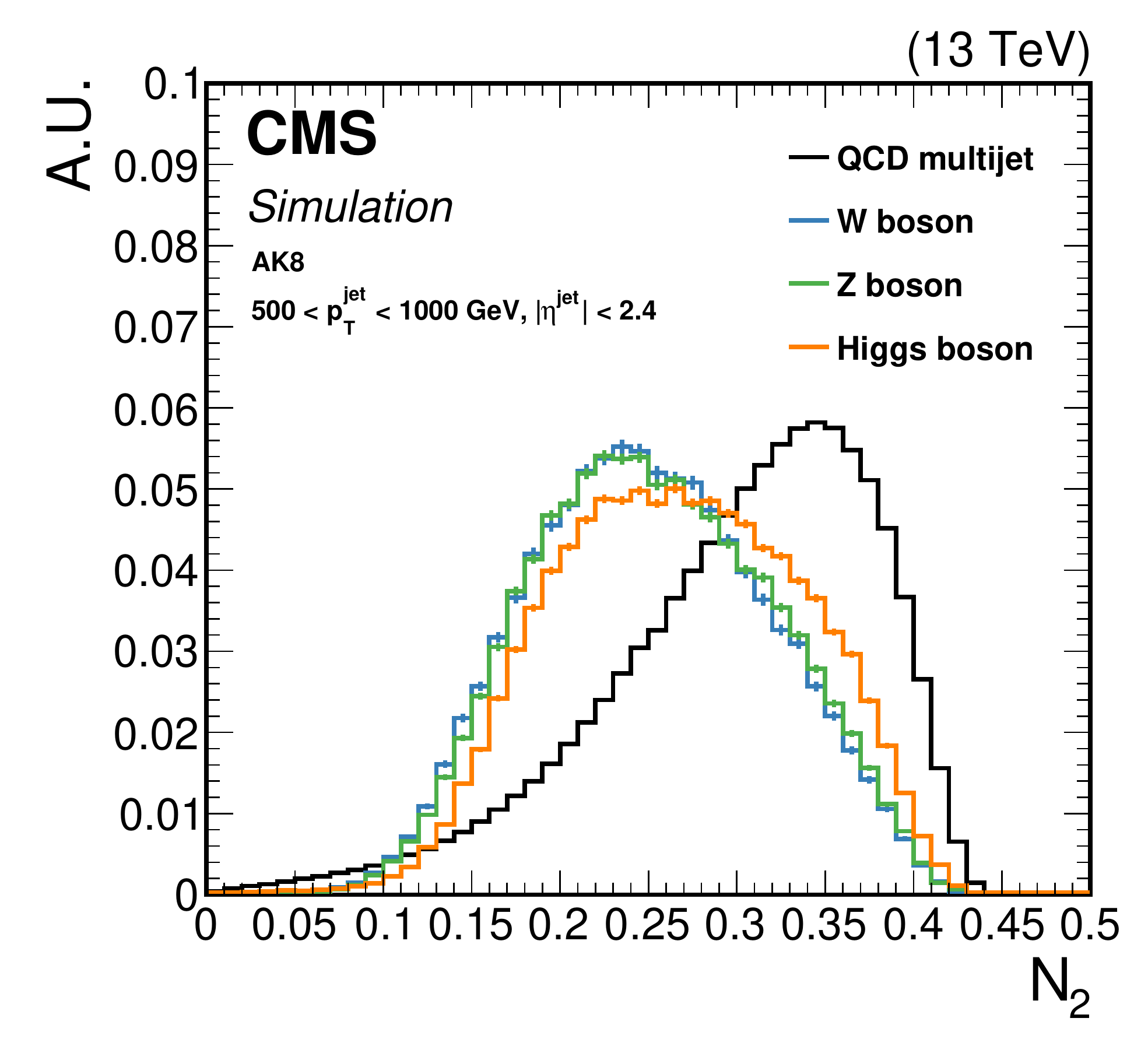}
  \includegraphics[width=0.45\textwidth]{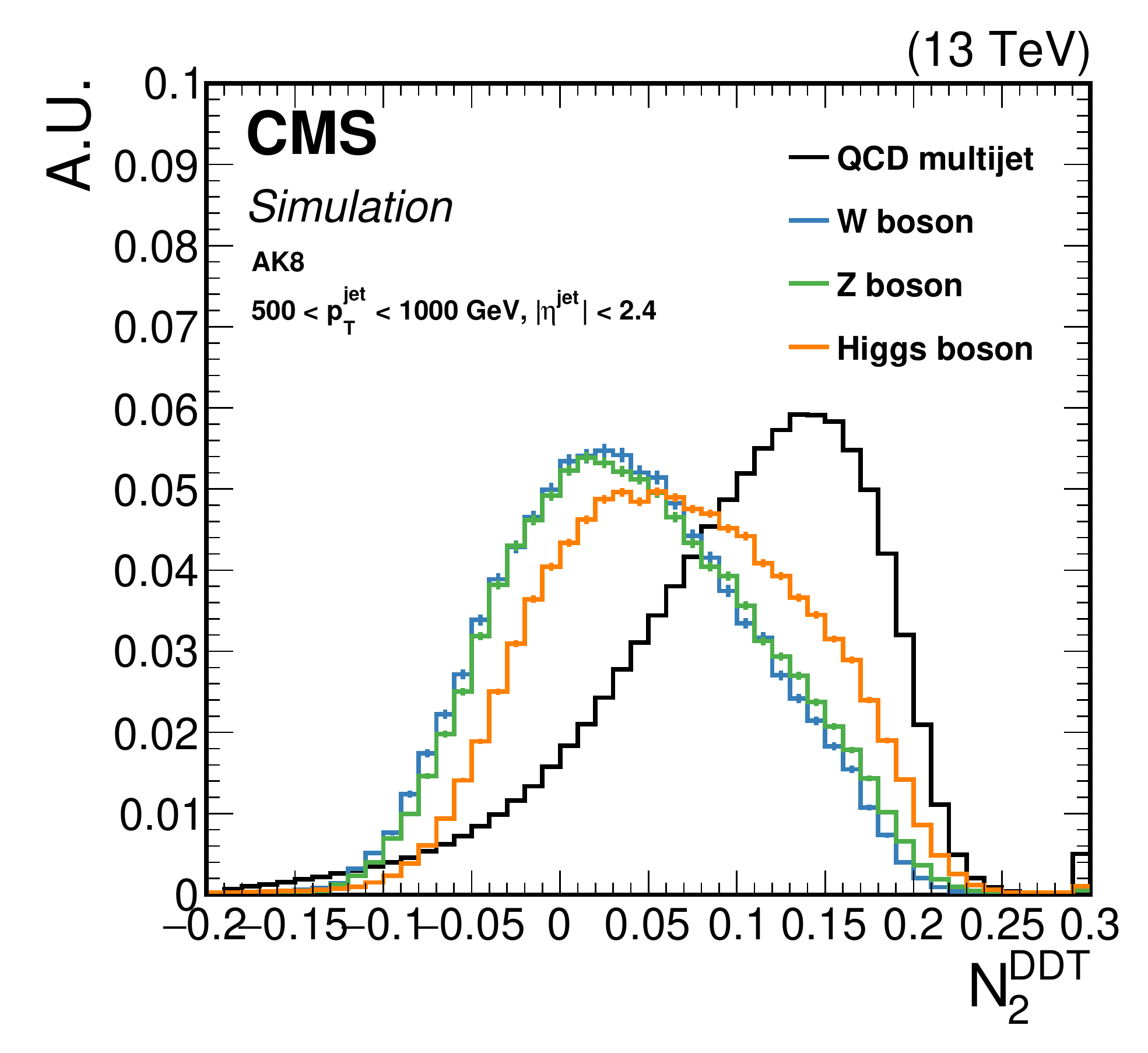}
  \caption{Distributions of the \ecfv (left) and \ecfvddt (right) in signal and background jets.}
  \label{fig:n2}
\end{figure}

A decorrelation procedure is further applied to avoid distorting the
jet mass distribution when a selection based on $N_{2}$ is made. We
design a transformation from $N_{2}$ to $N_{2}^{\text{DDT}}$, where
DDT stands for ``designed decorrelated tagger'' described in Ref.~\cite{Dolen:2016kst}.
The transformation is defined as a function of the dimensionless scaling variable
$\rho=\ln(\msd^2/\pt^2)$ and the jet \pt:
\begin{equation}
N_{2}^{\text{DDT}}(\rho, \pt) =  N_{2}(\rho, \pt) - N_{2}^{(X\%)}(\rho, \pt) ,
\end{equation}
where $N_{2}^{(X\%)}$ is
the $X$ percentile of the $N_{2}$ distribution in simulated QCD
events. This ensures that the selection $N_{2}^{\text{DDT}}  < 0$
yields a constant QCD background efficiency of $X\%$ across the mass and
\pt range considered with no loss in performance. The value $X=5$ is used
throughout this paper, following the choice in \cite{Sirunyan:2017nvi}. The distributions
of $N_{2}$ and $N_{2}^{\text{DDT}}$ in signal and background jets are shown
in Fig.~\ref{fig:n2}.  Signal jets have smaller values and background jets have larger values.
The $N_{2}^{\text{DDT}}$ is used for \PV tagging with \pt in excess
of 500\GeV in the search for light dijet resonances \cite{Sirunyan:2017nvi}.

The \ecfvddt observable was used and validated in several
analyses, including the ones described in Refs.\cite{Sirunyan:2017nvi,Sirunyan:2017dgc}.

\subsection{The double-\texorpdfstring{$\PQb$}{b} tagger}
\label{sec:doubleb}
The standard \PQb tagging tools, such as the CSVv2 discussed in
Section~\ref{sec:evtreco}, can be applied to the subjets returned by
the SD algorithm applied to AK8 jets. Characteristic examples are the
\msdtopbtag and \ecftop algorithms. However, these tools have limitations
 in certain topologies, for example when the two subjets become
very collimated. The ``double-\PQb'' tagger was developed to
specifically target Higgs decays to pairs of \PQb quarks in the
boosted regime~\cite{Sirunyan:2017ezt}. While it utilizes many of the
variables used in the standard CSVv2 \PQb tagging algorithm, it also
employs variables related to the track properties, such as the track
impact parameter and its significance, the positions
of secondary vertices, and information from the
two-secondary-vertex system, among others listed in
Ref.~\cite{Sirunyan:2017ezt}.
An important feature of the double-\PQb algorithm
is that it uses the $N$-subjettiness axes, defined in Eq.~(\ref{eq:njettiness}), for $N=2$, to
group the tracks to the direction of the partons giving rise to the two subjets.
The double-\PQb variables are then used as inputs to a
BDT. A key feature of the double-\PQb algorithm is that it is designed
to minimize the dependence of the BDT discriminant on the jet mass and
\pt, thus making it suitable for other topologies such as decays of boosted \PZ
bosons to bottom quarks~\cite{Sirunyan:2017dgc}.

The performance of the double-\PQb tagger in simulation is detailed
in Ref.~\cite{Sirunyan:2017ezt} using \PH boson jets as signal, and single-\PQb,
double-\PQb jets from gluon splitting to a pair of
\PQb quarks, and light-flavor quark or gluon jets. The
$\PH \to\bbbar$
identification efficiency is
$\sim$25\% ($\sim$70\%) for $\sim$1\% ($\sim$10\%) misidentification
rate~\cite{Sirunyan:2017ezt}.

The double-\PQb tagger performance in data is studied in~\cite{Sirunyan:2017ezt}
using data in a recent inclusive search for the Higgs boson in
the $\bbbar$ decay mode~\cite{Sirunyan:2017dgc}. In that
analysis, the \PZ boson was observed for the first time in the
single-jet topology and $\bbbar$ decay mode, with a rate
consistent within uncertainties with the SM expectation, validating
the double-\PQb tagging algorithm for the Higgs boson measurements and future
searches.

The double-\PQb tagger will serve as a reference for the performance
of the new methods explored in CMS.

\subsection{Boosted event shape tagger}

The boosted event shape tagger (BEST)~\cite{BEST} is a multi-classification
algorithm designed to discriminate hadronic decays of
high-\pt \PQt quarks and \PW{}/\PZ{}/\PH bosons from jets arising from \PQb
quarks, light flavor quarks,
and gluons.  The original algorithm was
demonstrated using generator-level particles and efficiently separated
jets originating from \PW{}/\PZ{}/\PH bosons, \PQt quarks, and \PQb jets.  The
algorithm has been extended and deployed for use in the CMS
experiment, adding an additional category to discriminate jets from
light-flavor quarks and gluons.

The BEST algorithm obtains discrimination on a jet-by-jet basis by
transforming the entire set of jet constituents four times, each
with a different boost vector.  The boost vectors are
obtained by assuming the jet originating from one of the heavy objects
under consideration (\PW{}/\PZ{}/\PH{}/\PQt).  The jet momentum is
held constant while the mass of the jet is adjusted to the theoretical
value of the corresponding particle.  This results in four
distributions of constituents that can be used to discriminate between
particle origins.  If a jet did originate from one of the hypothesized
heavy objects, its jet constituents will, in general, be more isotropic
in the rest frame of that particle.  By examining the differences
between heavy object hypotheses, discrimination is obtained between the
categories of interest (\PW{}/\PZ{}/\PH{}/\PQt{}/\PQb{}/other).

In total, 59 quantities are used to train a neural network (NN) and
classify the AK8 jets.  The variables are listed in Table
\ref{tab:input_vars}.  For each boost transformation, we calculate the
following observables: Fox--Wolfram moments~\cite{fwm};
aplanarity, sphericity, and isotropy quantities based on the
eigenvalues of sphericity tensor, as defined in
Ref. \cite{sphericity}; and jet thrust \cite{thrust}.
Additionally, in each boost hypothesis, AK4 subjets are clustered from
the constituents and used to compute pairwise subjet masses
for the leading three subjets, as well as the combined mass of the
leading four subjets $m_{1234}$.   These AK4 subjets are also used to
compute the longitudinal asymmetry $A_L$, defined as the ratio of the
sum of longitudinal components of the AK4 subjet momenta to the
sum of the total AK4 subjet momenta.  In addition to these quantities
evaluated for each set of jet constituents, the
$m_{\mathrm{SD}}$, rapidity, charge, $\tau_{32}$, $\tau_{21}$, and the
CSVv2 discriminant for each subjet provide additional inputs for each
set of boosted jet constituents.

The NN is trained with the \textsc{scikit-learn} package
\cite{scikitlearn} using the \textsc{MLPClassifier} module.  The network
architecture is fully connected and consists of 3 hidden layers with
40 nodes in each layer using a rectified linear unit
(ReLU)~\cite{Nair:2010:RLU:3104322.3104425} activation function.  The six output
nodes correspond to the 6 particle species of interest.  We use
500\,000 jets to train the network, split evenly between the 6
training samples.  The training is performed using the
\textsc{Adam}~\cite{kingma2014method} optimizer to minimize the cross entropy
loss with a constant learning rate of 0.001. Cross entropy is a 
measure of the difference (entropy) between two probability 
distributions and it is used for optimizing a classification model. 
The  BEST \PW{}/\PZ{}/\PH{}/\PQt{}/\PQb{}/other multi classification is currently used
for tagging high-\pt jets in the search for vector-like quark pair production \cite{Sirunyan:2019sza}.

\begin{table}
\centering
\caption{List of input quantities used for the training and evaluation
  of the BEST algorithm on AK8 jets.}
\begin{tabular}{ccc}
\hline
\multicolumn{3}{c}{ BEST training quantities} \\
\hline
Jet charge & Fox--Wolfram moment $H_1 / H_0$ (\PQt,\PW,\PZ,\PH) & $m_{12}$ (\PQt,\PW,\PZ,\PH) \\
Jet $\eta$ & Fox--Wolfram moment $H_2 / H_0$ (\PQt,\PW,\PZ,\PH) & $m_{23}$ (\PQt,\PW,\PZ,\PH) \\
Jet $\tau_{21}$ & Fox--Wolfram moment $H_3 / H_0$ (\PQt,\PW,\PZ,\PH) & $m_{13}$ (\PQt,\PW,\PZ,\PH) \\
Jet $\tau_{32}$ & Fox--Wolfram moment $H_4 / H_0$ (\PQt,\PW,\PZ,\PH) & $m_{1234}$ (\PQt,\PW,\PZ,\PH) \\
Jet soft-drop mass & Sphericity (\PQt,\PW,\PZ,\PH) & $A_L$ (\PQt,\PW,\PZ,\PH) \\
Subjet 1 CSV value & Aplanarity (\PQt,\PW,\PZ,\PH) & \\
Subjet 2 CSV value & Isotropy (\PQt,\PW,\PZ,\PH) & \\
Maximum subjet CSV value & Thrust (\PQt,\PW,\PZ,\PH) & \\
\hline
\end{tabular}
\label{tab:input_vars}

\end{table}

\subsection{Identification using particle-flow candidates: ImageTop}
Recent studies, e.g., in Ref.~\cite{Macaluso:2018tck}, have shown that
jet identification algorithms deploying ML methods directly on the jet
constituents yield significantly improved performance compared to
traditional algorithms.

To this end, the ``ImageTop'' \PQt quark identification algorithm was
developed. The ImageTop algorithm closely follows the  network
framework described in
Ref.~\cite{Macaluso:2018tck}, which is an optimization based on the
DeepTop framework described in Ref.~\cite{Kasieczka:2017nvn}. This
tagging approach uses
standard image recognition techniques based on two-dimensional
convolutional neural networks (CNNs) to discriminate  \PQt quark jets
from
QCD jets.  This is performed by pixelizing the jet energy deposits and define
different channels based on relevant detector information. Before
pixelization, the centroid of the jet
is shifted to the origin and then a rotation is
performed to make the major principal axis vertical.  The image
is then flipped along both
the horizontal and vertical axes as appropriate such that the maximum intensity is in
the lower-left quadrant. After this, the image intensity is
normalized and the image
is pixelized using $37{\times}37$ pixels with a total
$\Delta\eta=\Delta\phi=3.2$, with channels split into neutral \pt, track
\pt, number of muons, and of
tracks as an analogue to colors used in image recognition.  The network architecture uses a layer of 128 feature maps
with a $4{\times}4$ kernel followed by a second convolutional
layer of 64 feature maps each. Then a max-pooling layer with a $2{\times}2$
reduction factor is used, followed by two more consecutive
convolutional layers with 64 features maps
followed by another max-pooling layer.  A zero-padding in each
convolutional layer is used to correct for image-border effects.  In
the last pooling layer, the 64 maps are
flattened into a single one that is passed into a set of three fully
connected dense layers, one of 64 neurons, and two more with 256 neurons. The training
is performed using the Tensorflow \cite{tensorflow2015-whitepaper} software package
using the \textsc{AdaDelta} optimizer~\cite{DBLP:journals/corr/abs-1212-5701}
with a
learning rate of 0.3, a minibatch size of 128, and the binary cross
entropy loss function.

\begin{figure}[htb!]
  \centering
  \includegraphics[width=0.49\textwidth]{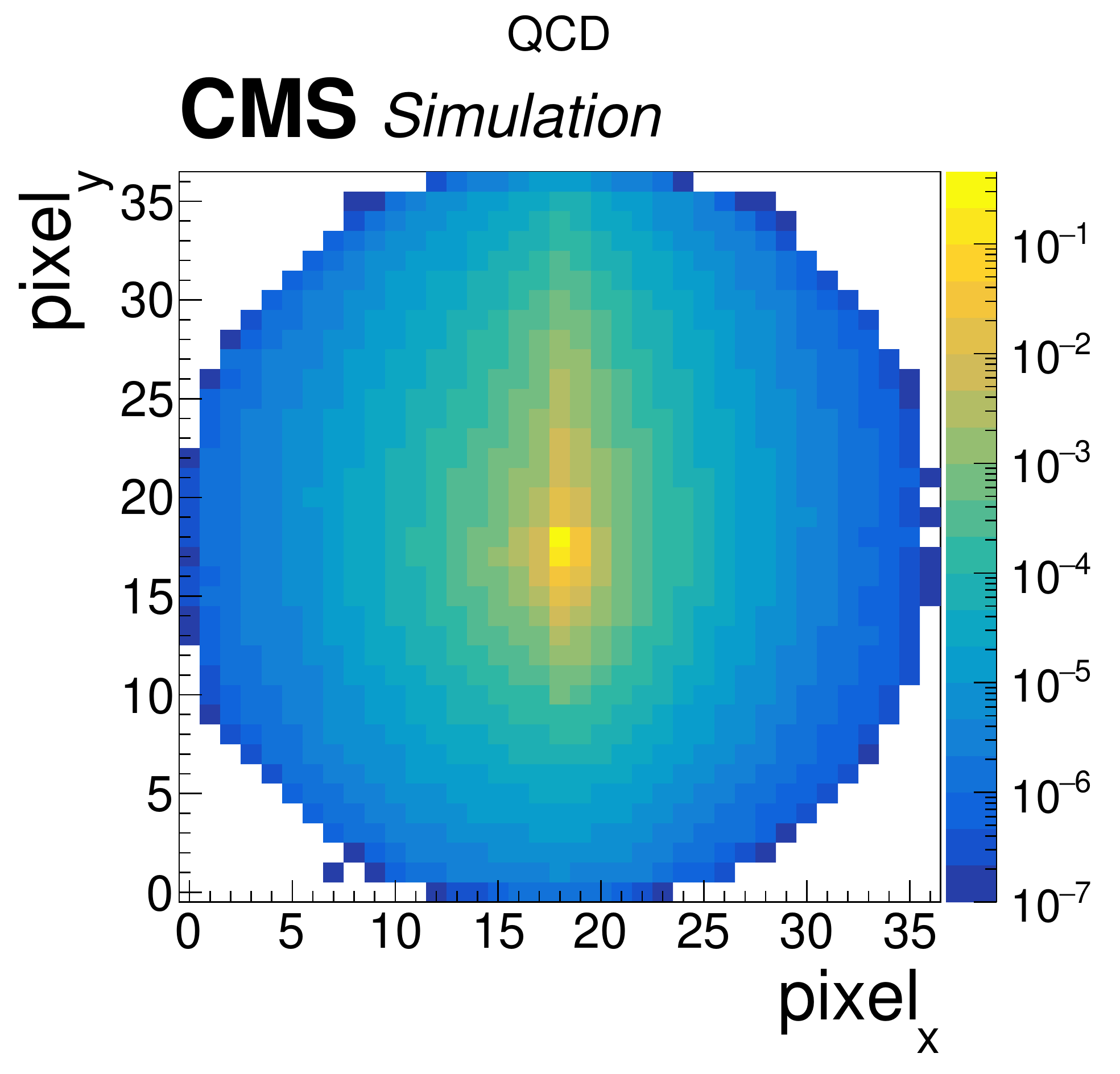}
  \includegraphics[width=0.49\textwidth]{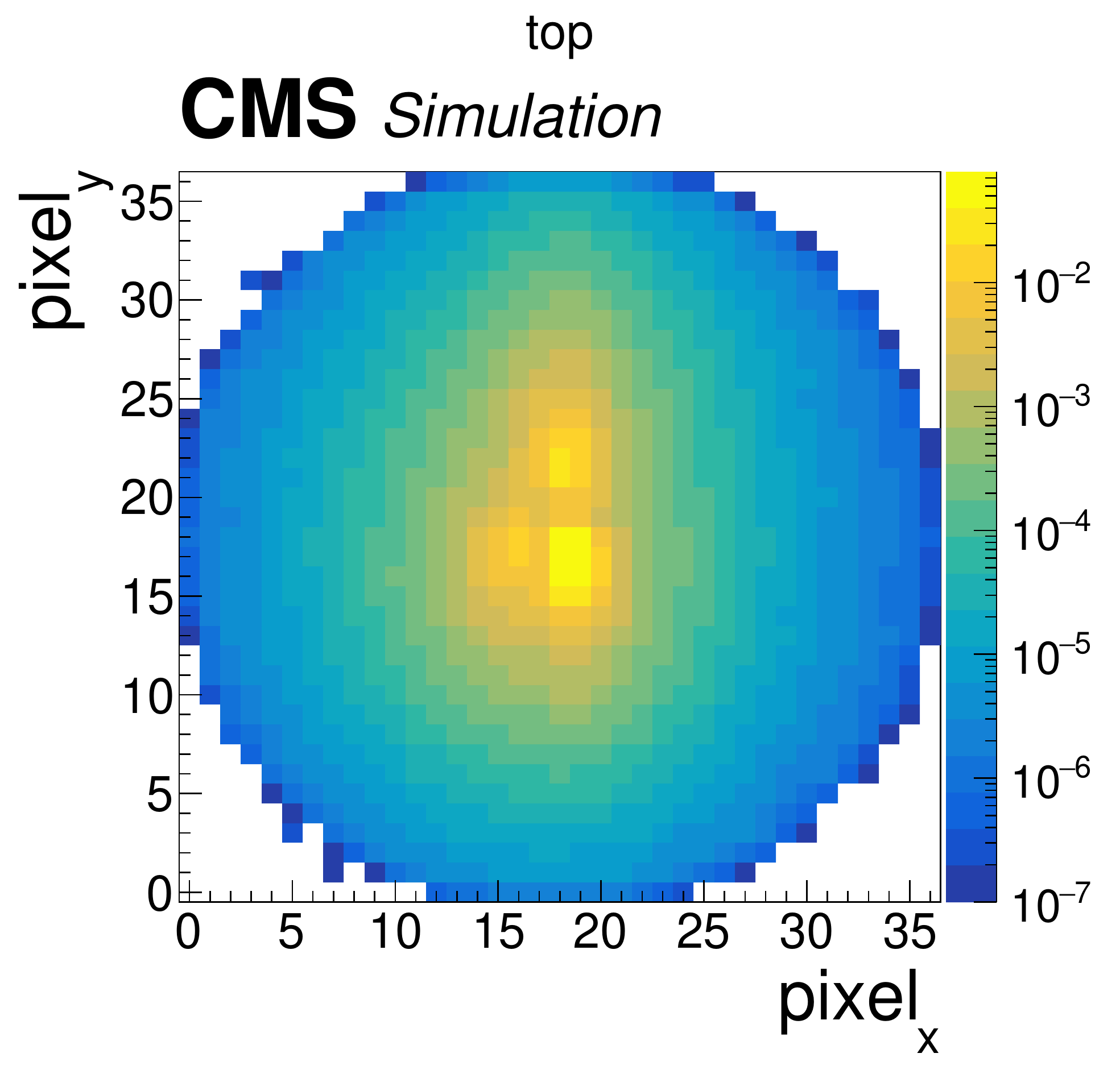}
  \caption{The pixelized images used in the ImageTop network with PF candidate colors summed together (``greyscale'')
    for QCD (left) and \PQt quark (right) jets.  The $x$ and $y$ axes are the pixel
    number, and roughly scale with $\Delta R$.
    The $\PZ$ axis is the intensity of the greyscale image in the given pixel, related to the PF candidate $\pt$, and
    has been normalized to unity.  This figure shows an ensemble of
    overlaid images after the image post processing;
    we can see clear differences between the QCD jet energy and \PQt quark    
    deposition patterns.}
  \label{fig:imagepix}
\end{figure}

\begin{figure}[htb!]
  \centering
  \includegraphics[width=0.98\textwidth]{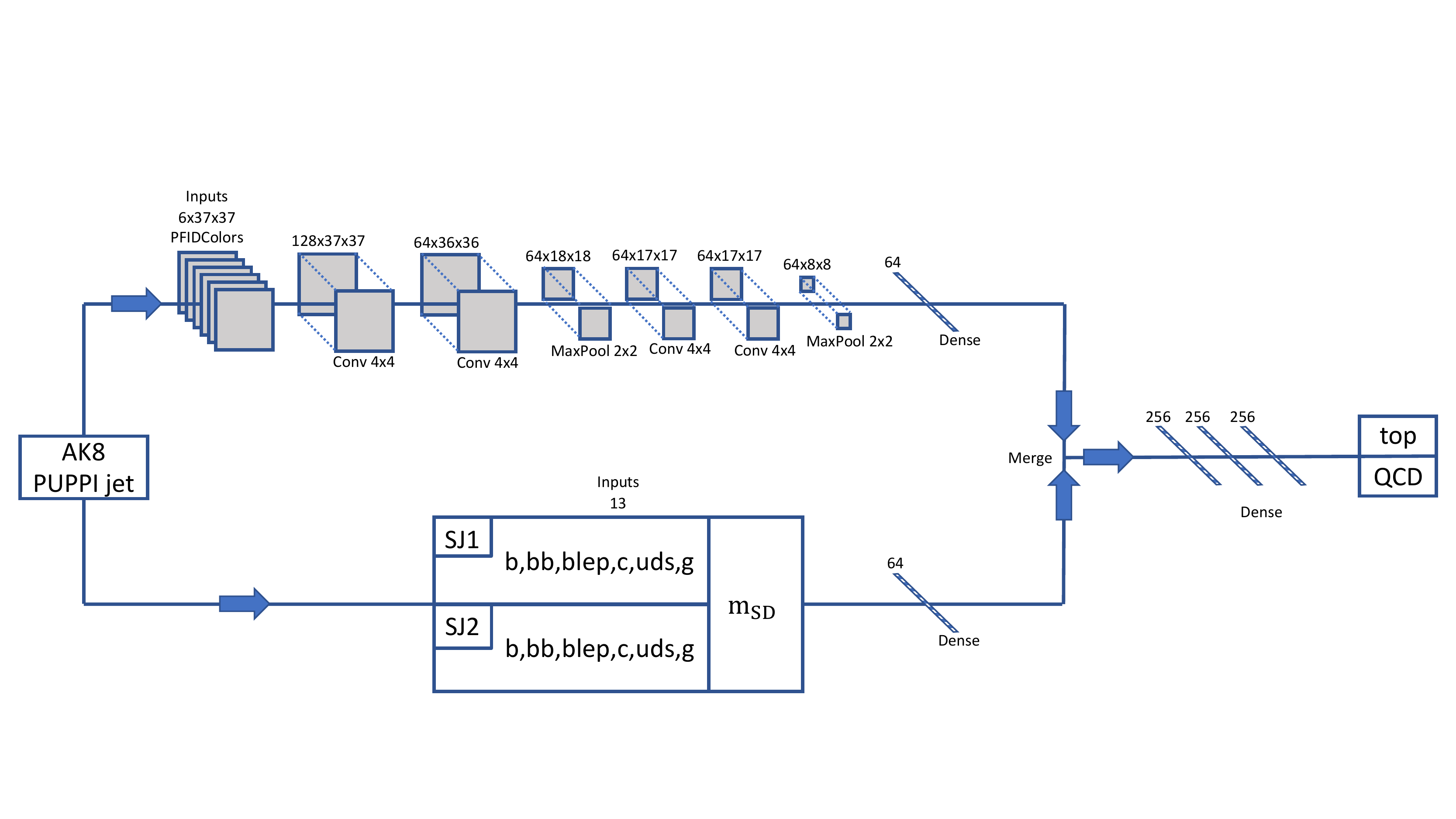}
  \caption{The ImageTop network architecture.  The neural network inputs are the
    37x37 pixelized PF candidate \pt map, which is split into colors
    based on the PF candidate flavor, and the DeepFlavor
    subjet \PQb tags applied to both subjets.  The pixelized images are
    sent through a two-dimensional CNN, and the subjet \PQb tags are
    inputs to a dense layer.  After flattening the CNN, the two networks are taken as input
    to three dense layers and finally to the two-node output, which is
    used as the top tagging discriminator. }
 \label{fig:imagediag}
\end{figure}

The tagger is modified to use the PF candidates contained in
the AK8 jets as inputs, with the
colors being the \pt of the PF candidates
for the full greyscale image, and a separate color for each PF candidate
flavor, namely charged and neutral hadrons, photons, electrons,
and muons.
The pixelized greyscale images used in the ImageTop network for QCD and \PQt quark
jets are shown in Fig.~\ref{fig:imagepix}.
The characteristic flavor of
the \PQt quark decay is included by applying the
DeepFlavor~\cite{CMS-DP-2018-033} \PQb tagging algorithm to the
SD subjets of the AK8 jet.  The subjet \PQb tagging outputs include
the probability of the jet to originate from the following six
sources: \PQb quark, $\bbbar$ pair,
leptonic \PQb decays, \PQc quark, light-flavor quark, or gluon. These output
probabilities calculated for both subjets along with \msd,
are used as  inputs (13 in total) into a 64-neuron dense layer and
merged with the previous flattened CNN layer and finally input into
three fully connected layers of 256 neurons each.
The factorization of the \PQb flavor discrimination is important for the
versatility
of the network, allowing for the flavor identification to be easily
removed or validated in parallel, which can be necessary for the
validation of objects with no SM analog.
The diagram of the CMS application of this NN can be seen in
Fig.~\ref{fig:imagediag}.

The training is performed for jets in the
$\pt>600\GeV$ region.
To sustain the ImageTop performance over a wide range of
$\pt(\text{jet})$,
the image is adaptively zoomed based on $\pt(\text{jet})$ to
account for the increased collimation of the \PQt quark
decay products at high Lorentz boosts and maintain a static pixel
size. The functional form of the zoom is extracted from the average
$\Delta R$ of the three
generator-level hadronic \PQt quark decay products, and the jet energy
deposits are corrected to make this constant on average, as evaluated
from a fit using the inverse jet \pt functional form $f(\pt)=0.066+264/\pt$.

A jet \pt bias is further reduced by
ensuring that the input \pt
distributions for signal and background jets are similarly shaped by
probabilistically removing QCD
events based on the ratio of \PQt quark and QCD jet \pt distributions when training the nominal ImageTop tagger.
The mass correlation of the tagger is reduced by additionally constraining
\msd in a similar manner to define a new discriminator, which will be referred to as ``ImageTop-MD''.
Since the inputs are relatively simple and do not exhibit secondary
mass correlation,
this passive approach for decorrelating the ImageTop network is
sufficient to remove the
mass bias in the fiducial training region ($\pt>600\GeV$ and $\abs{\eta}<2.4$).
This method of mass decorrelation also leads to a factorized
sensitivity where the sensitivity of the full ImageTop network in the
\PQt quark mass region is closely
approximated by the sensitivity of the mass-decorrelated version after
including a mass selection. 

\subsection{Identification using particle-flow candidates: DeepAK8}

An alternative approach to exploit particle-level information directly
with customized ML methods is the ``DeepAK8'' algorithm, a multiclass
classifier for the identification of hadronically decaying particles
with five main categories, \PW{}/\PZ{}/\PH/\PQt/other. To
increase the versatility of the algorithm, the main classes are
further subdivided into the minor categories corresponding to the
decay modes of each particle (e.g., $\PZ \to \bbbar$, 
     $\PZ\to \ccbar$ and $\PZ \to \qqbar$).

In the DeepAK8 algorithm, two lists of inputs are defined for each
jet. The first list (the ``particle'' list) consists of up to 100 jet
constituent particles, sorted by decreasing \pt. Typically less than 5\%
of the jets have more than 100 reconstructed particles, therefore
restricting to the 100 hardest particles results in a negligible loss of performance.
Measured properties of each particle, such as the \pt, the energy deposit, the
charge, the angular separation between the particle and the jet axis
or the subjet axes, etc., are included to help the algorithm extract
features related to the substructure of the jet. For charged particles,
additional information measured by the tracking detector is also
included, such as the displacement and quality of the tracks,
etc. These inputs are particularly useful to enable the algorithm to extract
features related to the presence of heavy-flavor (\PQb or \PQc)
quarks. In total, 42 variables are included for each particle in the
``particle'' list. A secondary vertex (SV) list consists of up to 7
SVs, each with 15 features, such as
the SV kinematics, the displacement, and quality criteria.
The SV list helps the network to extract features related to the
heavy-flavor content of the jet. The elements of the SV list as sorted based
on the two-dimensional impact parameter significance (S$_{\text{IP2D}}$).

\begin{figure}[ht]
\centering
\includegraphics[width=0.65\textwidth]{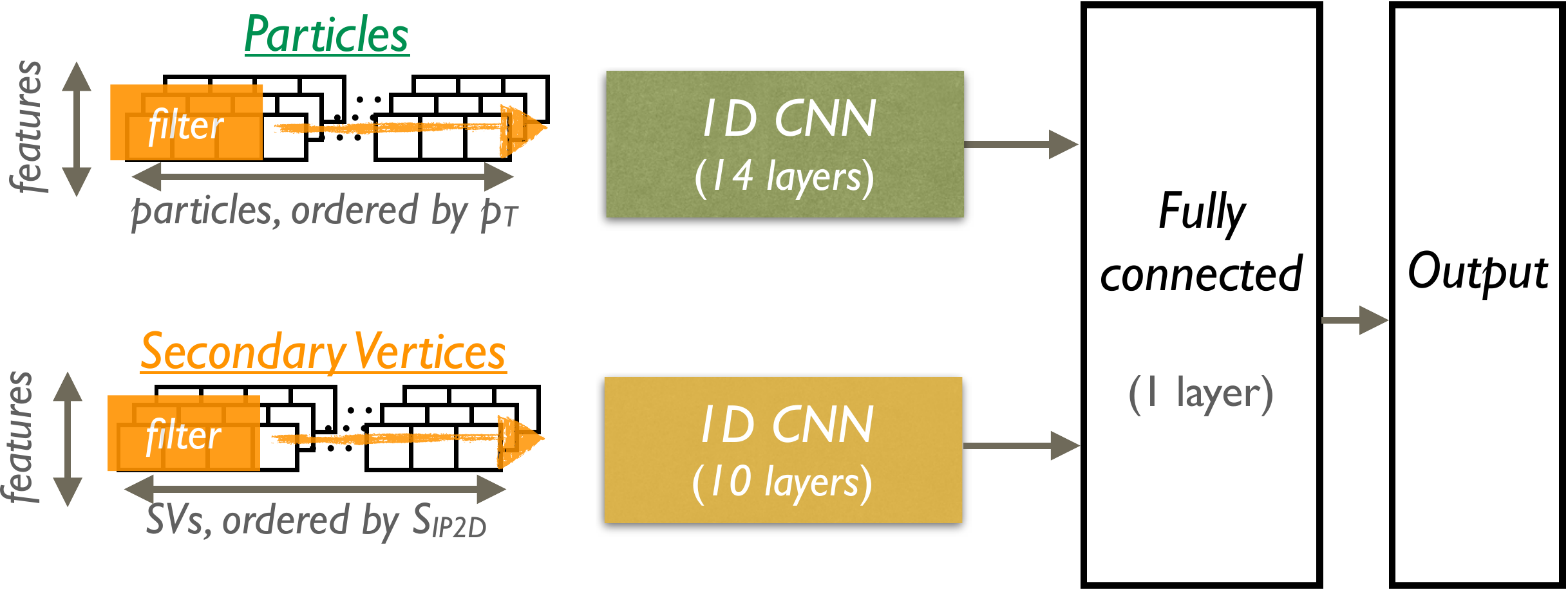}
\caption{\label{fig:deepAK8_arch} The network architecture of DeepAK8.}

\end{figure}

A significant challenge posed by the direct use of particle-level
information is a substantial increase in the number of
inputs. Additionally, the correlations between these inputs are of vital
importance. Therefore, an algorithm that can both process the inputs
efficiently and exploit the correlations effectively is required. A
customized DNN architecture is thus developed in DeepAK8 to fulfill
this requirement. As illustrated in Fig.~\ref{fig:deepAK8_arch}, the
architecture consists of two steps. In the first step, two one-dimensional CNNs are
applied to the particle list and the SV list in parallel to
transform the inputs and extract useful features. In the second
step, the outputs of these CNNs are combined and processed by a simple
fully connected network to perform the jet classification. The CNN
structure in the first step is based on the ResNet
model~\cite{DBLP:journals/corr/HeZRS15}, but adapted from
two-dimensional images to one-dimensional particle lists. The CNN for
the particle list has 14 layers, and the one for the SV list
has 10 layers. A convolution window of length 3 is used, and the
number of output channels in each convolutional layer ranges between
32 to 128. The ResNet architecture allows for an
efficient training of deep CNNs, thus leading to a better exploitation
of the correlations between the large inputs and improving the
performance. The CNNs in the first step already contain strong
discriminatory ability, so the fully connected network in the second
step consists of only one layer with 512 units, followed by a
ReLU activation function
and a Dropout \cite{JMLR:v15:srivastava14a} layer of 20\% drop rate. The
NN is implemented using the \textsc{MXNet}
package~\cite{DBLP:journals/corr/ChenLLLWWXXZZ15} and trained with the
\textsc{Adam} optimizer to minimize the cross-entropy loss. A minibatch 
size of 1024 is used, and the initial
learning rate is set to 0.001 and then reduced by a factor of 10 at
the 10th and 20th epochs to improve convergence. The training completes
 after 35 epochs. A sample of 50 million jets is used, of which
80\% are used for training and 20\% for
validation. Jets from different signal and background samples are reweighted
to yield flat distributions in \pt to avoid any potential bias in the training process.
The DeepAK8 algorithm is designed for jets with
$\pt>200\GeV$ and typical operating regions for which the misidentification rate
is greater than 0.1\%.

\subsubsection{A mass-decorrelated version of DeepAK8}

As will be discussed in Section~\ref{sec:performanceinmc},
background jets selected by the DeepAK8 algorithm exhibit a modified
mass distribution similar to that of the signal. The
mass of a jet is one of the most discriminating variables and, although
it is not directly used as an input to the algorithm, the CNNs are
able to extract features that are correlated to the mass to improve
the discrimination power. However, such modification of the mass
distribution may be undesirable (as described in
Ref.~\cite{Dolen:2016kst}) if the mass variable itself is used
for separating signal and background processes. Thus, an alternative
DeepAK8 algorithm, ``DeepAK8-MD'', is developed to be largely
decorrelated with the mass of a jet, while preserving the
discrimination power as much as possible using an adversarial
training approach~\cite{NIPS2017_6699}. Jets from different signal and
background samples are also weighted to yield flat distributions in both
\pt and \msd to aid the training.

\begin{figure}[ht]
\centering
\includegraphics[width=0.7\textwidth]{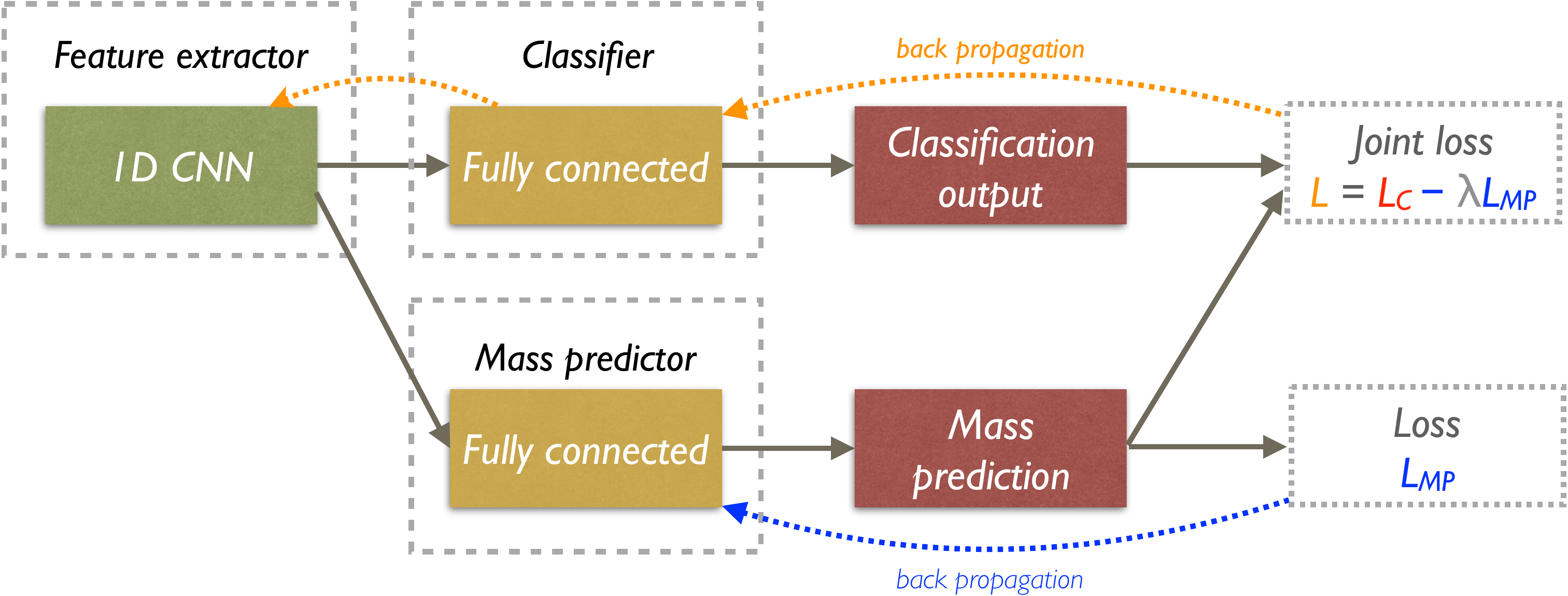}
\caption{\label{fig:deepAK8_adv_arch} The network architecture of DeepAK8-MD.}
\end{figure}

The architecture of DeepAK8-MD is shown in Fig.~\ref{fig:deepAK8_adv_arch}.
Compared to the nominal version of DeepAK8, a mass prediction
network is added with the goal of predicting the mass of a background jet from
the features extracted by the CNNs. 
The mass prediction network consists of 3 fully-connected layers, each with 
256 units and a SELU activation function \cite{Klambauer:2017:SNN:3294771.3294864}. 
It is trained to predict the \msd of background jets to the closest 10\GeV value 
between 30 and 250\GeV by minimizing the cross-entropy loss. 
When properly trained, the mass
prediction network becomes a good indicator of how strongly the
features extracted by the CNNs are correlated with the mass of a jet,
because the stronger the correlation is, the more accurate the mass
prediction will be. With the introduction of the mass prediction
network, the training target of the algorithm can be modified to
include the accuracy of the mass prediction for the background jets as a penalty, therefore
preventing the CNNs from extracting features that are correlated with
the mass. In this way, the final prediction of the algorithm also
becomes largely independent of the mass. As the features extracted by
the CNNs evolve during the training process, the mass prediction
network itself needs to be updated regularly to adapt to the
changes of its inputs and remain as an effective indicator of mass
correlation. Therefore, for each training step of the DeepAK8 
network (the Particle and SV CNNs and the 1-layer fully-connected network), 
the mass prediction network is trained for 10 steps. 
Each training step corresponds to a minibatch of 6000 jets. 
A large minibatch size is used to reduce statistical fluctuation 
on the mass correlation penalty evaluated by the mass prediction network, 
since only background jets are used in the evaluation. Both the DeepAK8 network 
and the mass prediction network are trained with the \textsc{Adam} optimizer.
A constant learning rate of 0.001 (0.0001) is used for the training of 
the DeepAK8 (mass prediction) network.

Forcing the algorithm to be decorrelated with the jet mass, inevitably
leads to a loss of discrimination power, and the resulting algorithm is
a balance between performance and mass independence.
Because the training of DeepAK8-MD is carried out only on jets with $30<\msd<250\GeV$,
jets with \msd outside this range should be removed when using DeepAK8-MD.

\section{Performance in simulation}
\label{sec:performanceinmc}

As presented in Section~\ref{sec:hrtalgorithms}, a variety of
algorithms have been developed by the CMS Collaboration to identify the
hadronic decays of
\PW{}/\PZ{}/\PH/ bosons and \PQt quarks. To gain an initial
understanding of the tagging performance and the complementarity
between the different approaches, the algorithms were studied in
simulated events. The performance of the algorithms is evaluated using
the signal and background efficiencies, $\epsilon_{\text{S}}$ and
$\epsilon_{\text{B}}$, respectively, as a figure of merit. The efficiencies \esig~and
\ebkg~are defined as:
\begin{equation}
\esig = \frac{N^{\text{tagged}}_\text{S}} {N^{\text{total}}_\text{S}} ~~~\text{and}~~~
\ebkg = \frac{N^{\text{tagged}}_\text{B}} {N^{\text{total}}_\text{B}},
\end{equation}

where $N^{\text{tagged}}_{S}$ ($N^{\text{tagged}}_{B}$) is the number of
signal (background) jets satisfying the identification criteria of each algorithm,
and $N^{\text{total}}_{S}$ ($N^{\text{total}}_{B}$) is
the total number of generated particles considered to be signal (background).
Hadronically decaying \PW{}/\PZ{}/\PH bosons or \PQt quarks are
signal,
whereas quarks (excluding \PQt~quarks) and gluons from the QCD multijet process
are  background.

\begin{figure}[!p]
\centering
\includegraphics[width=0.45\textwidth]{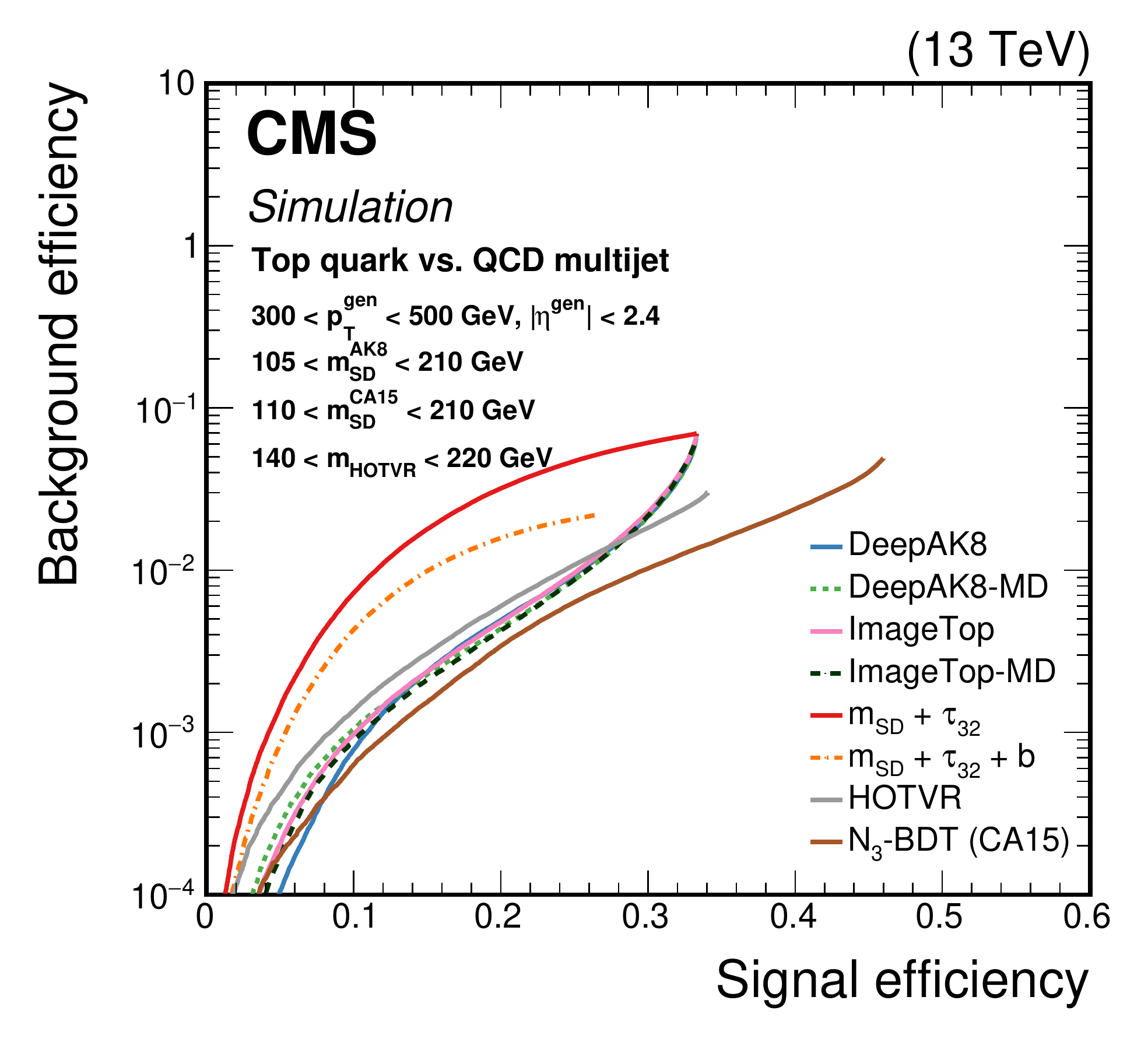}
\includegraphics[width=0.45\textwidth]{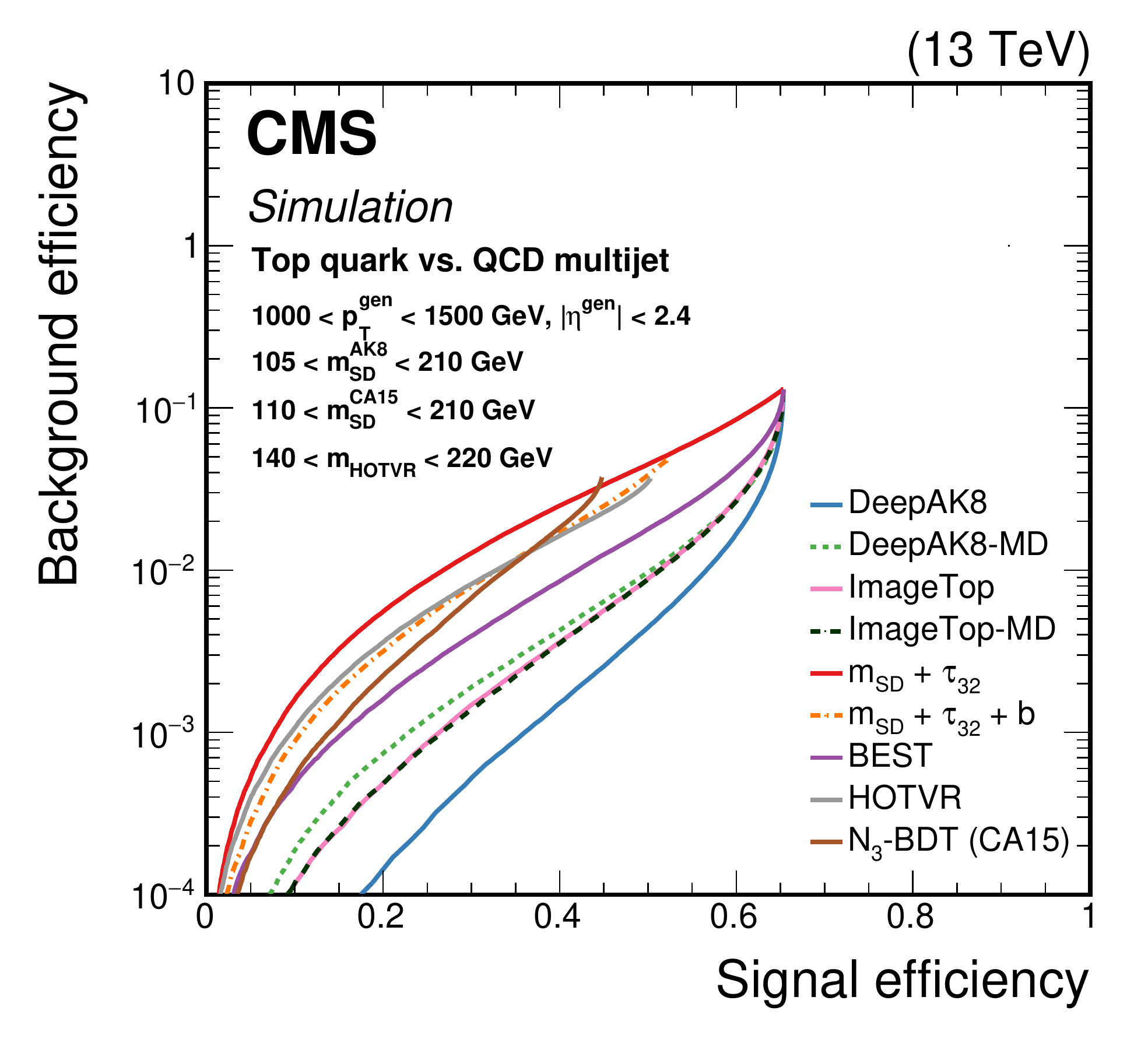}
\caption{\label{fig:roc_top_vs_qcd} Comparison of the identification algorithms for
  hadronically decaying \PQt~quark in terms
  of ROC curves in two regions
  based on the \pt~of the generated particle; Left: $300<\pt<500\GeV$, and
  Right: $1000<\pt<1500\GeV$. Additional fiducial selection criteria applied to
  the jets are listed on the plots.}
\end{figure}

\begin{figure}
\includegraphics[width=0.45\textwidth]{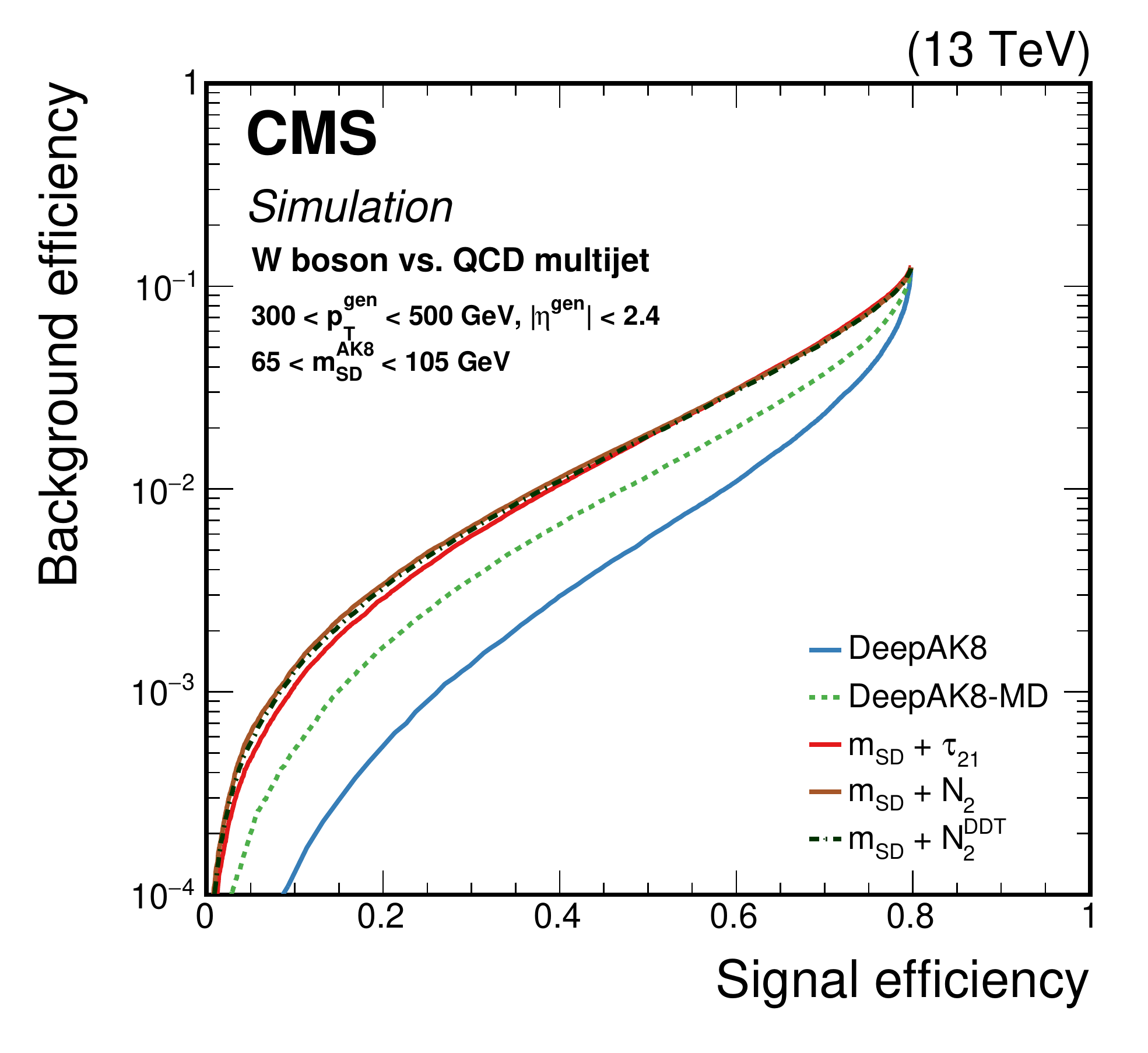}
\includegraphics[width=0.45\textwidth]{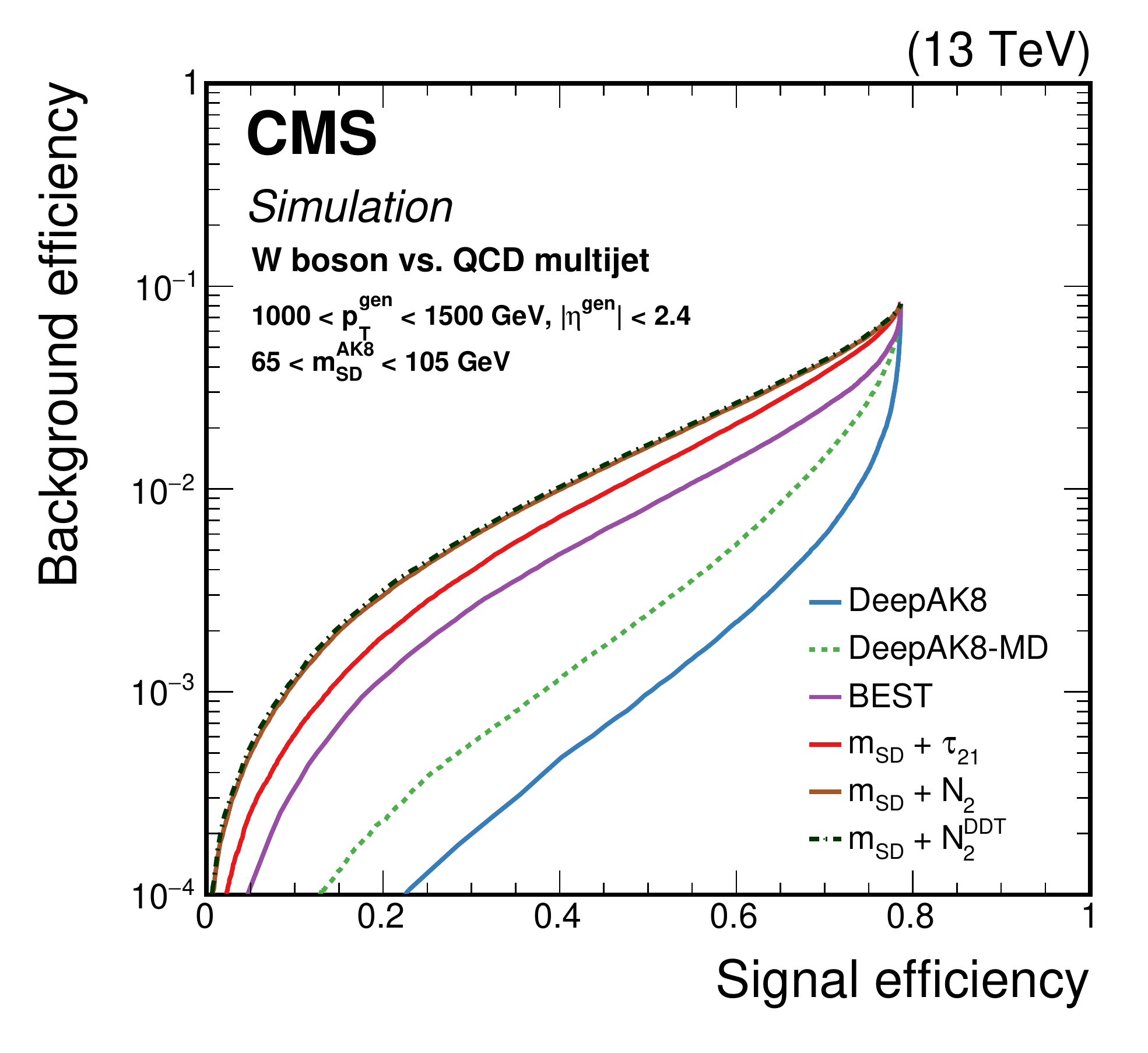}
\caption{\label{fig:roc_w_vs_qcd} Comparison of the identification algorithms for
  hadronically decaying \PW~boson  in terms of
  ROC curves in two regions based
  on the \pt~of the generated particle; Left: $300<\pt<500\GeV$, and
  Right: $1000<\pt<1500\GeV$.  Additional fiducial selection criteria applied to
  the jets are listed on the plots.}
\end{figure}

\begin{figure}[!p]
\centering
\includegraphics[width=0.45\textwidth]{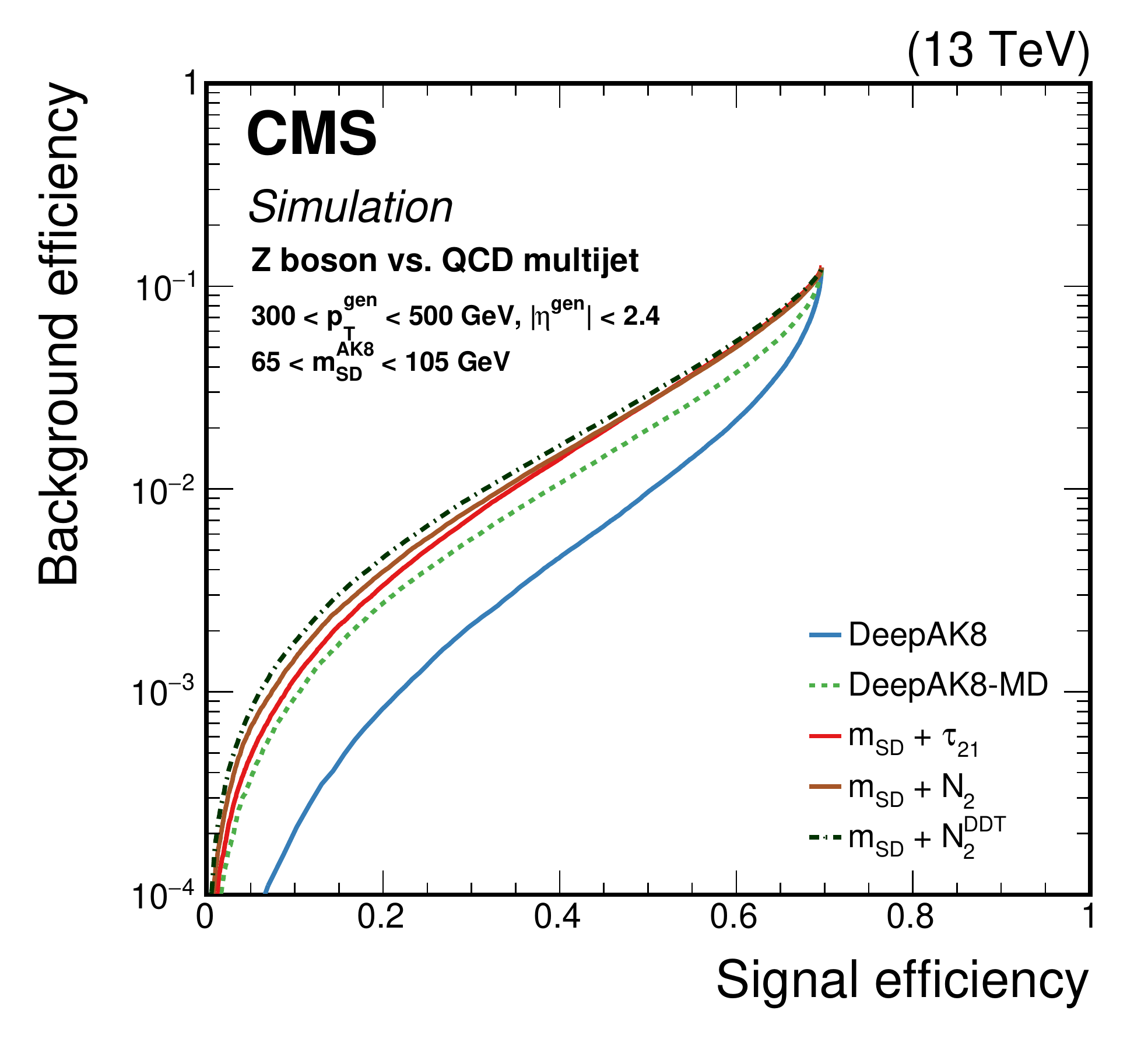}
\includegraphics[width=0.45\textwidth]{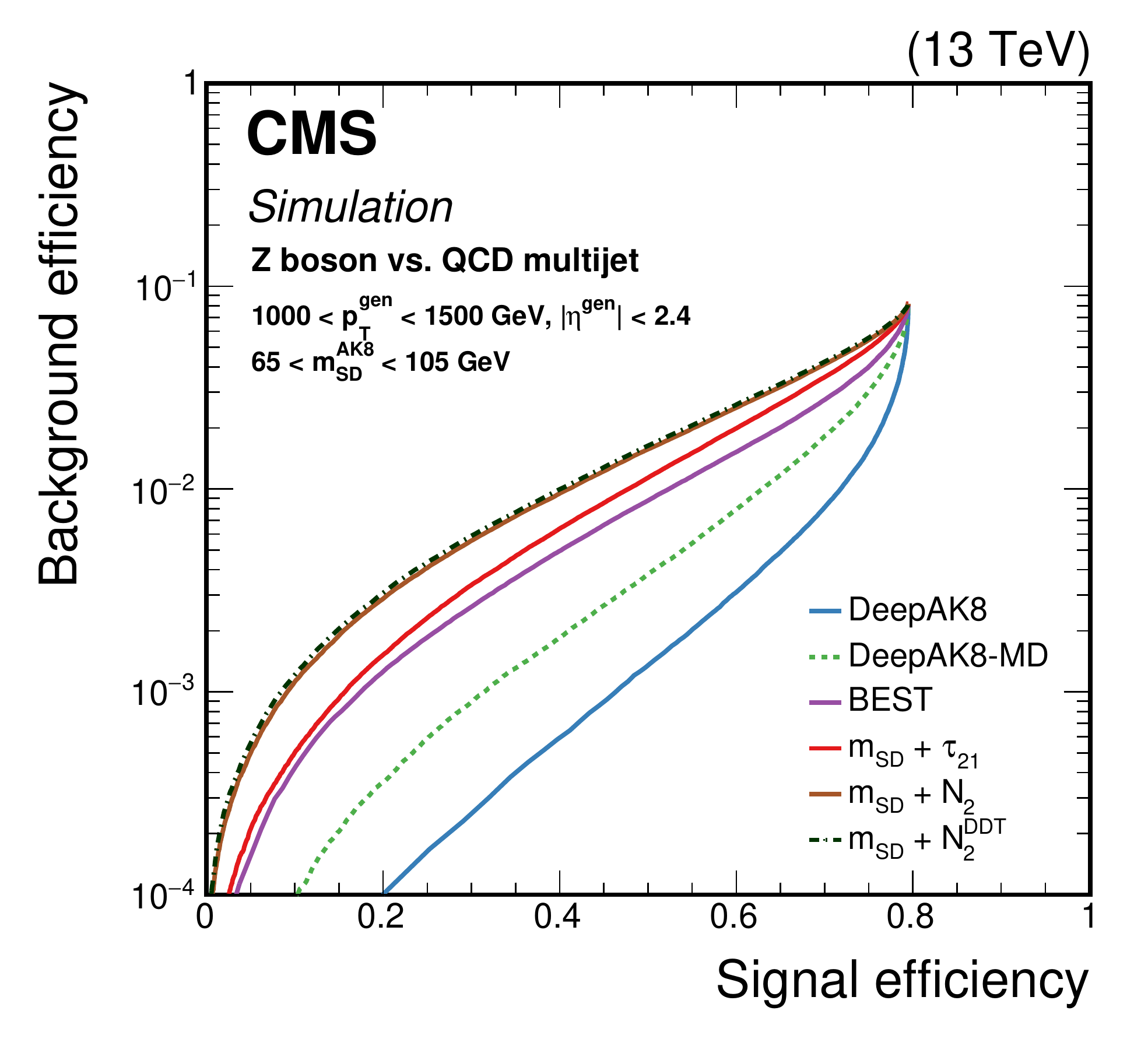}
\caption{\label{fig:roc_z_vs_qcd} Comparison of the identification algorithms for
  hadronically decaying \PZ~boson  in terms of
  ROC curves in two regions based
  on the \pt~of the generated particle; Left: $300<\pt<500\GeV$, and
  Right: $1000<\pt<1500\GeV$.  Additional fiducial selection criteria applied to
  the jets are listed on the plots.}
\end{figure}

\begin{figure}
\includegraphics[width=0.45\textwidth]{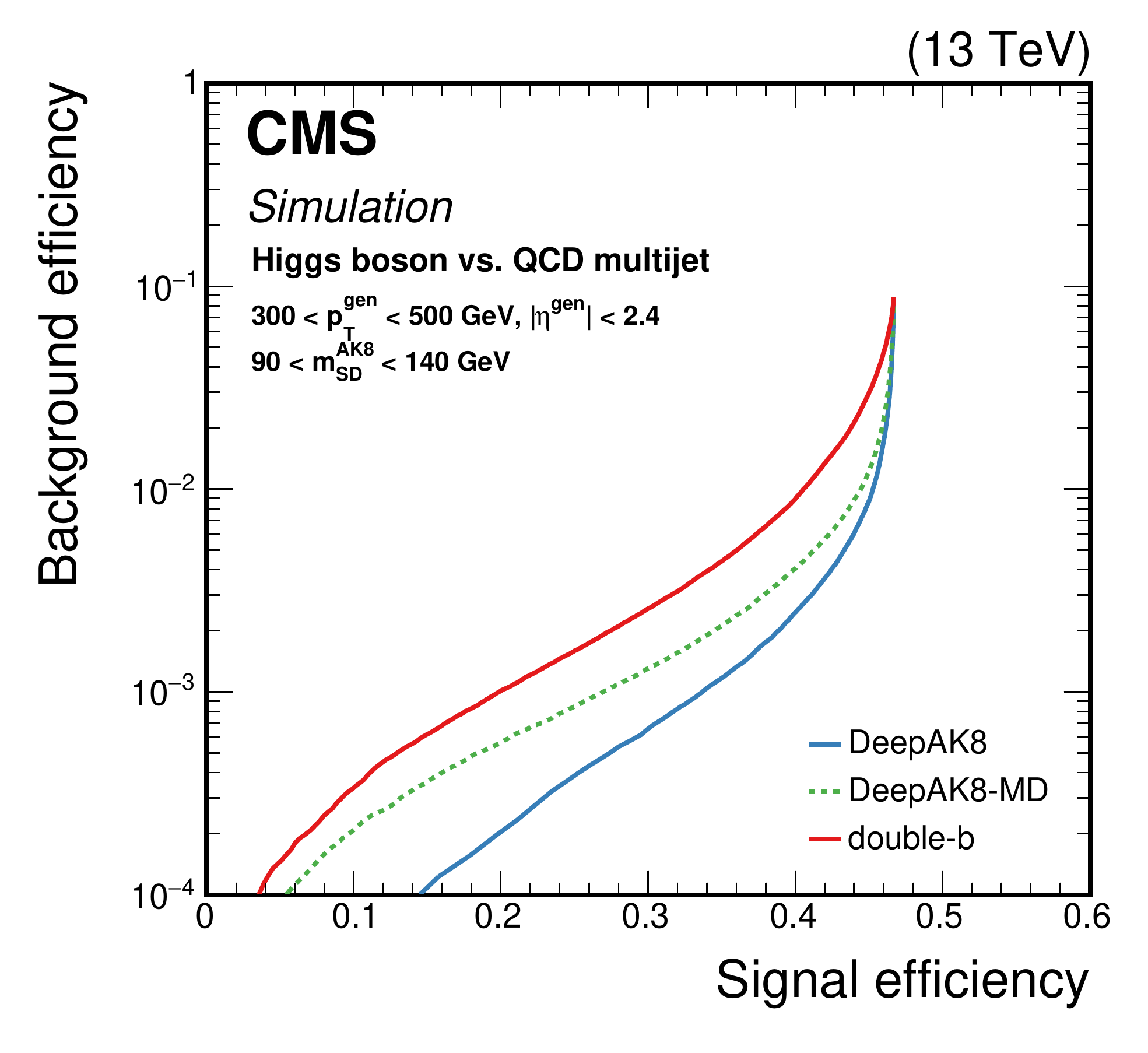}
\includegraphics[width=0.45\textwidth]{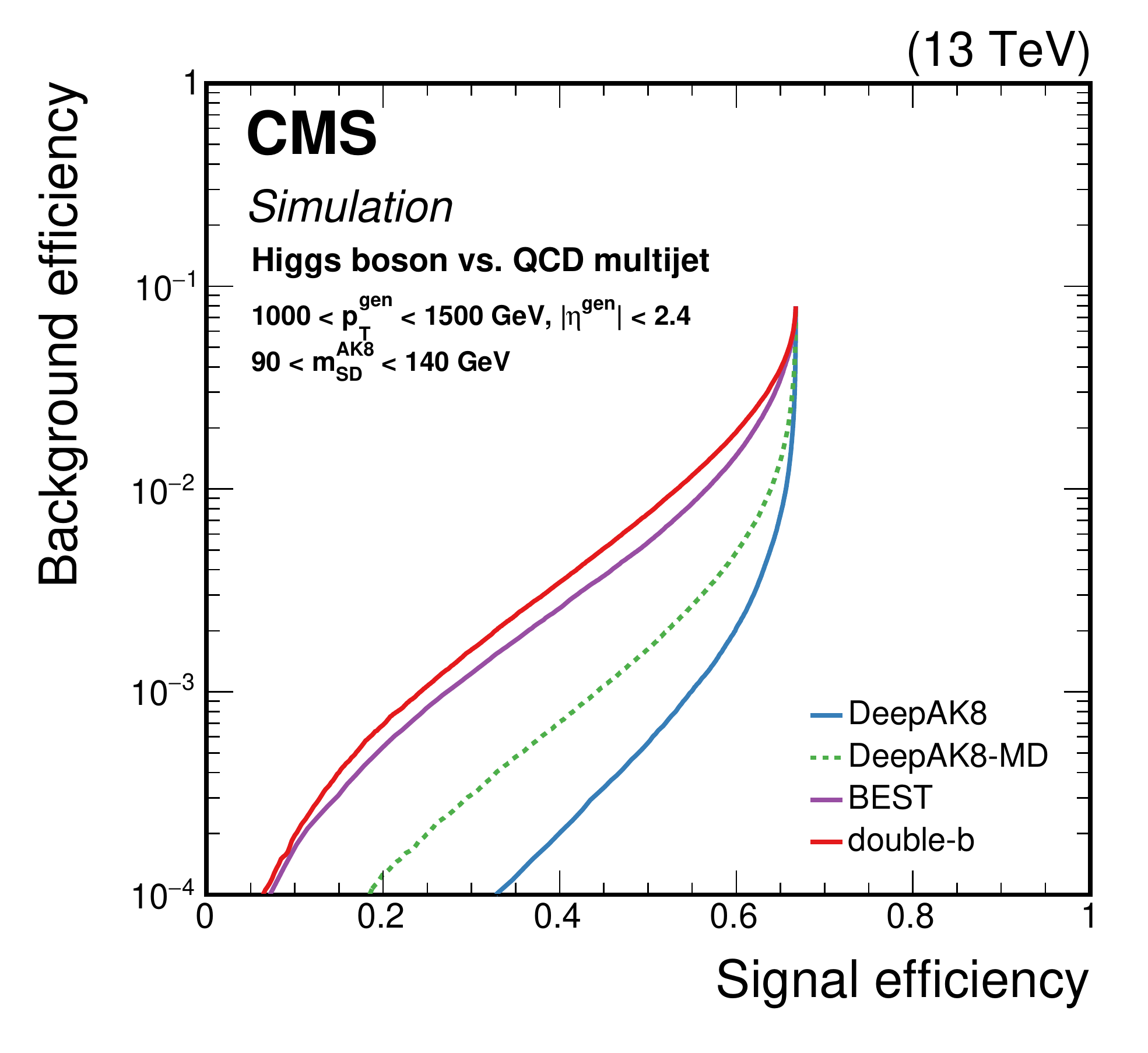}
\caption{\label{fig:roc_higgs_vs_qcd} Comparison of the identification algorithms for
  hadronically decaying \PH boson  in terms
  of ROC curves in two regions
  based on the \pt~of the generated particle; Left: $300<\pt<500\GeV$, and
  Right: $1000<\pt<1500\GeV$. The \PH boson decays to a
  pair of \PQb~quarks.  Additional fiducial selection criteria applied to
  the jets are listed on the plots.}
\end{figure}

First, for each algorithm, the \ebkg~as a function of \esig~is
evaluated in terms of a receiver operating characteristic (ROC)
curve. Figures~\ref{fig:roc_top_vs_qcd}--\ref{fig:roc_higgs_vs_qcd}
summarize the ROC curves of all algorithms for the identification of
\PQt~quarks, and \PW, \PZ, and \PH bosons, respectively. The comparisons
are performed at low and high values of the generated particle \pt. The
fiducial selection criteria applied to the generator-level particles are
displayed in the plots. For the cutoff-based algorithms, namely \msdtop,
\msdtopbtag, \msdv, \ecfv, and \ecfvddt, all selections except the
selection on $\tauthreetwo$, $\tautwoone$, or $N_{2}$, are applied, as
described in Sections~\ref{sec:subsec_softdrop} and~\ref{sec:ecfv}.

In \PQt tagging, the addition of the subjet $\PQb$ tagging
in the \msdtop algorithm reduces the misidentification probability
for \PQt~quarks by up to $\sim$50\% depending on the \pt. The
performance of the HOTVR algorithm lies between \msdtop and
\msdtopbtag, and the \ecftop algorithm shows improved performance
compared to these algorithms, particularly in the low-\pt range.
 The improved performance stems from the usage of the ECFs,
which provide complementary information to \tauthreetwo. Particularly
in the low-\pt region, the gain is mainly due to the use of
larger-cone jets (i.e., jets clustered with $R=1.5$). The BEST
algorithm targets the high-\pt regime and shows similar performance to the ECF
algorithm in this regime. The best discrimination is achieved with
algorithms based on lower-level information, namely the ImageTop and
DeepAK8 algorithms. ImageTop and DeepAK8-MD yield comparable
performance in the low and high \pt regions. The best performance
in terms of ROC curves is achieved with the nominal
version of DeepAK8 over the entire \pt region.


Various arguments contribute to the significantly improved performance
of ImageTop and DeepAK8 with respect to the other algorithms. 
First, the usage of lower-level variables as inputs to the network 
exploits the high granularity of the CMS detector. Second,
the architectures
of these algorithms provide quark-gluon discrimination information.
 Moreover,
information about the jet flavor content is extracted, which is
particularly important for \PQt~quark and \PZ/\PH boson identification. The
flavor identification in jets from boosted object decay is very
challenging because the
decay products overlap and traditional \PQb~tagging algorithms
perform significantly less well. The usage of the type of the PF
candidates, and the secondary vertices in the case of DeepAK8, provides
 a more precise description of the flavor content inside the jet.

Similar conclusions hold for the identification of hadronically
decaying \PW and \PZ~bosons. The BEST, DeepAK8, and DeepAK8-MD
algorithms show enhanced performance compared with the simpler
\msdv algorithm. The gain in terms of misidentification rate can be
as large as an order of magnitude in the case of DeepAK8. The smaller
relative gain of DeepAK8 over BEST for discriminating between \PW or \PZ~bosons, 
and \PQt~quarks occurs because flavor information for the \PW and \PZ bosons 
is not as critical as for \PQt~quarks.
The \ecfv and \ecfvddt show weaker performance compared with 
the \msdv algorithm.

The double-\PQb, BEST, DeepAK8, and DeepAK8-MD
algorithms are used to identify hadronic decays of the \PH
boson. In Fig.~\ref{fig:roc_higgs_vs_qcd}, the \PH boson decays
to a pair of \PQb~quarks. The performance of the BEST algorithm
lies between the double-\PQb algorithm and
DeepAK8. The gain with DeepAK8 is expected just as in
 \PQt~quark identification for similar arguments.

\begin{figure}[!phtb]
\centering
\includegraphics[width=0.45\textwidth]{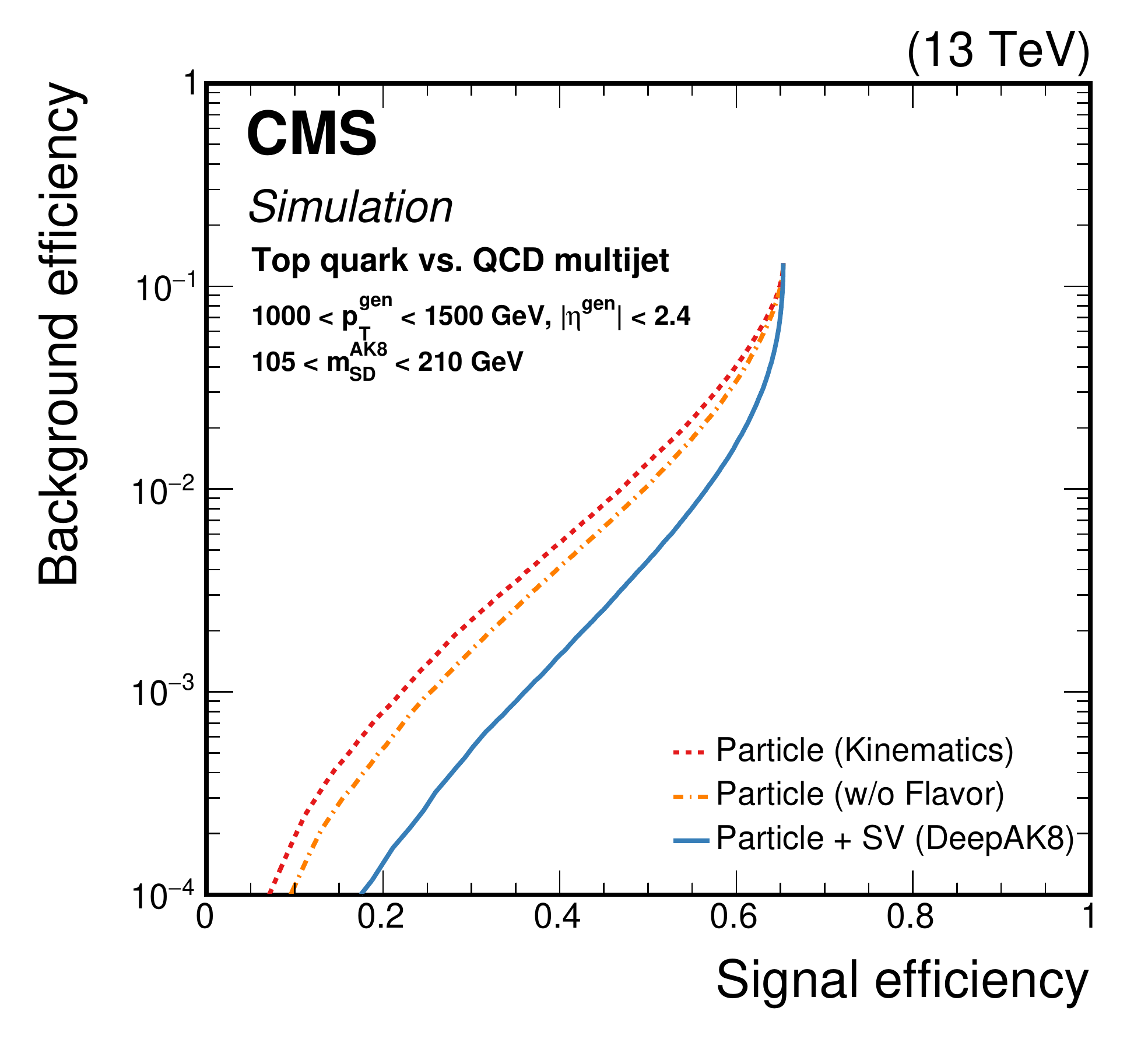}\\
\includegraphics[width=0.45\textwidth]{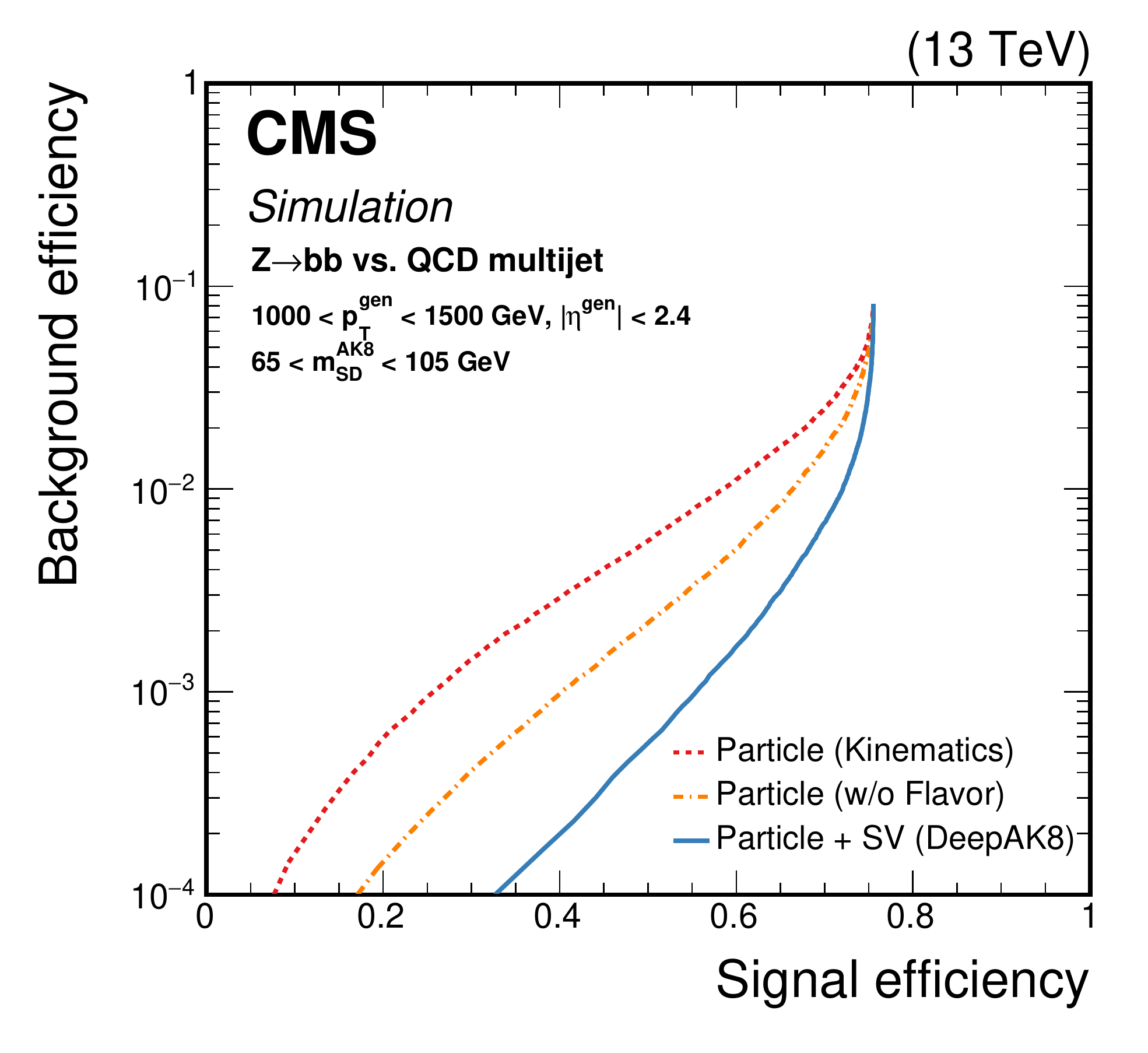}
\includegraphics[width=0.45\textwidth]{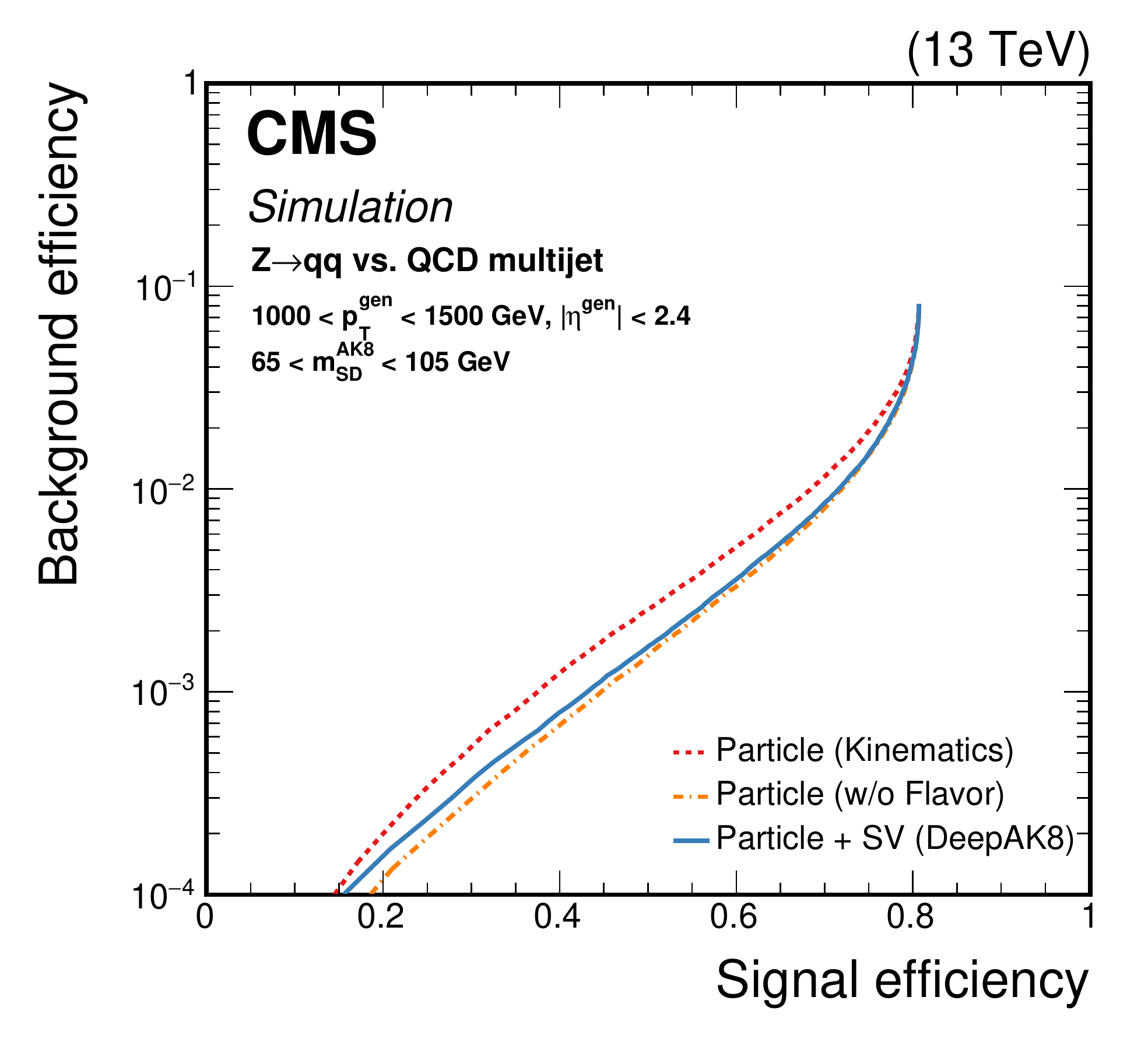}
\caption{\label{fig:deepak8_roc_decomp}Alternative versions of DeepAK8
  trained using a subset of the input features. The details about each
  version are discussed in the text. The performances of the three
  versions of DeepAK8 are compared for \PQt quark (upper) and \PZ boson
  (lower) identification. For the latter, the left plot
  corresponds to \PZ bosons decaying to a pair of \PQb quarks, and the
  right plot to a pair of light-flavor quarks. }
\end{figure}

To gain a deeper understanding of the DeepAK8 performance, two alternative
versions of DeepAK8 were trained using a subset of the input
features. Three sets of input features were studied and compared.
The ``Particle (kinematics)'' set
consists of only the kinematic information on the PF candidates, e.g.,
the four-momenta and the distances to the jet and subjet axes. This set
serves as a baseline to evaluate the performance using only
substructure of the jets. The ``Particle (w/o Flavor)'' set includes
additional experimental information for each PF candidate, such as the
electric charge, particle identification, and track quality
information. Compared with the nominal DeepAK8 algorithm, input features
that contribute to the identification of heavy-flavor quarks, such as
the displacement of the tracks, the association of tracks to the
reconstructed vertices, and the SV features, are not included
in the ``Particle (w/o Flavor)'' set. The performances of the three
versions of DeepAK8 are compared in Fig.~\ref{fig:deepak8_roc_decomp}
for \PQt~quark and \PZ boson identification. In both cases, the addition of
experimental information brings sizable improvement in
performance. Although the additional features contributing to heavy-flavor
 identification lead to no improvement for the
identification of \PZ bosons decaying to a pair of light-flavor quarks, a significant
improvement is observed for \PZ bosons decaying to a pair of \PQb quarks, as well
as \PQt quark decays, showing the strong complementarity between heavy-flavor
identification and jet substructure for heavy-resonance identification
where heavy-flavor quarks are involved in the decay.

\subsection{Robustness of tagging algorithms}
\label{sec:robustnessstudies}
In addition to the performance of the algorithms in pure
discrimination, an important ingredient is their robustness to changes in
jet kinematics and data-taking conditions. To quantify this, we study the
\esig and \ebkg of the algorithms as a function of the \pt~of the
generated particle and the number of reconstructed vertices (\nvtx) in the event.
 For these studies, a common working point is defined,
corresponding to $\esig=30$ (50)\% for \PQt quark
(\PW{}/\PZ{}/\PH boson) with $500<\pt(\text{generated particle})<600\GeV$.
Working points used in CMS analyses vary from analysis to analysis,
since they are optimized to achieve the best sensitivity for the targeted signal processes.
For example, CMS employs a \PQt quark tagging working point
at approximately 40\% signal efficiency in the search for BSM \ttbar production~\cite{Sirunyan:2018ryr},
a \PW tagging working point at approximately 20\% signal efficiency in the search
for BSM diboson production~\cite{Sirunyan:2019jbg}, and an \PH tagging working point at
approximately 30\% signal efficiency in the search for \PH boson pair production~\cite{Sirunyan:2018qca}.

\begin{figure}[!phtb]
\centering
\includegraphics[width=0.45\textwidth]{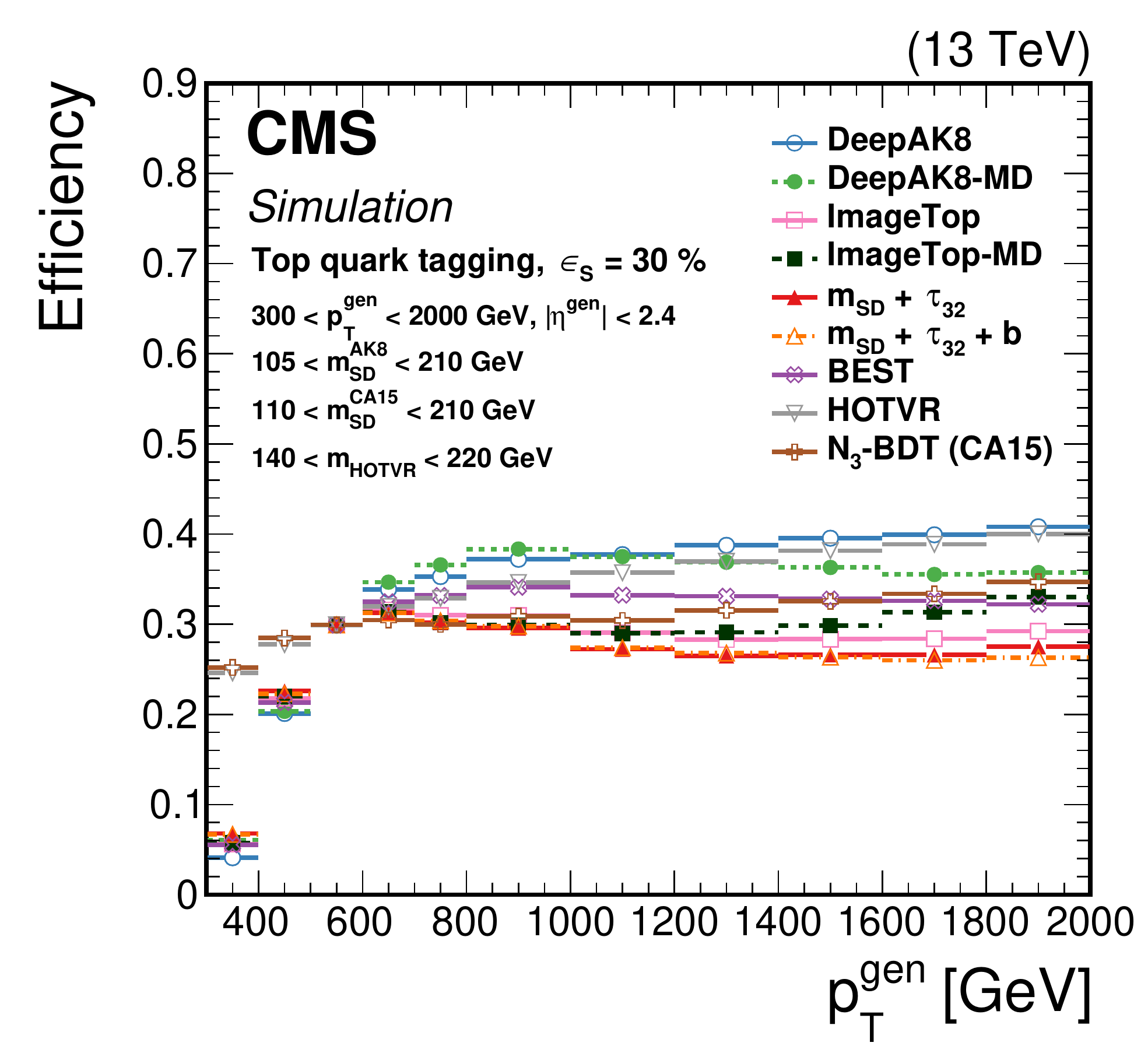}
\includegraphics[width=0.45\textwidth]{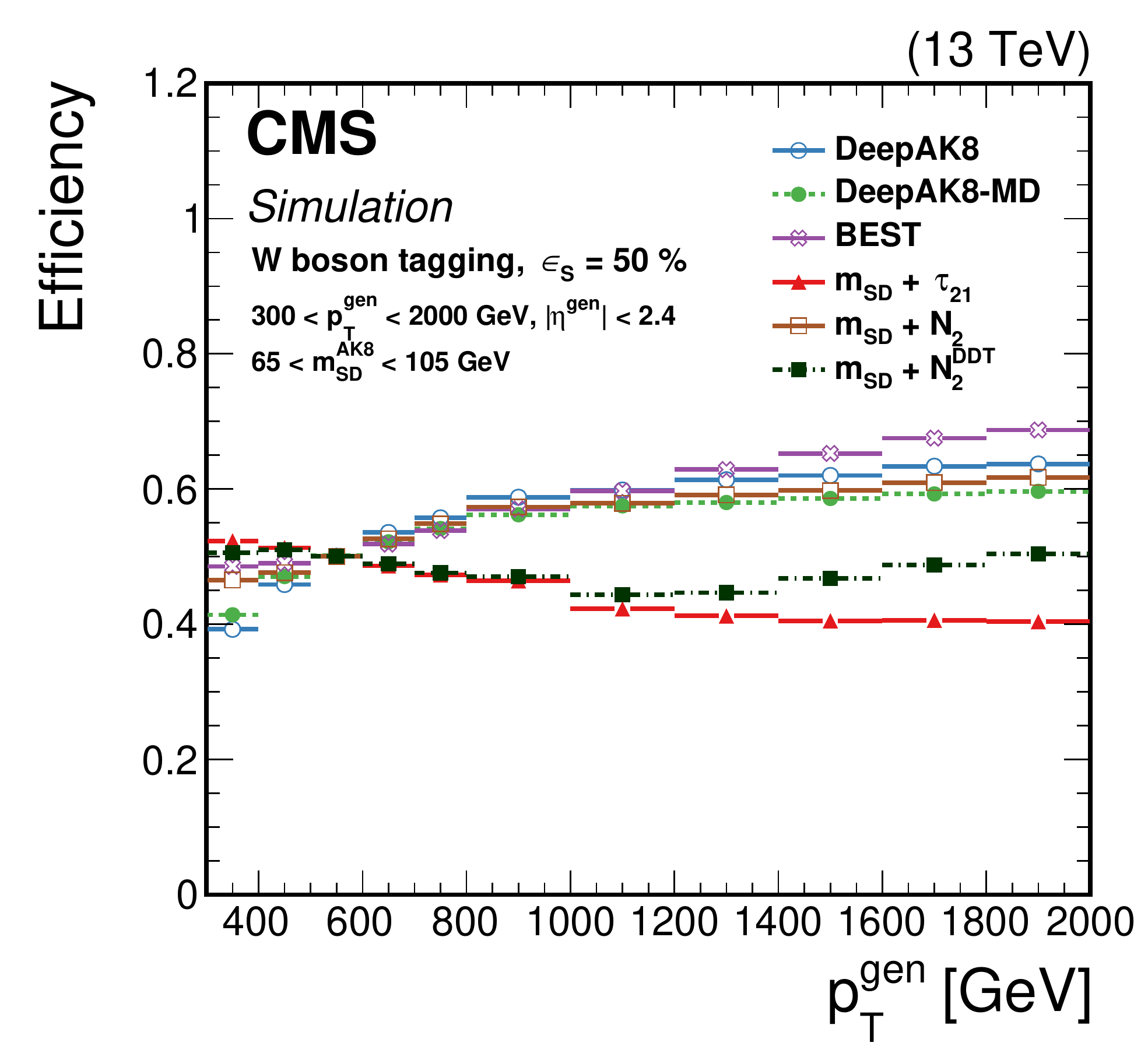}
\includegraphics[width=0.45\textwidth]{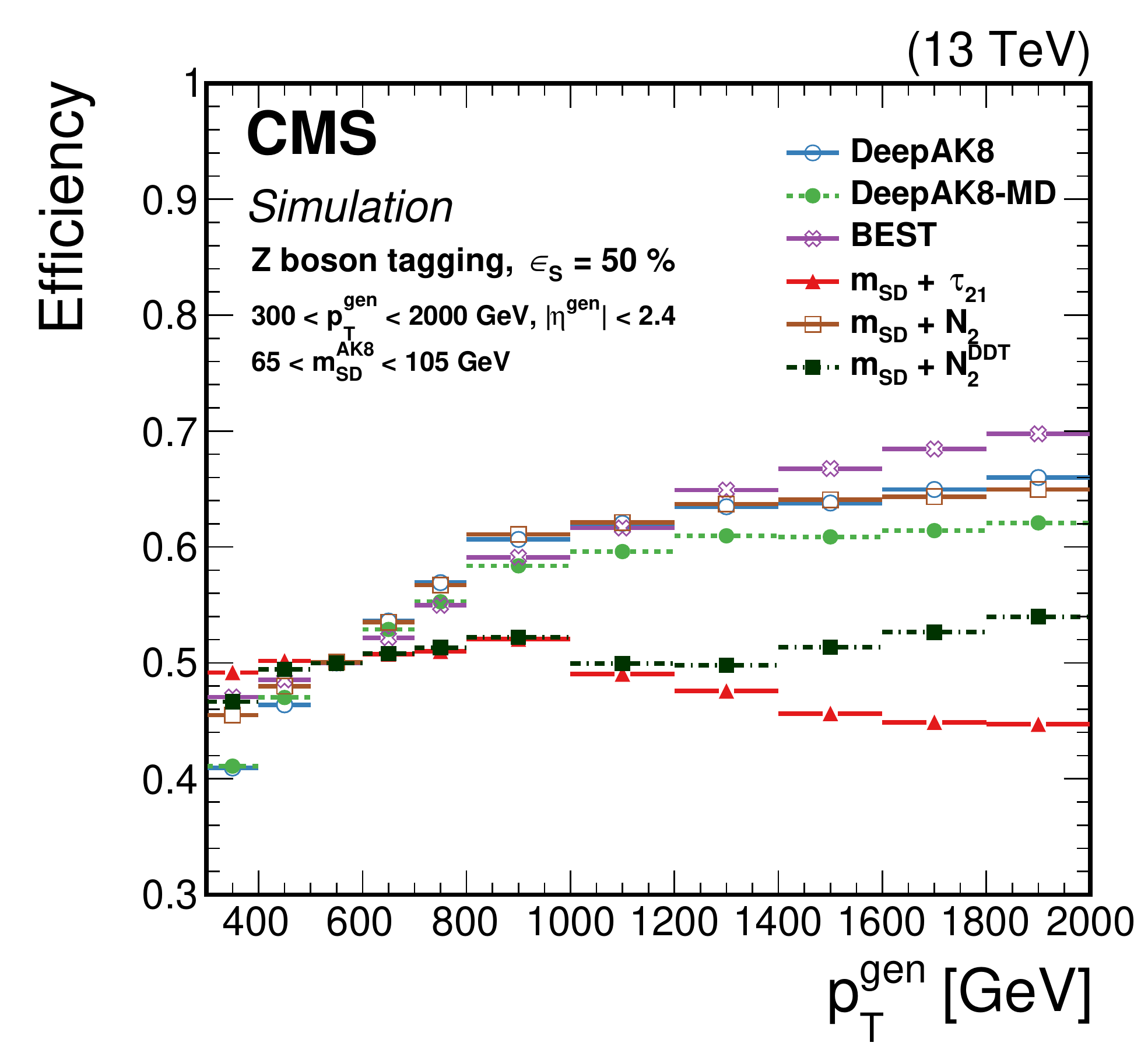}
\includegraphics[width=0.45\textwidth]{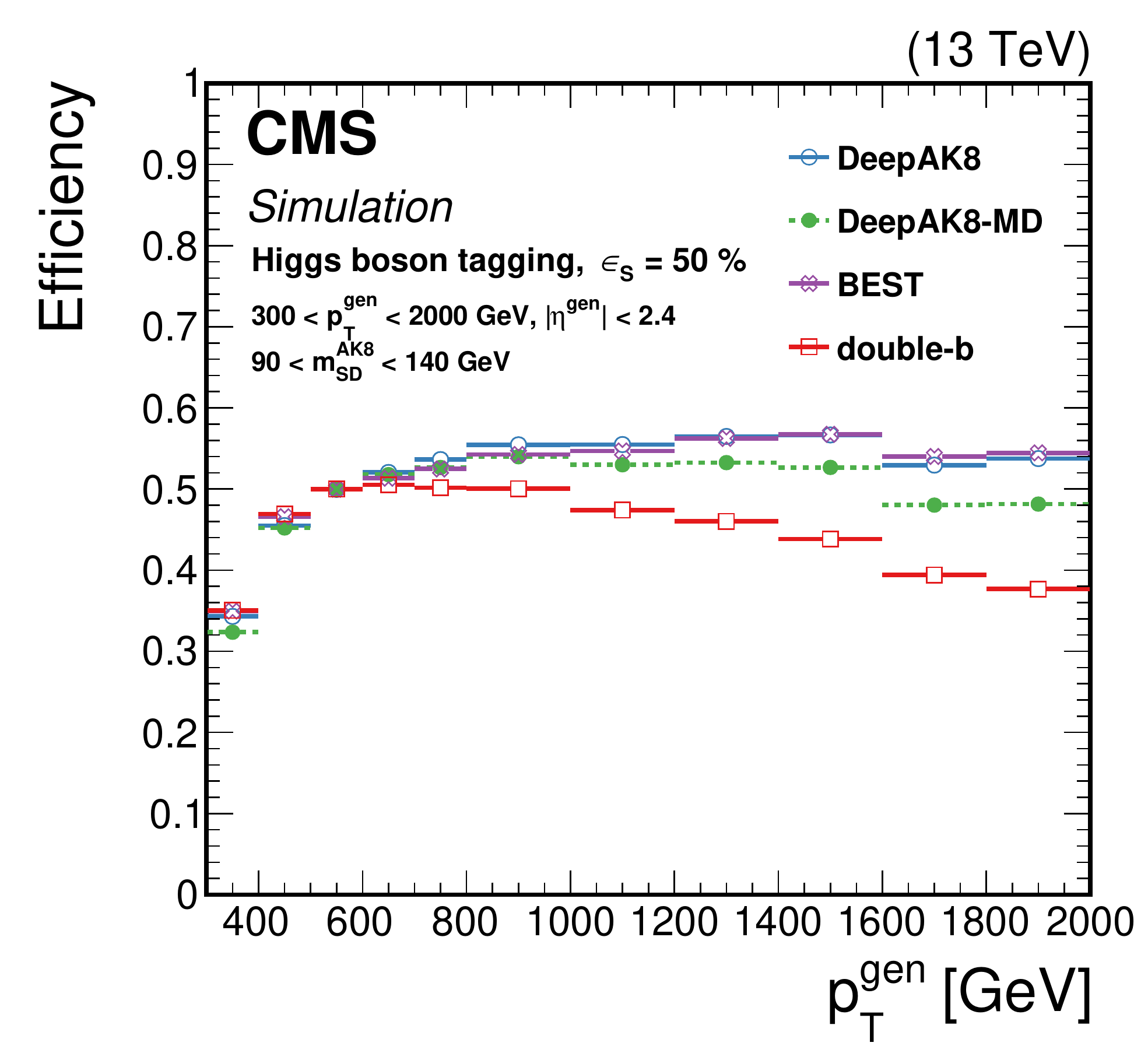}
\caption{\label{fig:esig_vs_pt} The efficiency \esig as a
  function of the generated particle \pt for a working point
  corresponding to $\esig = 30\,(50)$\% for \PQt~quark
 (\PW{}/\PZ{}/\PH boson) identification. Upper left:
  \PQt~quark, upper right: \PW~boson, lower left: \PZ~boson, lower
  right: \PH boson. The error bars represent the
  statistical uncertainty in each specific bin, due to the limited
  number of simulated events.  Additional fiducial selection criteria applied to
  the jets are listed in the plots.}
\end{figure}

\begin{figure}[!phtb]
\centering
\includegraphics[width=0.45\textwidth]{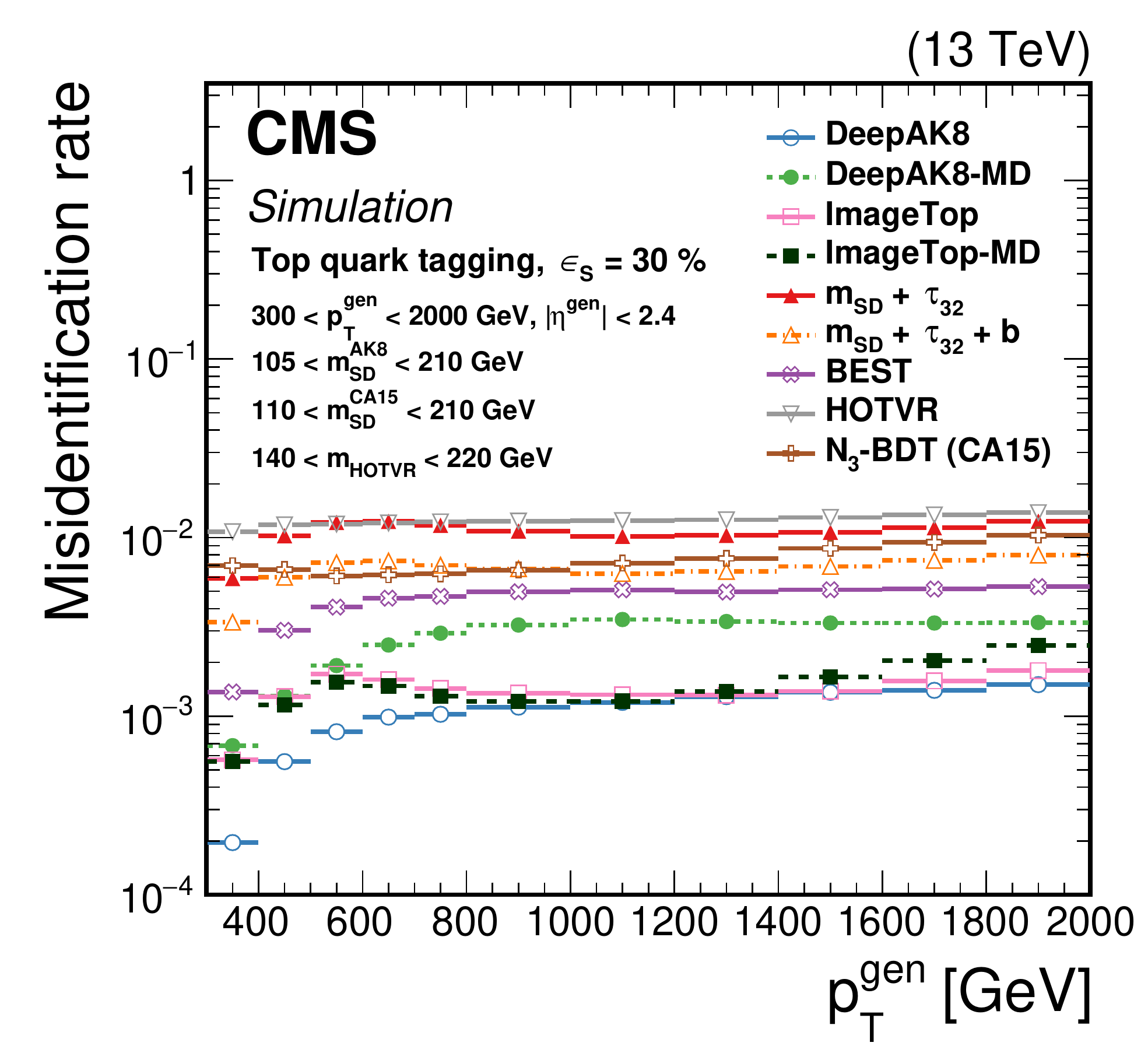}
\includegraphics[width=0.45\textwidth]{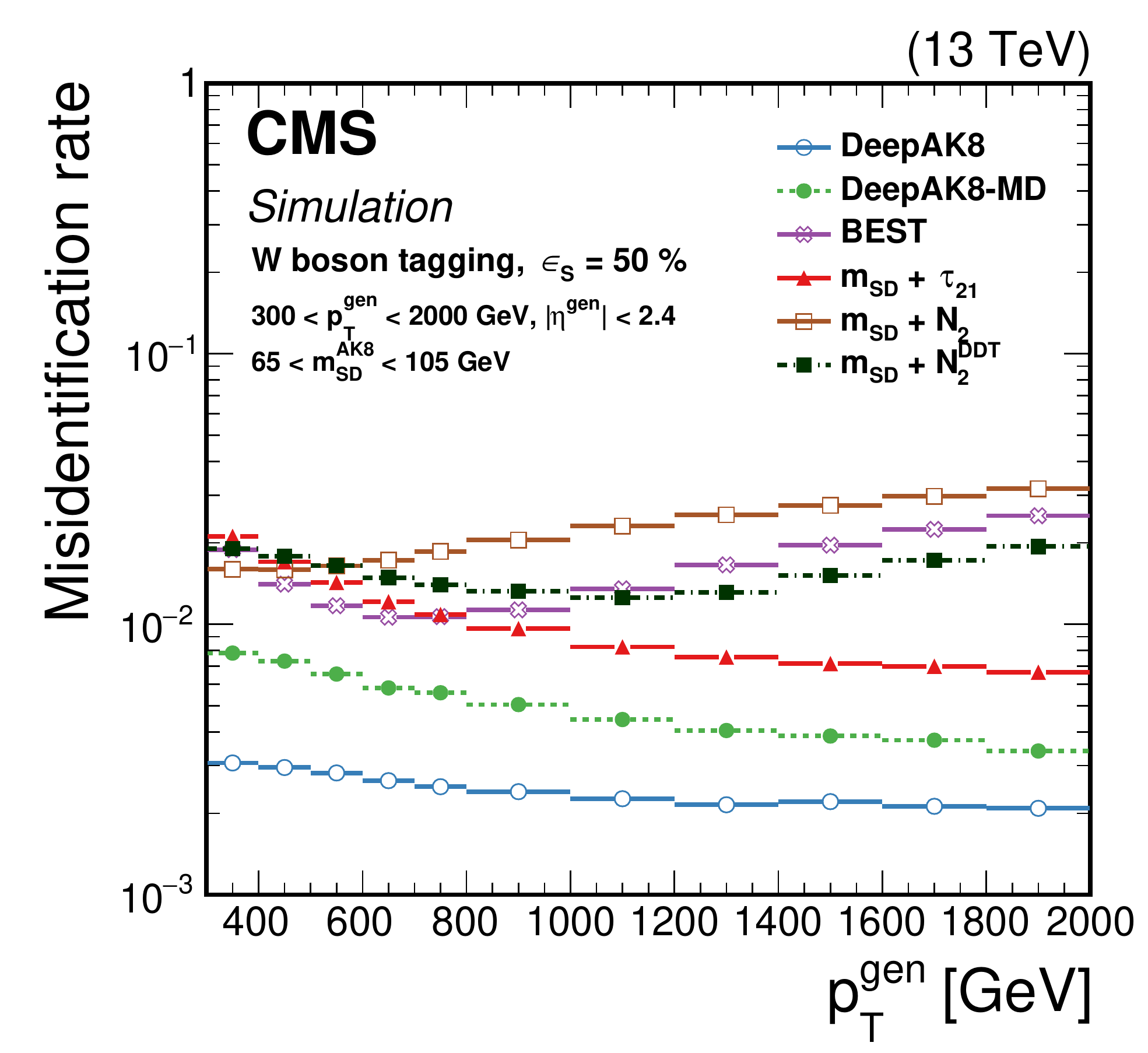}
\includegraphics[width=0.45\textwidth]{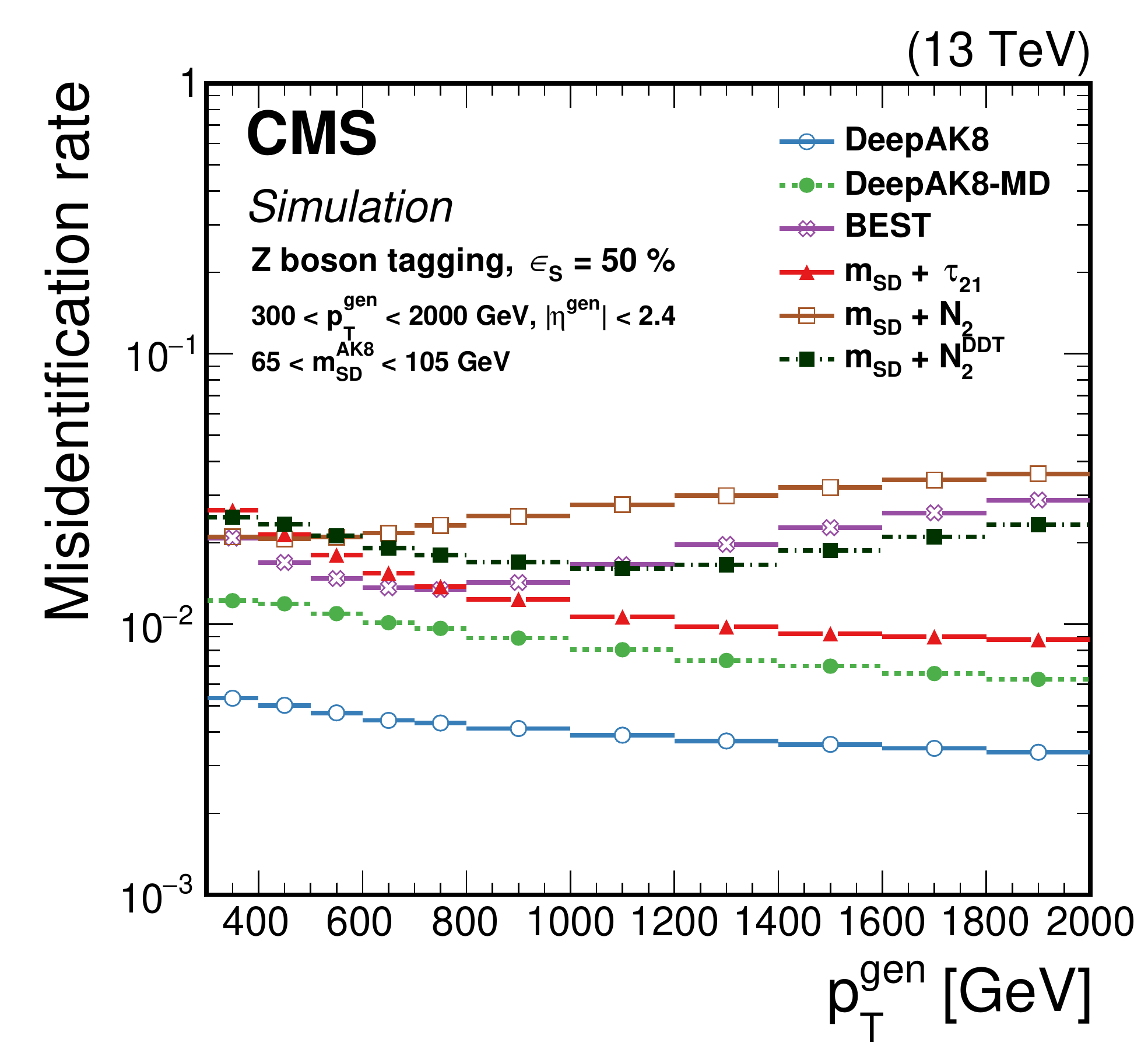}
\includegraphics[width=0.45\textwidth]{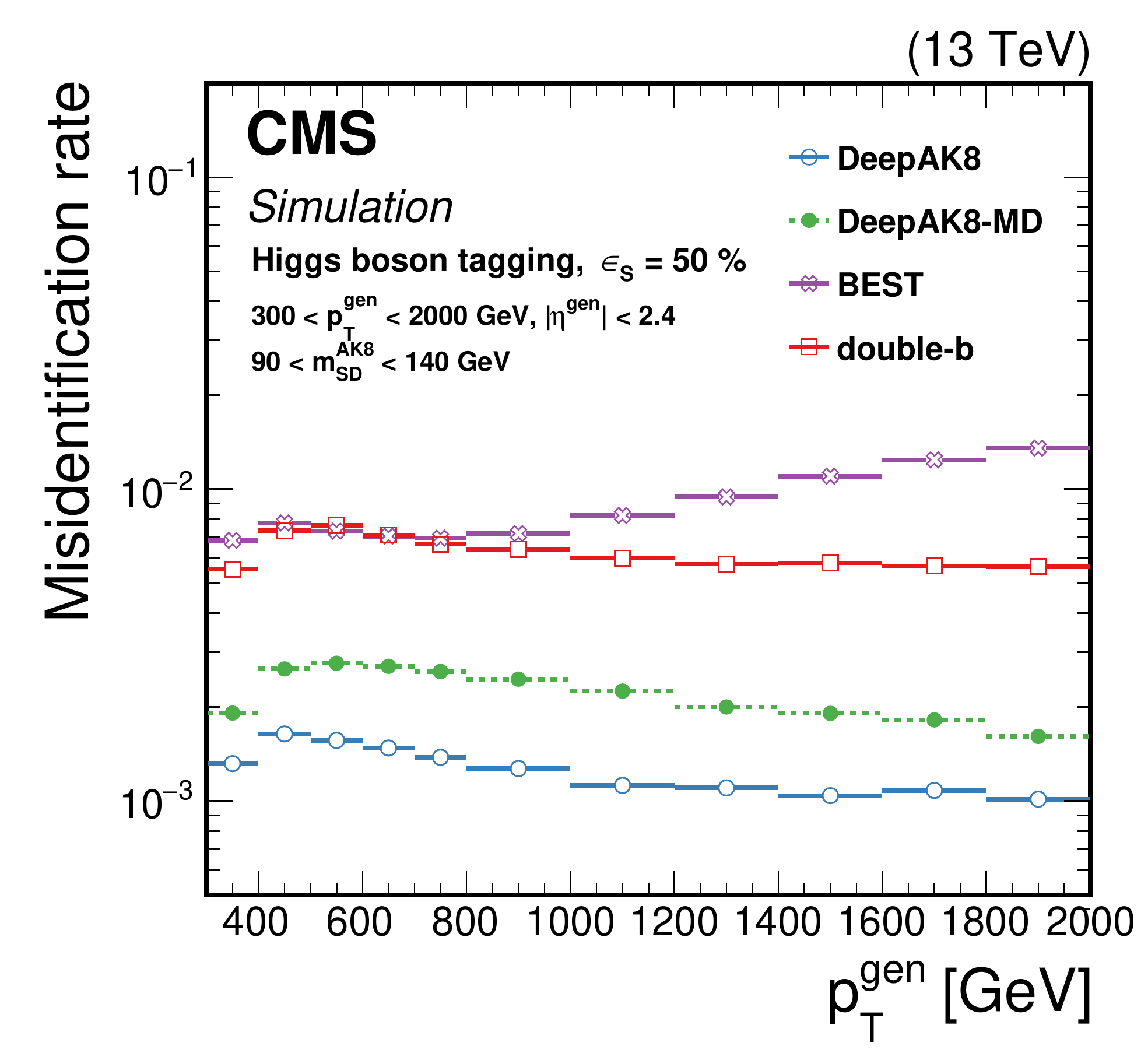}
\caption{\label{fig:ebkg_vs_pt} The distribution of \ebkg as a
  function of the generated particle \pt for a working point
  corresponding to $\esig = 30\,(50)$\% for \PQt~quark
 (\PW{}/\PZ{}/\PH boson) identification. Upper left:
  \PQt~quark, upper right: \PW~boson, lower left: \PZ~boson, lower
  right: \PH boson. The error bars represent
  the statistical uncertainty in each specific bin, due to the limited
  number of simulated events.  Additional fiducial selection criteria applied to
  the jets are listed in the plots.}
\end{figure}

The distributions of the \esig and \ebkg as a function of \pt~of
the generated particle for the different particle identification
scenarios are displayed in Figs.~\ref{fig:esig_vs_pt} and
\ref{fig:ebkg_vs_pt}, respectively. In the low-\pt range for the
\PQt tagging case, the \esig for the algorithms using AK8 jets
increases
rapidly until $\pt \gtrsim 600\GeV$, where a large fraction of jets
contain all the \PQt decay products. As expected, the \ecftop and HOTVR
algorithms have a stable \esig as a function of the generator-level particle
\pt. Similar behavior is observed for the \PQt quark
misidentification rate.

In the case of the \PW and \PZ boson tagging, the \esig for the
\msdv algorithm decreases as a function of $\pt(\text{generated particle})$,
whereas the BEST, DeepAK8, and DeepAK8-MD
algorithms exhibit improvements in \esig as a function of $\pt(\text{generated particle})$.
The drop in \esig for \msdv is a result of the correlation between \msdv and
 the jet \pt, leading to a shift in the jet mass distribution to higher values.
The \ecfv algorithm shows similar behavior to BEST and DeepAK8 algorithms,
whereas the \esig in the case of \ecfvddt is
stable as a function of $\pt(\text{generated particle})$. In contrast
to $N$-subjettiness, the ECF observable uses an axis-free approach, which is more
efficient in the case of highly collimated decay products.

The misidentification rate has a nontrivial behavior for most
algorithms. In the case of DeepAK8 and DeepAK8-MD the \ebkg value decreases
with $\pt(\text{generated particle})$, which is mainly a result of the use of low-level
features as inputs to the algorithm. For \ecfv, the \ebkg increases with
$\pt(\text{generated particle})$, whereas for \ecfvddt, it is, by design, significantly more stable.
In the case of \msdv, the decrease of \ebkg as a function of $\pt(\text{generated particle})$
is mainly caused by the strong shift of the \msd shape of the background jets to larger values as a result of the selection on \tautwoone.
This will be discussed in more detail in Section~\ref{sec:mc_mass}. Finally,
for the BEST, the \ebkg decreases up to $\pt(\text{generated particle})\sim 1000\GeV$, and
then increases again. This is a feature of the training of the BEST algorithm,
stemming from an imbalance in the relative fraction of jets between the low- and high-\pt regimes.

{\tolerance=38400
In the case of \PH tagging, the BEST and DeepAK8 algorithms have stable
\esig for $\pt(\text{generated particle})\gtrsim600\GeV$, whereas for
the double-\PQb algorithm the \esig starts to decrease around this \pt
regime. There are two main reasons for this behavior. First, the double-\PQb algorithm exploits
axis-dependent observables, similar to \tautwoone, which are less efficient at high \pt
where the decay products become highly collimated. Second, the selection on the
tracks used to construct the variables used for the training of the double-\PQb algorithm, discussed
in Section~\ref{sec:doubleb}, is suboptimal in the very high-\pt regime. The efficiency
 \ebkg for both double-\PQb and DeepAK8 decreases as a
function of $\pt(\text{generated particle})$, whereas for BEST it shows a
modest increase for $\pt(\text{generated particle}) \gtrsim 1000\GeV$, for the same
reasons as in the \PW and \PZ boson tagging case.\par}

\begin{figure}[!phtb]
\centering
\includegraphics[width=0.45\textwidth]{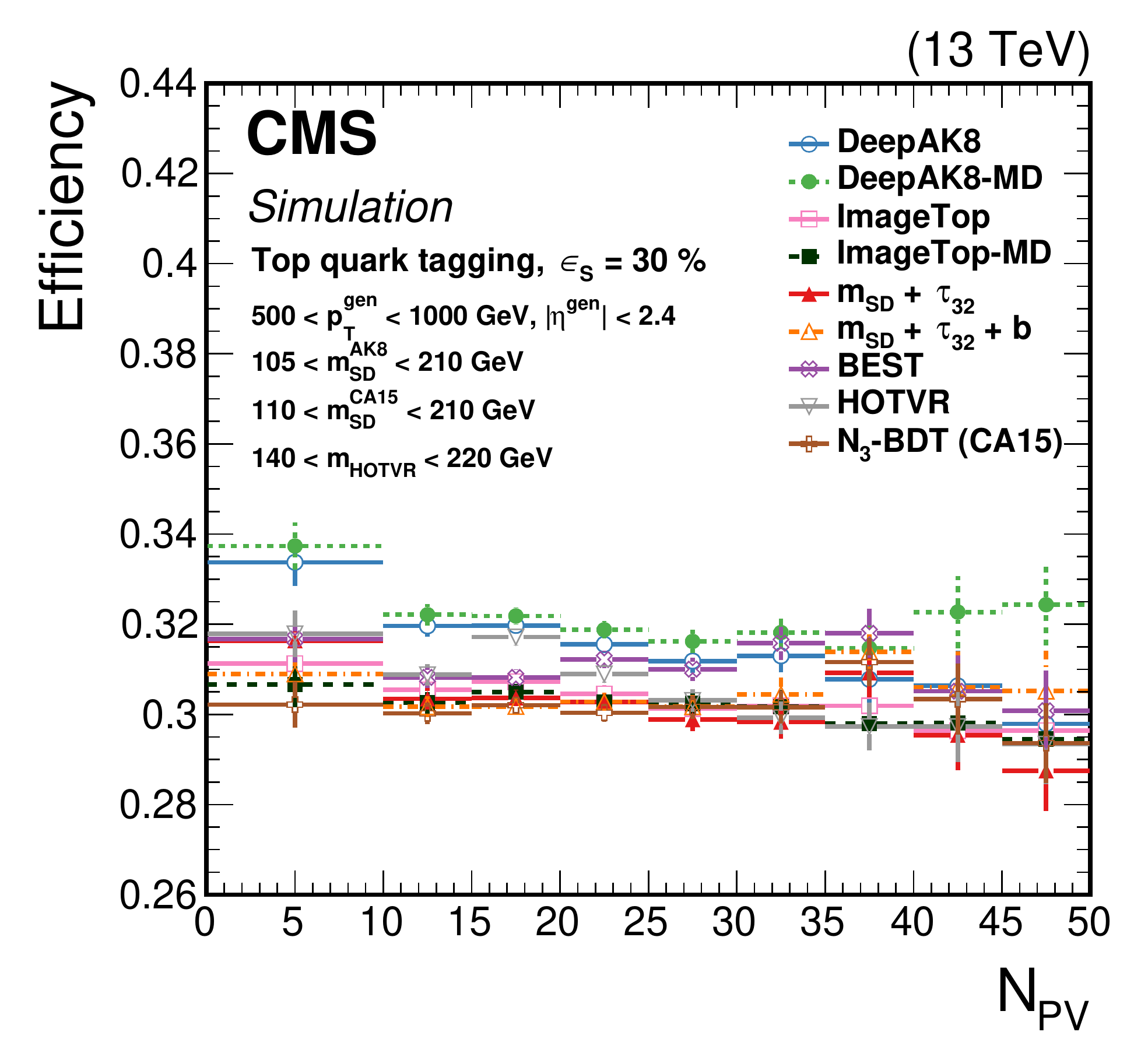}
\includegraphics[width=0.45\textwidth]{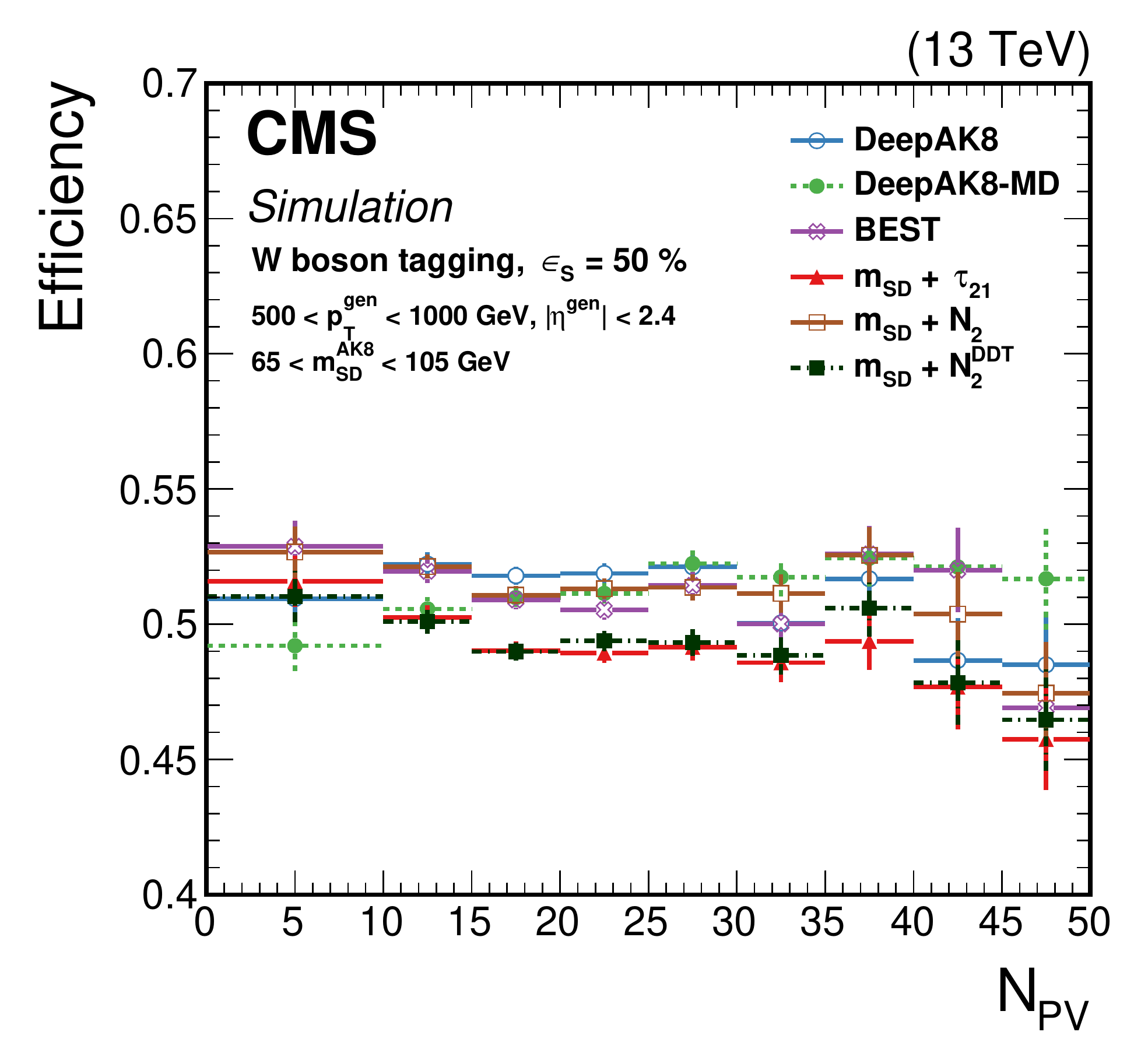}
\includegraphics[width=0.45\textwidth]{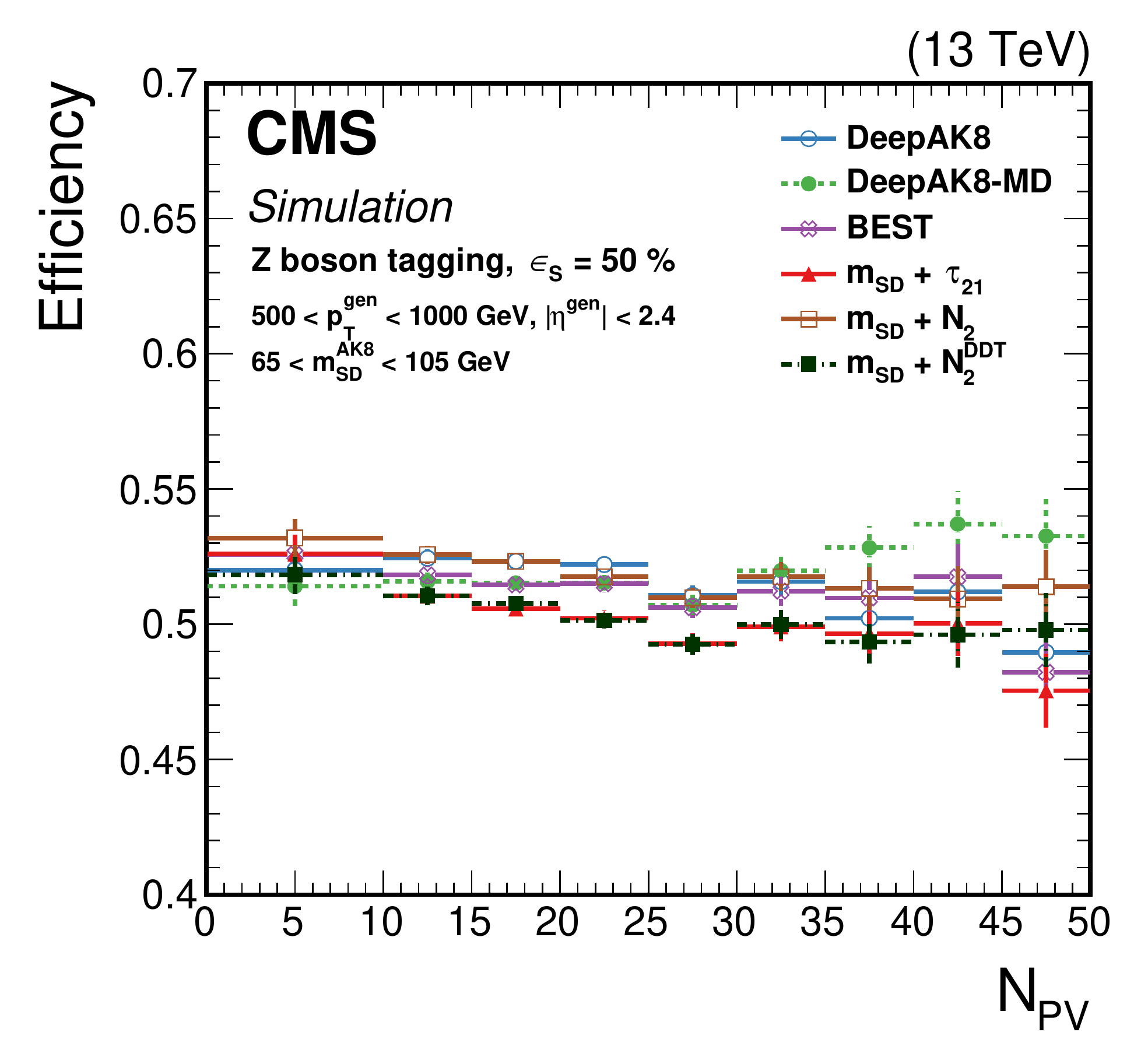}
\includegraphics[width=0.45\textwidth]{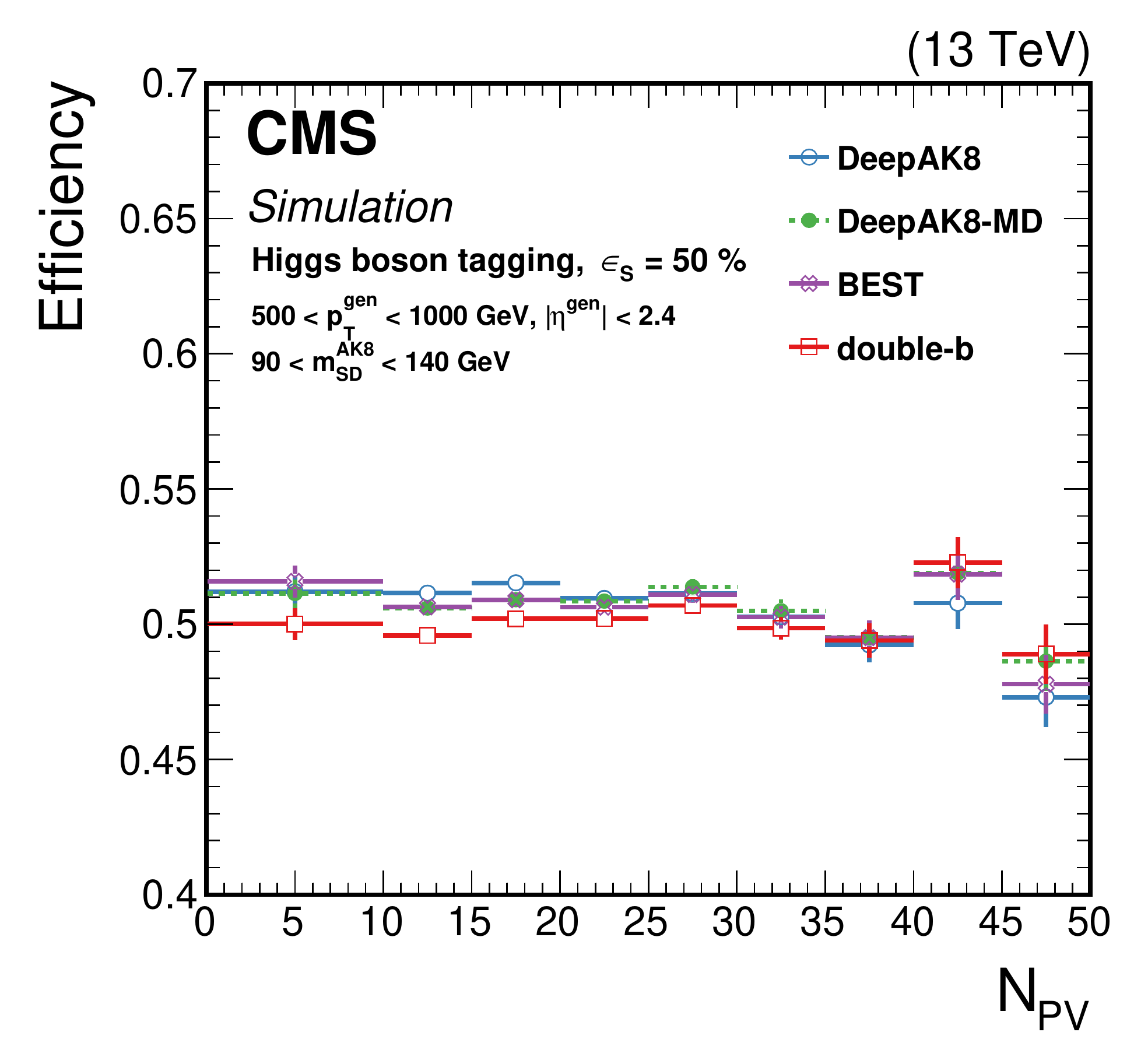}
\caption{\label{fig:esig_vs_npv_lowpt}The efficiency \esig~as a function of
                                          the number of primary vertices
  (\nvtx) for generated particles with $500<\pt<1000\GeV$
  at a working point corresponding to $\esig = 30\,(50)$\%
  for \PQt~quark (\PW{}/\PZ{}/\PH boson) identification.
  Upper left: \PQt~quark, upper right: \PW~boson, lower left: \PZ~boson,
  lower right: \PH boson. The error bars
  represent the statistical uncertainty in each specific bin, due to a
  limited number of simulated events.  Additional fiducial selection criteria applied to
  the jets are listed in the plots.}
\end{figure}

\begin{figure}[!phtb]
\centering
\includegraphics[width=0.45\textwidth]{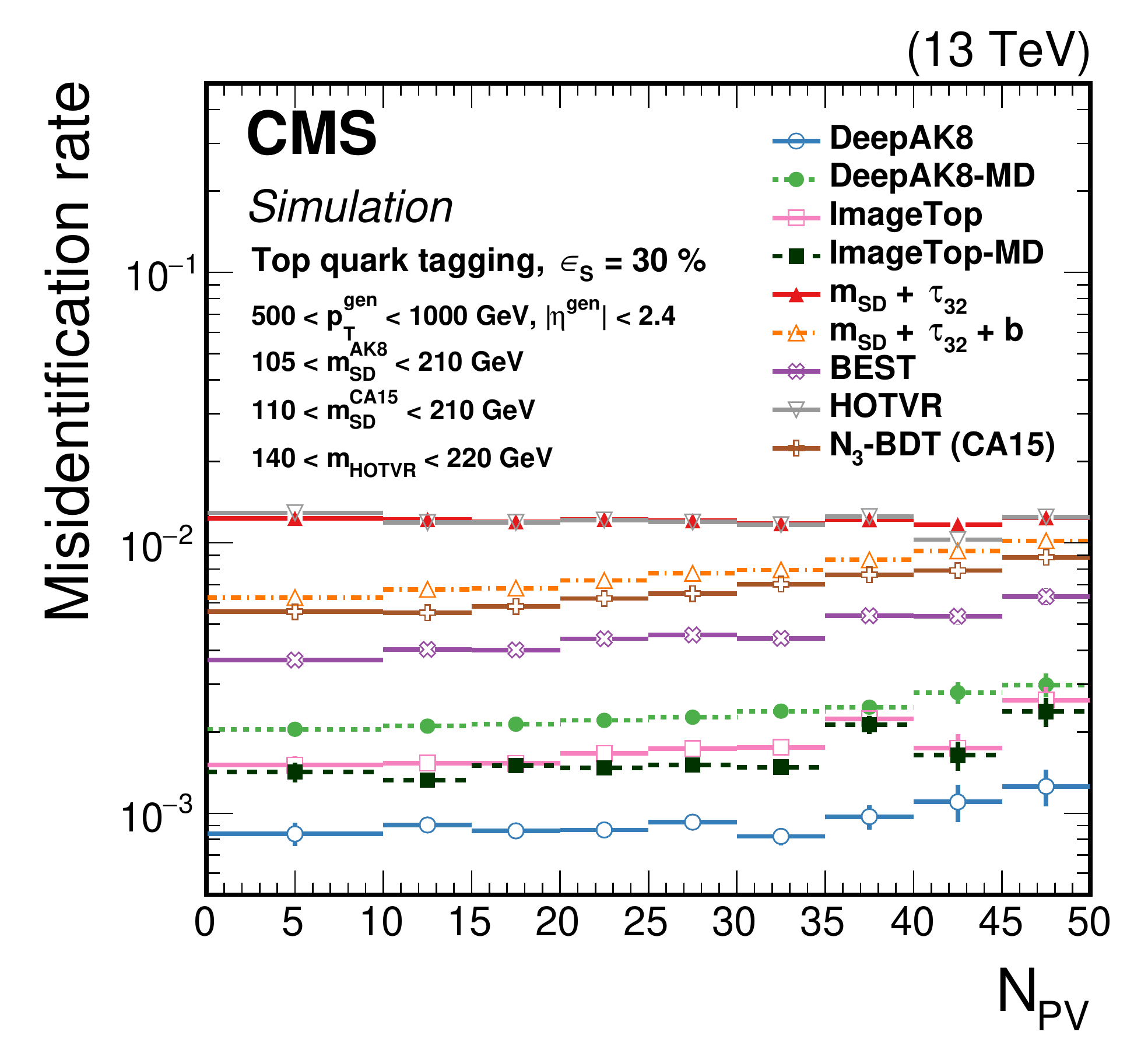}
\includegraphics[width=0.45\textwidth]{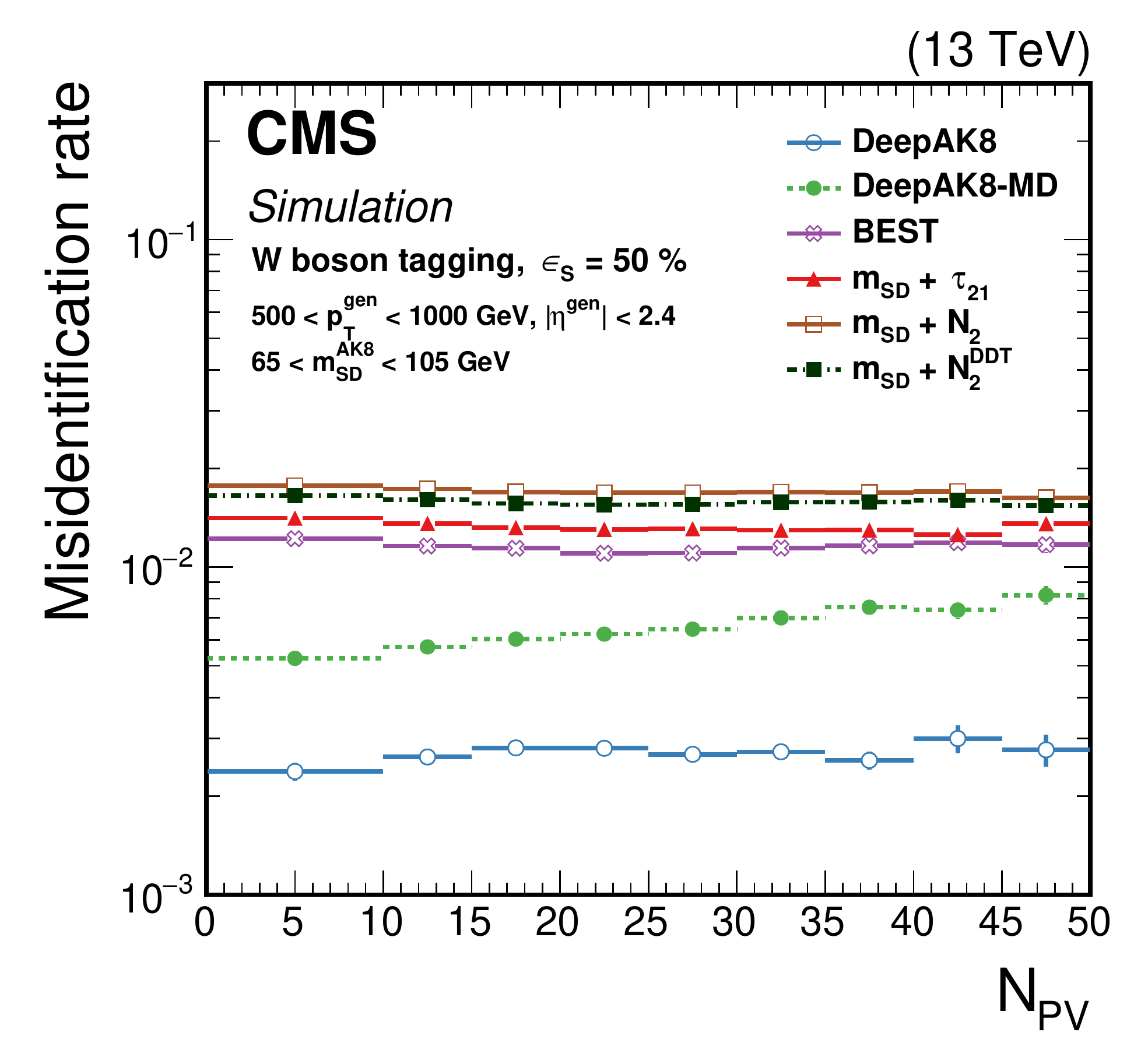}
\includegraphics[width=0.45\textwidth]{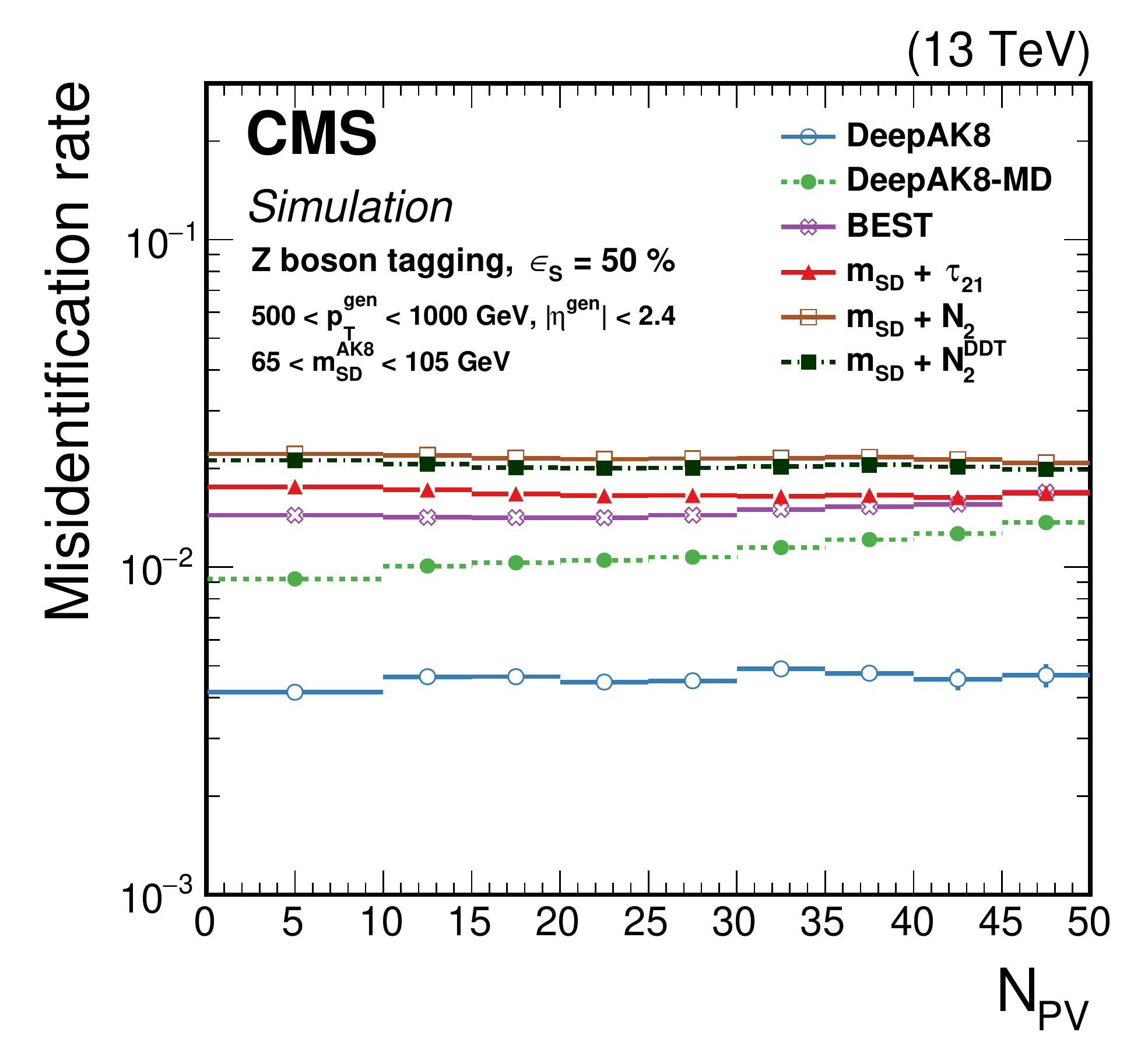}
\includegraphics[width=0.45\textwidth]{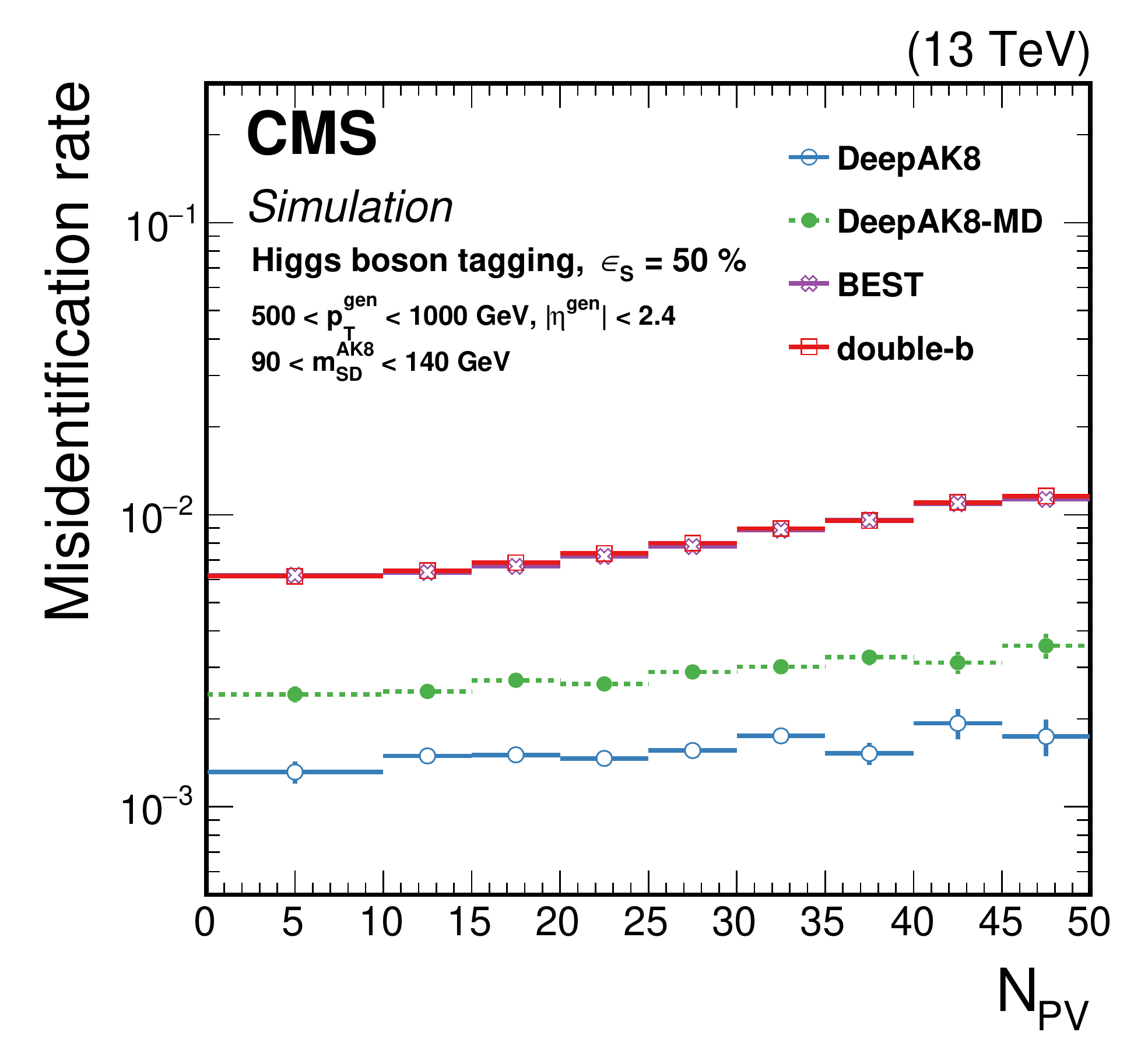}
\caption{\label{fig:ebkg_vs_npv_lowpt}The efficiency \ebkg~as a function of
  the number of primary  vertices (\nvtx)
 for generated particles with $500<\pt<1000\GeV$
  at a working point corresponding to $\esig = 30\,(50)$\%
  for \PQt~quark (\PW{}/\PZ{}/\PH boson) identification.
  Upper left: \PQt~quark, upper right:
  \PW~boson, lower left: \PZ~boson, lower right: \PH
  boson. The error bars represent the
  statistical uncertainty in each specific bin, due to the limited
  number of simulated events.  Additional fiducial selection criteria applied to
  the jets are listed in the plots.}
\end{figure}

The dependence of the algorithms on \nvtx is also examined
using simulated events. Figure~\ref{fig:esig_vs_npv_lowpt}
 shows the distribution of \esig, and Fig.~\ref{fig:ebkg_vs_npv_lowpt}
that of \ebkg, as a function of \nvtx for generated particles with $500<\pt<1000\GeV$,
operating at a working point with $\esig=30$
($50$)\% for \PQt quark (\PW{}/\PZ{}/\PH boson)
identification as defined above. The algorithms make use of jets that employ PUPPI for pileup
mitigation, which results in a roughly constant \esig and
\ebkg for all different pileup scenarios.

\subsection{Correlation with jet mass}
\label{sec:mc_mass}
\begin{figure}[!htbp]
\centering
\includegraphics[width=0.45\textwidth]{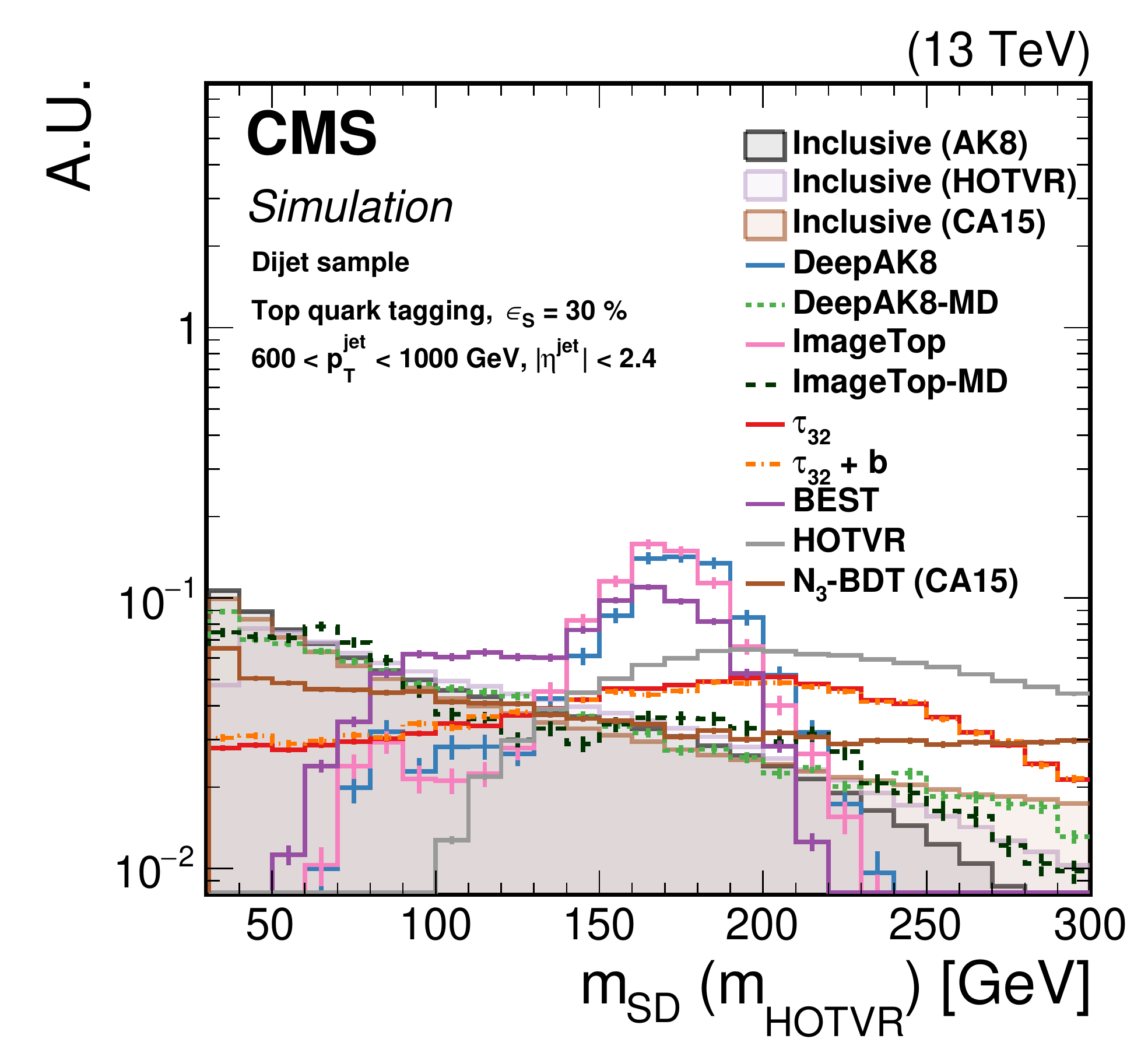}
\includegraphics[width=0.45\textwidth]{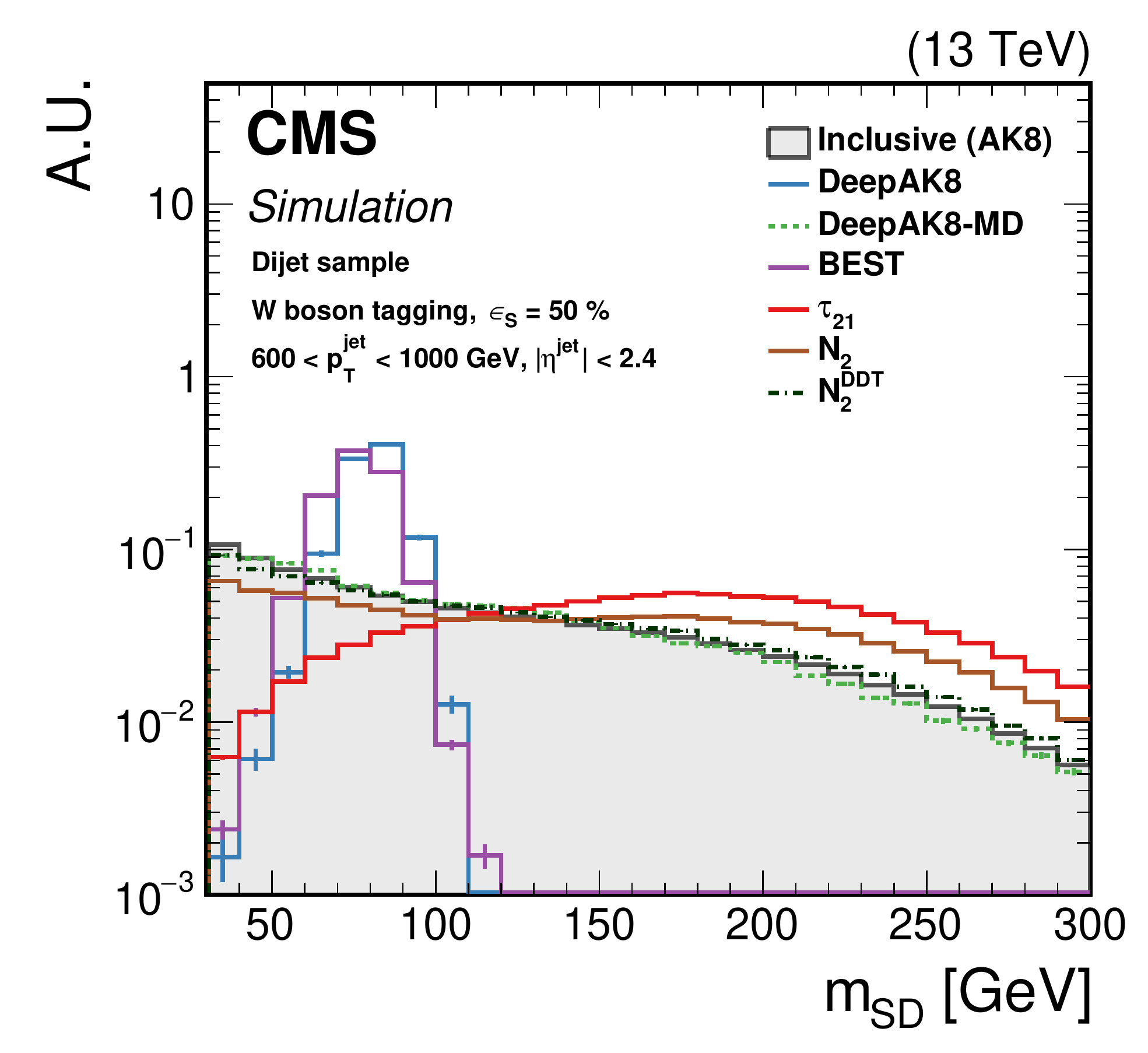}
\includegraphics[width=0.45\textwidth]{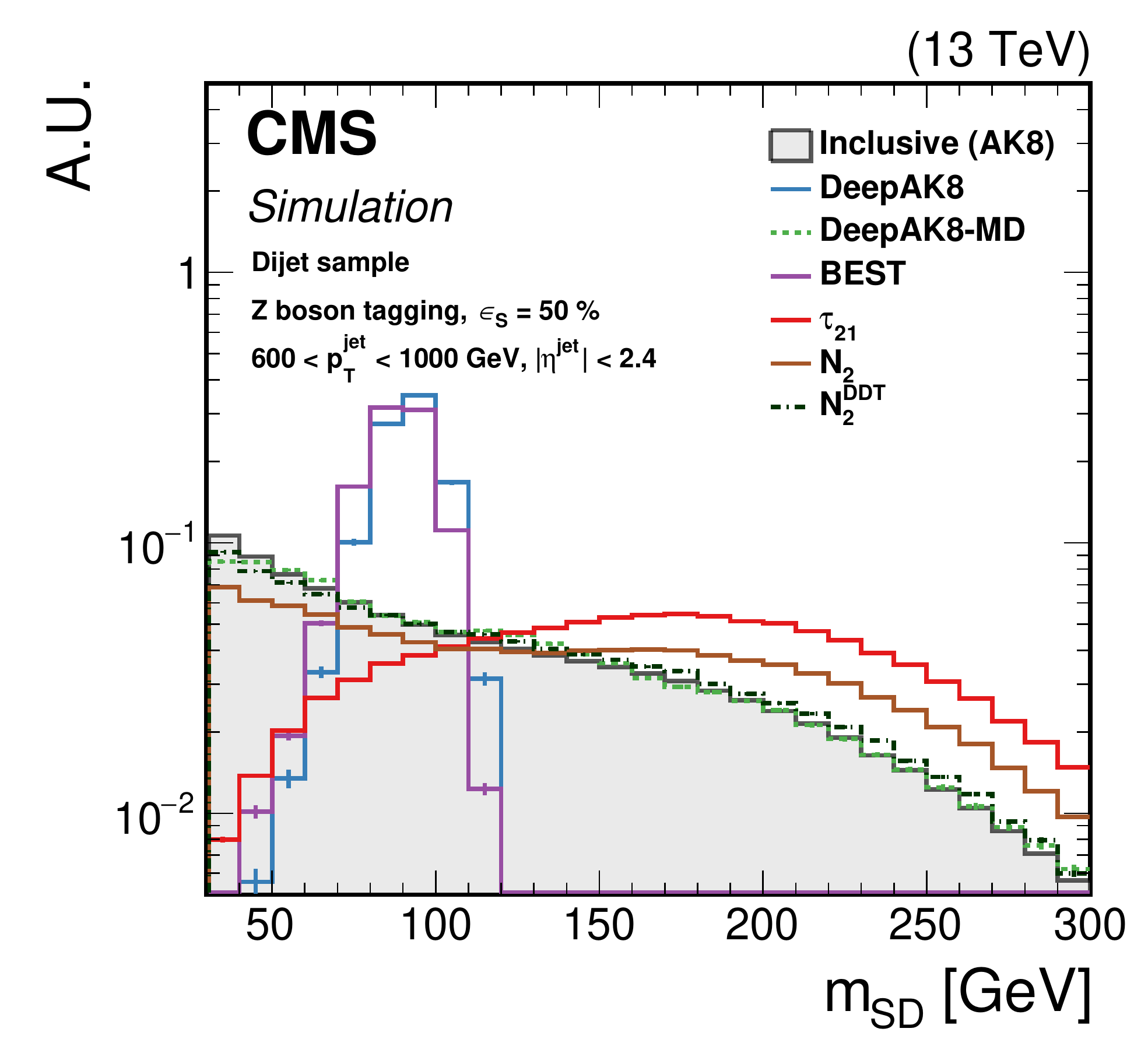}
\includegraphics[width=0.45\textwidth]{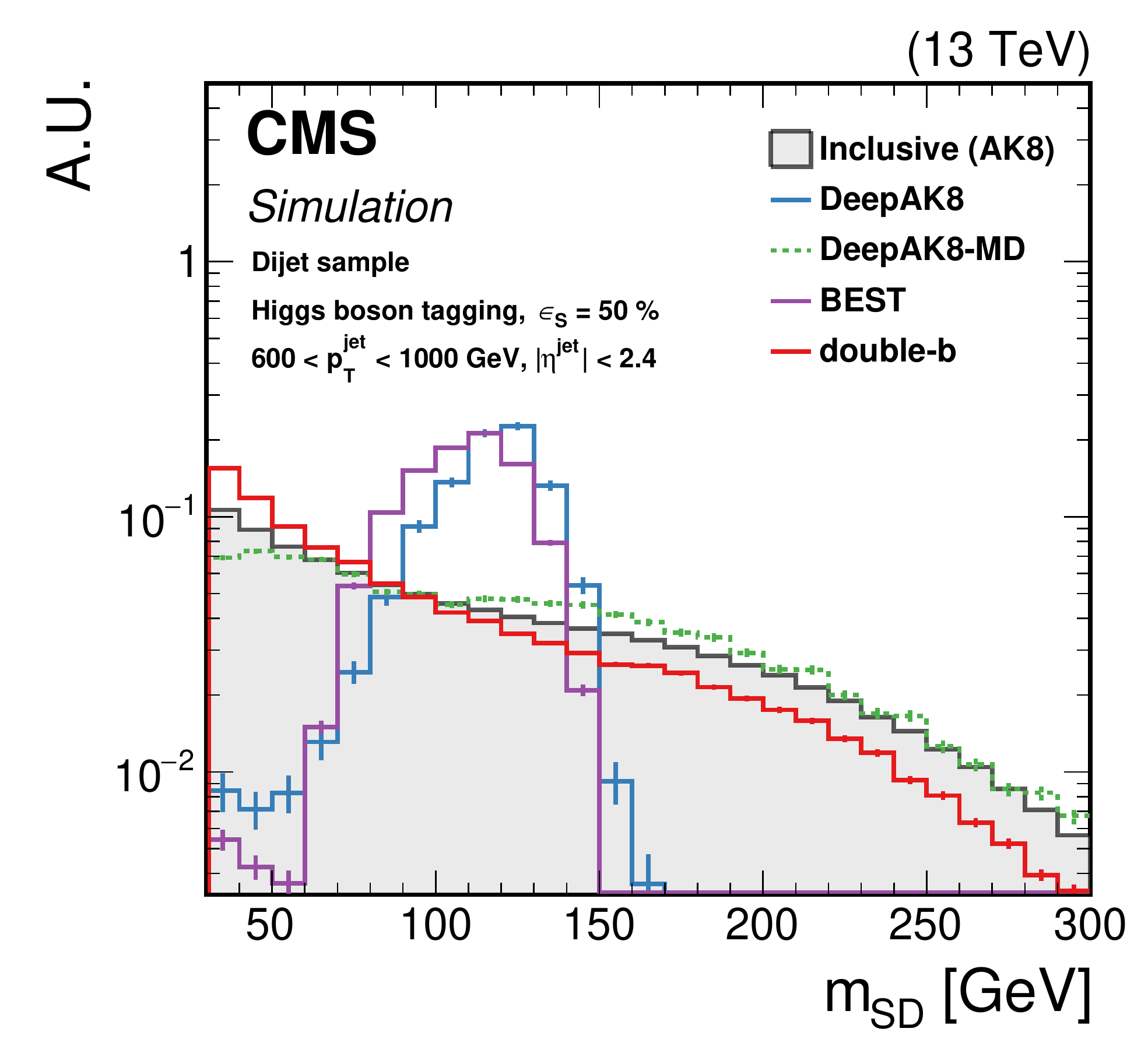}
\caption{\label{fig:mass_shape} The normalized \msd\
  distribution for background QCD jets with $600<\pt<1000\GeV$, inclusively and after selection
  by each algorithm. The working point chosen corresponds to
  $\esig=30$ ($\esig=50$)\% for \PQt~quark (\PW{}/\PZ{}/\PH boson) identification.
  Upper left: \PQt~quark, upper right: \PW~boson, lower left:
  \PZ~boson, lower right: \PH boson. The error
  bars represent the statistical uncertainty in each specific bin, which is related to
  the limited number of simulated events.  Additional fiducial selection criteria applied to
  the jets are listed on the plots.}
\end{figure}

\begin{figure}[!phtb]
\centering
\includegraphics[width=0.45\textwidth]{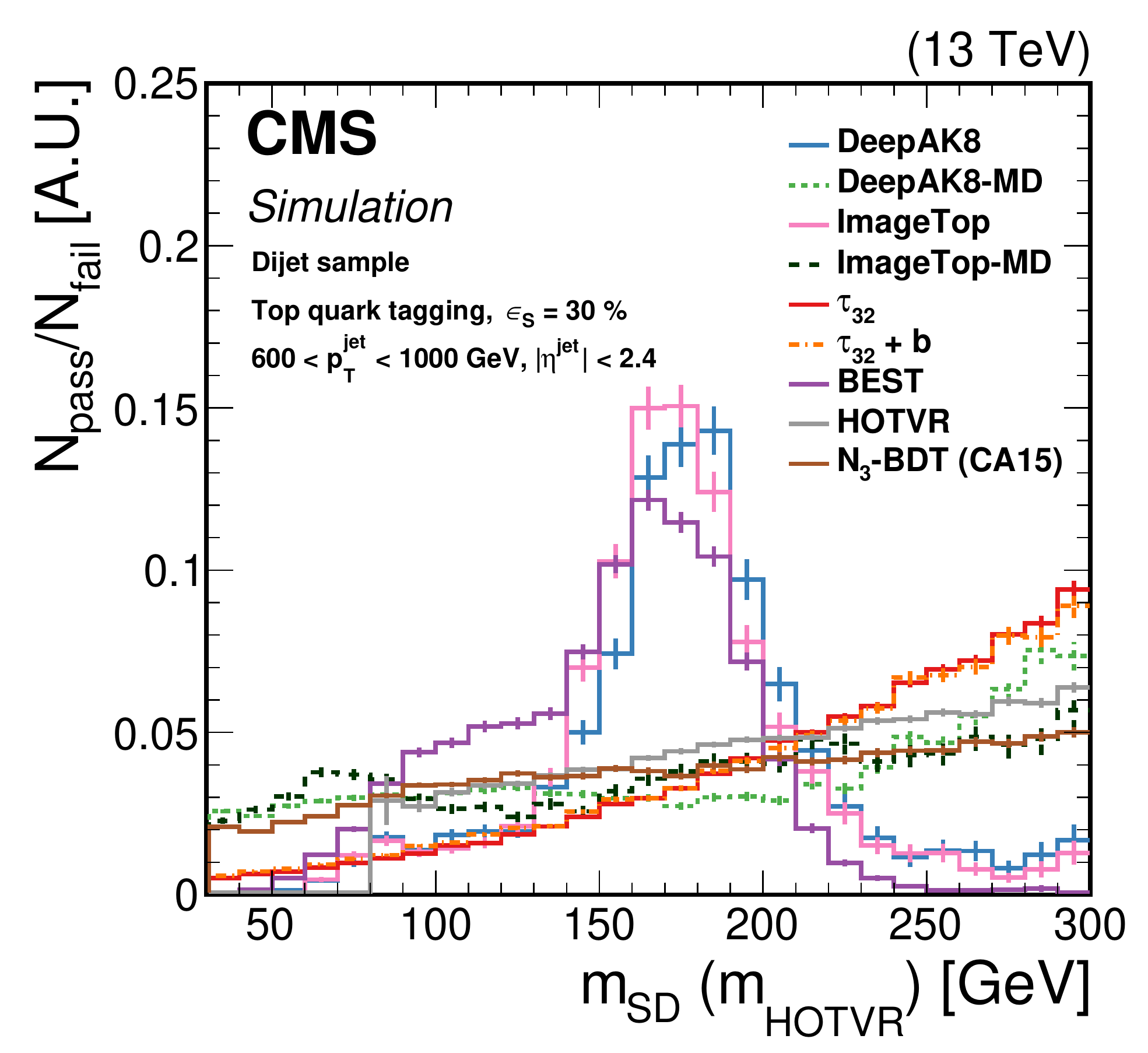}
\includegraphics[width=0.45\textwidth]{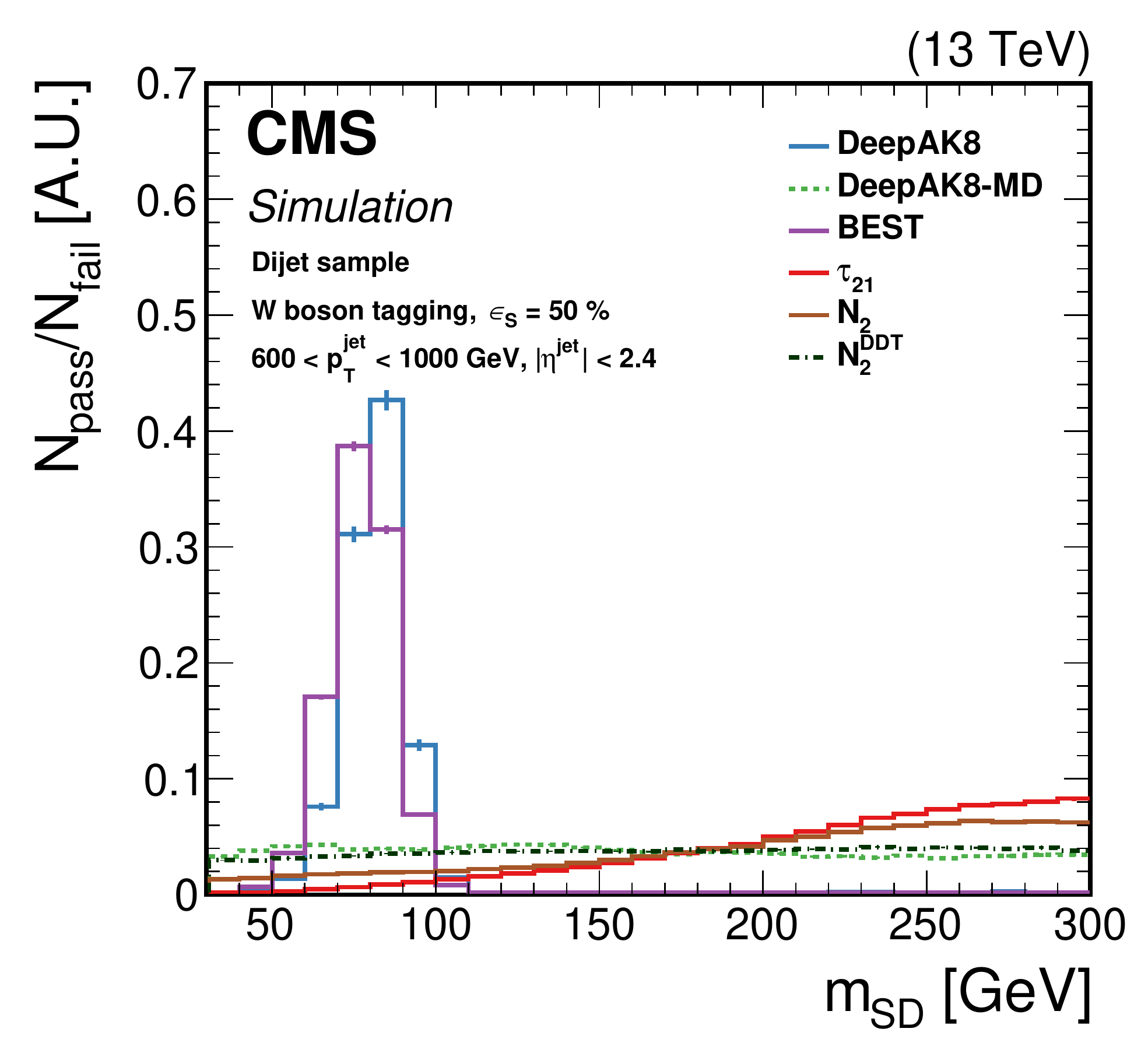}
\includegraphics[width=0.45\textwidth]{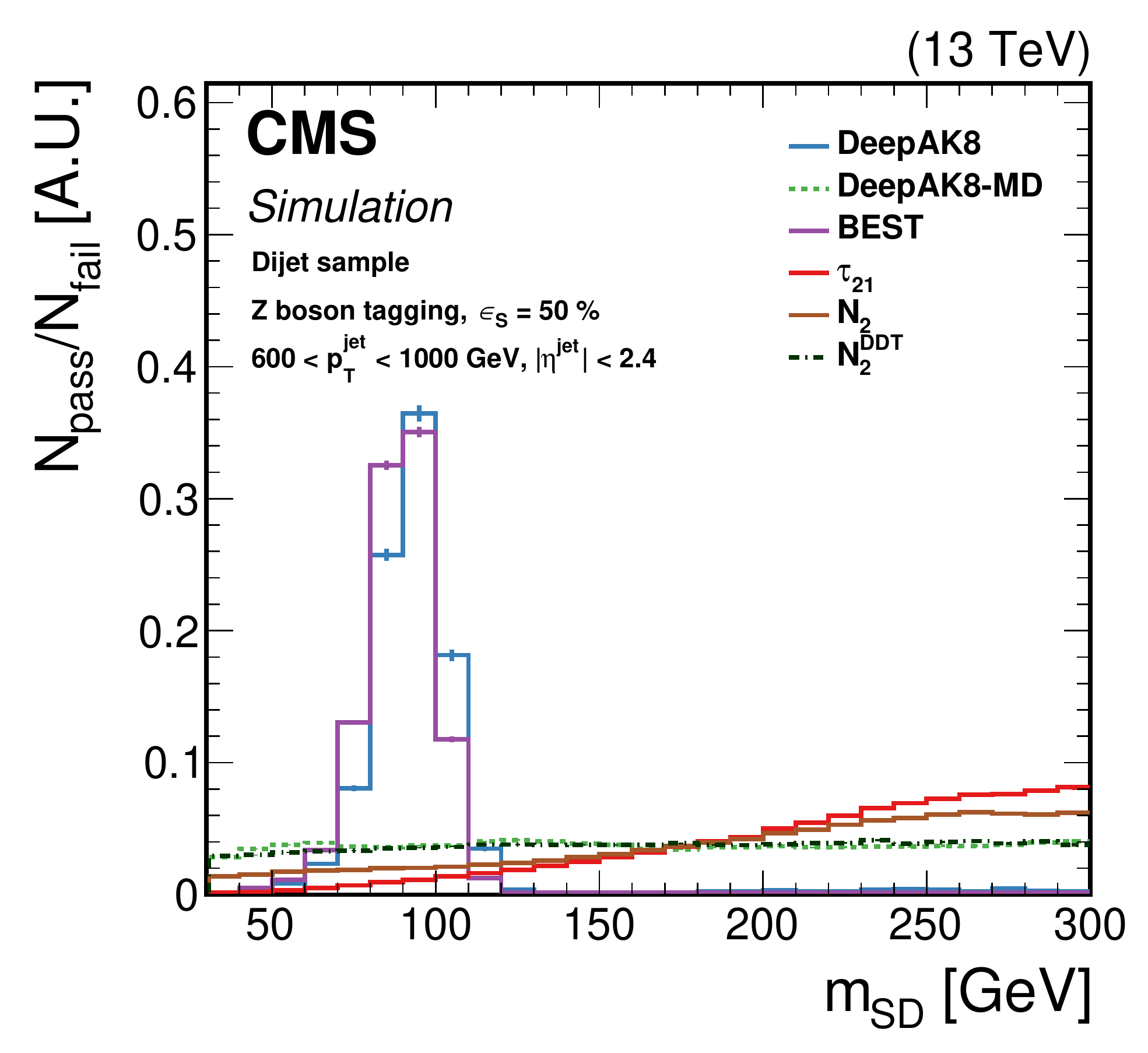}
\includegraphics[width=0.45\textwidth]{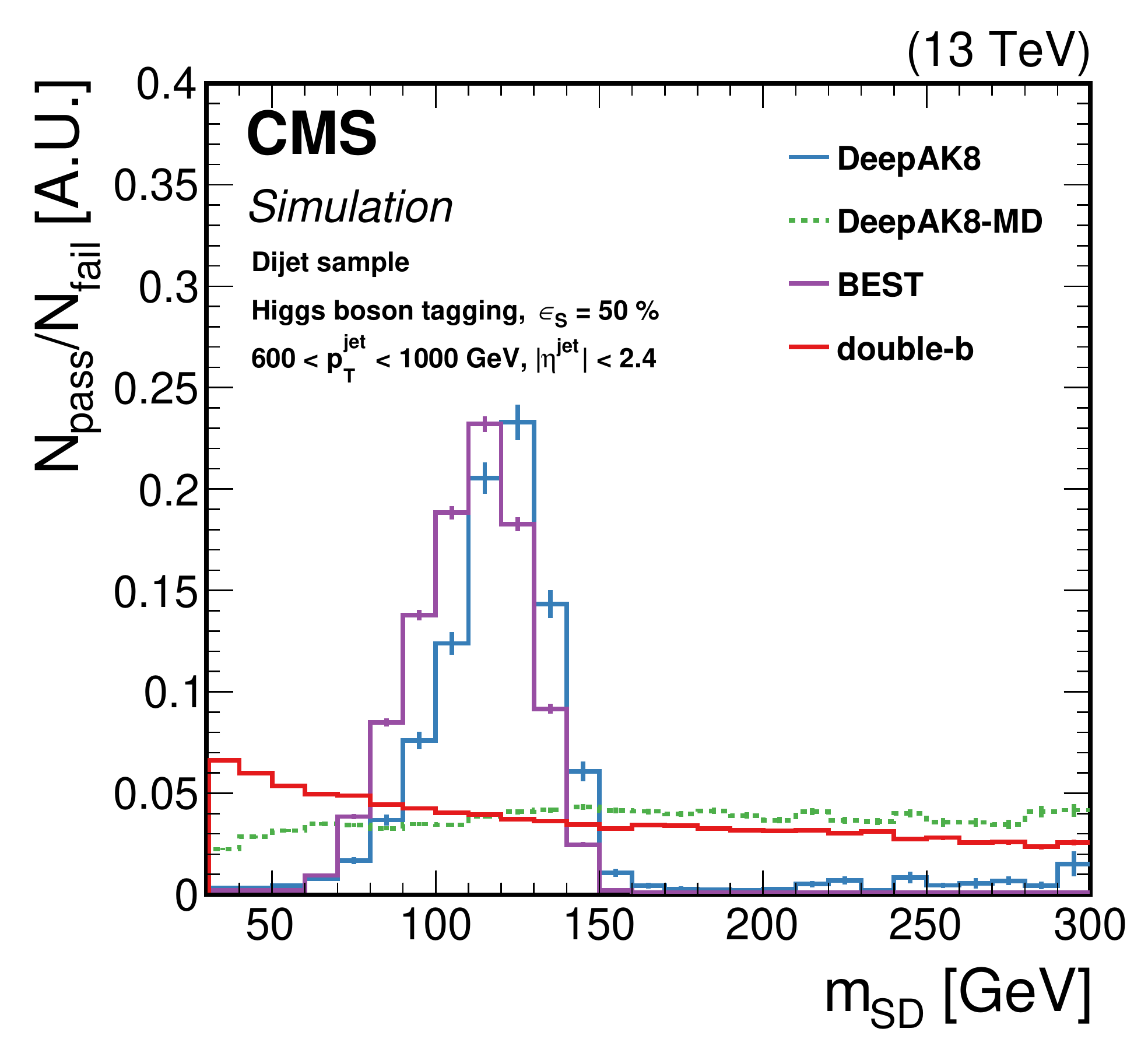}
\caption{\label{fig:mass_passfail_shape} Normalized ratio of the QCD background jet mass distribution 
                                         for the passing and failing jets with $600<\pt<1000\GeV$,    
  by each algorithm. The working point chosen corresponds to 
  $\esig=30$ ($\esig=50$)\% for \PQt~quark (\PW{}/\PZ{}/\PH boson) identification.
  Upper left: \PQt~quark, upper right: \PW~boson, lower left:
  \PZ~boson, lower right: \PH boson. The error
  bars represent the statistical uncertainty in each specific bin, which is related to
 the limited number of simulated events.  Additional fiducial selection criteria applied to
  the jets are listed on the plots.}
\end{figure}

A set of studies was performed to understand the
correlation of the algorithms with the jet mass.
This understanding benefits from
 the theoretical progress made in jet
substructure studies~\cite{Larkoski:2017jix}, which can result in reduced
systematic uncertainties~\cite{Dolen:2016kst}. The jet
mass is one of the most discriminating
variables, and many analyses require a smoothly falling background jet mass
spectrum under a signal peak (e.g., in
Ref.~\cite{Sirunyan:2018ikr}).
Figure~\ref{fig:mass_shape} displays the normalized  \msd\
distribution for jets obtained from the QCD multijet sample,
inclusively and after applying a selection with each algorithm. The
working point chosen corresponds to $\esig=30$ ($50$)\% for
\PQt~quark (\PW{}/\PZ{}/\PH boson). The results are shown for
one region of the generated particle \pt distribution, but similar behavior is seen for
 other \pt~regions as well. By design, the BEST and the nominal
version of the DeepAK8 algorithms lead to significant sculpting of
the background jet mass shape, but this does not affect analyses
unless the jet mass distribution is explicitly used in signal extraction, 
e.g., Ref.~\cite{Sirunyan:2017wif}.
An alternative way of presenting the sculpting of the background jet mass introduced by each tagging
 algorithm is displayed in Fig.~\ref{fig:mass_passfail_shape}. The figure shows 
the normalized ratio of the background jet mass distributions for the passing and failing jets
 for each algorithm, after selecting a working point corresponding to $\esig=30$\,(50)\% for
\PQt~quark (\PW{}/\PZ{}/\PH boson). For the mass decorrelated versions of the algorithms, the ratio 
typically shows very little dependence on  \msd.

To quantify the level of mass sculpting we use the Jensen--Shannon divergence (JSD)~\cite{jsdref},
which is a symmetrized version of the Kullback--Leibler divergence (KLD)~\cite{Kullback51klDivergence},
and provides a metric for the similarity of the shape between distributions. The KLD is defined as:
\begin{equation}
\label{eq:kl}
\text{KLD} (P || Q) = \sum_{i} P(i) \text{log}_{10} \frac{P(i)}{Q(i)},
\end{equation}
where $P(i)$ and $Q(i)$ are the normalized mass distributions of the
background jets that fail and pass a selection with a given algorithm, respectively,
and the symbol $||$ represents the divergence of $P$ from $Q$.
The index $i$ runs over the bins of the distributions.

The JSD metric is defined as:
\begin{equation}
\label{eq:jsd}
\text{JSD} (P || Q) = \frac{1}{2} ( \text{KLD}(P||M)) + \text{KLD}(Q||M) ), ~~\text{where}~~M = \frac{P+Q}{2}.
\end{equation}
Lower values of JSD indicate larger similarity between the mass distributions of
jets passing and failing a selection on a given algorithm.

\begin{figure}[!phtb]
\centering
\includegraphics[width=0.45\textwidth]{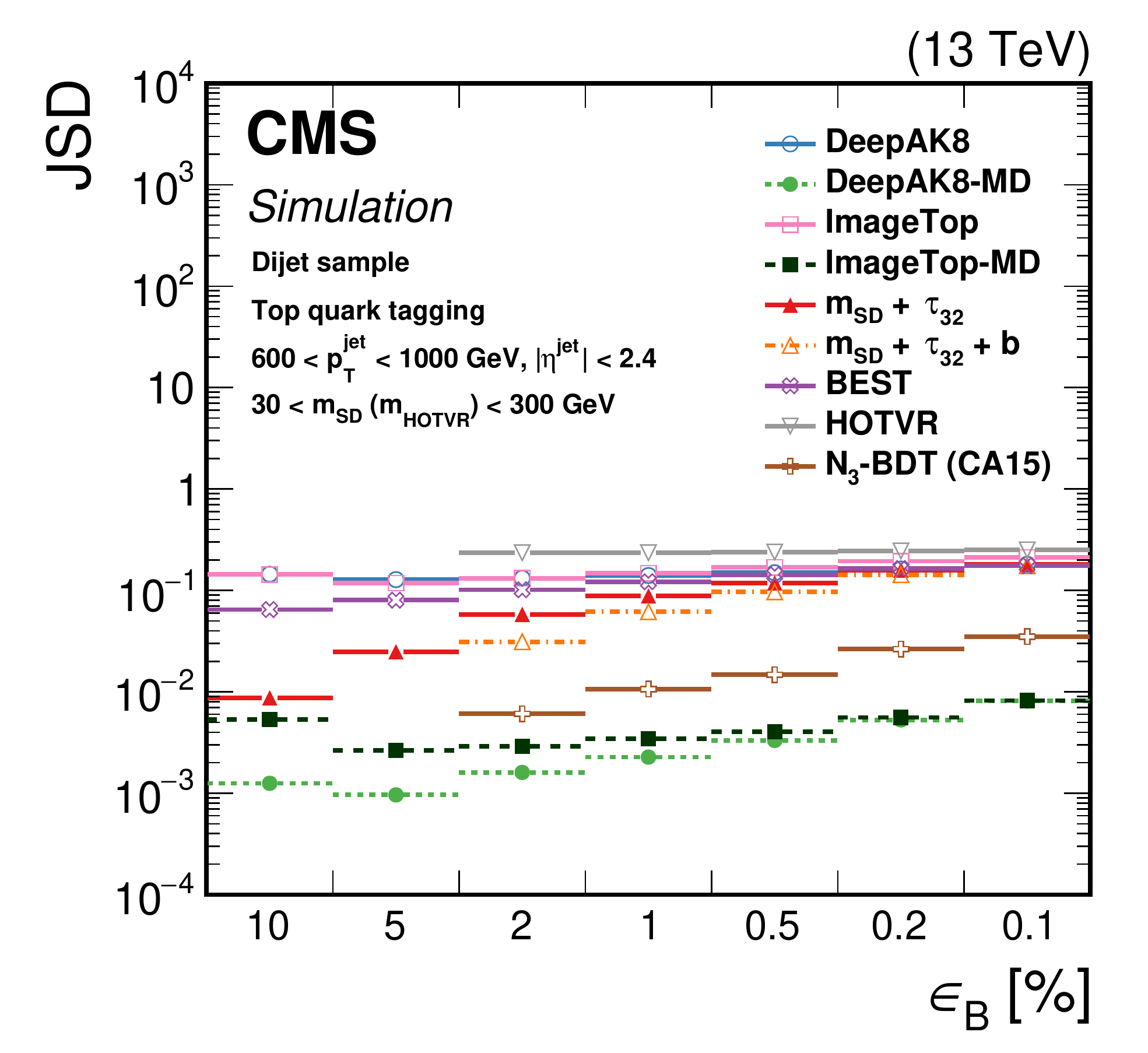}
\includegraphics[width=0.45\textwidth]{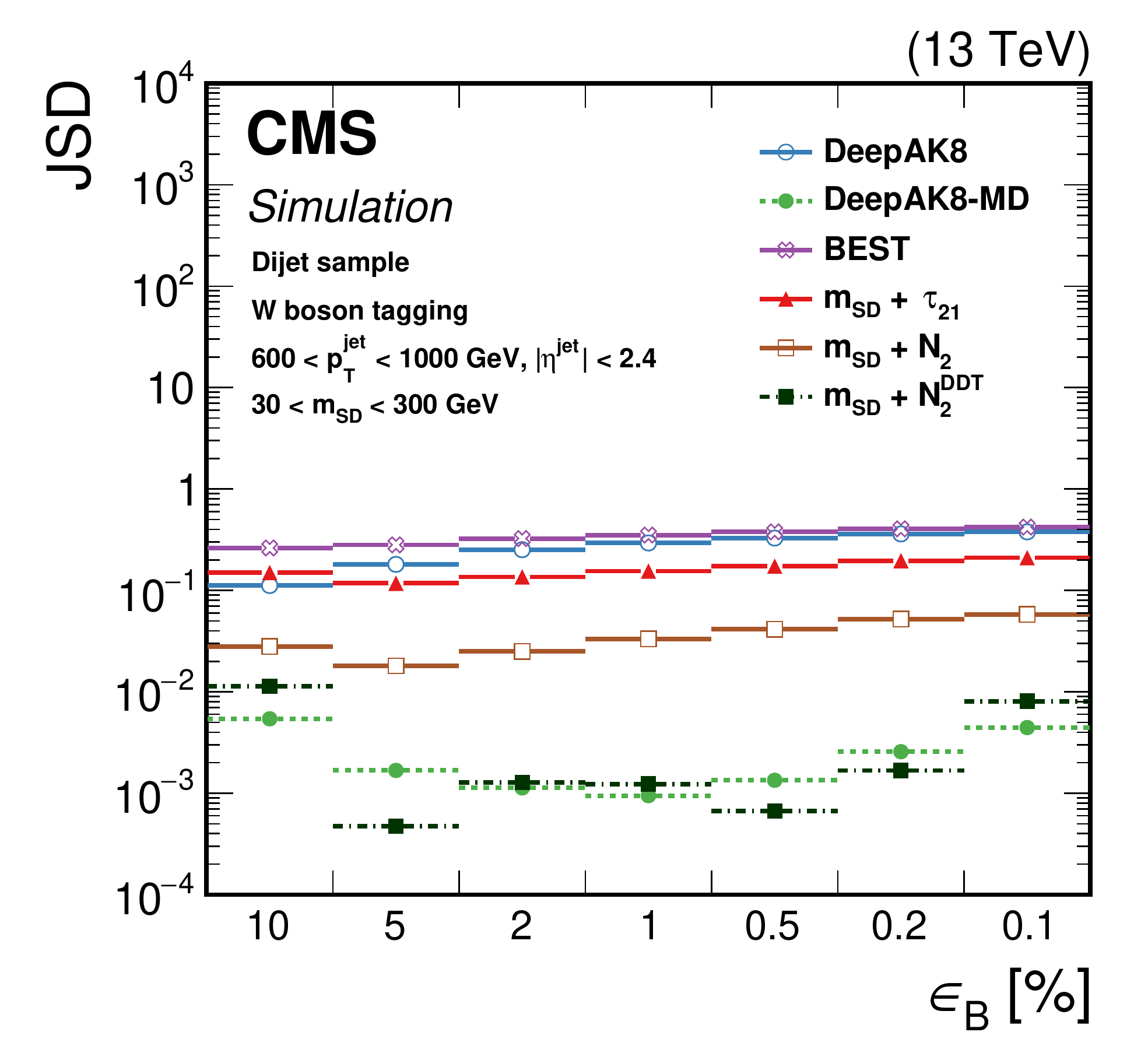}
\caption{\label{fig:KL_vs_score} The JSD as a function of
  successively tighter selections (expressed in terms of \ebkg) for
  the various \PQt (left) and \PW (right) tagging algorithms. Lower
  values of JSD indicate larger similarity of the \sdmass in QCD
  multijet events passing and failing the selection on the tagging
  algorithm.  Additional fiducial selection criteria applied to
  the jets are listed in the plots.}
\end{figure}

In our studies, the jet mass
distributions lay between 30 and 300\GeV with a bin size of
10\GeV. The JSD values for successively tighter selections (expressed in terms of decreasing
\ebkg) for the various \PQt quark and \PW boson tagging algorithms are shown in
Fig.~\ref{fig:KL_vs_score}. The best decorrelation for the \PQt tagging
cases is achieved with the DeepAK8-MD algorithm, which
exploits an adversarial network to reduce the correlation of the
tagging score with the jet mass.
For \PW tagging, \ecfvddt and DeepAK8-MD achieve similar levels of mass decorrelation.
As expected, tighter selection on the
tagging score results in an increase of the mass sculpting. A similar
behavior is observed for all algorithms.

\begin{figure}[phtb!]
\includegraphics[width=0.45\textwidth]{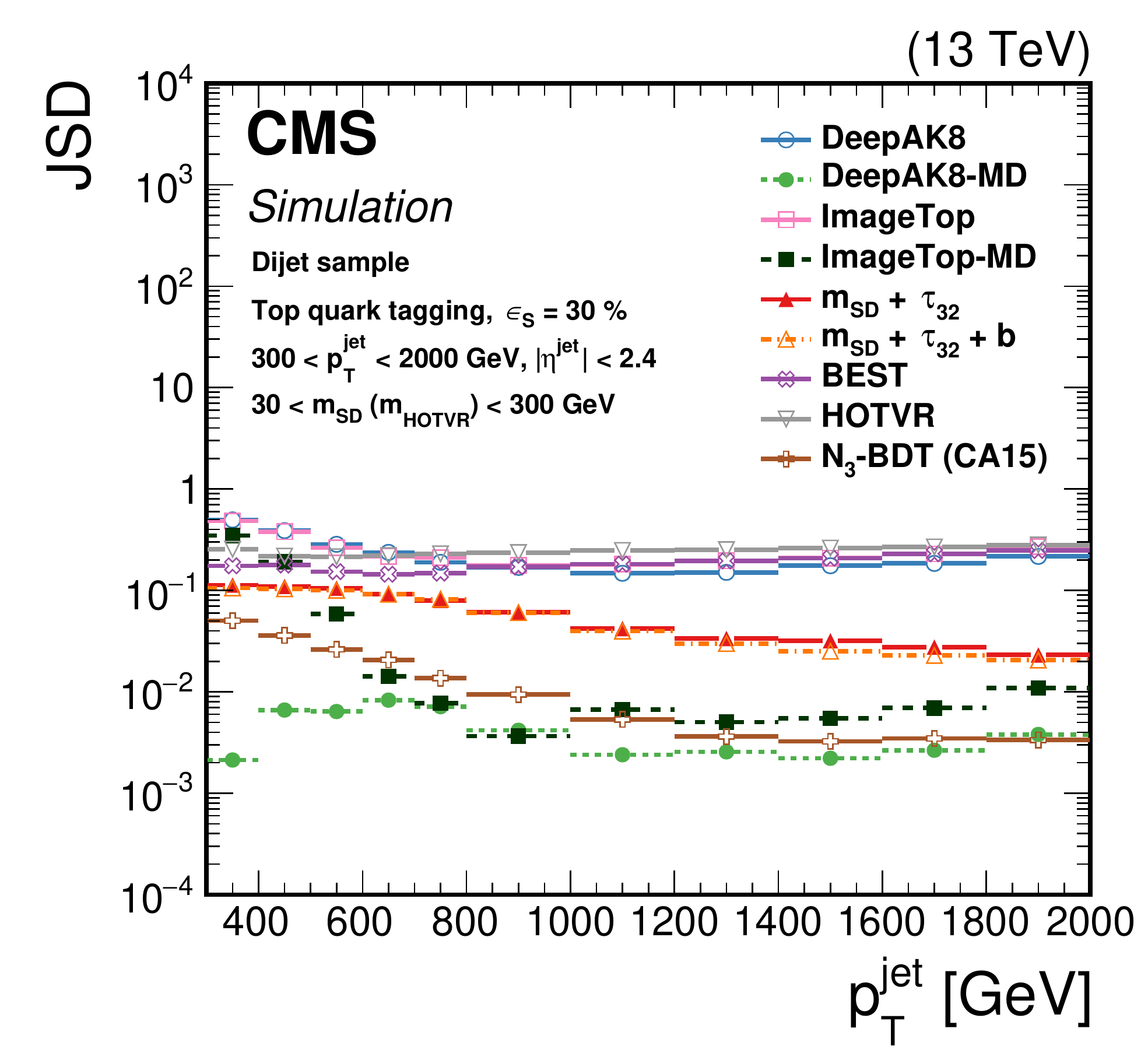}
\includegraphics[width=0.45\textwidth]{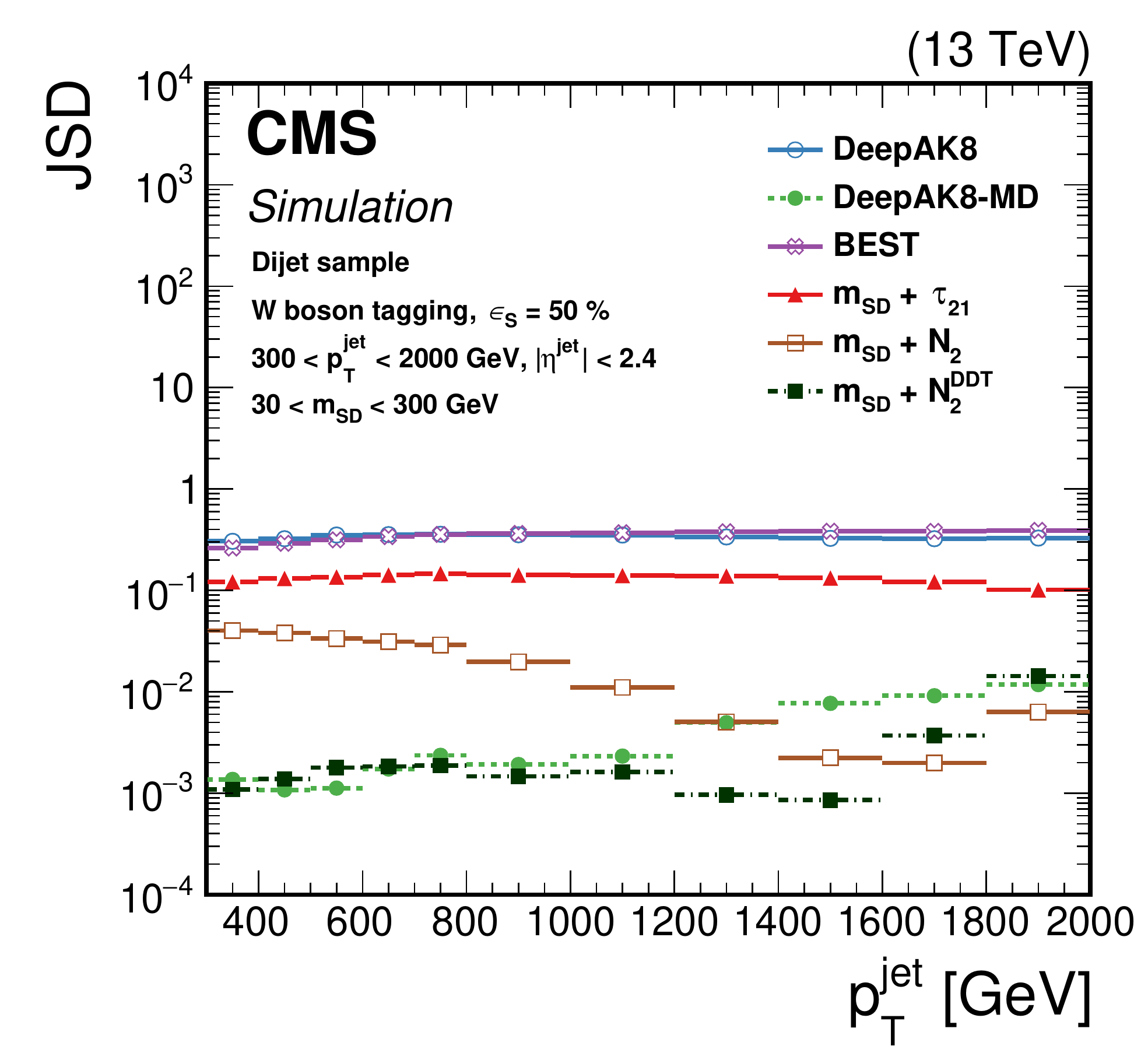}
\caption{\label{fig:KL_vs_jetpt}The JSD, as a function of the jet
  \pt for the various \PQt (left) and \PW (right) tagging
  algorithms. Lower values of JSD indicate larger similarity of the
  \sdmass in QCD multijet events passing and failing the selection on
  the tagging algorithm.  Additional fiducial selection criteria applied to
  the jets are listed in the plots.}
\end{figure}

The robustness of the mass decorrelation techniques is further
studied as a function of jet \pt and \nvtx.
These studies are performed for a working point corresponding
to $\esig=30$  (50)\% for \PQt (\PW) tagging.
Figure~\ref{fig:KL_vs_jetpt} shows the JSD values as a
function of the jet \pt for jets from QCD multijet
events. Most algorithms show modest dependence on jet
\pt, except for ImageTop-MD, where the mass dependence increases rapidly
when $\pt\lesssim600\GeV$ as the training was only performed  for jets with $\pt>600\GeV$.
The DeepAK8-MD and \ecfvddt algorithms for \PW tagging
also show modestly increased mass dependence in the \pt range of 1200
and $1600\GeV$, respectively.
The dependence of the mass mitigation techniques on \nvtx was also
studied and was small.

\section{Performance in data and systematic uncertainties}
\label{sec:performanceindata}

In this section, the validation of the algorithms using
data is presented. The validation is performed in two steps. In the
first step, we focus on  studying the overall modeling of key
variables in simulation and their agreement with data, as well as the
dependence on the simulation details. The second step is to use these results to
extract corrections to the simulation so that 
the algorithms perform similarly in simulation and data.
Differences in the performance between data and
simulation are corrected by scale factors (SF)
extracted by comparing the efficiencies in data and simulation. To account for
effects not captured in the SF, multiple sources of systematic
uncertainties are considered. The data and simulated samples used for
these studies are described in Section~\ref{sec:evtselection}.

In this paper, we focus on the calibration of the \PQt quark
and \PW/\PZ boson tagging algorithms.
The calibration of tagging algorithms
where \PZ and \PH bosons decay to a pair of  bottom or charm quarks
requires alternative methods that go beyond the scope of this paper.
Since it is challenging to obtain a pure \PZ or \PH boson sample,
the calibration of such taggers relies on the use of a proxy jet,
i.e., a jet obtained in the dijet sample
with characteristics similar to signal jets. Data-to-simulation correction factors
are extracted based on these proxy jets, which are then applied to signal jets.
Therefore, the proxy jets should be selected
 to have similar characteristics to the signal jets.
To this end, jets arising from gluon splitting to $\bbbar$ or $\ccbar$ are
used as proxy jets from a sample dominated by QCD multijet events.
Such approaches have been followed in Refs.~\cite{Sirunyan:2017dgc,Sirunyan:2017ezt,CMS:2019tbh}.

\subsection{Systematic uncertainties}
\label{sec:systematics}
A number of sources of systematic effects can affect the
modeling of the performance of the algorithms in data by the simulation.
These include systematic uncertainties in the parton showering model, renormalization and
factorization scales, PDFs, jet energy scale and resolution, \ptmiss
unclustered energy, trigger and lepton identification, pileup modeling, and integrated luminosity,
as well as statistical uncertainties of simulated samples.

Parton shower uncertainties for signal jets are evaluated using samples with the
same event generators but a different choice for the modeling of the
parton showering. For background jets, a sample produced using an alternative generator
for both the hard scattering and the parton shower is used.
 The details of the samples are discussed in Section~\ref{sec:samples}.
Changes in renormalization ($\mu_{R}$) and factorization ($\mu_{F}$) scales are
estimated by varying the scales separately by a factor of two up and down,
relative to the choices of the scale values used
in the sample generation. The uncertainty related to the choice of the
PDFs is obtained from the standard deviation in 100 replicas of the
NNPDF3.0 PDF set~\cite{Ball:2012cx}.
The jet energy scale and resolution are changed within their $\pt$-
and $\eta$-dependent uncertainties, based on the studies presented in
Ref.~\cite{Khachatryan:2016kdb}. Their effects are also propagated to
\ptmiss.
The effect of the uncertainty in the measurement of the unclustered
energy (i.e., contribution of PF candidates not associated to any of
the physics objects) is evaluated based on the momentum resolution of
each PF candidate, which depends on the type of the
candidate~\cite{Sirunyan:2019kia}.
Uncertainties in the measurement of the trigger efficiency and in the
energy/momentum scale and resolution of the leptons are propagated in the SF
extraction.
The uncertainty in the pileup weighting procedure is determined by
varying the minimum bias cross section used to produce the pileup
profile by $\pm$5\% from the measured central value of 69.2 mb
\cite{Sirunyan:2018nqx,Aaboud:2016mmw}.
The limited size of the simulated samples and the size of the data
control samples are also considered.

The uncertainties described above contribute in different
ways to the modeling of jet kinematics and the extraction of 
SF. Because many of the algorithms detailed in this paper use jet substructure
and jet constituent information, either directly or as input to
multivariate techniques, the uncertainties in the choice of parton shower are significant.
Different parton showers directly affect the
number, momentum, and distribution of jet constituents, influencing
the observables used as inputs to the multivariate techniques, and
eventually propagating to the outputs of those algorithms. The magnitude of
this source of systematic uncertainty is from 10--30\%. The
uncertainty in the value of $\mu_{R}$ and $\mu_{F}$
chosen for event generation also has a sizable impact (5--15\%), because this
changes the amount of radiation that can enter into a reconstructed
jet. These dominant components contribute a total combined
uncertainty of 10--50\%, depending on the specific jet
kinematics of interest.

Additional sources of systematic uncertainties, with smaller impact, are also considered.
 For example, the trigger and lepton
identification uncertainties are a few percent,
and do not include uncertainties in the kinematic distributions.  The identification of leptons,
especially muons, is nearly fully efficient, and the trigger
is selected to ensure full efficiency in the regime of
interest.
The jet energy scale and resolution uncertainties are similar, including shape components,
and are between 1 and 5\% for the high-\pt jets
studied here. Uncertainties related to pileup modeling and the integrated luminosity measurement
have an effect smaller than $3\%$.

These uncertainties partially cancel in the SF measurement,
as will be discussed in Section~\ref{sec:sfmeasurement}.

\subsection{The \texorpdfstring{\PQt}{t} quark and \texorpdfstring{\PW}{W} boson identification performance in data}
\label{perf_in_data_1mu}
The single-$\mu$ event selection discussed in
Section~\ref{sec:evtselection_tt1l} provides a sample dominated by
semileptonic \ttbar events. One of the \PQt~quarks decays to a \PW\~boson 
that decays leptonically (to pass the selection), and the other provides
a hadronic decay to be used in validating the algorithms.

To study possible dependence of the tagging efficiency
on the parton showering scheme, we consider two alternative simulated \ttbar\
samples. As discussed in Section~\ref{sec:samples}, both samples are generated
with the same generator (i.e., \POWHEG), but one uses \PYTHIA for the modeling
of the parton showering, whereas the other uses \HERWIGpp. The total SM expectation from
simulation using the latter \ttbar sample will be referred to
as ``SM (Herwig)''. As we will see, the choice of the
parton showering generator has only a small impact on the overall agreement between
data and simulation in signal jets.

To account for the differences in the design of the algorithms, the large-$R$
jets discussed in Section~\ref{sec:evtselection_tt1l} are either AK8,
CA15, or HOTVR jets. For brevity we focus mainly on
results using AK8 jets, unless otherwise stated, but similar conclusions
can be drawn from all three jet collections.

\begin{figure}[hp!]
\centering
\includegraphics[width=0.38\textwidth]{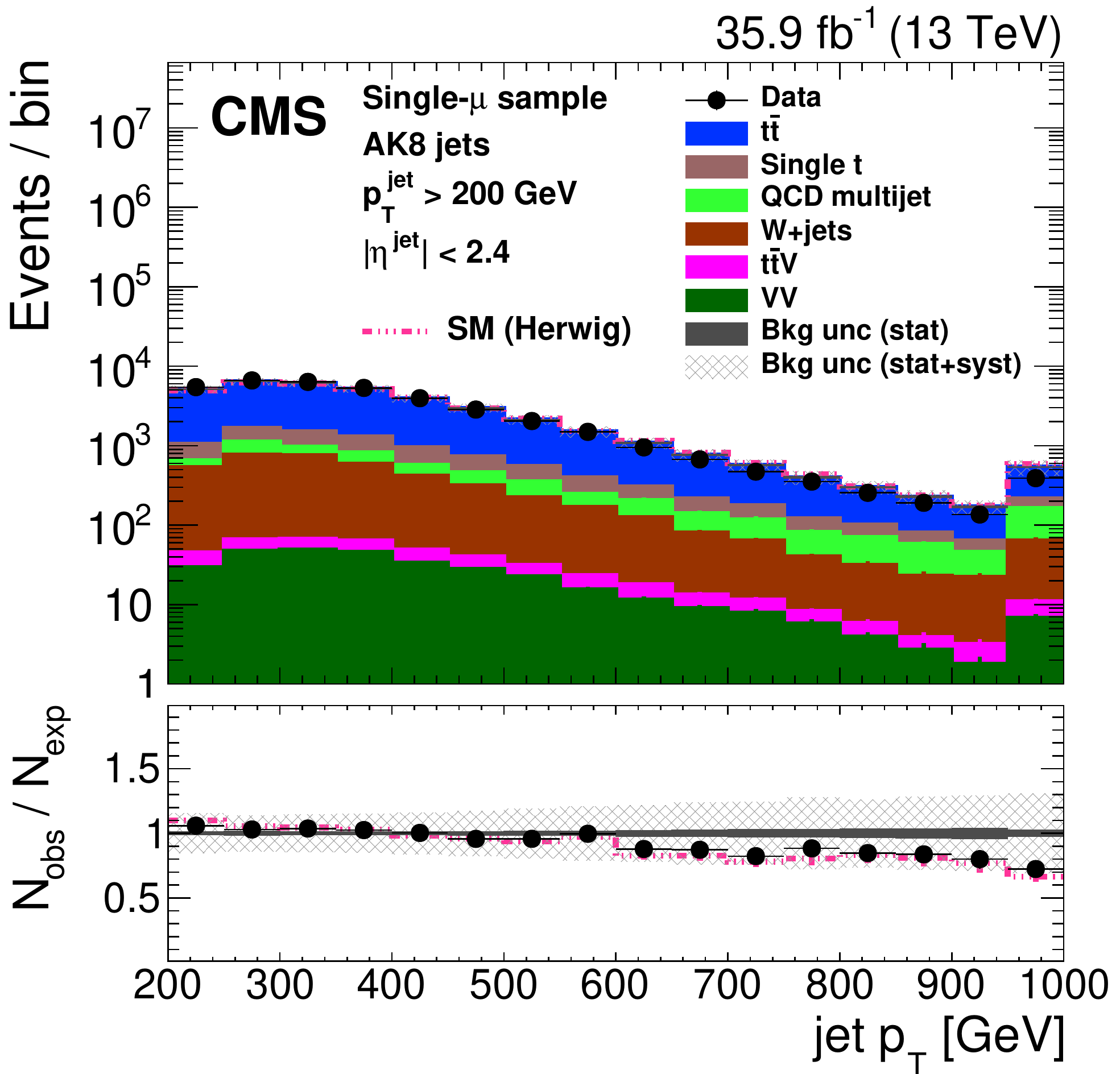}
\includegraphics[width=0.38\textwidth]{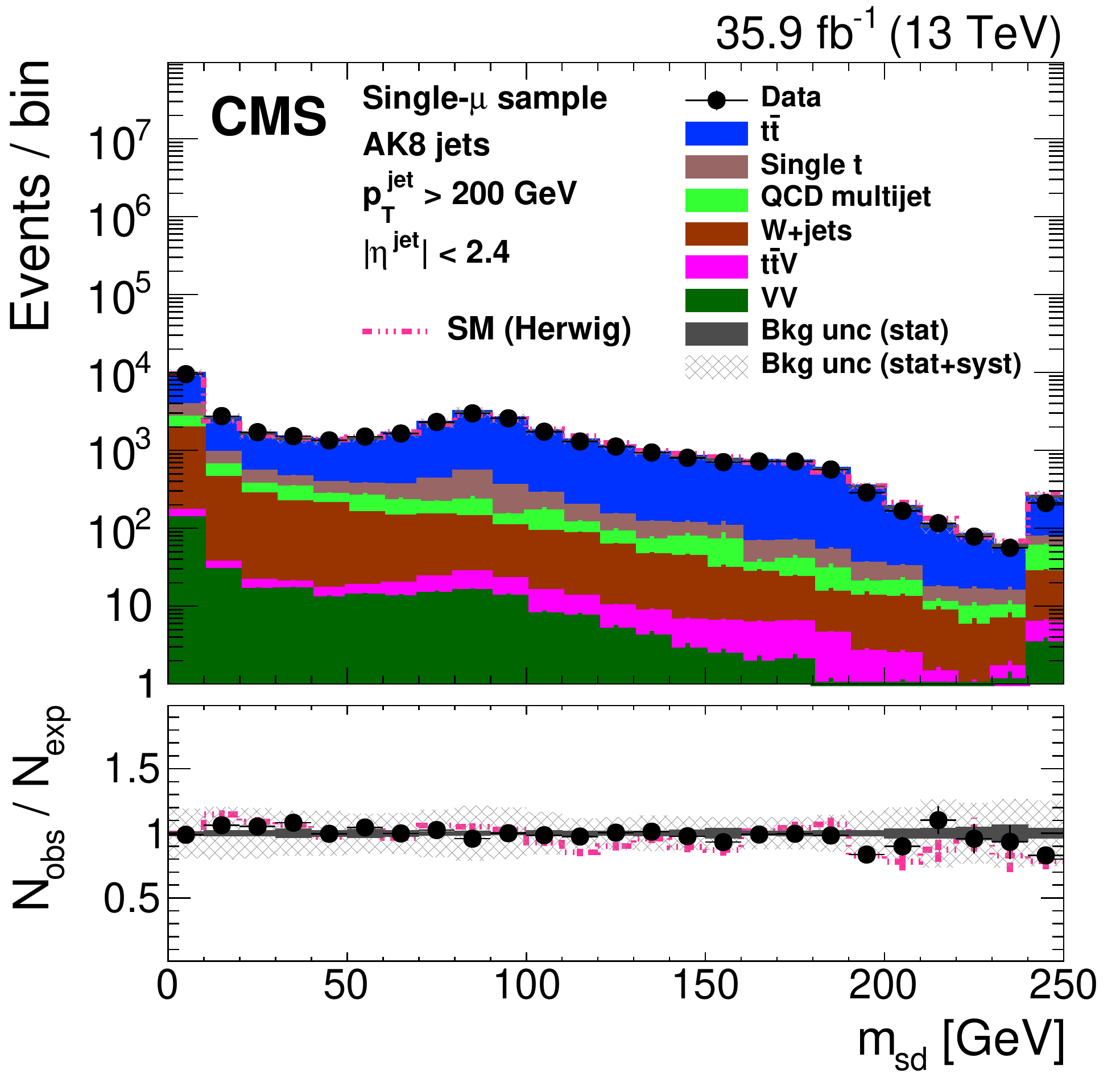}
\includegraphics[width=0.38\textwidth]{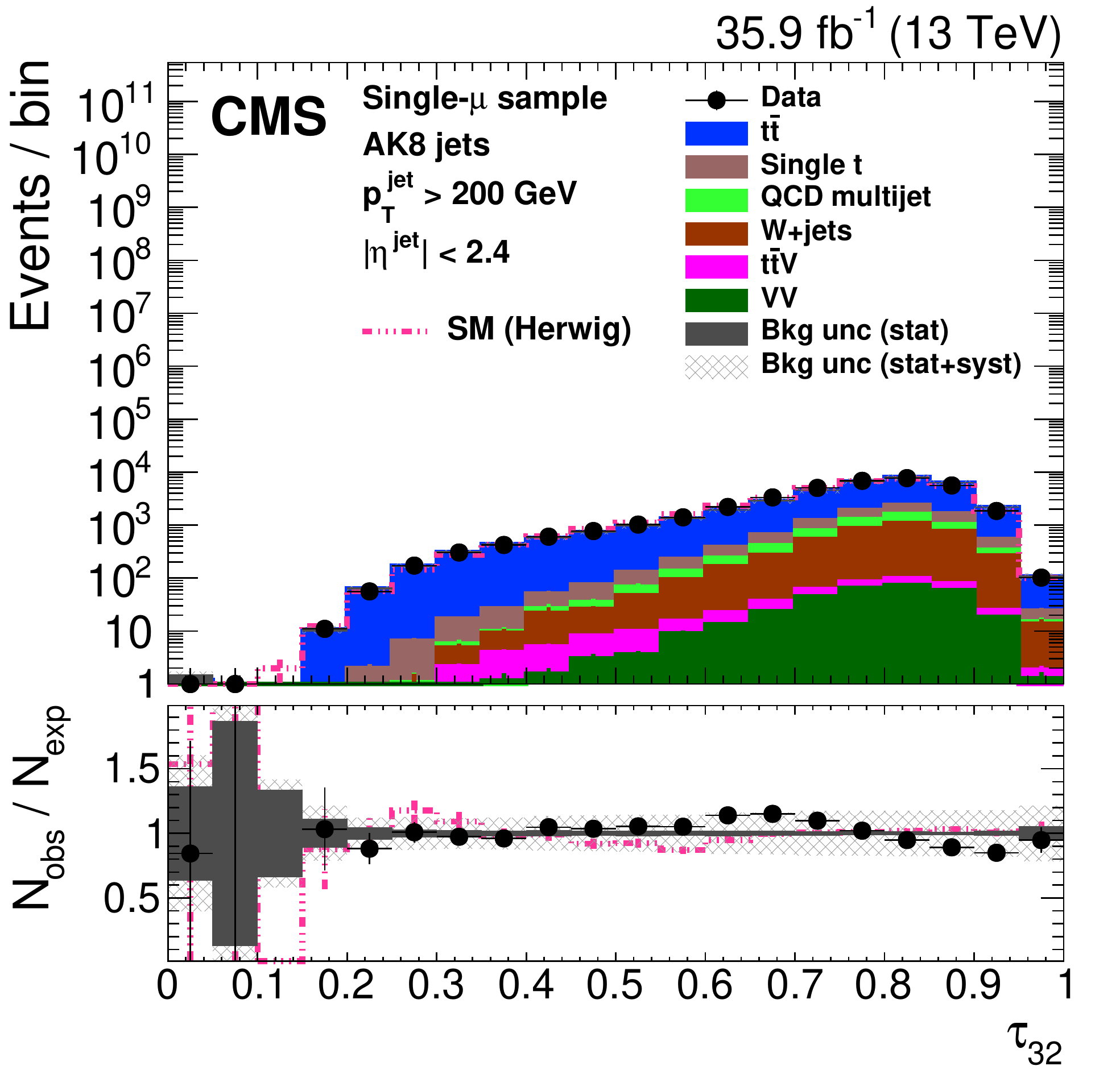}
\includegraphics[width=0.38\textwidth]{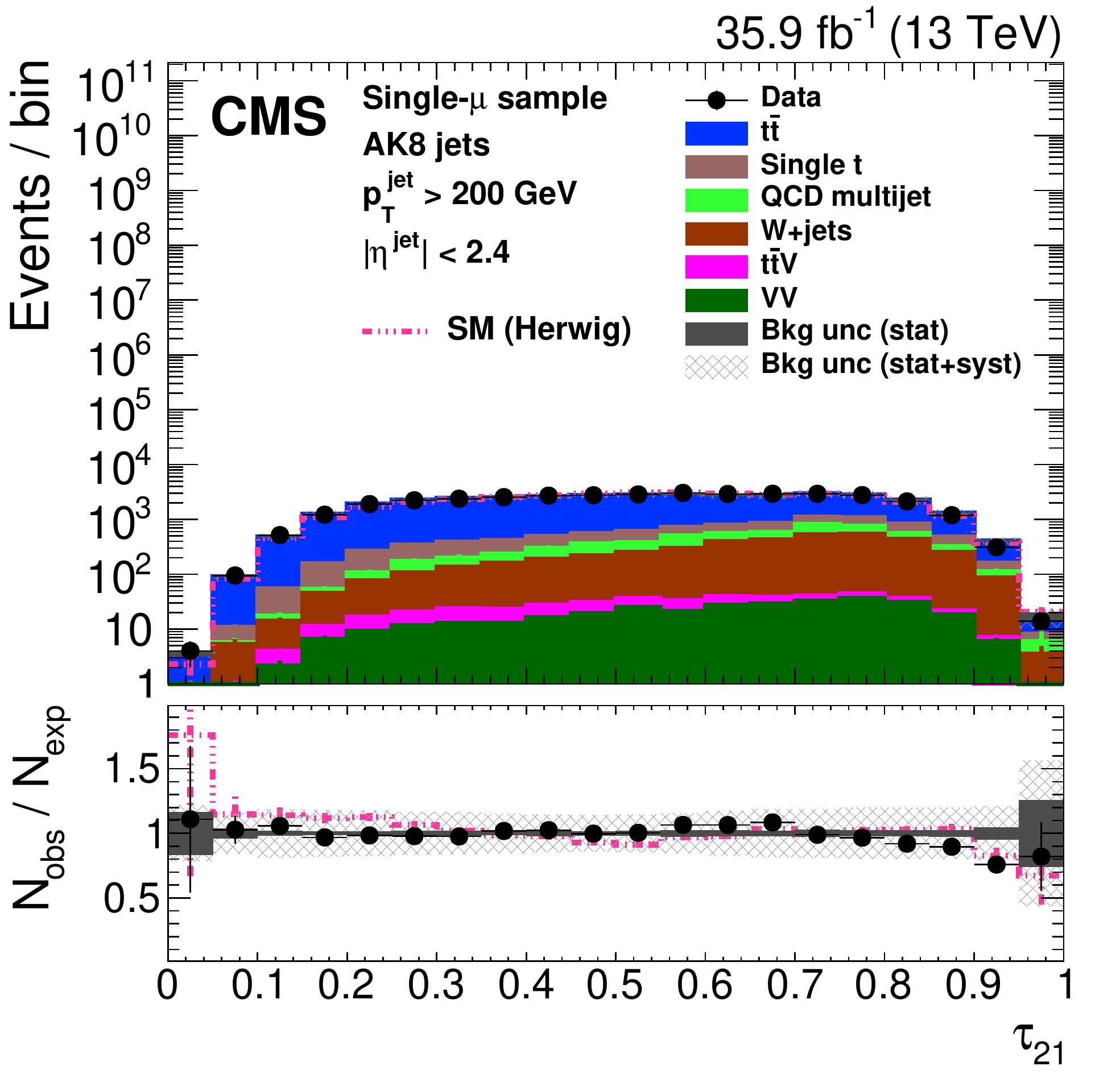}
\includegraphics[width=0.38\textwidth]{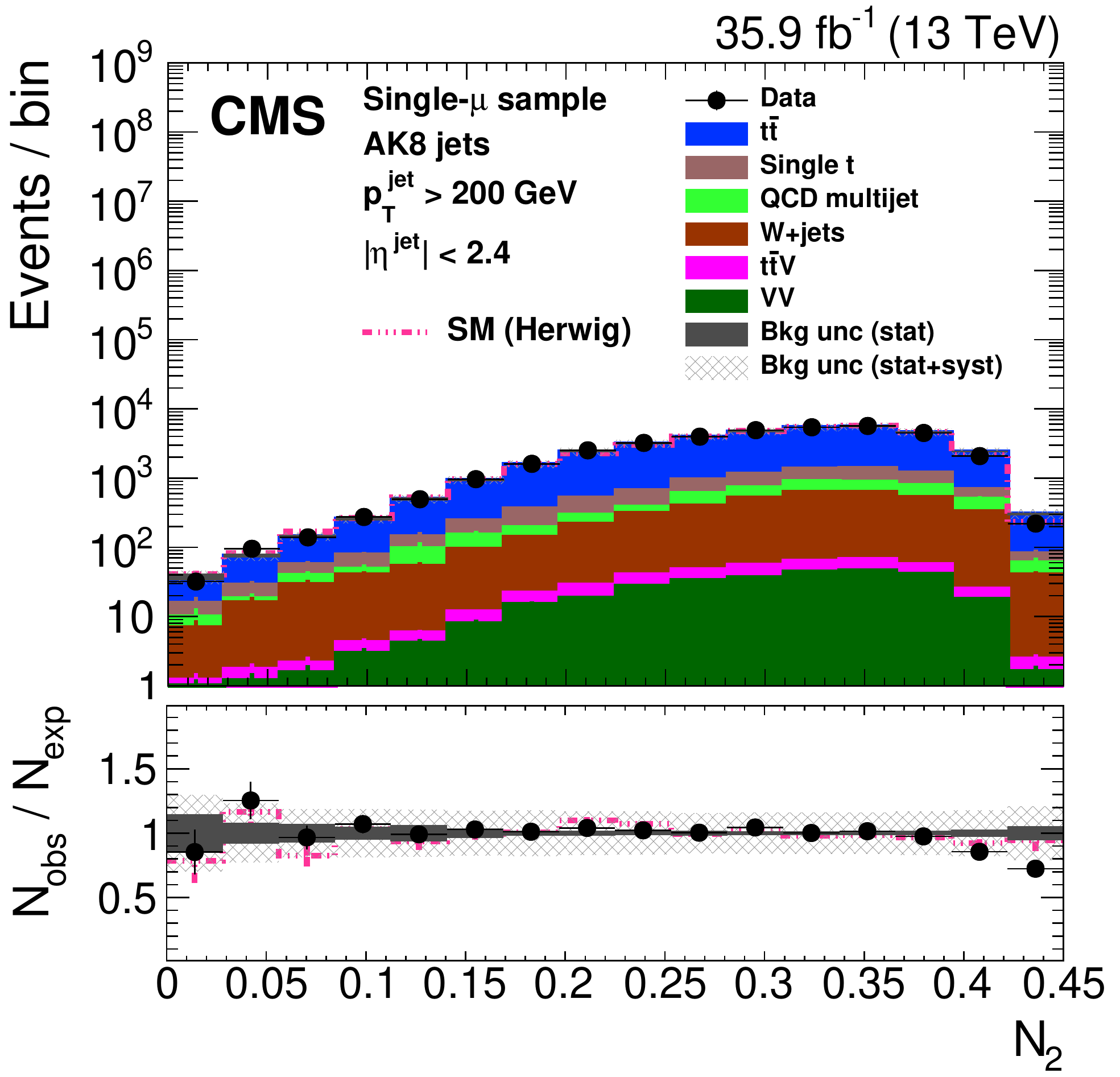}
\includegraphics[width=0.38\textwidth]{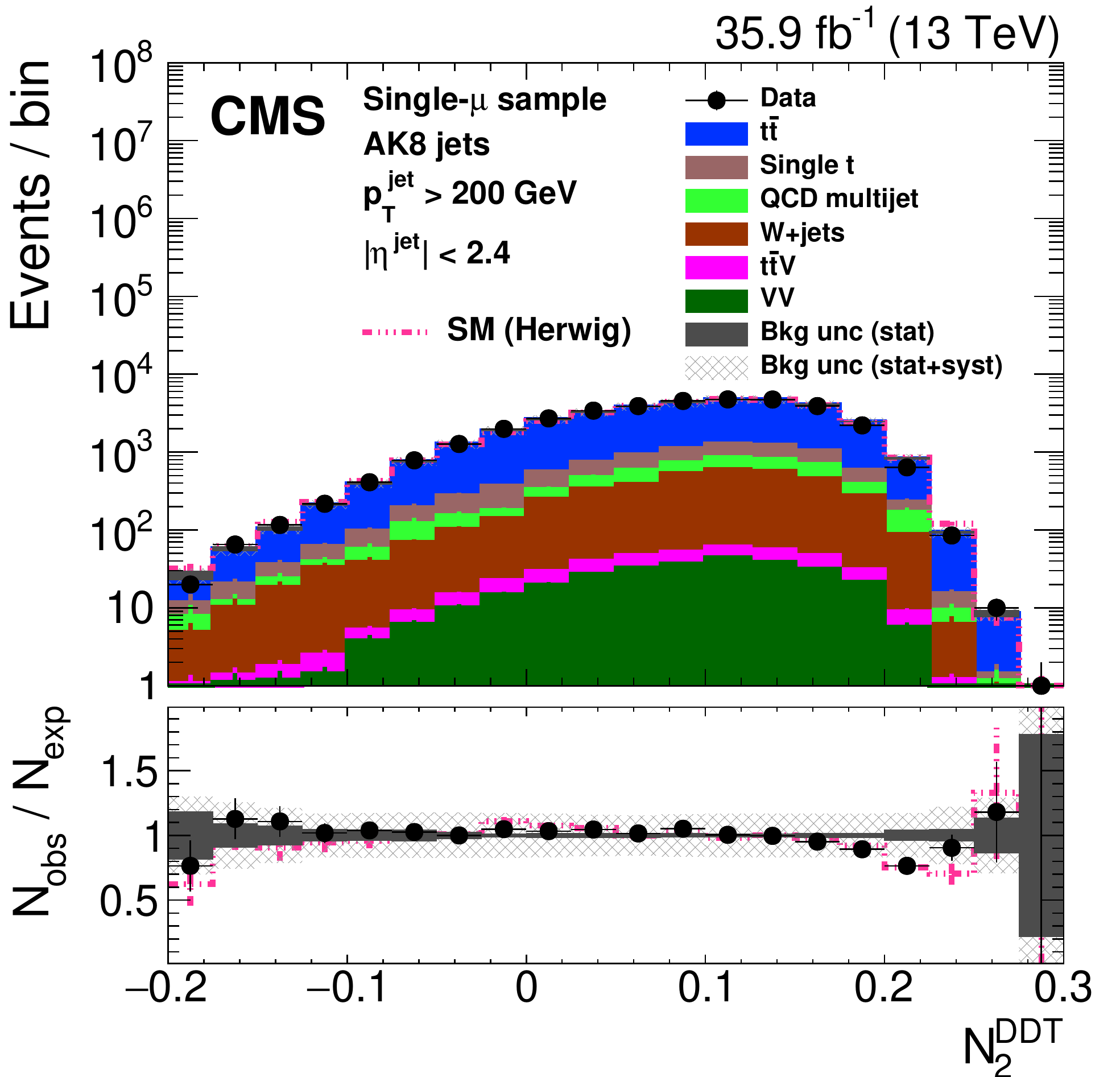}
\caption{\label{fig:tt1l_presel}Distribution of the jet \pt (upper left), jet mass, \msd\
  (upper right), the $N$-subjettiness ratios
  $\tauthreetwo$ (middle left) and $\tautwoone$ (middle right), and
  the $N_2$ (lower left) and $N_2^{\text{DDT}}$ (lower right) in data and
  simulation in the single-$\mu$ signal sample. The pink line corresponds to
  the simulation distribution obtained using the alternative \ttbar sample.
  The background event yield is normalized to the total observed data yield.
  The lower panel shows the data to simulation ratio. The solid dark-gray (shaded light-gray)
  band corresponds to the total uncertainty (statistical uncertainty of the simulated samples),
  the pink line to the data to simulation ratio using the alternative \ttbar sample,
  and the vertical black lines correspond to the statistical
  uncertainty of the data. The vertical pink lines correspond to the 
  statistical uncertainty of the alternative \ttbar sample. 
  The distributions are weighted according to
  the top quark \pt weighting procedure described in the text. }
\end{figure}

\begin{figure}[hp!]
\centering
  \includegraphics[width=0.45\textwidth]{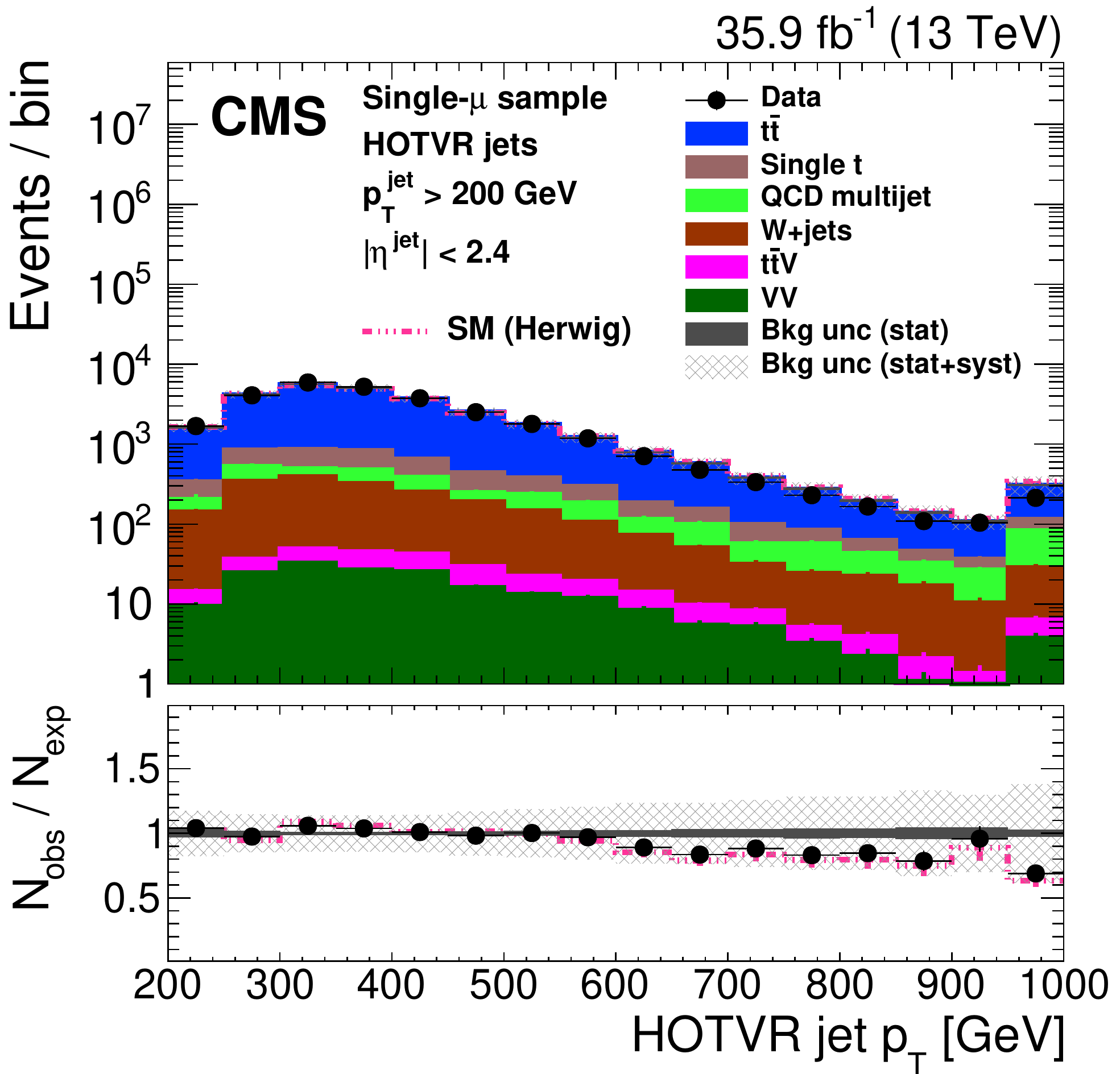}
  \includegraphics[width=0.45\textwidth]{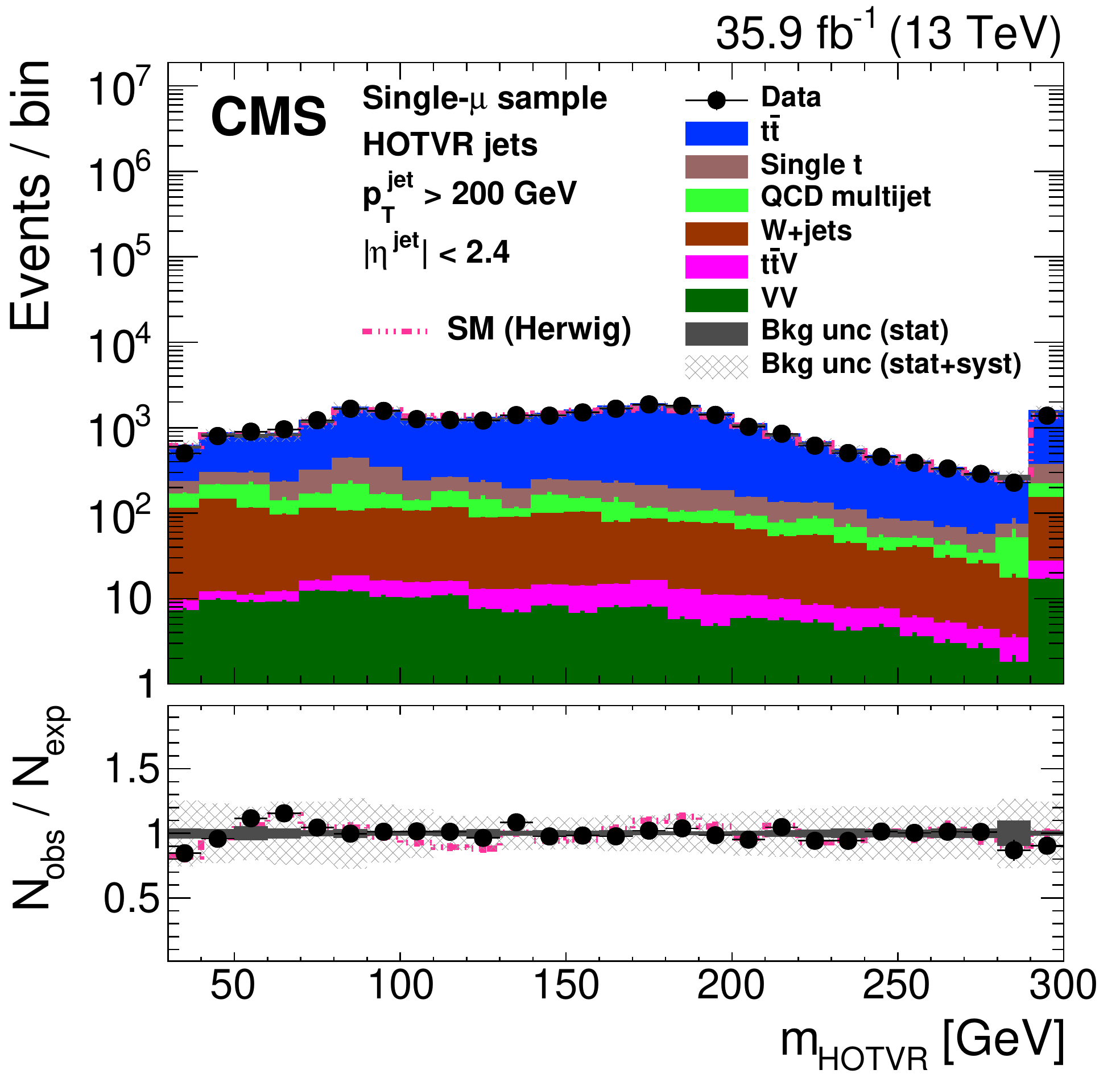}\\
  \includegraphics[width=0.45\textwidth]{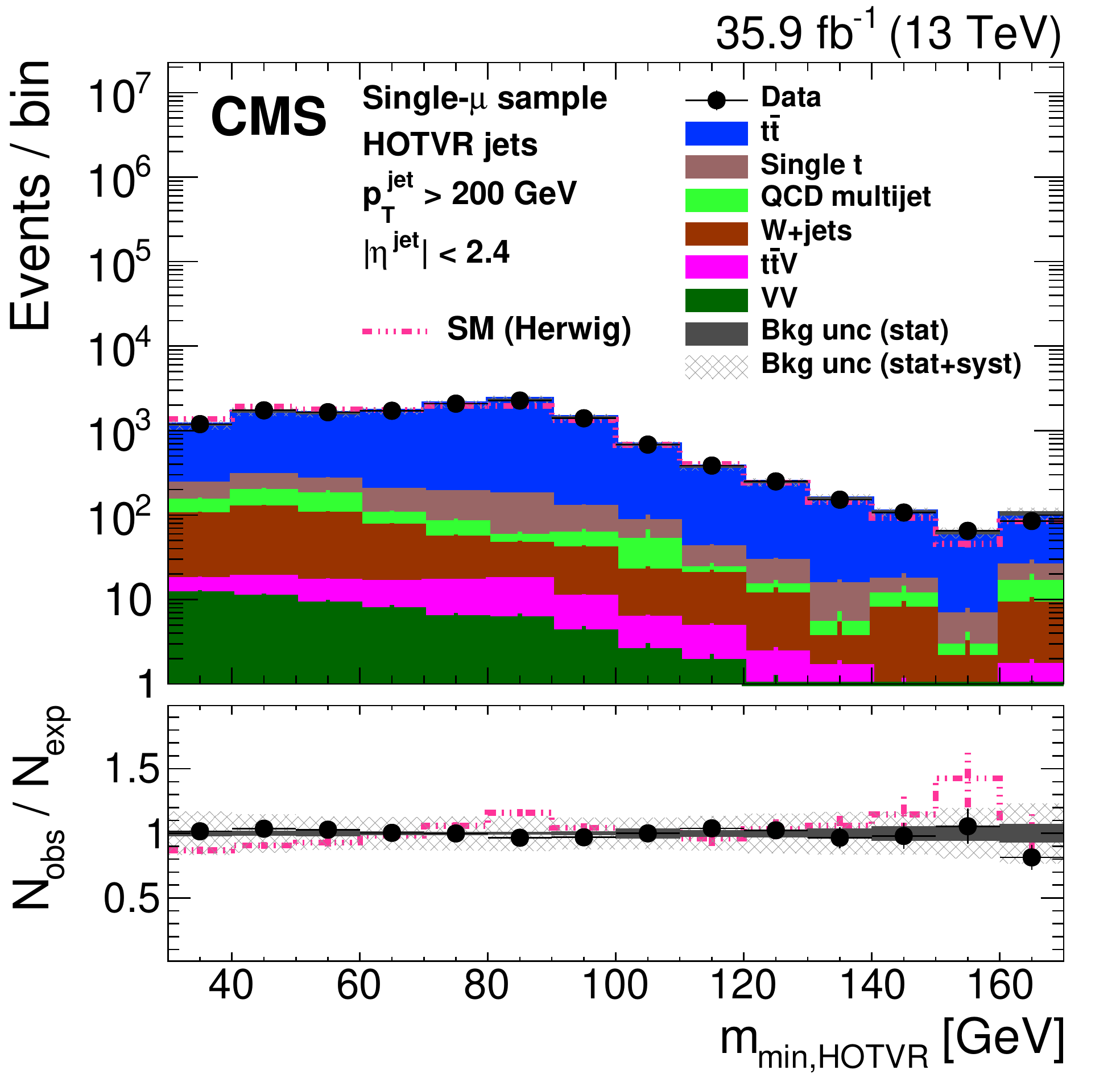}
  \includegraphics[width=0.45\textwidth]{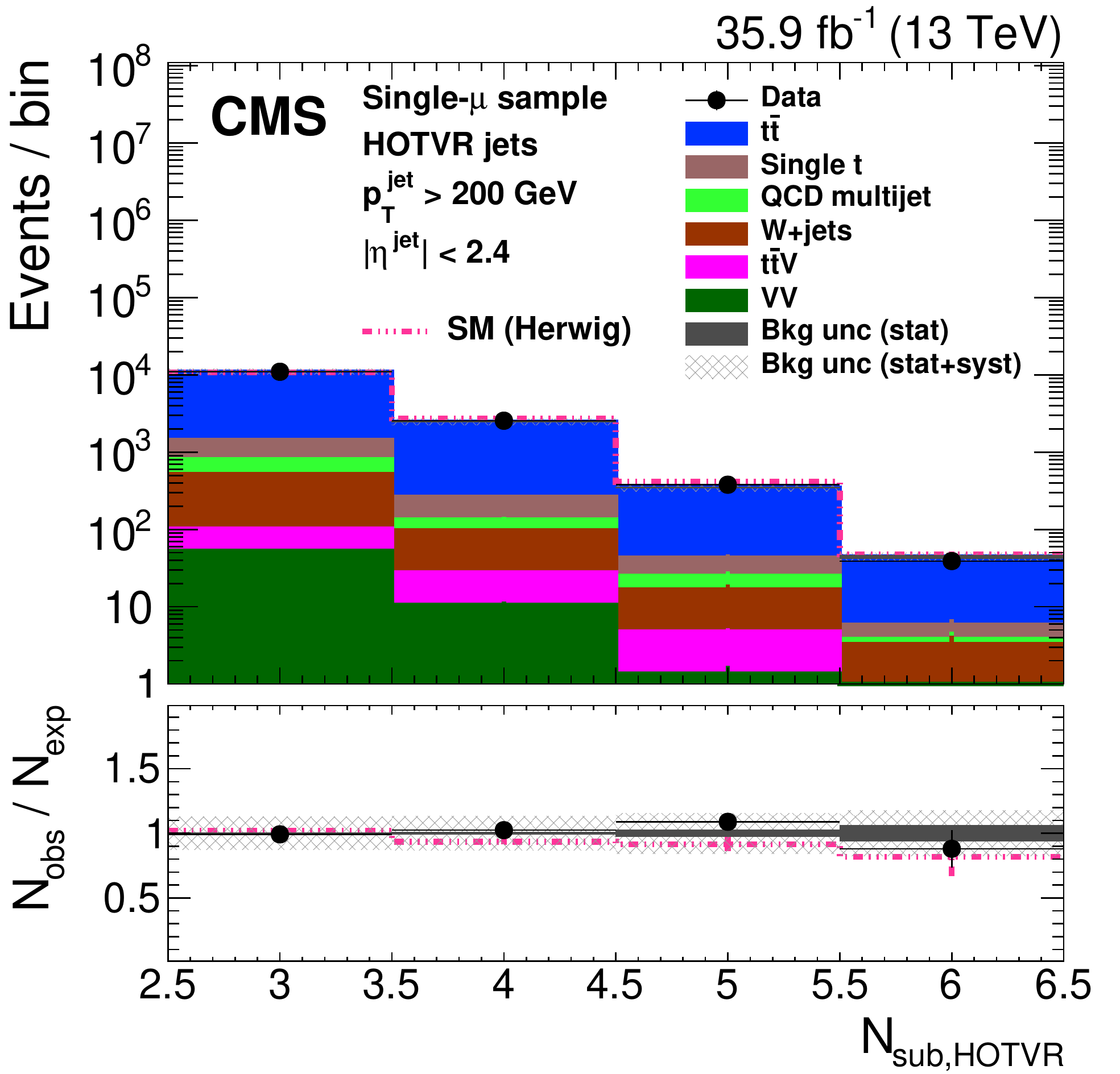}
\caption{\label{fig:tt1l_hotvr}Distribution of the main observables of
  the HOTVR algorithm, HOTVR jet \pt (upper left), $m_{\text{HOTVR}}$ (upper right),
  $m_{\text{min,HOTVR}}$ (lower left), and $N_{\text{sub,HOTVR}}$
  (lower right) in data and simulation in the single-$\mu$ signal sample. The pink line corresponds to
  the simulation distribution obtained using the alternative \ttbar sample.
  The background event yield is normalized to the total observed data yield.
  The lower panel shows the data to simulation ratio. The solid dark-gray (shaded light-gray)
  band corresponds to the total uncertainty (statistical uncertainty of the simulated samples),
  the pink line to the data to simulation ratio using the alternative \ttbar sample,
  and the vertical black lines correspond to the statistical
  uncertainty of the data. The vertical pink lines correspond to the
  statistical uncertainty of the alternative \ttbar sample.
  The distributions are weighted according to  the top quark \pt weighting procedure described in the text. }
\end{figure}

\begin{figure}[hp!]
\centering
\includegraphics[width=0.45\textwidth]{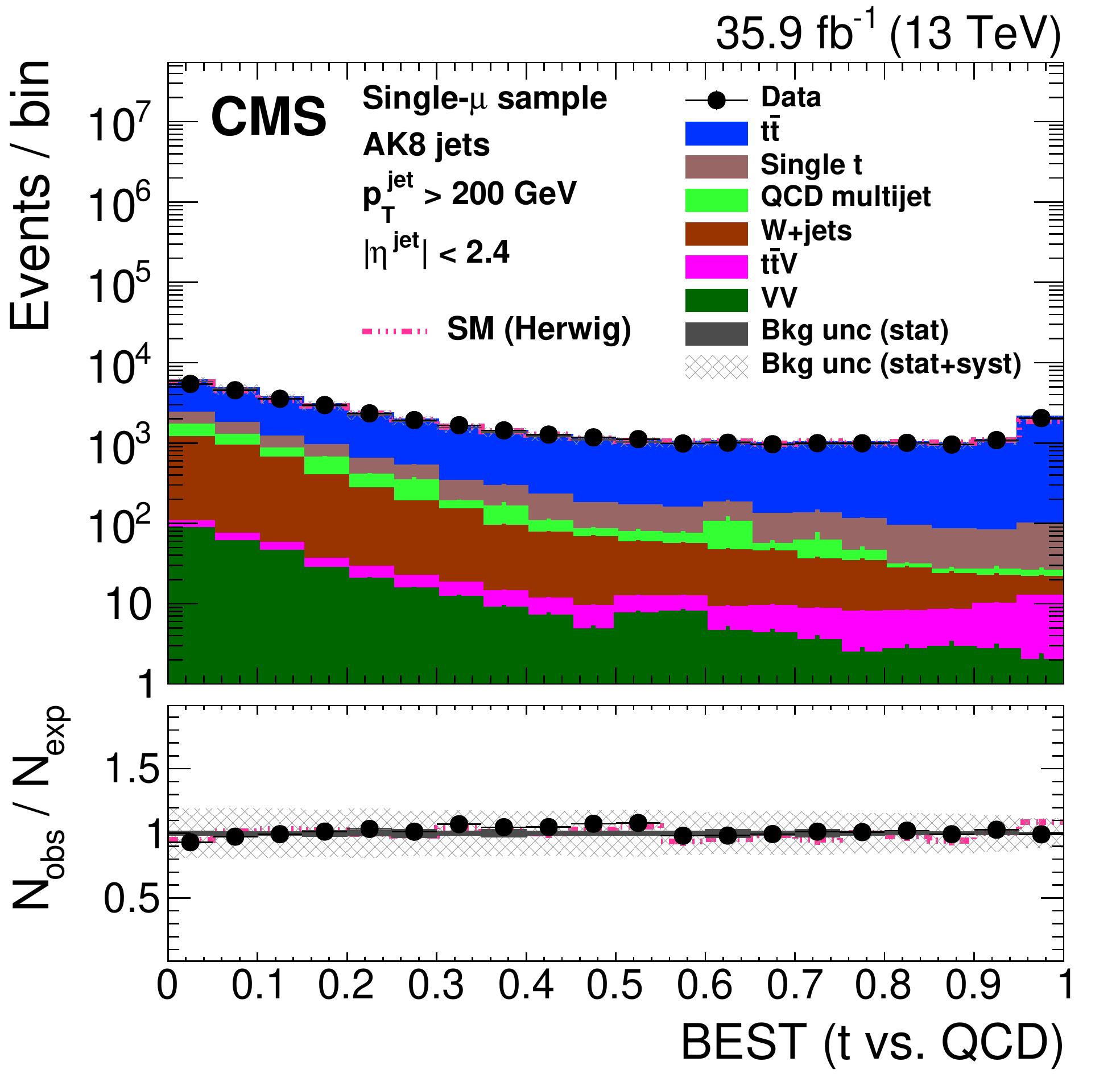}
\includegraphics[width=0.45\textwidth]{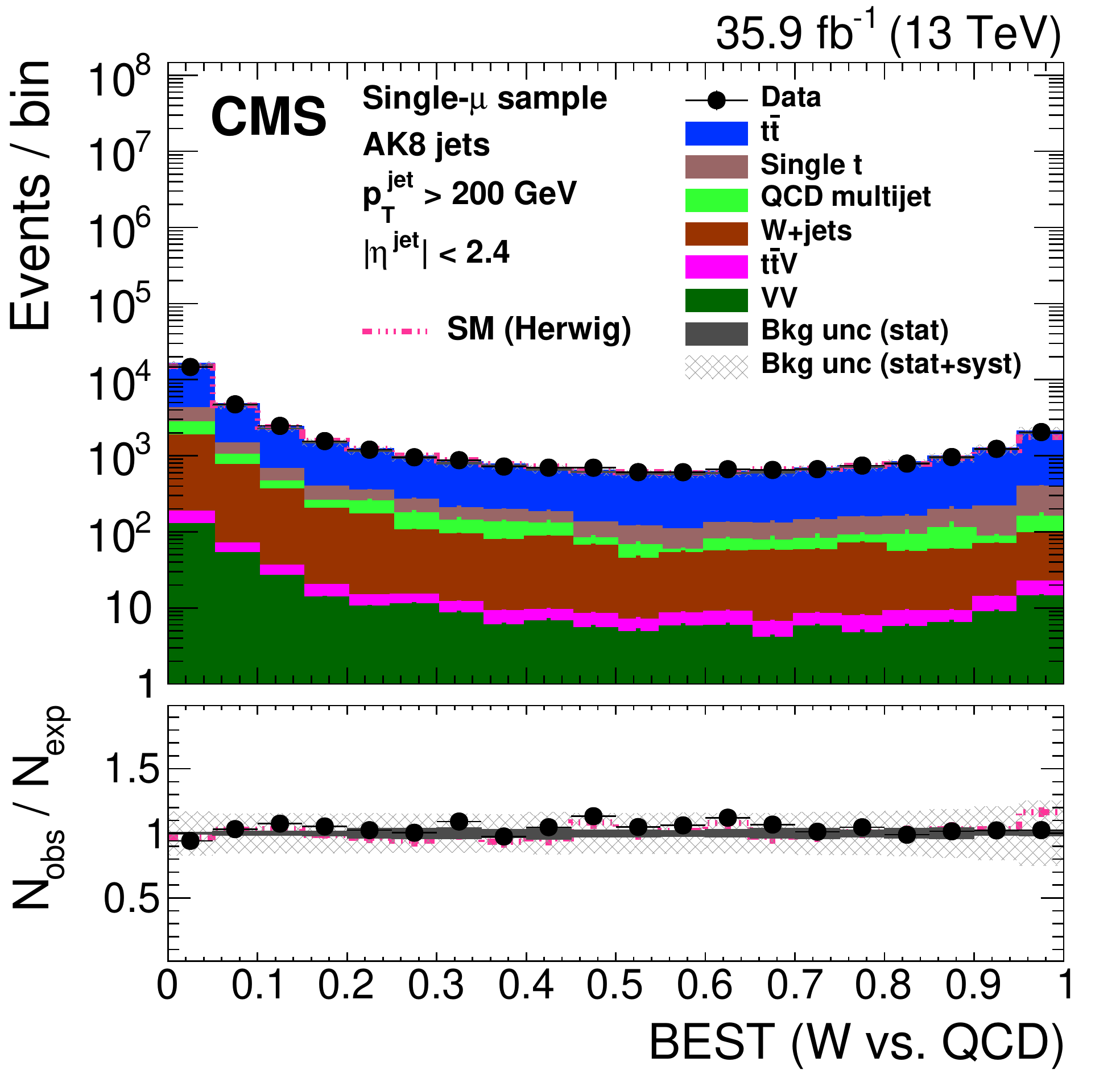}
\includegraphics[width=0.45\textwidth]{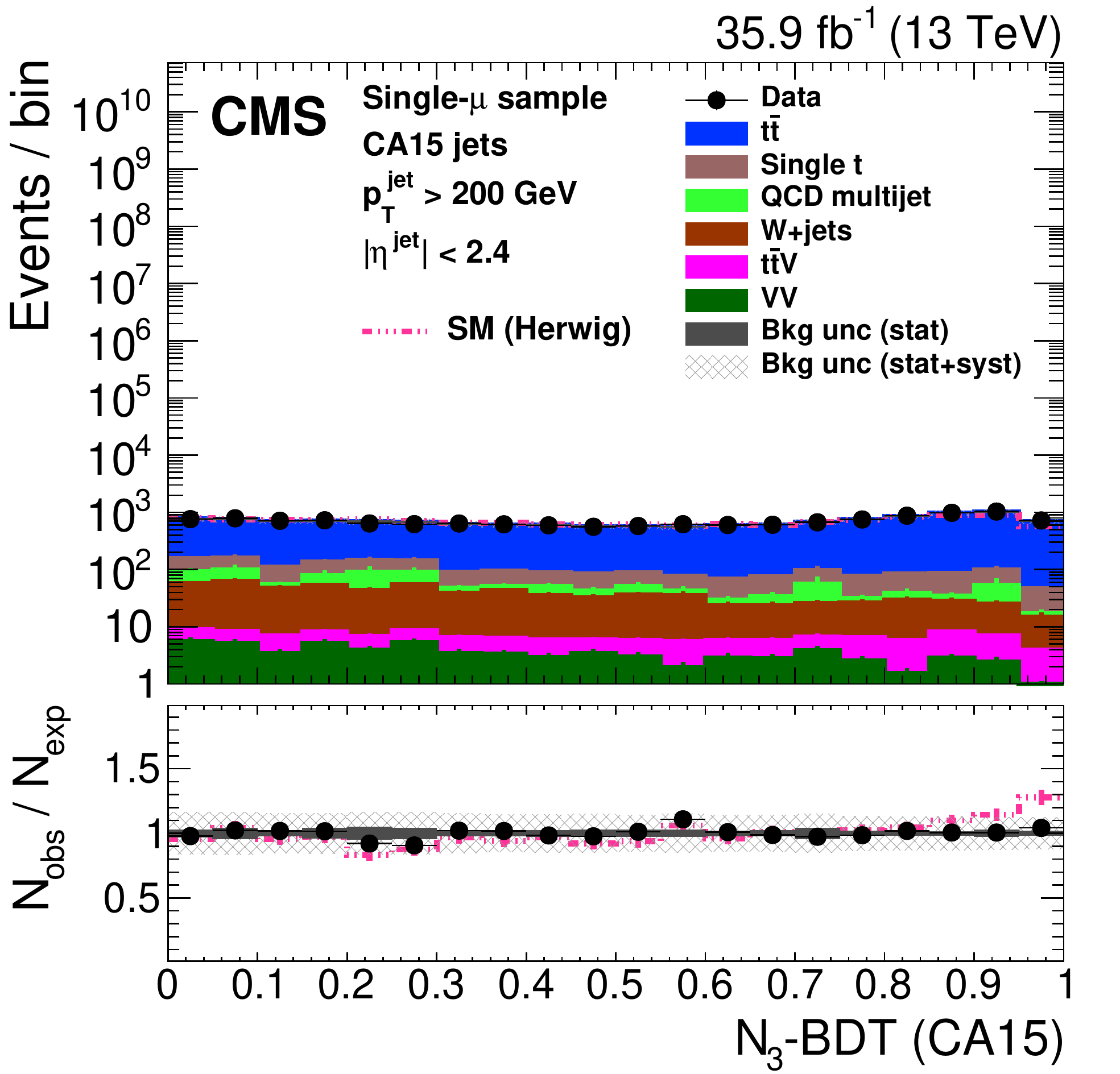}
\caption{\label{fig:tt1l_hlv}Distribution of the \PQt~quark (upper
  left) and \PW~boson (upper right) identification probabilities for
  the BEST algorithm, and the \ecftop discriminant in data and simulation
  in the single-$\mu$ signal sample. The pink line corresponds to
  the simulation distribution obtained using the alternative \ttbar sample.
  The background event yield is normalized to the total observed data yield.
  The lower panel shows the data to simulation ratio. The solid dark-gray (shaded light-gray)
  band corresponds to the total uncertainty (statistical uncertainty of the simulated samples),
  the pink line to the data to simulation ratio using the alternative \ttbar sample,
  and the vertical black lines correspond to the statistical
  uncertainty of the data. The vertical pink lines correspond to the
  statistical uncertainty of the alternative \ttbar sample.
  The distributions are weighted according to
  the top quark \pt weighting procedure described in the text.  }
\end{figure}

\begin{figure}[hp!]
\centering
\includegraphics[width=0.38\textwidth]{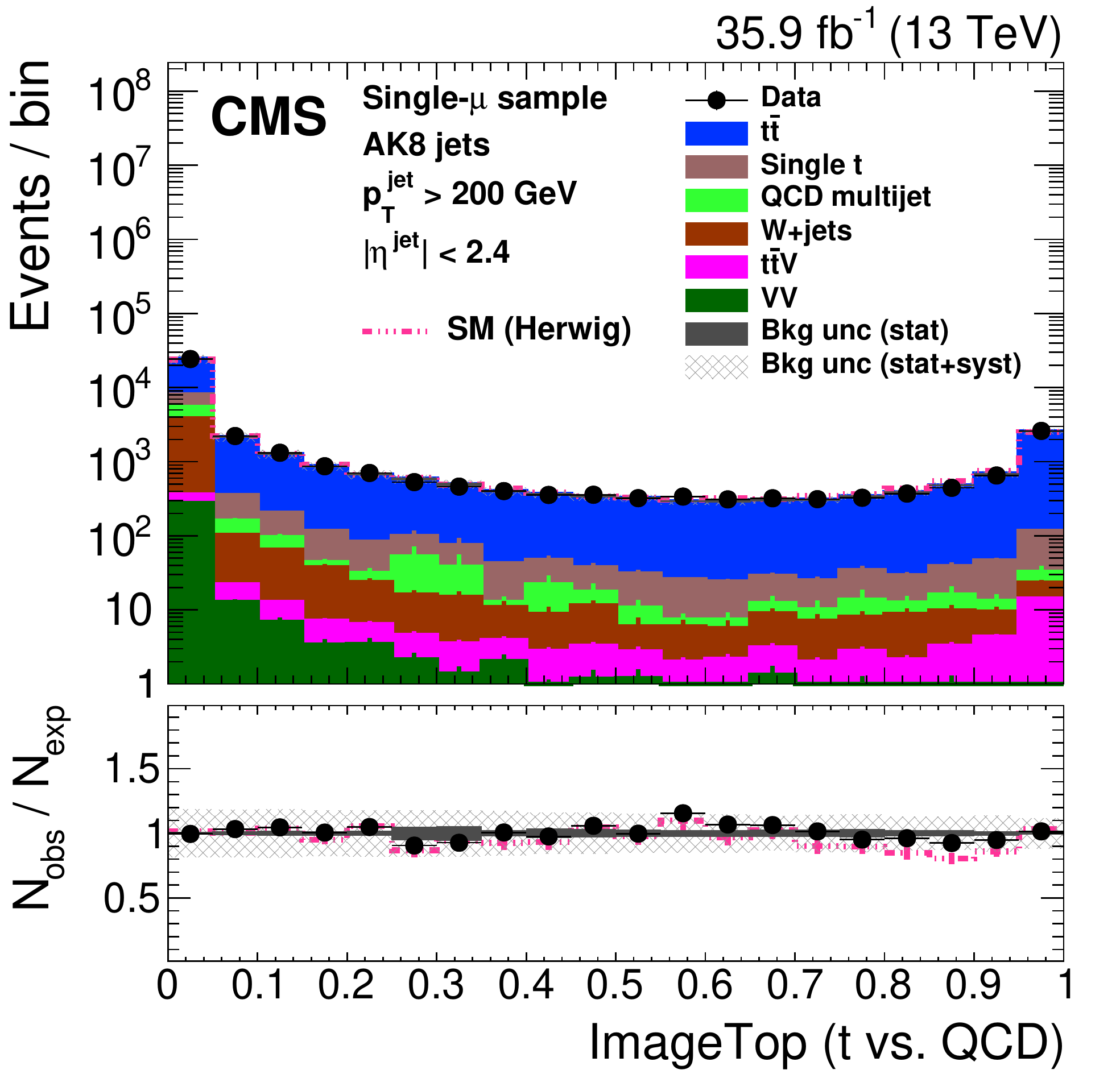}
\includegraphics[width=0.38\textwidth]{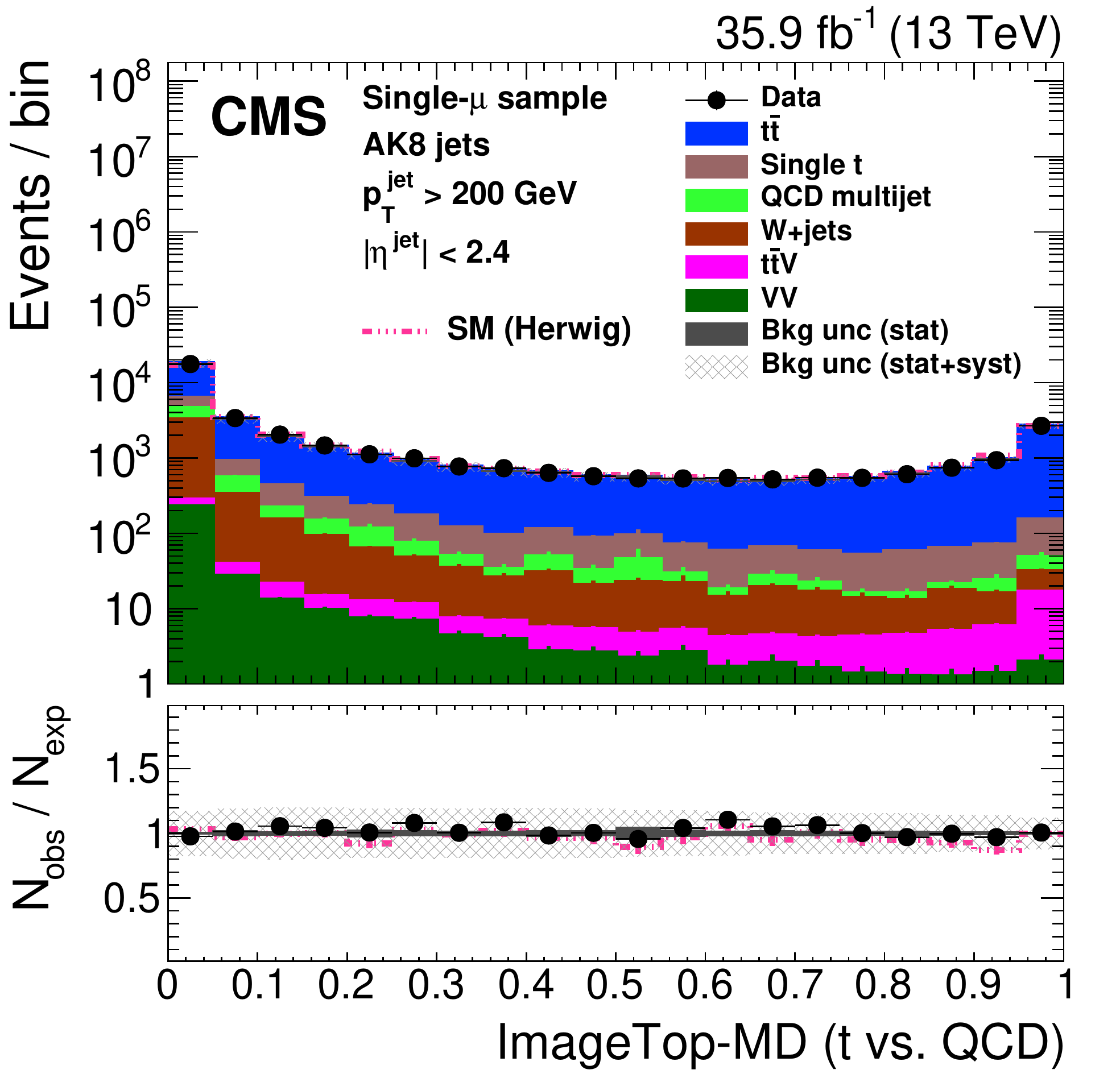}\\
\includegraphics[width=0.38\textwidth]{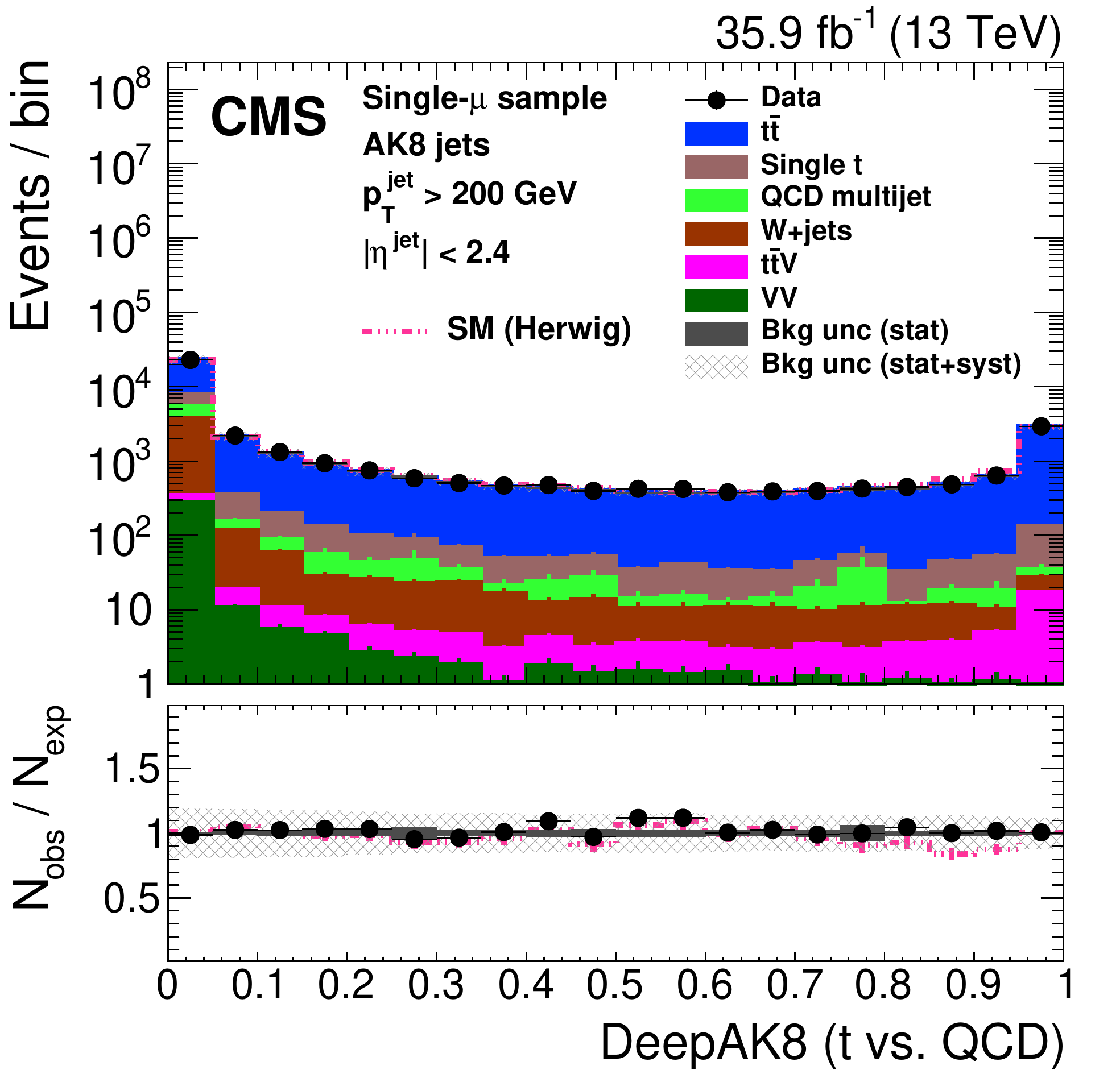}
\includegraphics[width=0.38\textwidth]{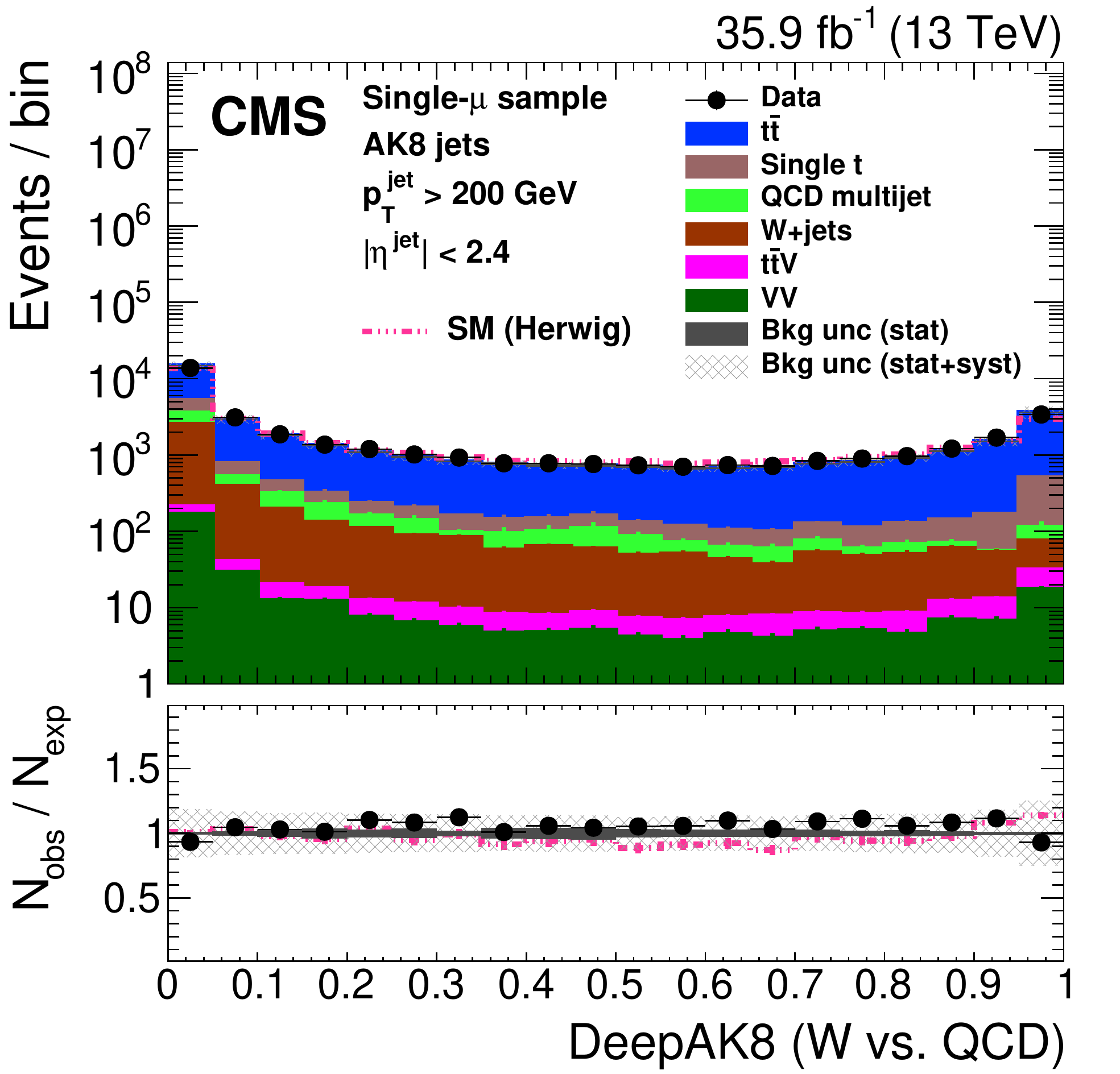} \\
\includegraphics[width=0.38\textwidth]{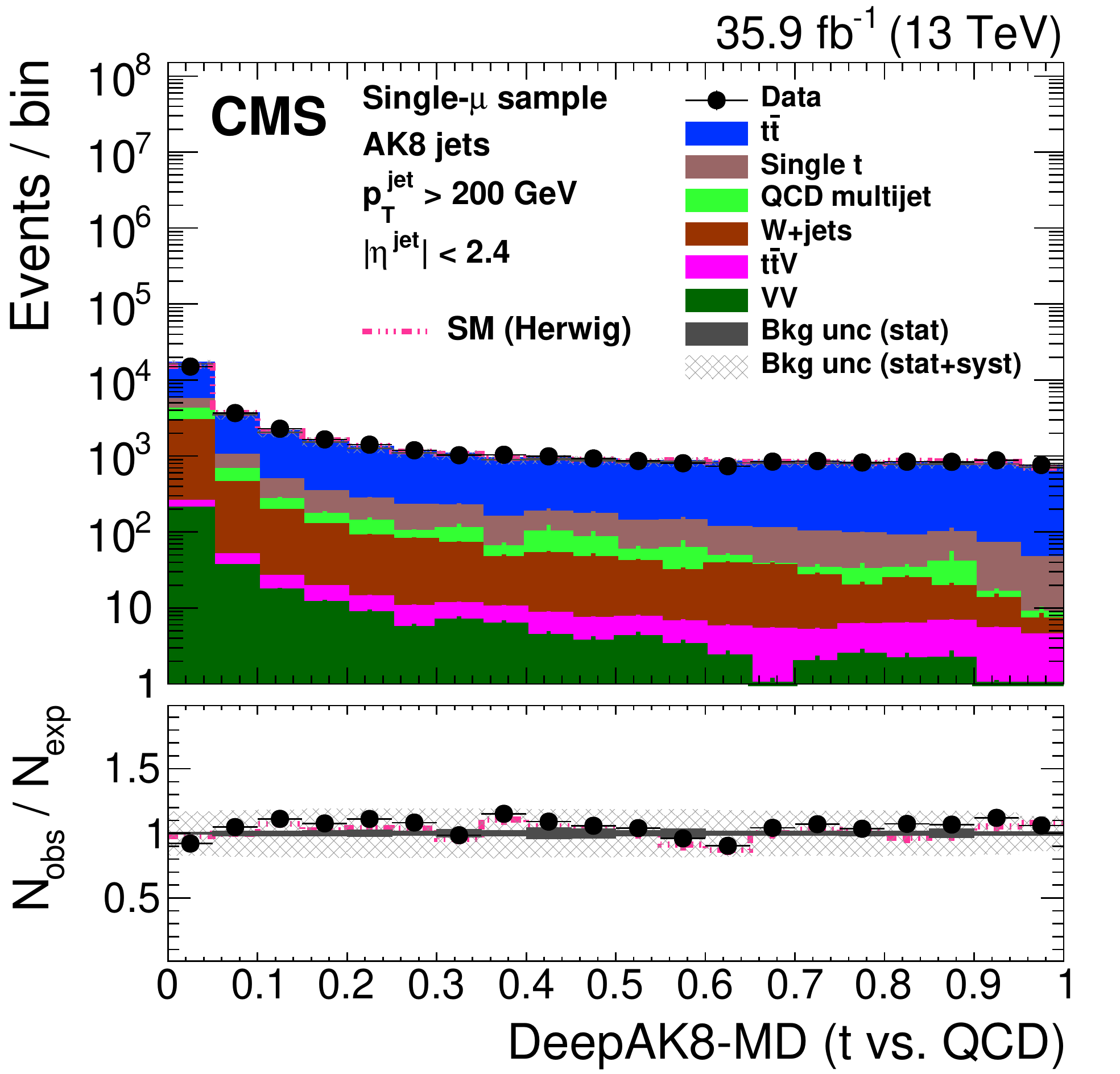}
\includegraphics[width=0.38\textwidth]{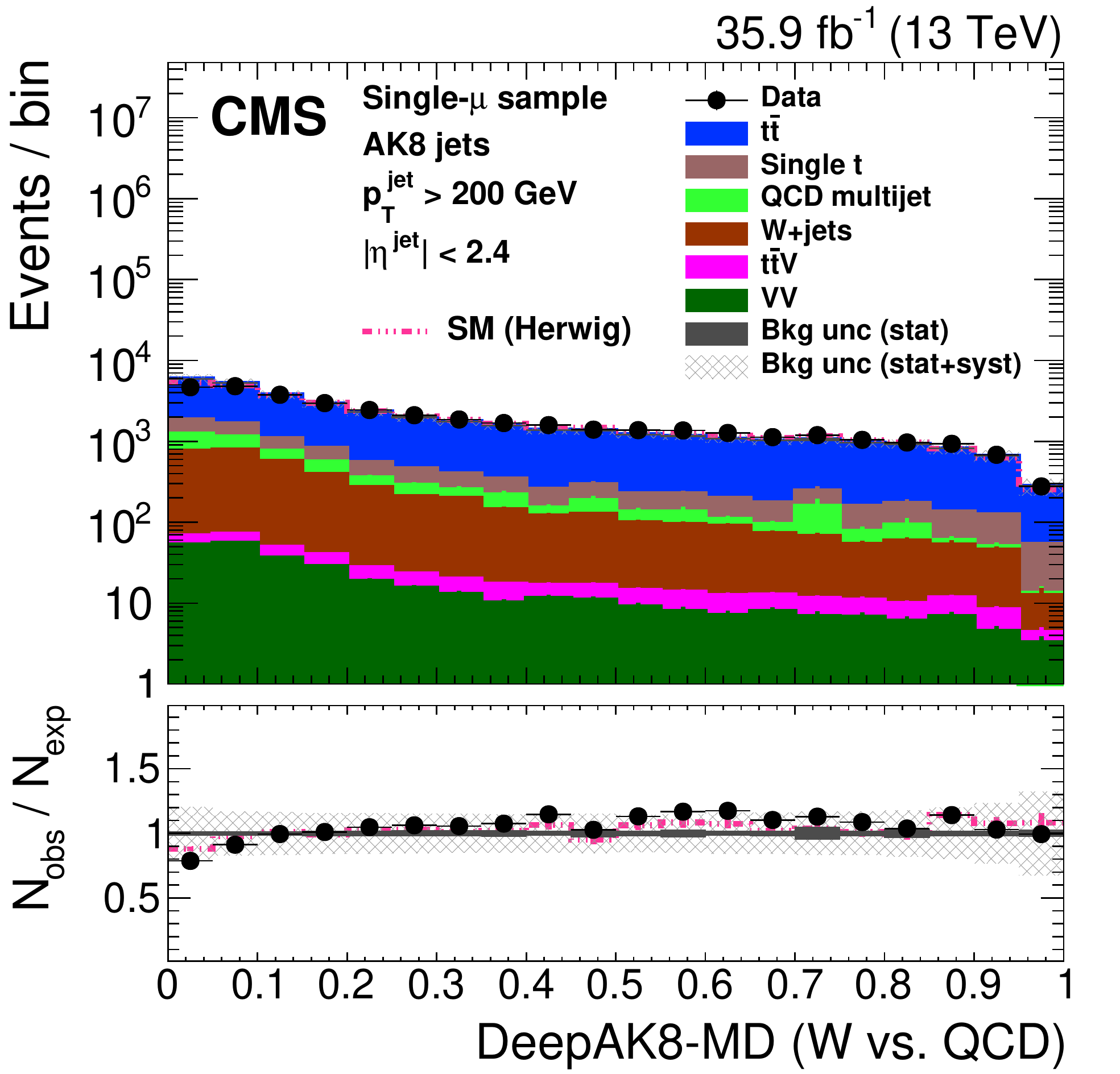} \\
\caption{\label{fig:tt1l_low}Distribution of the ImageTop (upper left)
  and ImageTop-MD (upper right) discriminant in data and simulation in the
  single-$\mu$ sample. The plots in the middle row show the
  \PQt~quark (left) and \PW~boson (right)  identification
  probabilities in data and simulation for the DeepAK8 algorithm. The corresponding plots for
  DeepAK8-MD are displayed in the lower row. The pink line corresponds to
  the simulation distribution obtained using the alternative \ttbar sample.
  The background event yield is normalized to the total observed data yield.
  The lower panel shows the data to simulation ratio. The solid dark-gray (shaded light-gray)
  band corresponds to the total uncertainty (statistical uncertainty of the simulated samples),
  the pink line to the data to simulation ratio using the alternative \ttbar sample,
  and the vertical black lines correspond to the statistical
  uncertainty of the data. The vertical pink lines correspond to the
  statistical uncertainty of the alternative \ttbar sample. 
  The distributions are weighted according to
  the top quark \pt weighting procedure described in the text. }

\end{figure}

The data-to-simulation comparisons of basic jet kinematic and substructure variables:
 $\pt(\text{jet})$, \msd, the $N$-subjettiness ratios
$\tauthreetwo$ and $\tautwoone$, and the $N_{2}$ and $N_{2}^{\mathrm{DDT}}$, are
shown in Fig.~\ref{fig:tt1l_presel}.
Figure~\ref{fig:tt1l_hotvr} displays the main observables of the HOTVR algorithm,
$m_{\text{HOTVR}}$, $m_{\text{min,HOTVR}}$ and $N_{\text{sub,HOTVR}}$,
in data and simulation. The next set of comparisons includes tagging
algorithms that are based on high-level jet substructure observables
and explore ML techniques to improve performance, namely the BEST and
the \ecftop algorithms. Figure~\ref{fig:tt1l_hlv} shows the \PQt\
quark and \PW boson identification probabilities of BEST and the
\PQt tagging discriminant for the \ecftop, in data and
simulation. The last set of comparisons is related to
the ImageTop and the DeepAK8 algorithms,
which both explore lower-level observables. Figure~\ref{fig:tt1l_low}
displays the distributions of the \PQt~quark identification
probability for the two versions of ImageTop, and the \PQt~quark and
\PW boson identification probabilities for DeepAK8 algorithms.

\begin{figure}[hp!]
\centering
\includegraphics[width=0.38\textwidth]{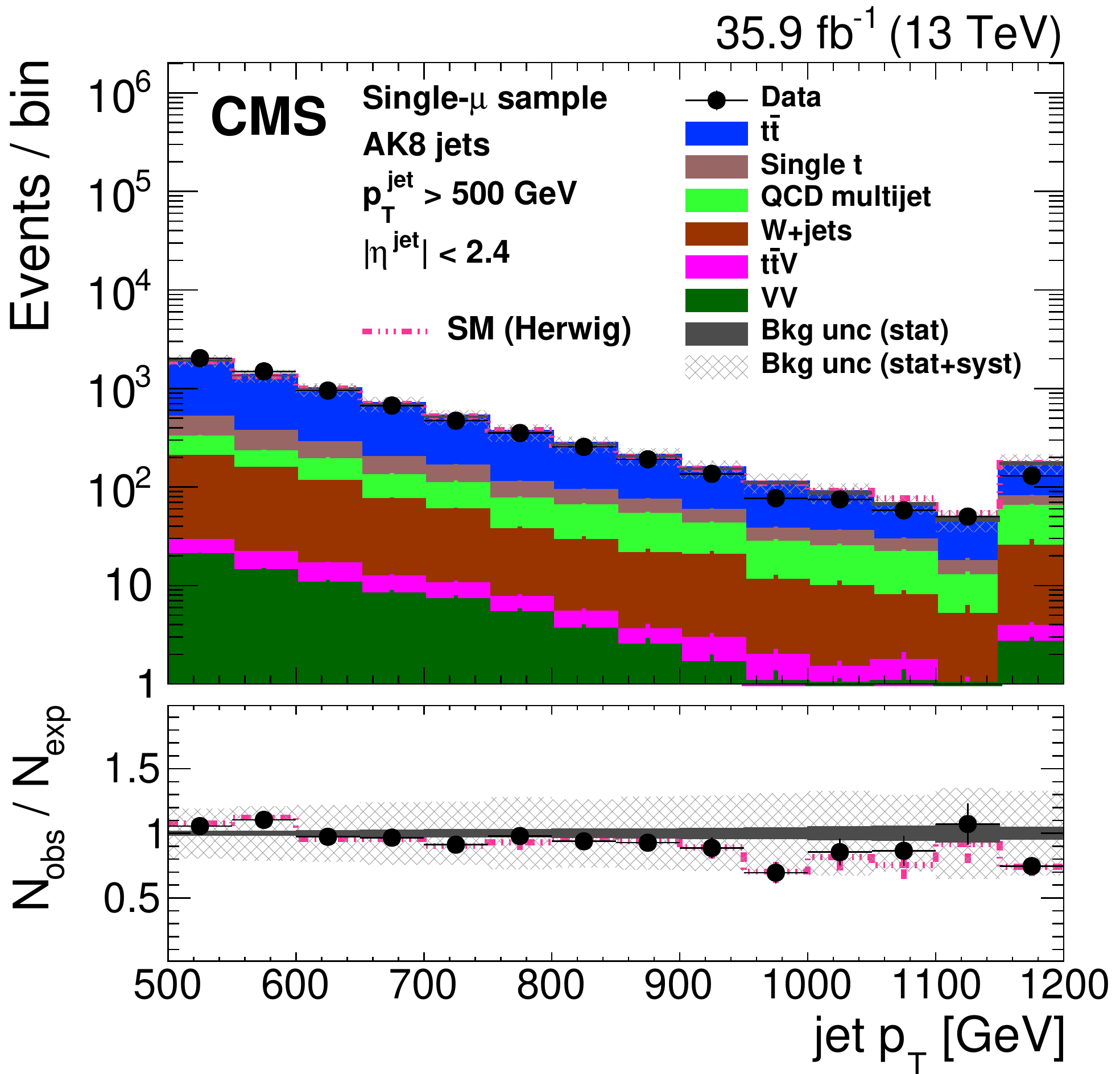}
\includegraphics[width=0.38\textwidth]{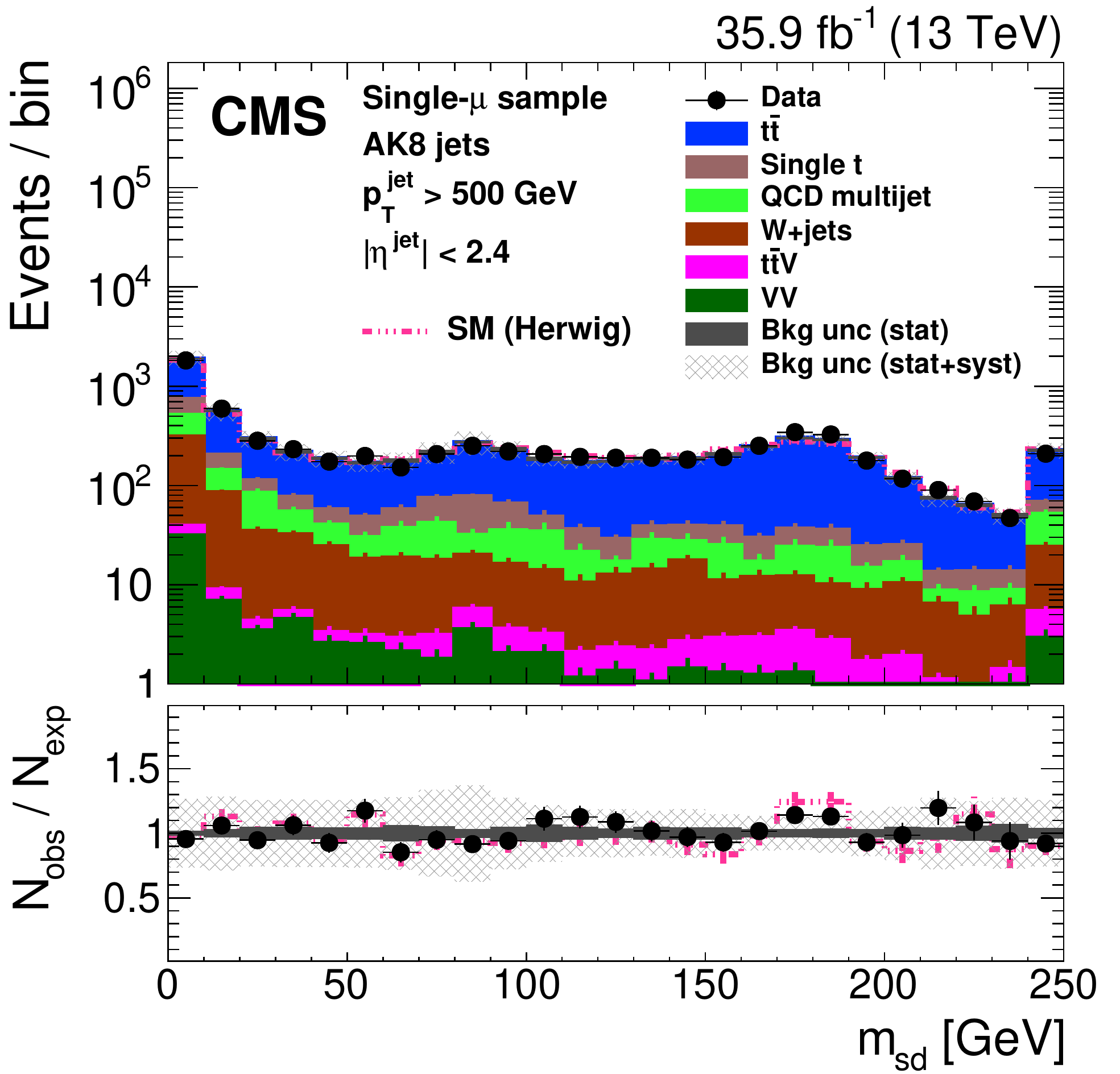}
\includegraphics[width=0.38\textwidth]{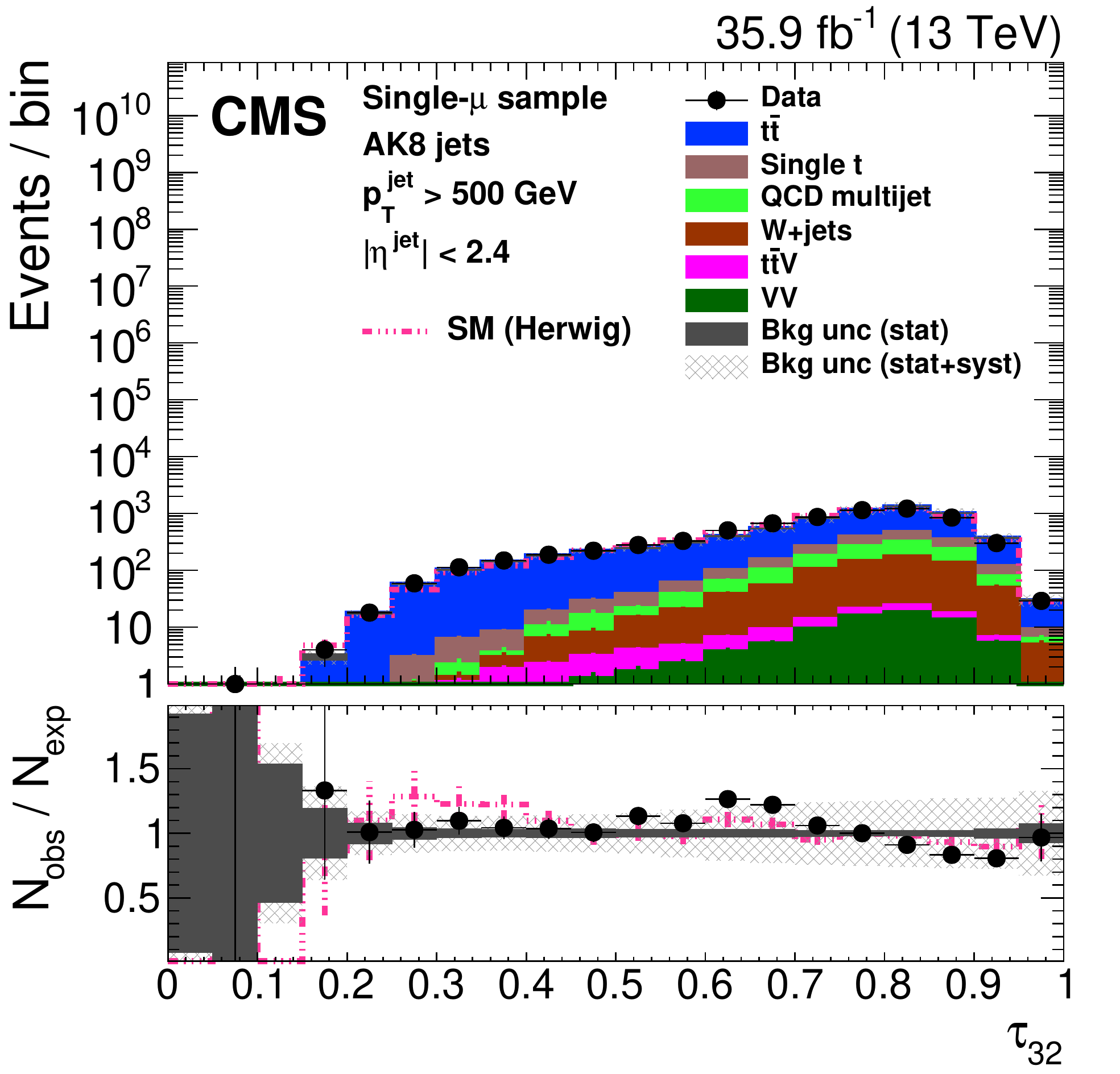}
\includegraphics[width=0.38\textwidth]{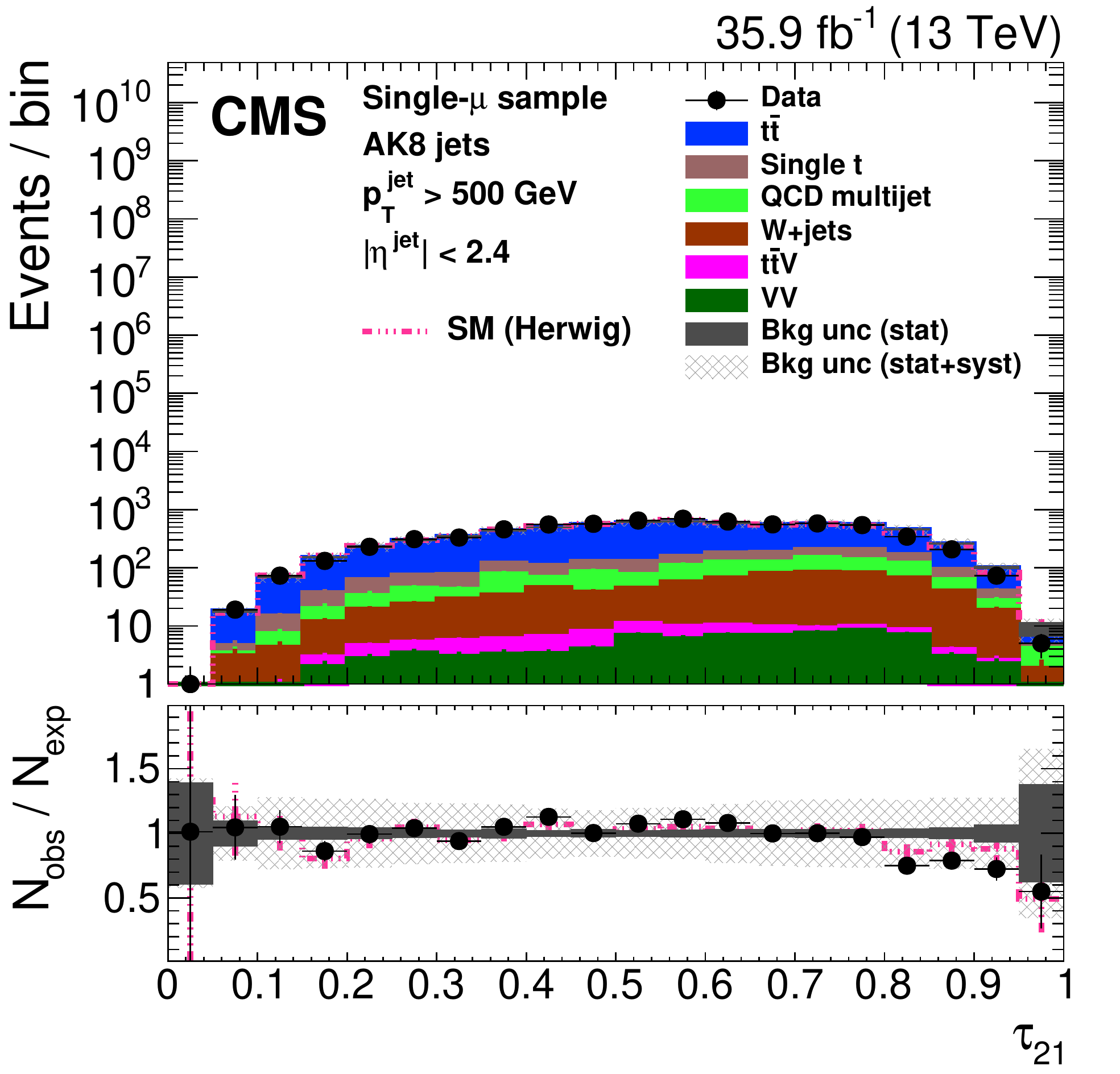}
\includegraphics[width=0.38\textwidth]{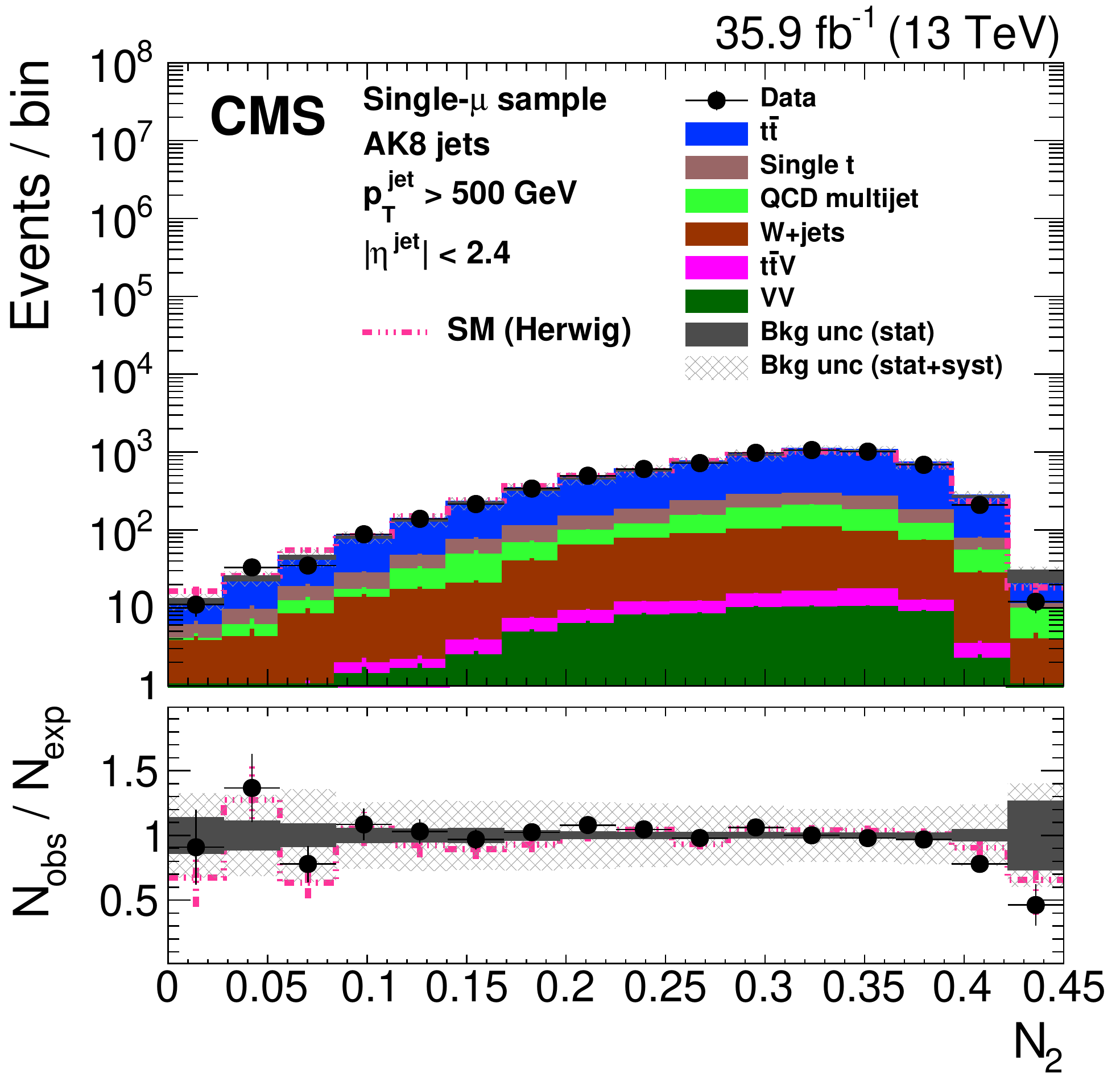}
\includegraphics[width=0.38\textwidth]{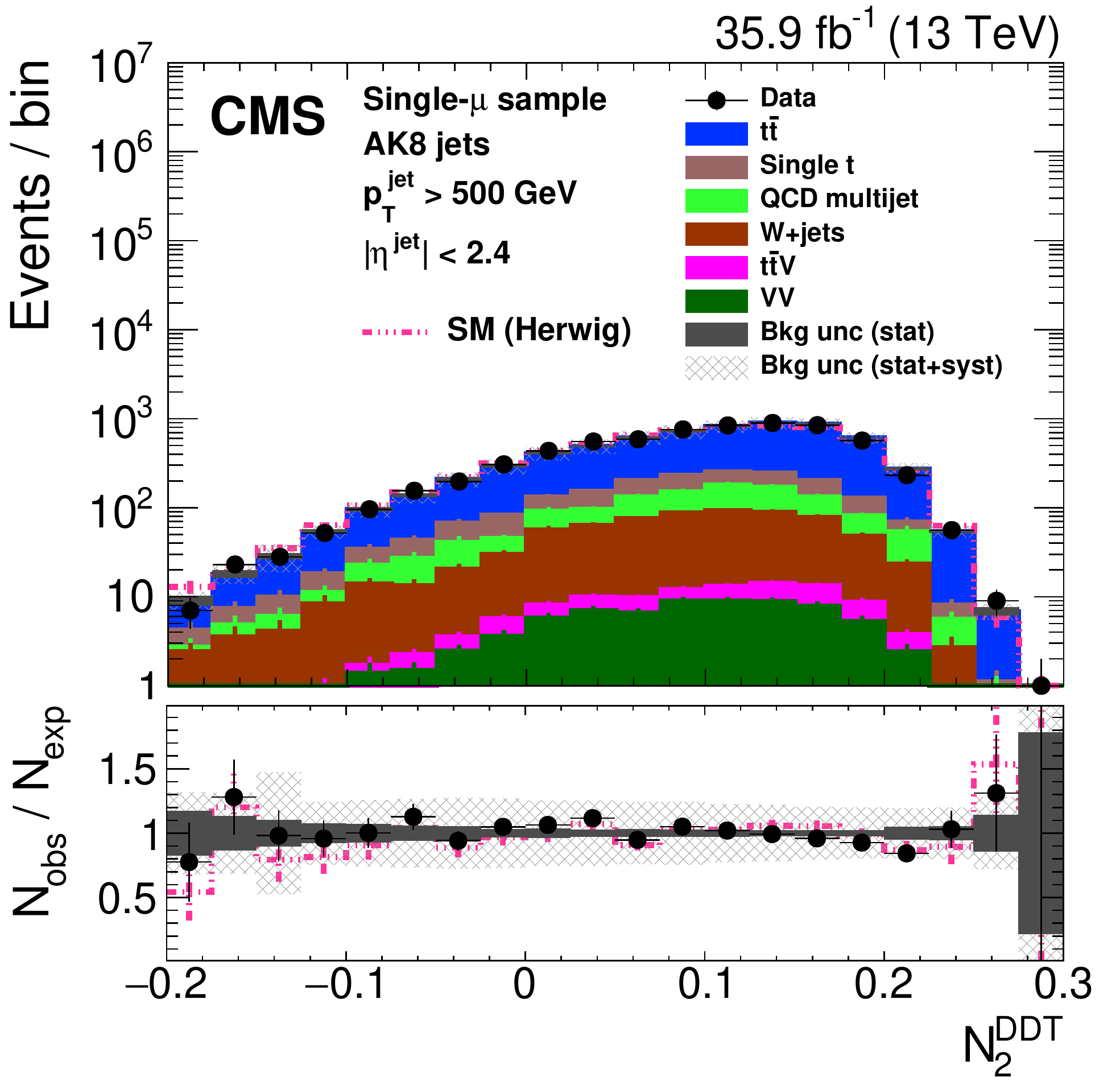}
\caption{\label{fig:tt1l_presel_pt500}Distribution of the jet \pt (upper left), the jet mass,
  \msd (upper right), the $N$-subjettiness
  ratios $\tauthreetwo$ (middle left) and $\tautwoone$
  (middle right), and the $N_2$ (lower left) and $N_2^{\text{DDT}}$
  (lower right) in data and simulation in the single-$\mu$ signal sample
  after applying a jet momentum cut $\pt > 500\GeV$.  The pink line corresponds to
  the simulation distribution obtained using the alternative \ttbar sample.
  The background event yield is normalized to the total observed data yield.
  The lower panel shows the data to simulation ratio. The solid dark-gray (shaded light-gray)
  band corresponds to the total uncertainty (statistical uncertainty of the simulated samples),
  the pink line to the data to simulation ratio using the alternative \ttbar sample,
  and the vertical black lines correspond to the statistical
  uncertainty of the data. The vertical pink lines correspond to the
  statistical uncertainty of the alternative \ttbar sample.
  The distributions are weighted according to
  the top quark \pt weighting procedure described in the text.}
\end{figure}

\begin{figure}[hp!]
\centering
  \includegraphics[width=0.38\textwidth]{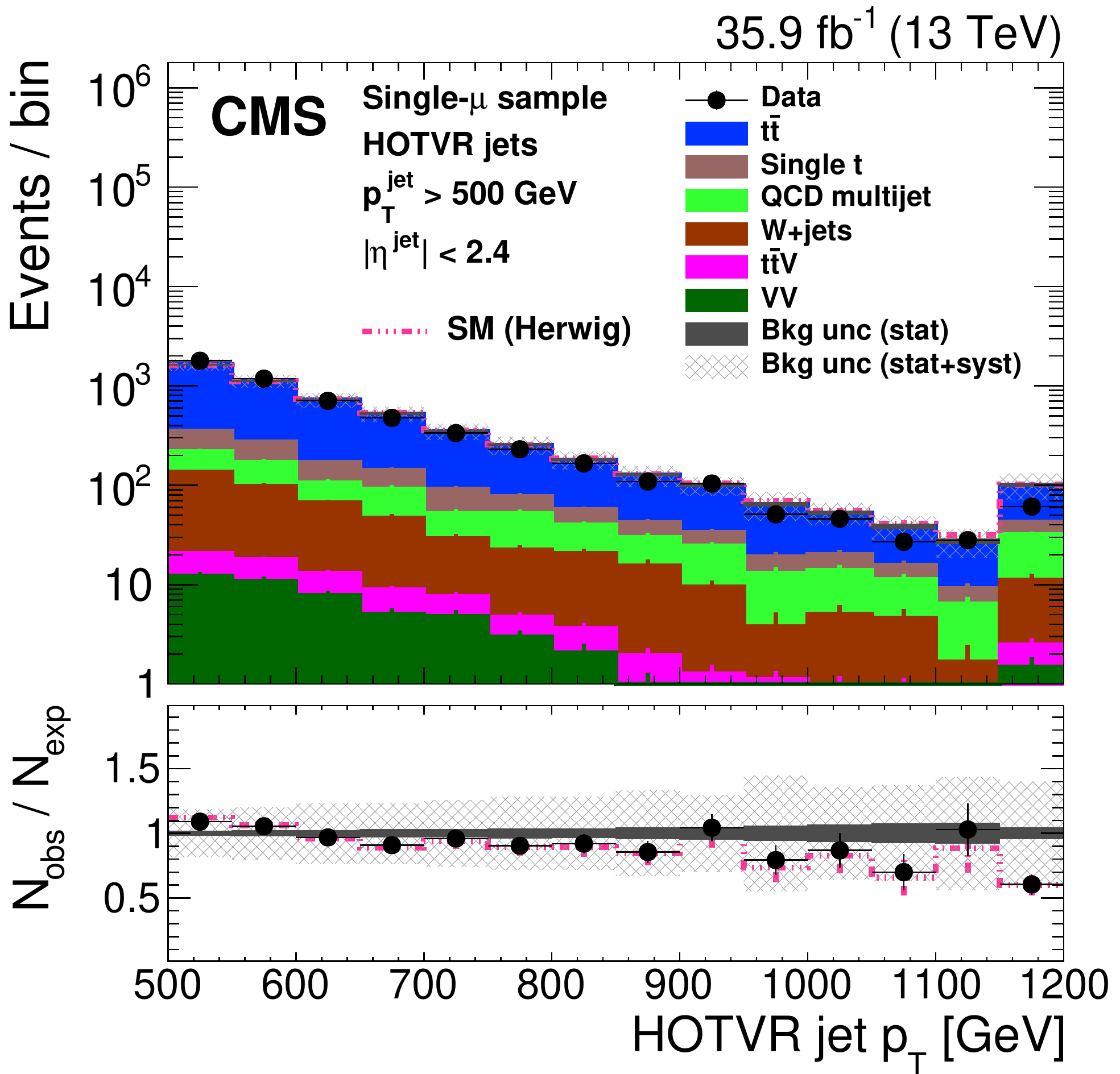}
  \includegraphics[width=0.38\textwidth]{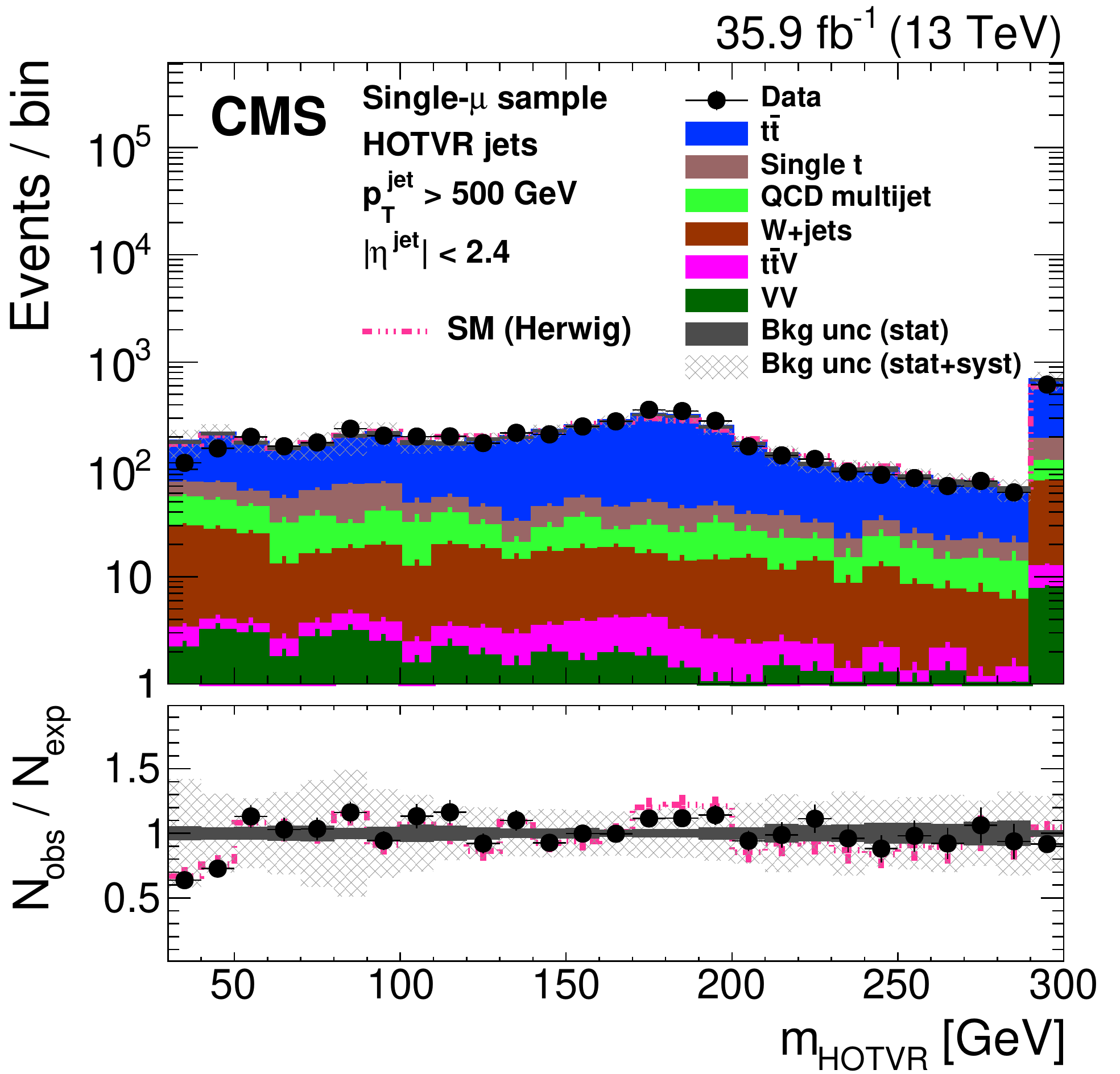}\\
  \includegraphics[width=0.38\textwidth]{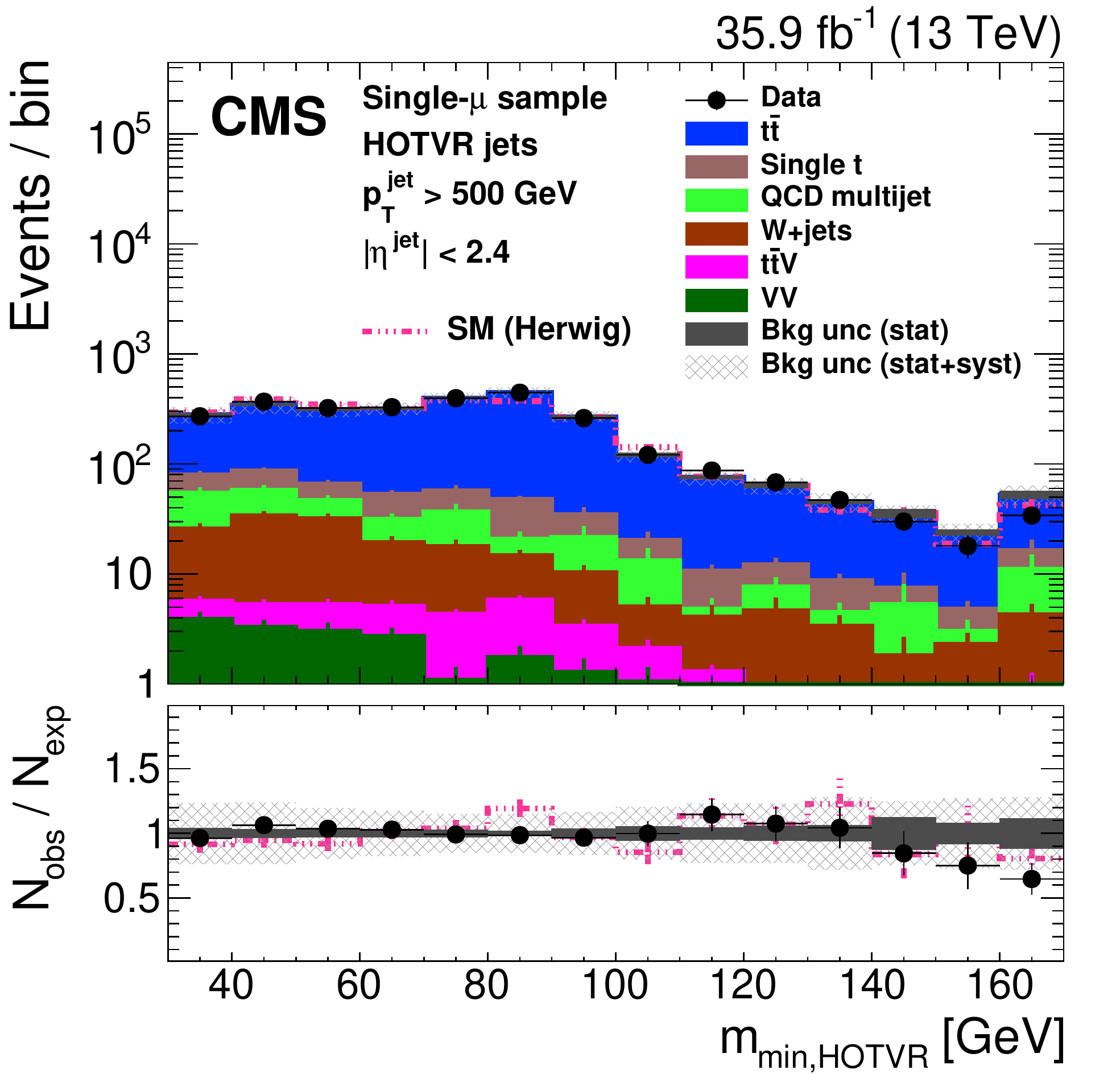}
  \includegraphics[width=0.38\textwidth]{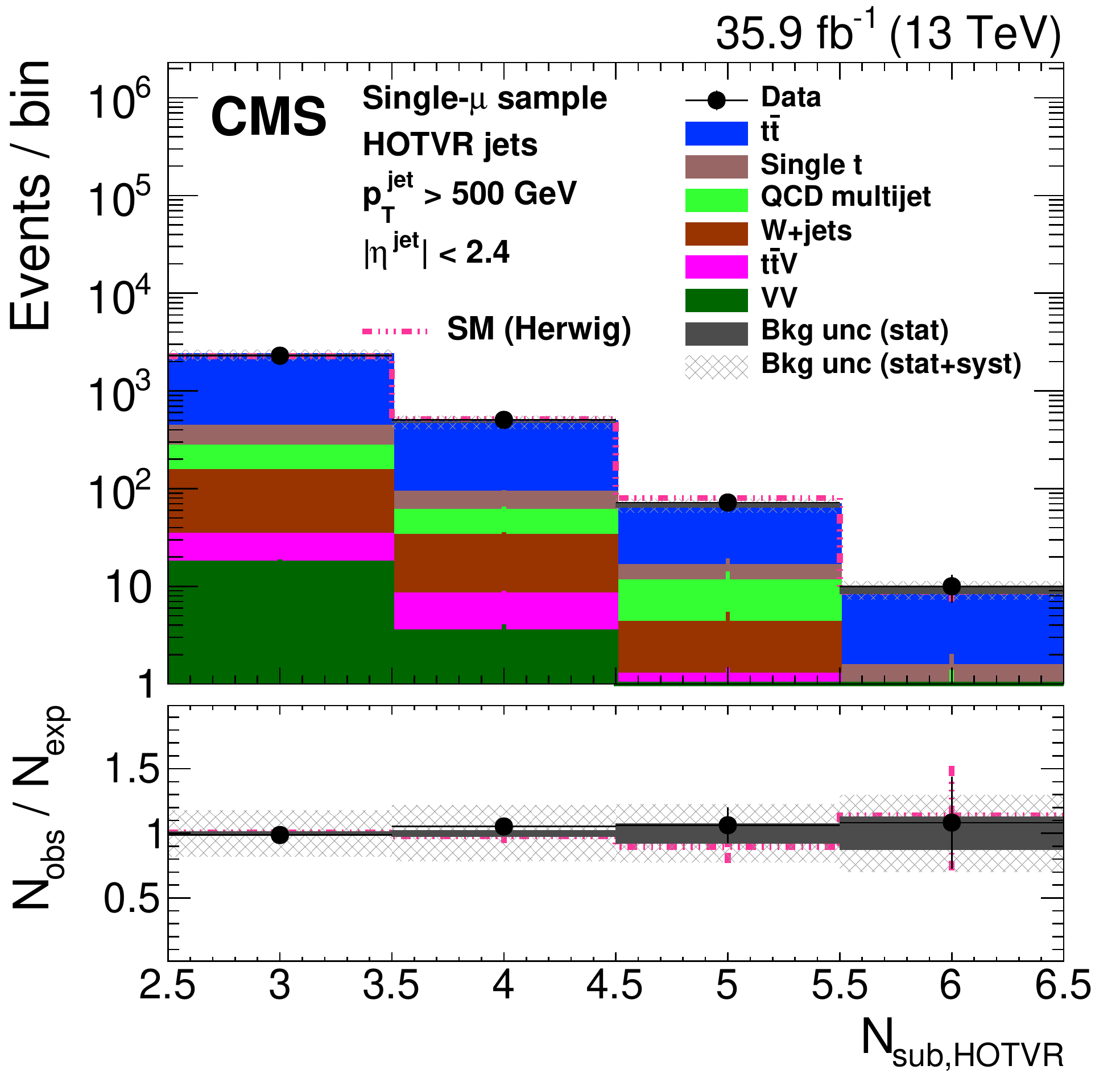}
  \caption{\label{fig:tt1l_hotvr_pt500}Distribution of the main
  observables of the HOTVR algorithm, HOTVR jet \pt (upper left), $m_{\text{HOTVR}}$ (upper right),
  $m_{\text{min,HOTVR}}$ (lower left) and $N_{\text{sub,HOTVR}}$
  (lower right) in data and simulation in the single-$\mu$ signal sample, after
  applying a jet momentum cut $\pt > 500\GeV$. The pink line corresponds to
  the simulation distribution obtained using the alternative \ttbar sample.
  The background event yield is normalized to the total observed data yield.
  The lower panel shows the data to simulation ratio. The solid dark-gray (shaded light-gray)
  band corresponds to the total uncertainty (statistical uncertainty of the simulated samples),
  the pink line to the data to simulation ratio using the alternative \ttbar sample,
  and the vertical black lines correspond to the statistical
  uncertainty of the data. The vertical pink lines correspond to the
  statistical uncertainty of the alternative \ttbar sample.
  The distributions are weighted according to
  the top quark \pt weighting procedure described in the text.}
\end{figure}

\begin{figure}[hp!]
\centering
\includegraphics[width=0.45\textwidth]{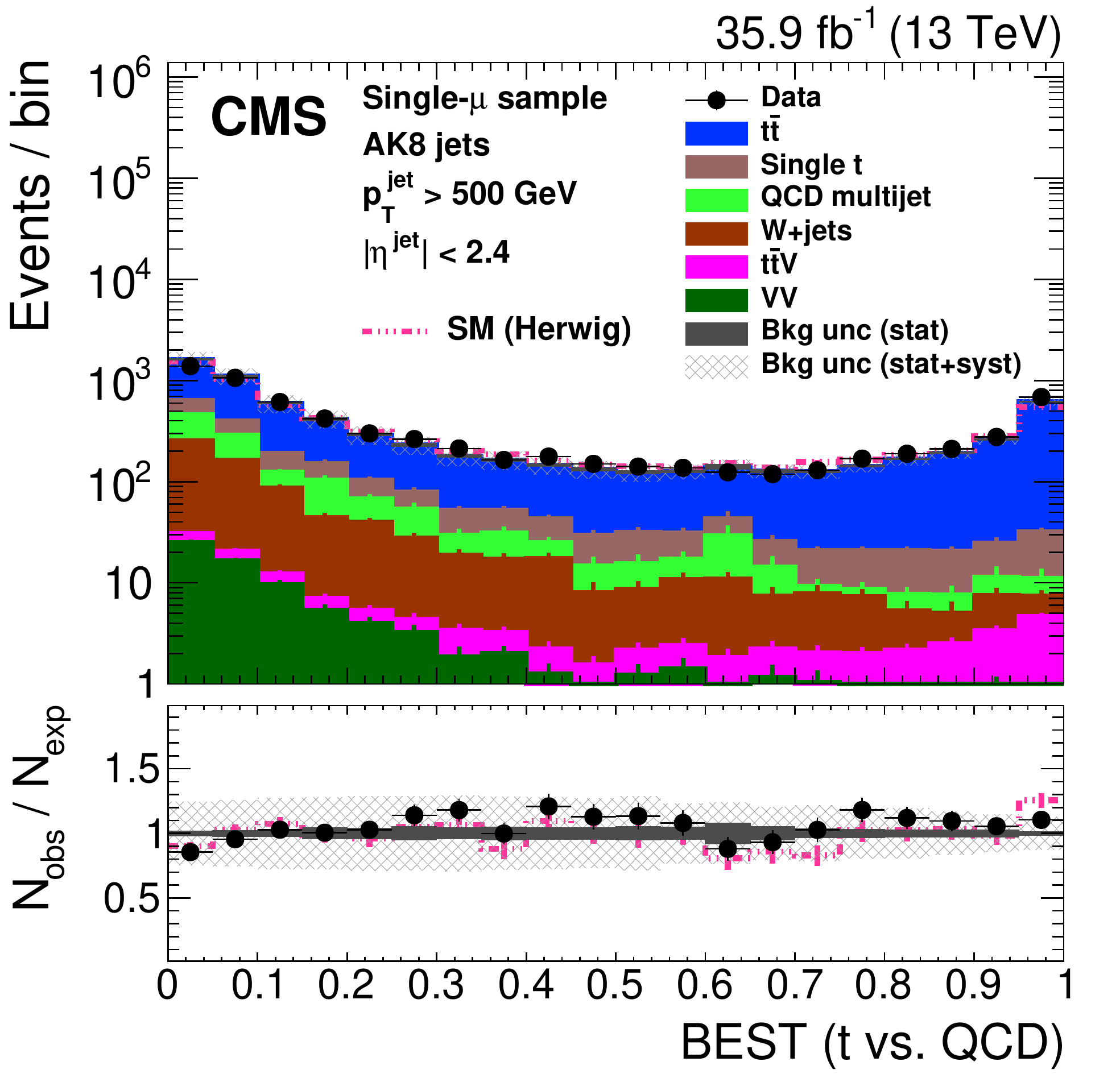}
\includegraphics[width=0.45\textwidth]{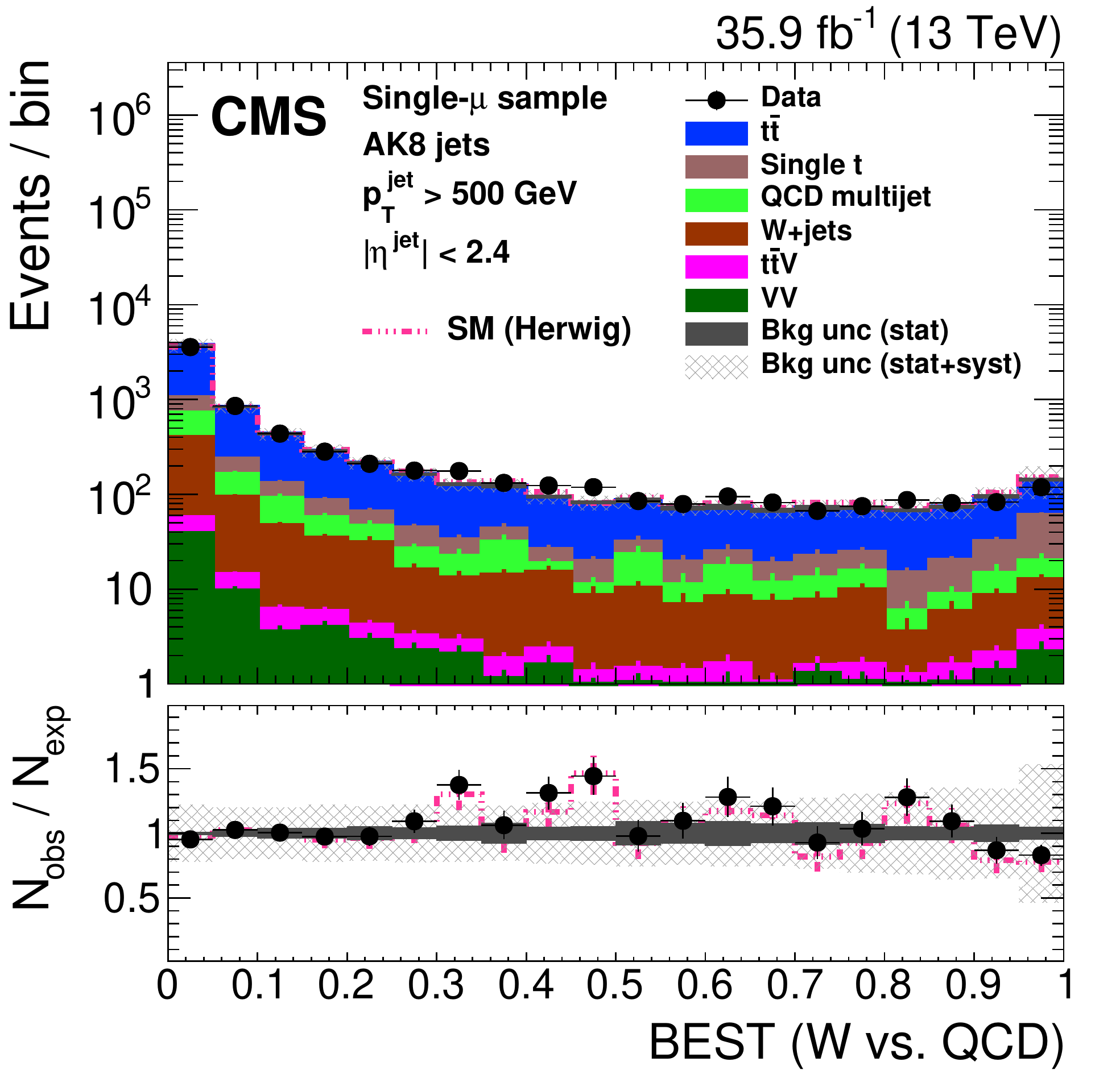}
\includegraphics[width=0.45\textwidth]{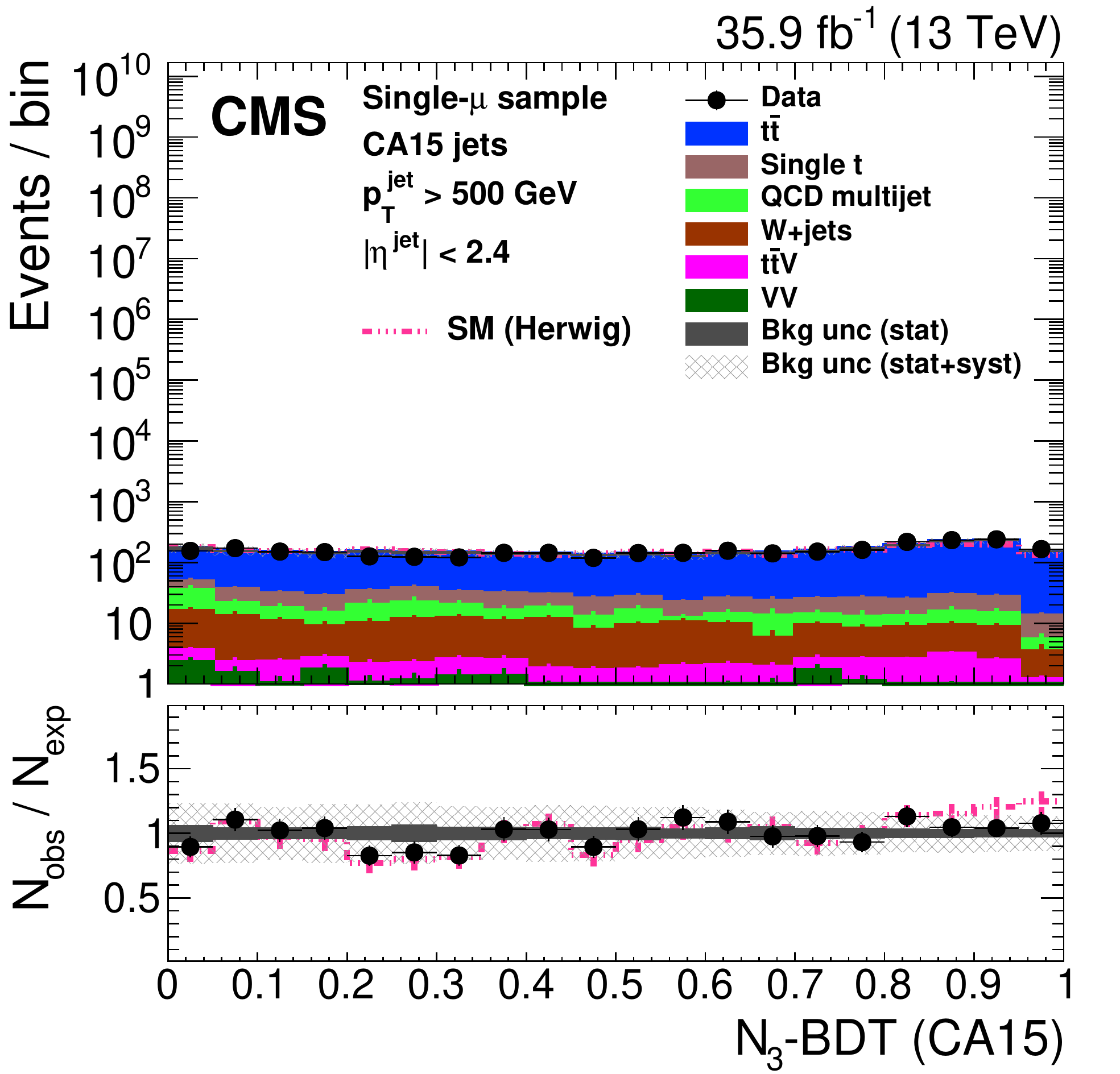}
\caption{\label{fig:tt1l_hlv_pt500}Distribution of the \PQt~quark
  (upper left) and \PW~boson (upper right) identification probabilities
  for the BEST algorithm, and the \ecftop discriminant in data and
  simulation in the single-$\mu$ signal sample, after applying a jet
  momentum cut $\pt > 500\GeV$. The pink line corresponds to
  the simulation distribution obtained using the alternative \ttbar sample.
  The background event yield is normalized to the total observed data yield.
  The lower panel shows the data to simulation ratio. The solid dark-gray (shaded light-gray)
  band corresponds to the total uncertainty (statistical uncertainty of the simulated samples),
  the pink line to the data to simulation ratio using the alternative \ttbar sample,
  and the vertical black lines correspond to the statistical
  uncertainty of the data. The vertical pink lines correspond to the
  statistical uncertainty of the alternative \ttbar sample.  
  The distributions are weighted according to 
  the top quark \pt weighting procedure described in the text.}
\end{figure}

\begin{figure}[hp!]
\centering
\includegraphics[width=0.38\textwidth]{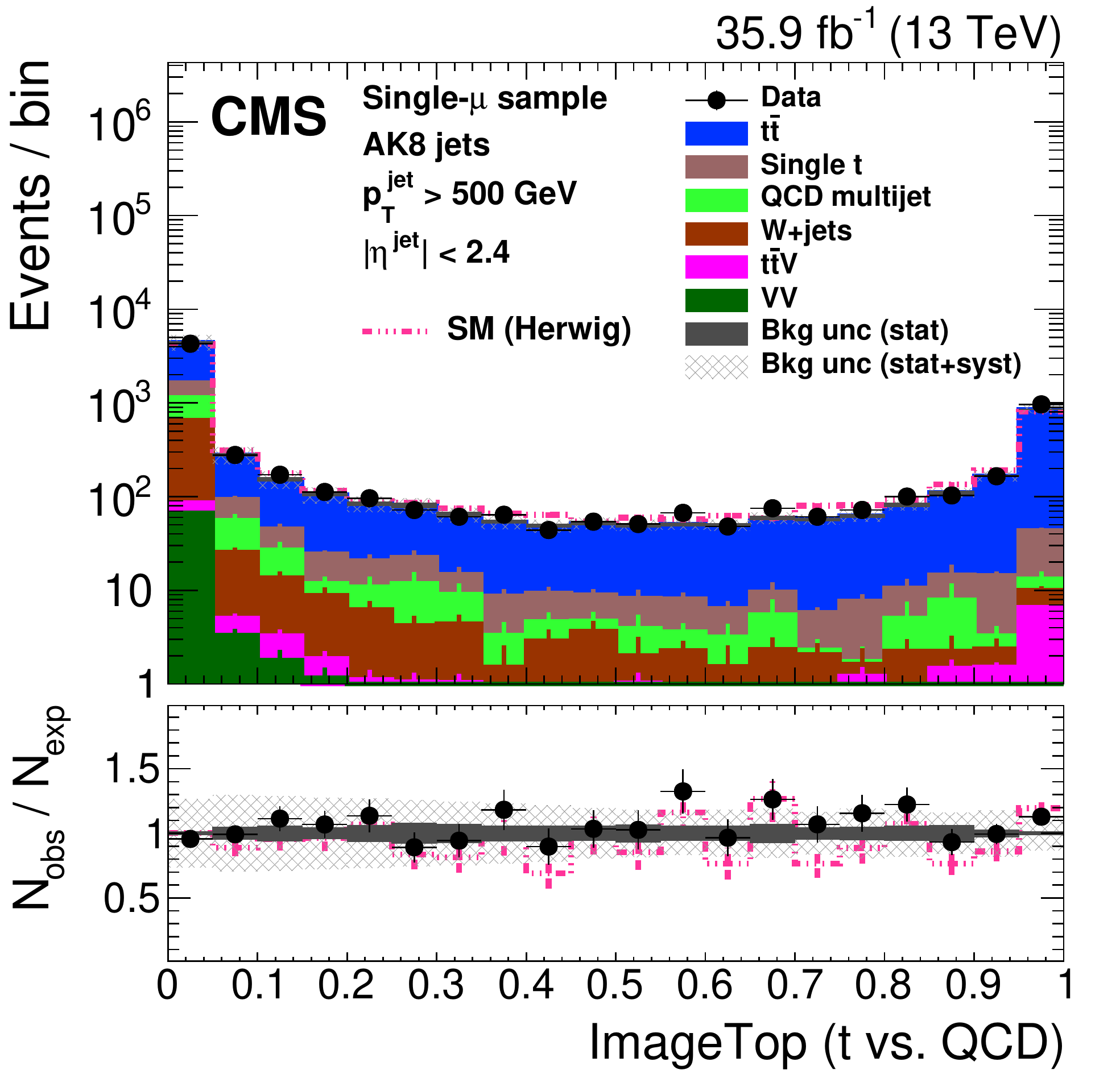}
\includegraphics[width=0.38\textwidth]{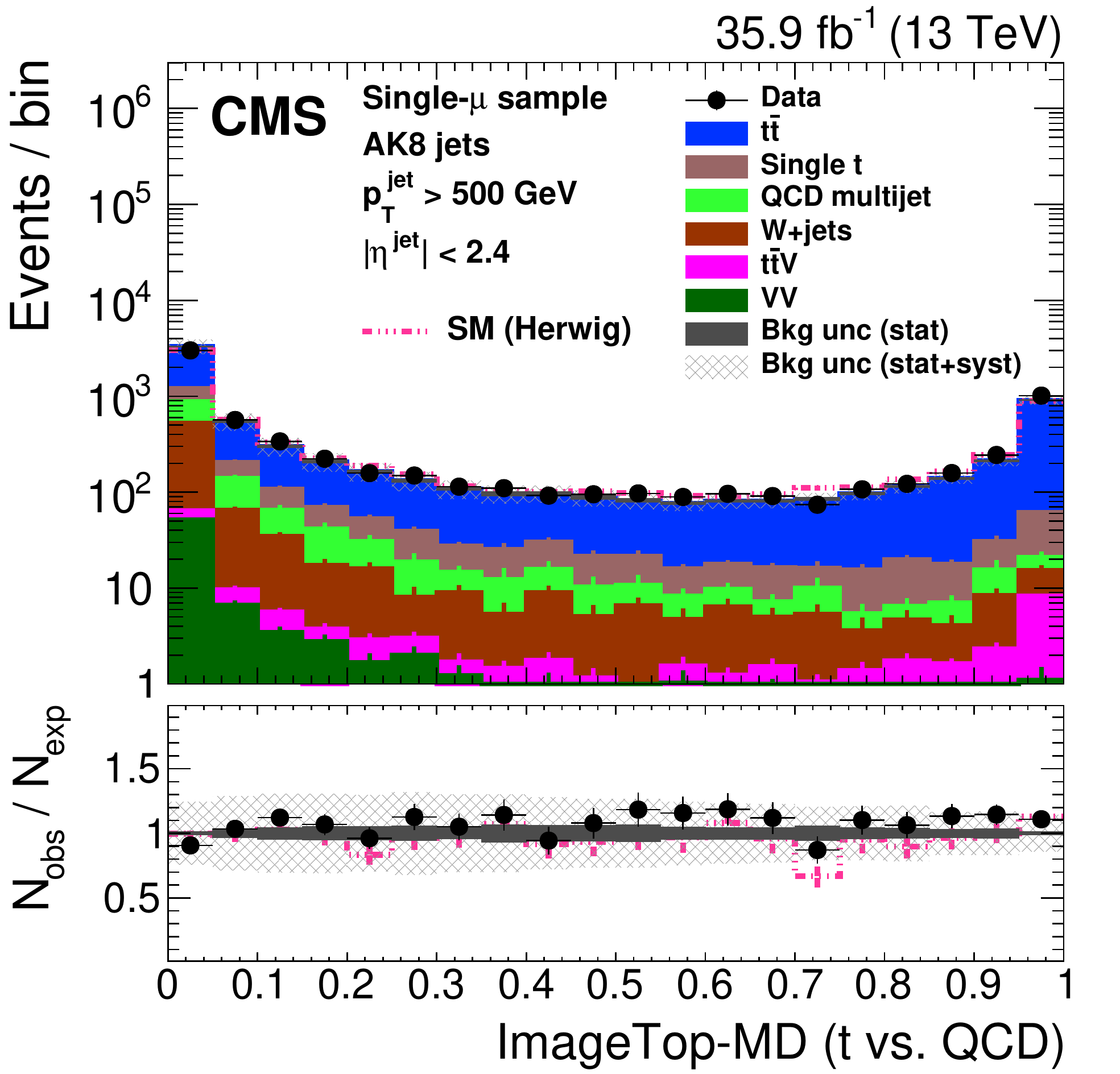}\\
\includegraphics[width=0.38\textwidth]{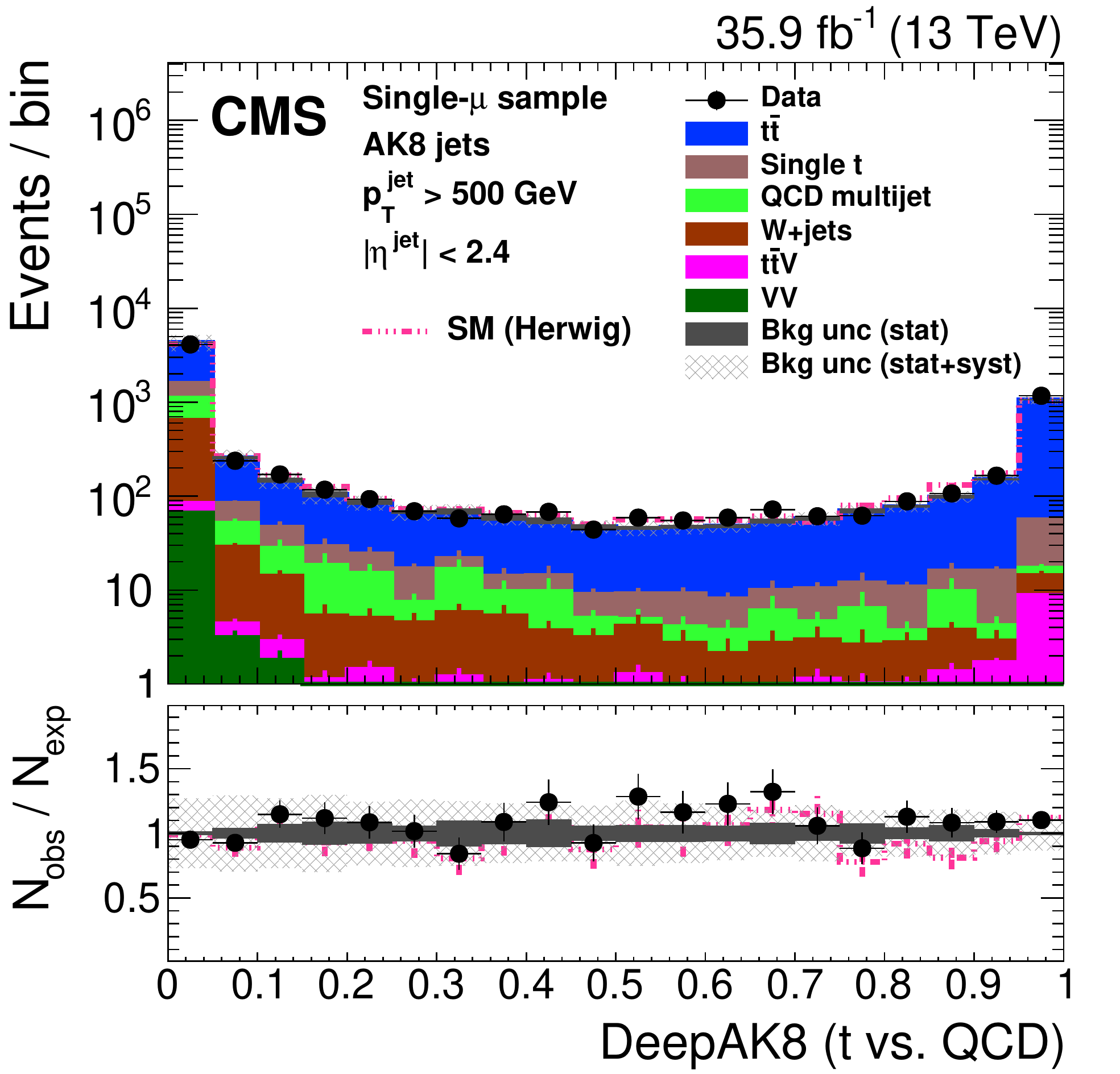}
\includegraphics[width=0.38\textwidth]{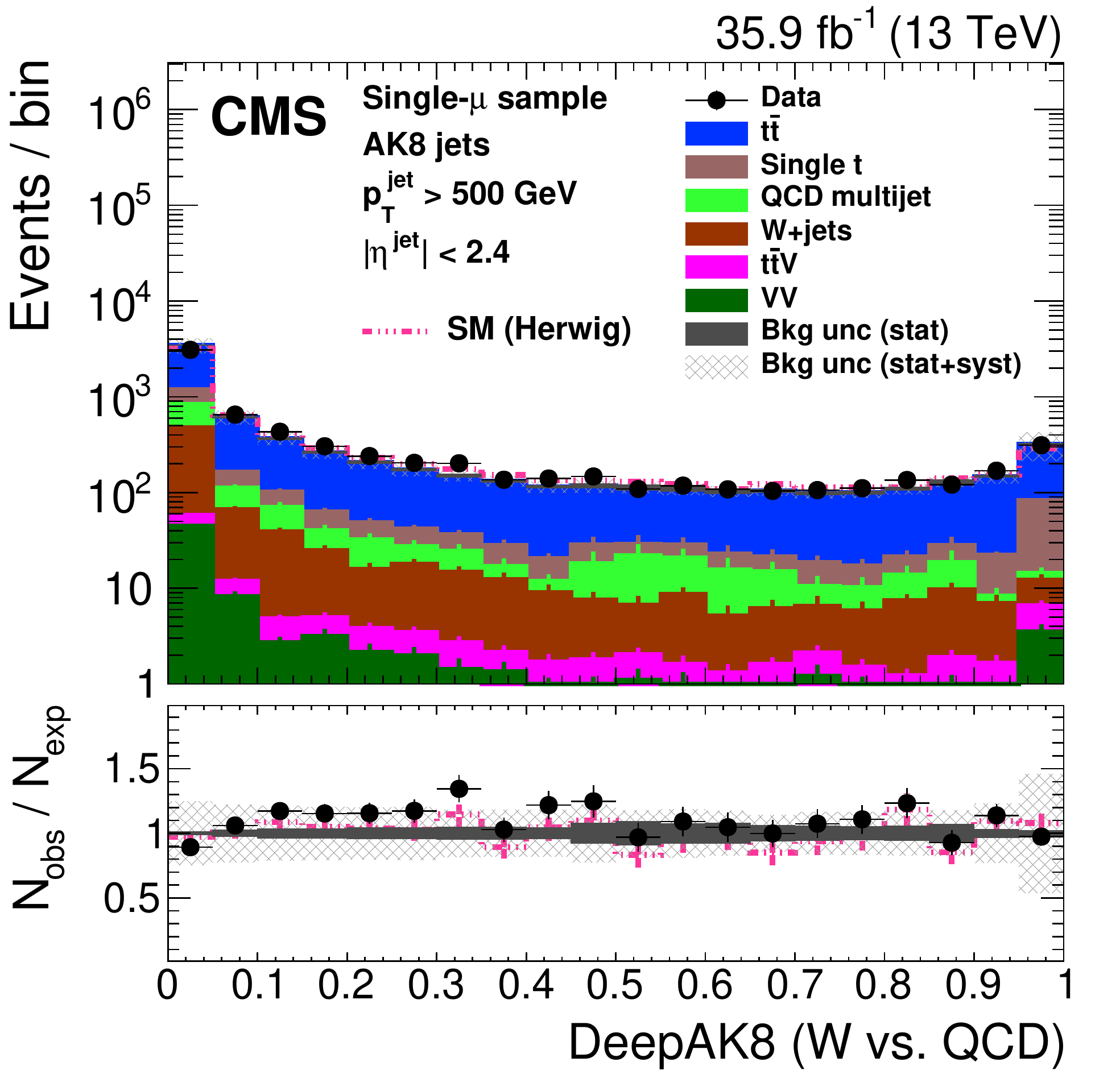} \\
\includegraphics[width=0.38\textwidth]{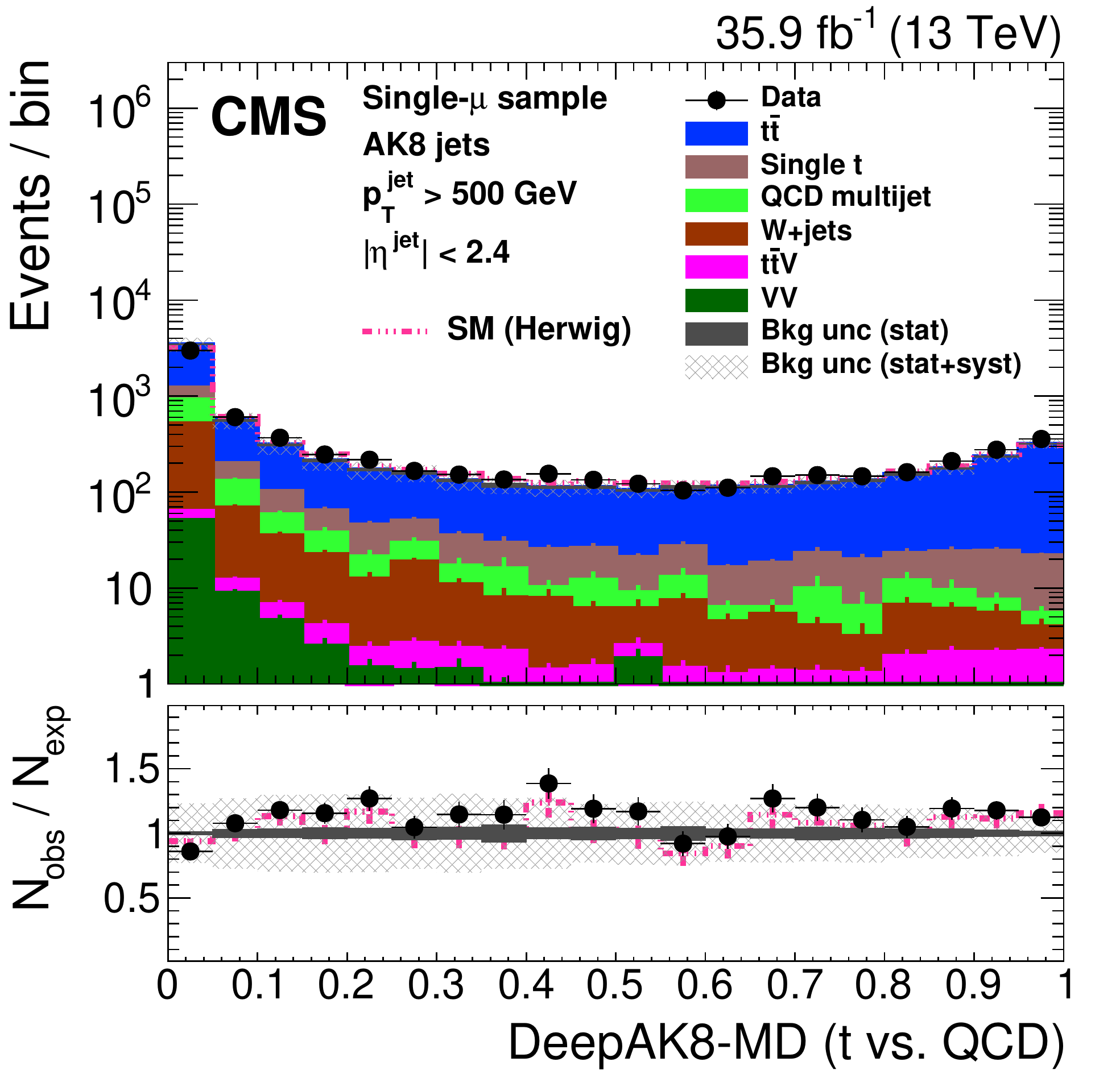}
\includegraphics[width=0.38\textwidth]{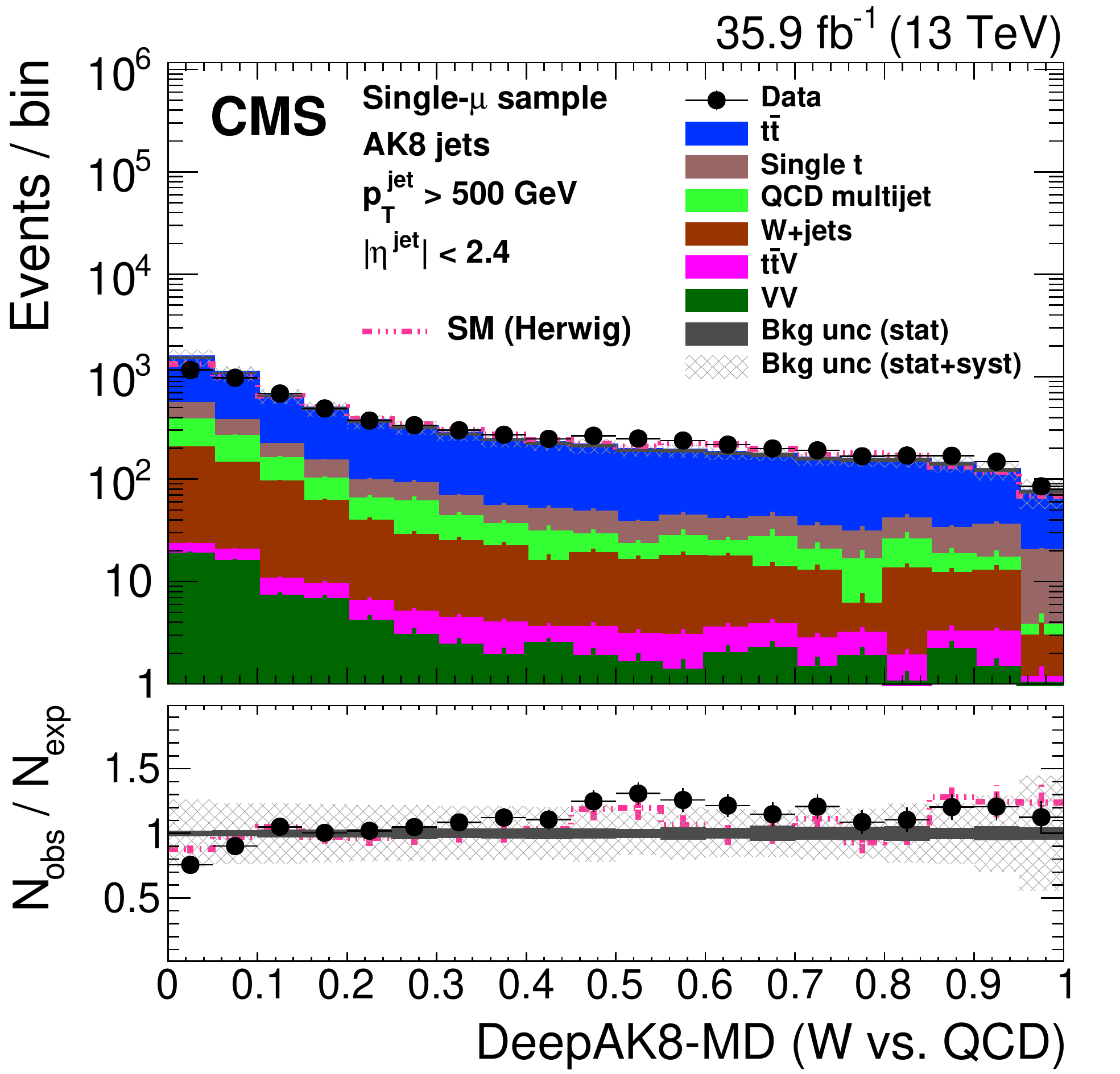} \\
\caption{\label{fig:tt1l_low_pt500}Distribution of the ImageTop (upper
  left) and ImageTop-MD (upper right) discriminant in data and simulation in
  the single-$\mu$ sample. The plots in the middle row show the
  \PQt~quark (left) and \PW~boson (right)  identification
  probabilities in data and simulation for the DeepAK8 algorithm after
  applying a jet momentum cut $\pt > 500\GeV$. The corresponding plots
  for DeepAK8-MD are displayed in the lower row. The pink line corresponds to
  the simulation distribution obtained using the alternative \ttbar sample.
  The background event yield is normalized to the total observed data yield.
  The lower panel shows the data to simulation ratio. The solid dark-gray (shaded light-gray)
  band corresponds to the total uncertainty (statistical uncertainty of the simulated samples),
  the pink line to the data to simulation ratio using the alternative \ttbar sample,
  and the vertical black lines correspond to the statistical
  uncertainty of the data. The vertical pink lines correspond to the
  statistical uncertainty of the alternative \ttbar sample.
The distributions are weighted according to
  the top quark \pt weighting procedure described in the text.}
\end{figure}

Because the selection applied to events shown in
Figs.~\ref{fig:tt1l_presel}--\ref{fig:tt1l_low} results in a sample
with low purity of fully merged \PQt quark decay products, we also study the
same distributions after applying a tighter requirement on the jet
momenta: $\pt > 500\GeV$.  This selection results in a sample
consisting of a higher fraction of fully merged \PQt quark jets,
relative to the jet component from the decay of a boosted \PW boson jet.
Figures~\ref{fig:tt1l_presel_pt500}-\ref{fig:tt1l_low_pt500} show the same
distributions for this high-\pt selection.

The total background yield
is normalized to the observed number of data events. The systematic
uncertainties discussed in Section~\ref{sec:systematics} are also
considered and are shown via the shaded dark-grey band in the
figures. Overall, the shapes in data are compatible with the
expectation from simulation within uncertainties for all the
algorithms.

\subsection{Misidentification probability in data}

The misidentification probability of the algorithms is studied in the
dijet and single-$\gamma$ data samples. The two samples differ
in the relative fraction of light-flavor quarks and gluons in the final
state. To study
the dependence of the misidentification probability on the choice of
the event generator and the parton showering scheme, we consider two
different simulated samples to model the QCD multijet background. The
nominal sample uses \MADGRAPH for the event generation and \PYTHIA (P8)
for the parton showering and hadronization, whereas the alternative
sample uses \HERWIGpp for event generation and the
modeling of the parton showering. More information on the generation
of these samples is discussed in Section~\ref{sec:samples}.
The total SM contribution  estimated using the
\HERWIGpp QCD multijet sample is referred to
as ``SM (Herwig)''.
As in Section~\ref{perf_in_data_1mu}, we will focus on results
using jets with $R=0.8$, unless otherwise stated. To account for
possible differences in the \pt distribution of the QCD multijet and
$\gamma+$jet simulated events, the total background yield is
weighted to match the \pt distribution in data, following the
procedure discussed in Section~\ref{sec:samples}.

\begin{figure}[hp!]
\centering
\includegraphics[width=0.38\textwidth]{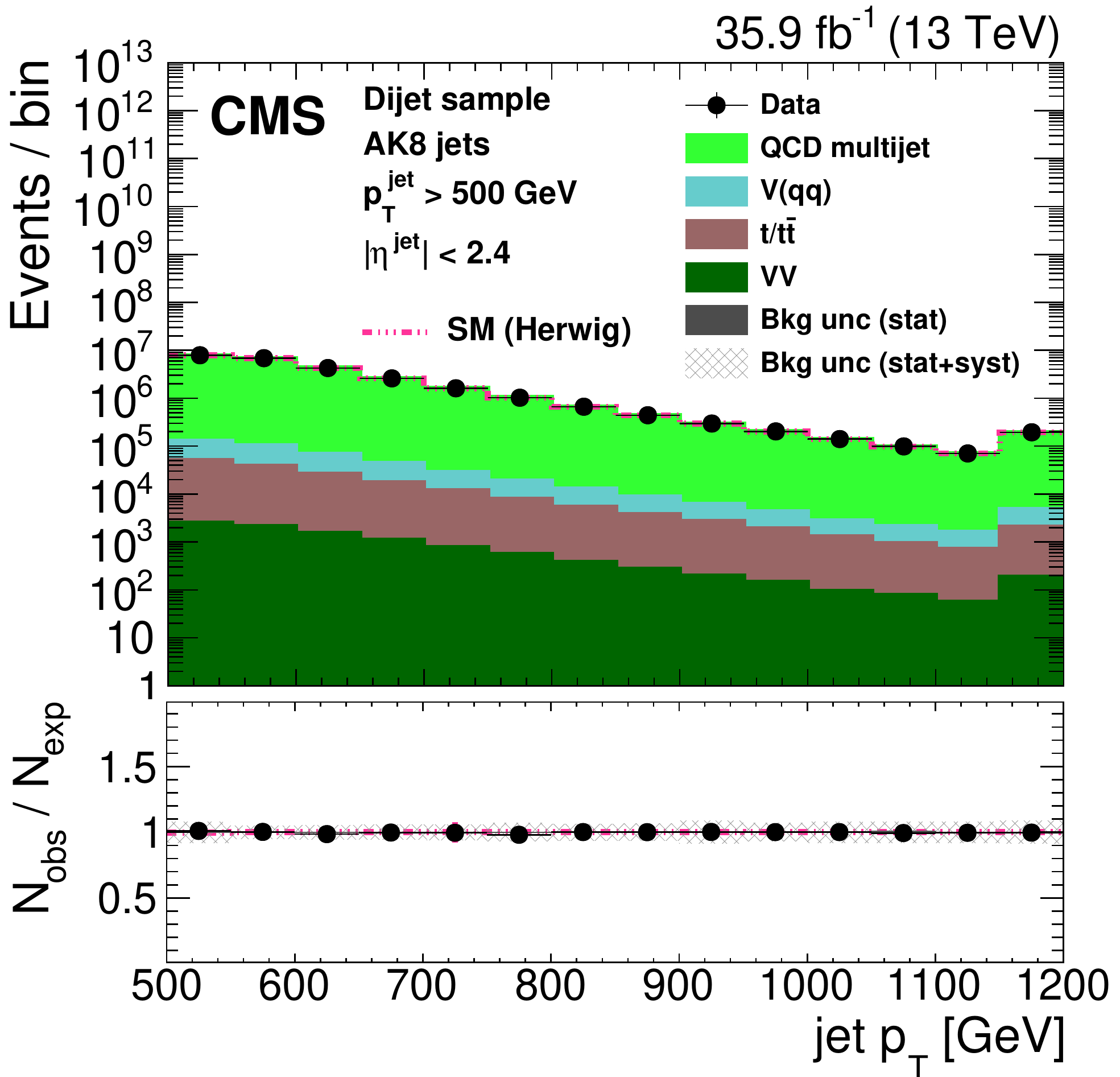}
\includegraphics[width=0.38\textwidth]{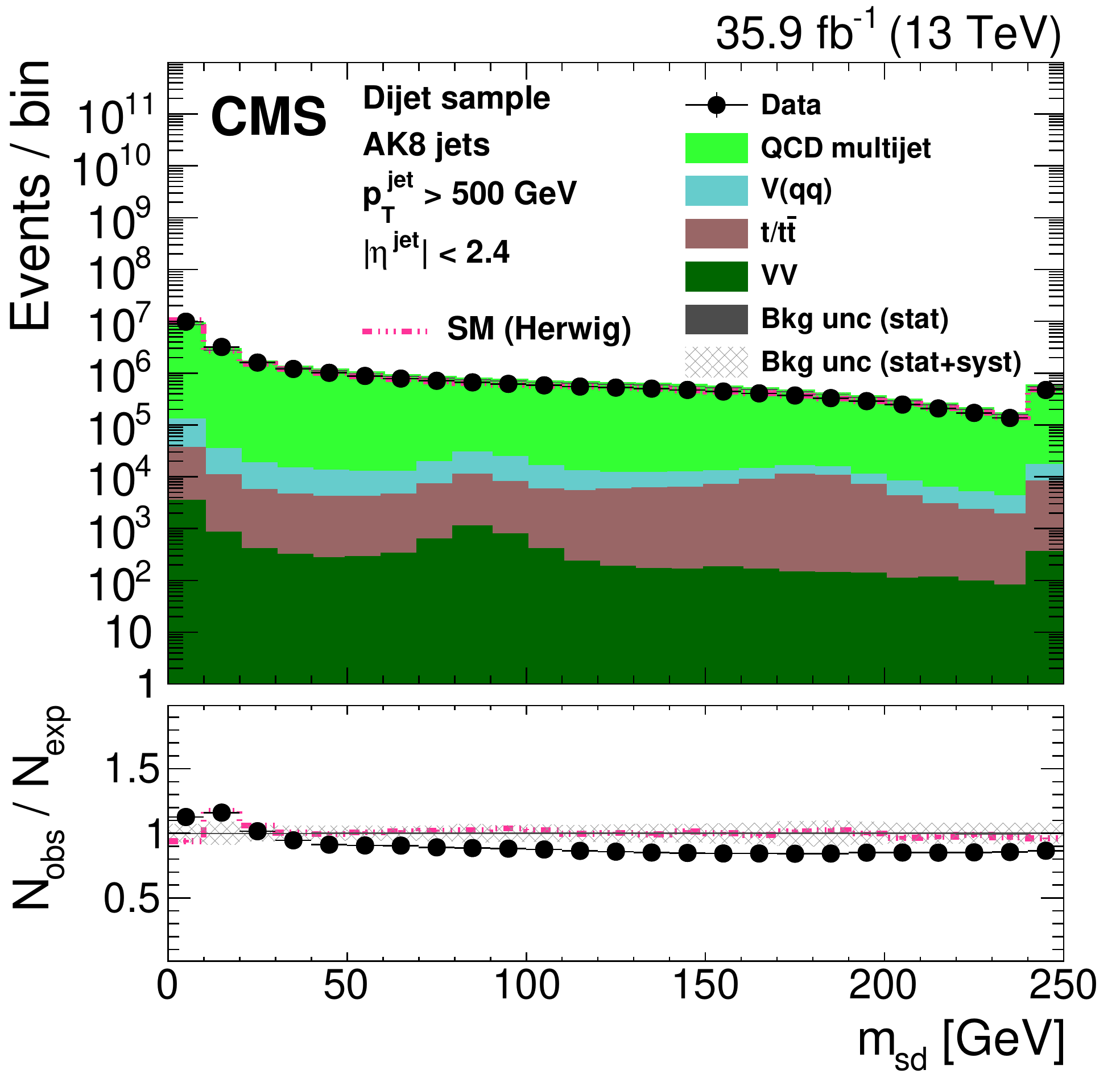}
\includegraphics[width=0.38\textwidth]{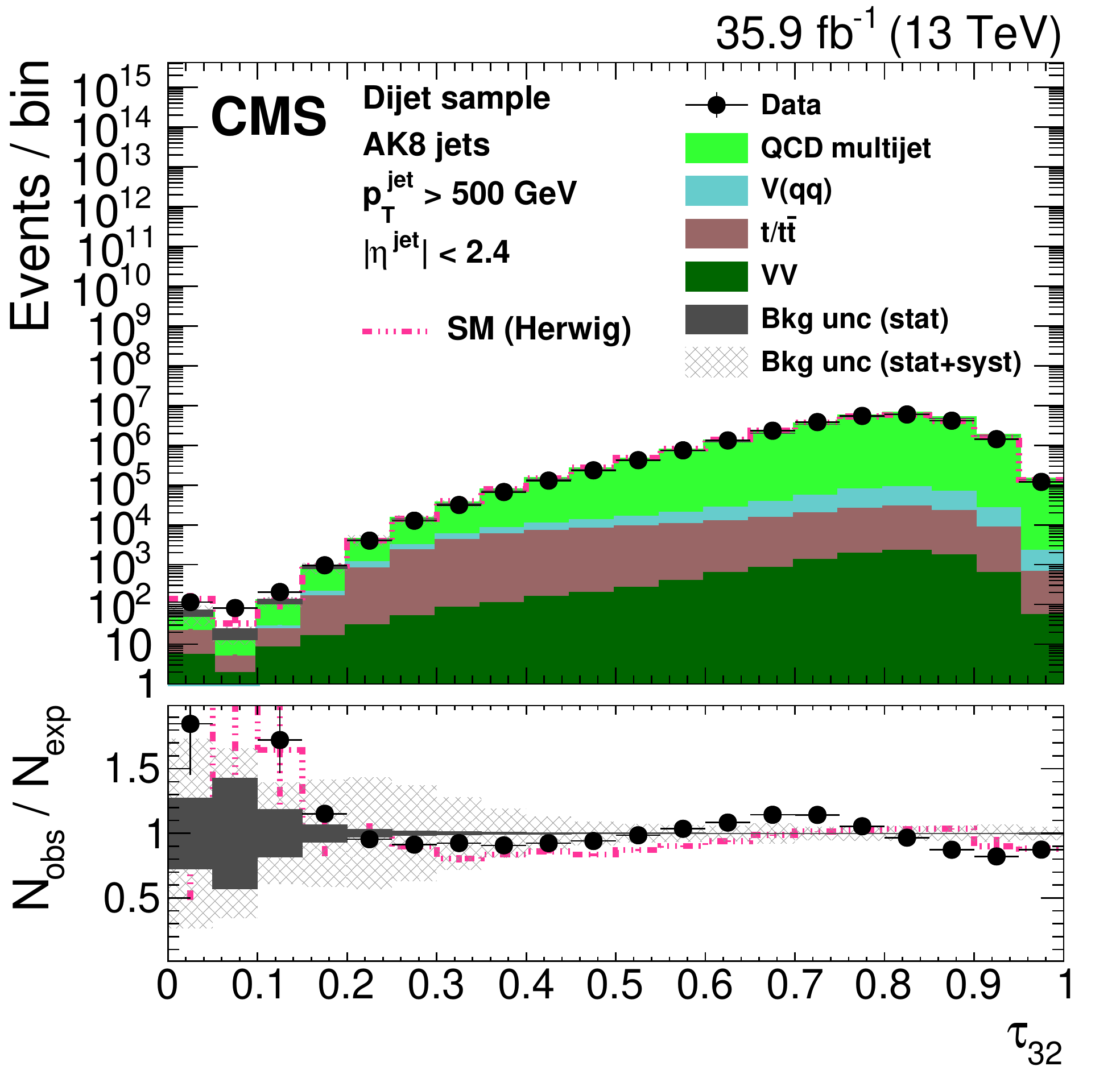}
\includegraphics[width=0.38\textwidth]{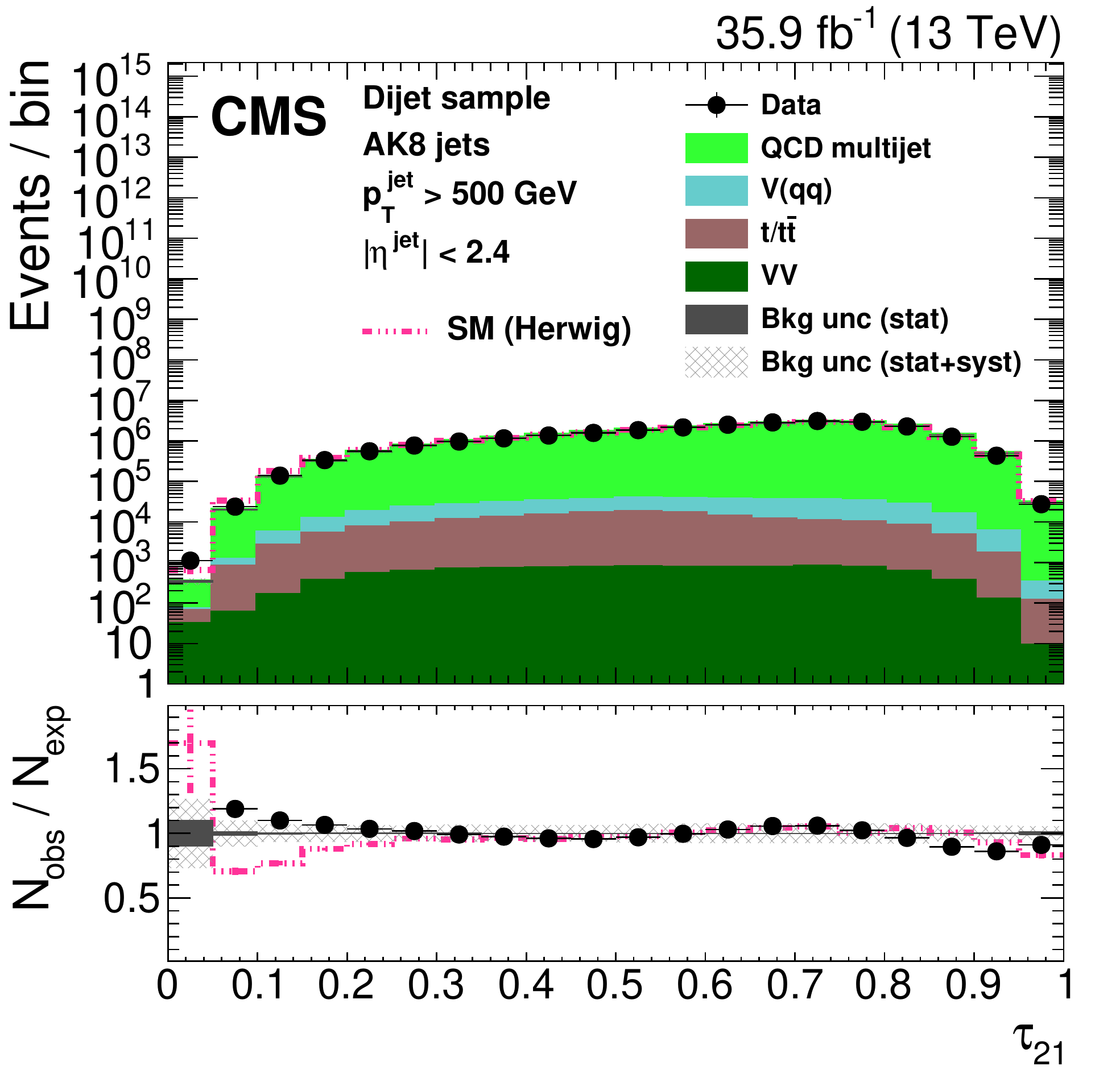}
\includegraphics[width=0.38\textwidth]{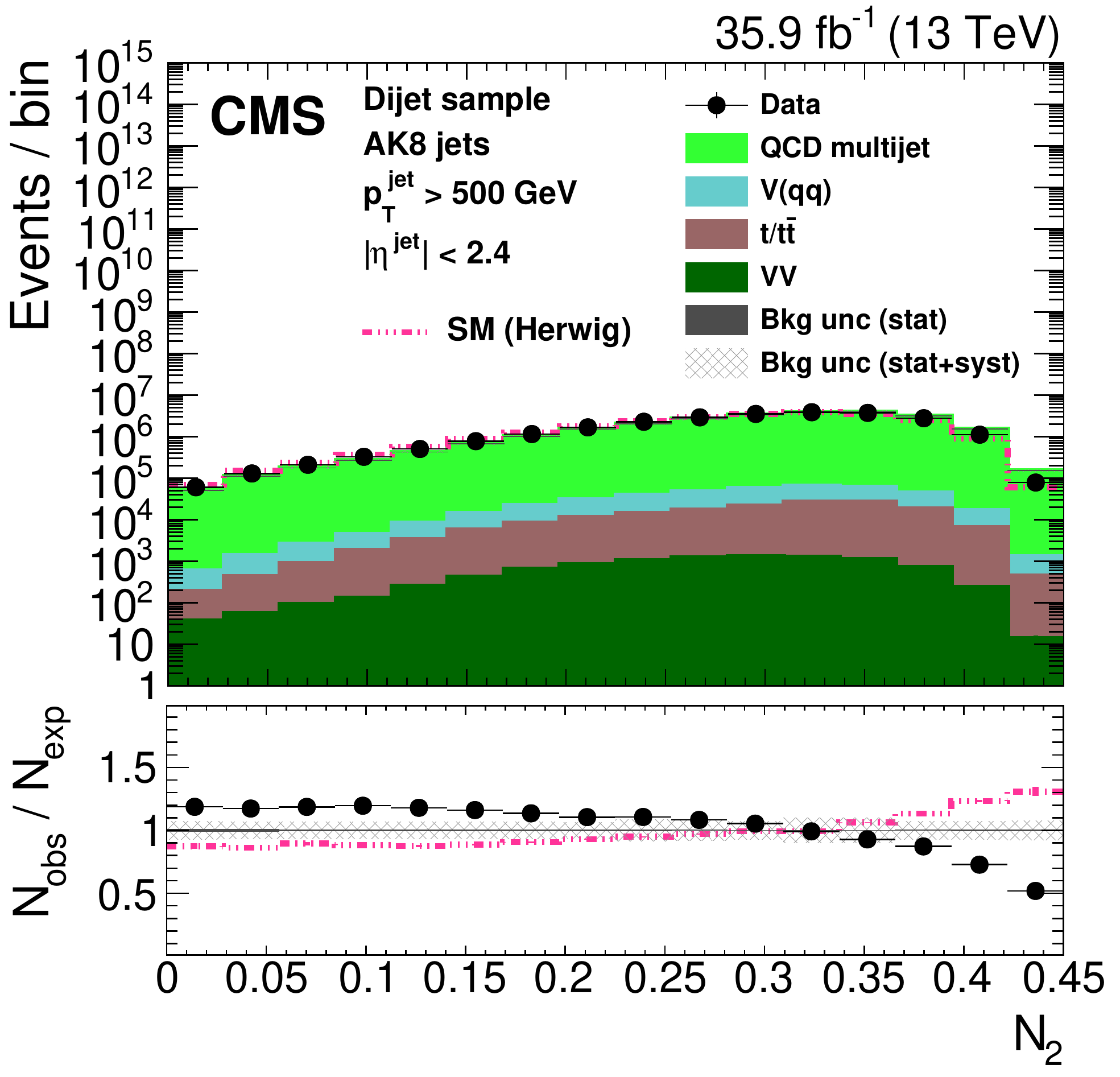}
\includegraphics[width=0.38\textwidth]{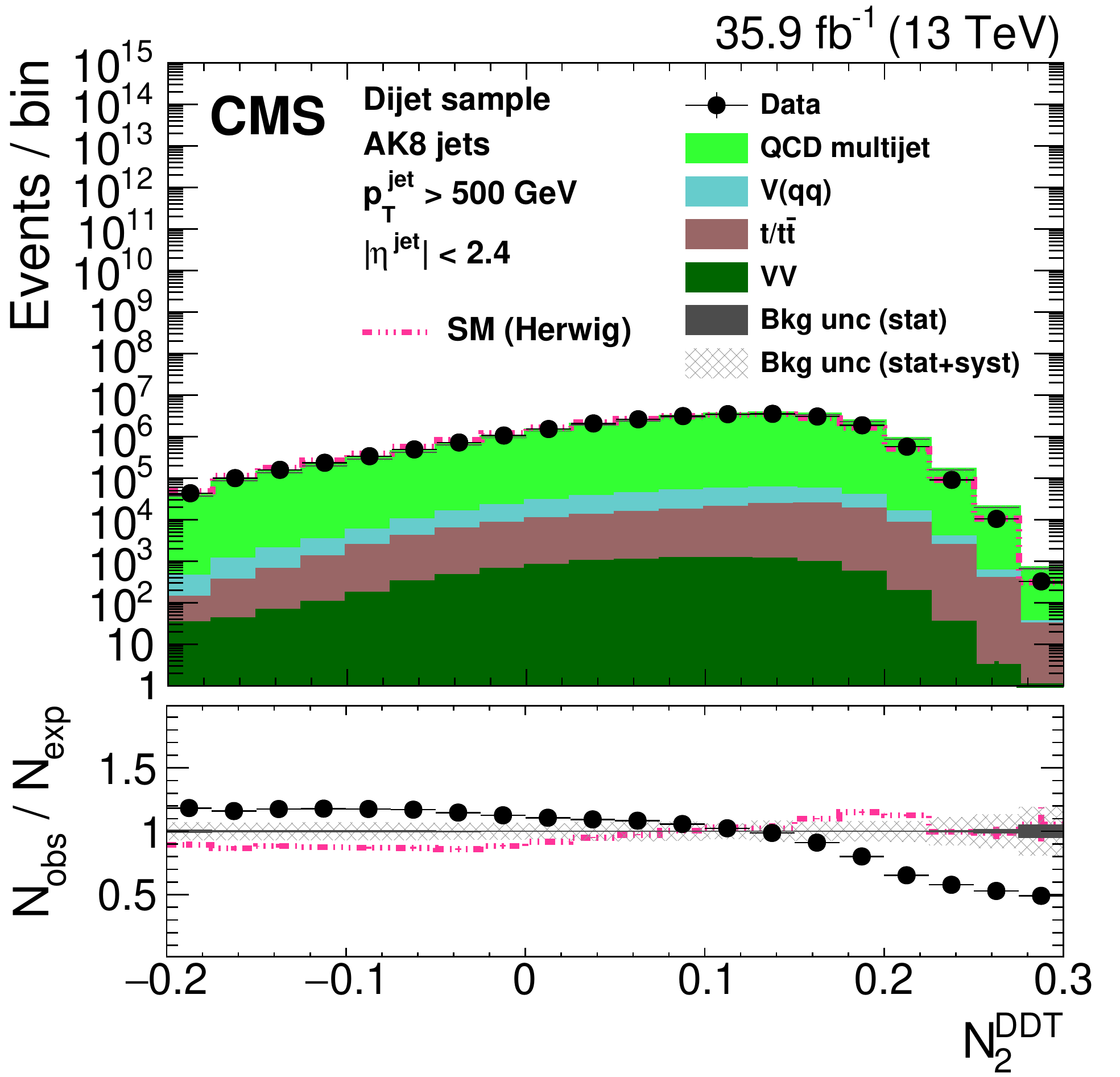}
\caption{\label{fig:qcd_presel}Distribution of the jet \pt (upper left),
  the jet mass, \msd (upper right), the $N$-subjettiness ratios
  $\tauthreetwo$ (middle left) and $\tautwoone$ (middle right), and
  the $N_2$ (lower left) and $N_{2}^{\text{DDT}}$ (lower right) in data and
  simulation in the dijet sample. The pink 
line corresponds to the simulation distribution obtained using the
  alternative QCD multijet sample. The background event yield is normalized to the total observed data yield.
  The lower panel shows the data to simulation ratio. The solid dark-gray (shaded light-gray)
  band corresponds to the total uncertainty (statistical uncertainty of the simulated samples),
  the pink line to the data to simulation ratio using the alternative QCD multijet sample,
  and the vertical black lines correspond to the statistical
  uncertainty of the data. The vertical pink lines correspond to the
  statistical uncertainty of the alternative QCD multijet sample.
 The distributions are weighted so that the jet \pt
  distribution of the simulation matches the data. }
\end{figure}

The distributions of \msd, jet \pt, the $N$-subjettiness ratios
$\tauthreetwo$ and $\tautwoone$, and the $N_2$ and $N_{2}^{\text{DDT}}$,
in the dijet sample are displayed in
Fig.~\ref{fig:qcd_presel}. For this event selection, the shapes of
the \msd and the $N$-subjettiness ratios are described
well by simulation, whereas there is disagreement between data
and simulation for high values of $N_2$ and $N_{2}^{\text{DDT}}$.
A better description of the data, particularly for $N_{2}^{\text{DDT}}$,
is achieved with the \HERWIGpp QCD multijet sample, which hints that the
disagreement is related to the description of the parton shower.
For the other observables we observe
similar level of agreement between the two generators.

\begin{figure}[hp!]
\centering
\includegraphics[width=0.38\textwidth]{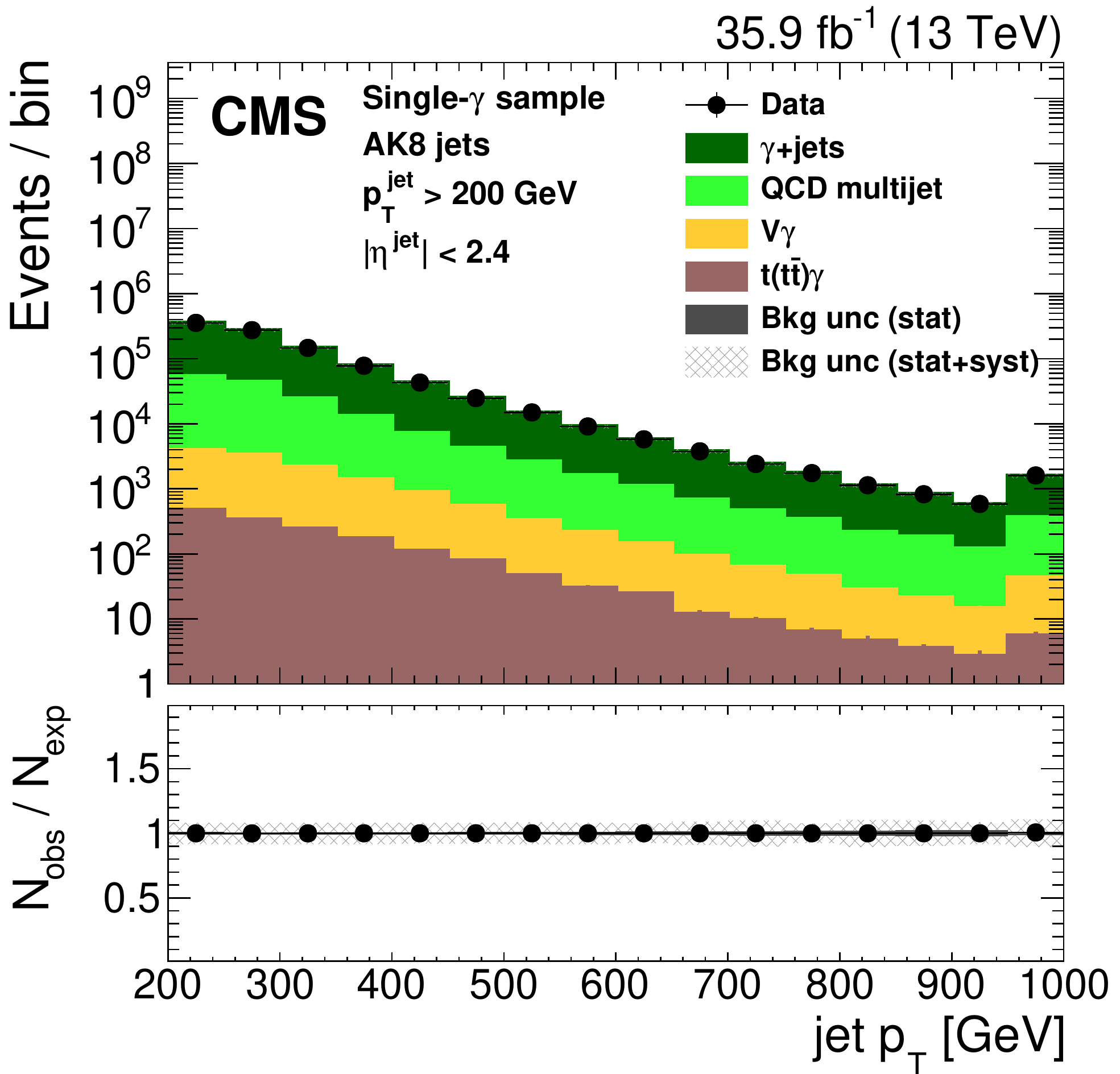}
\includegraphics[width=0.38\textwidth]{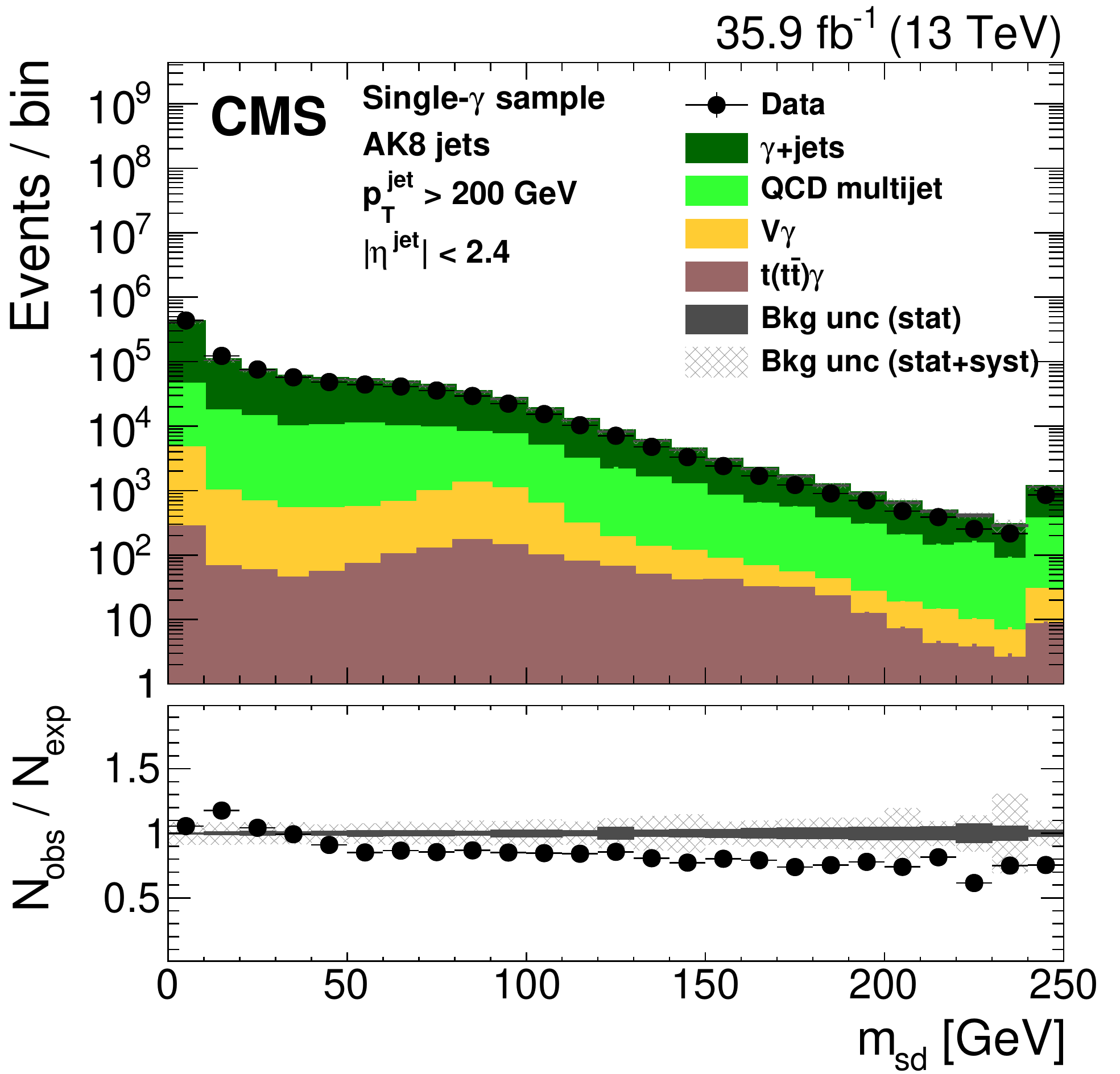}
\includegraphics[width=0.38\textwidth]{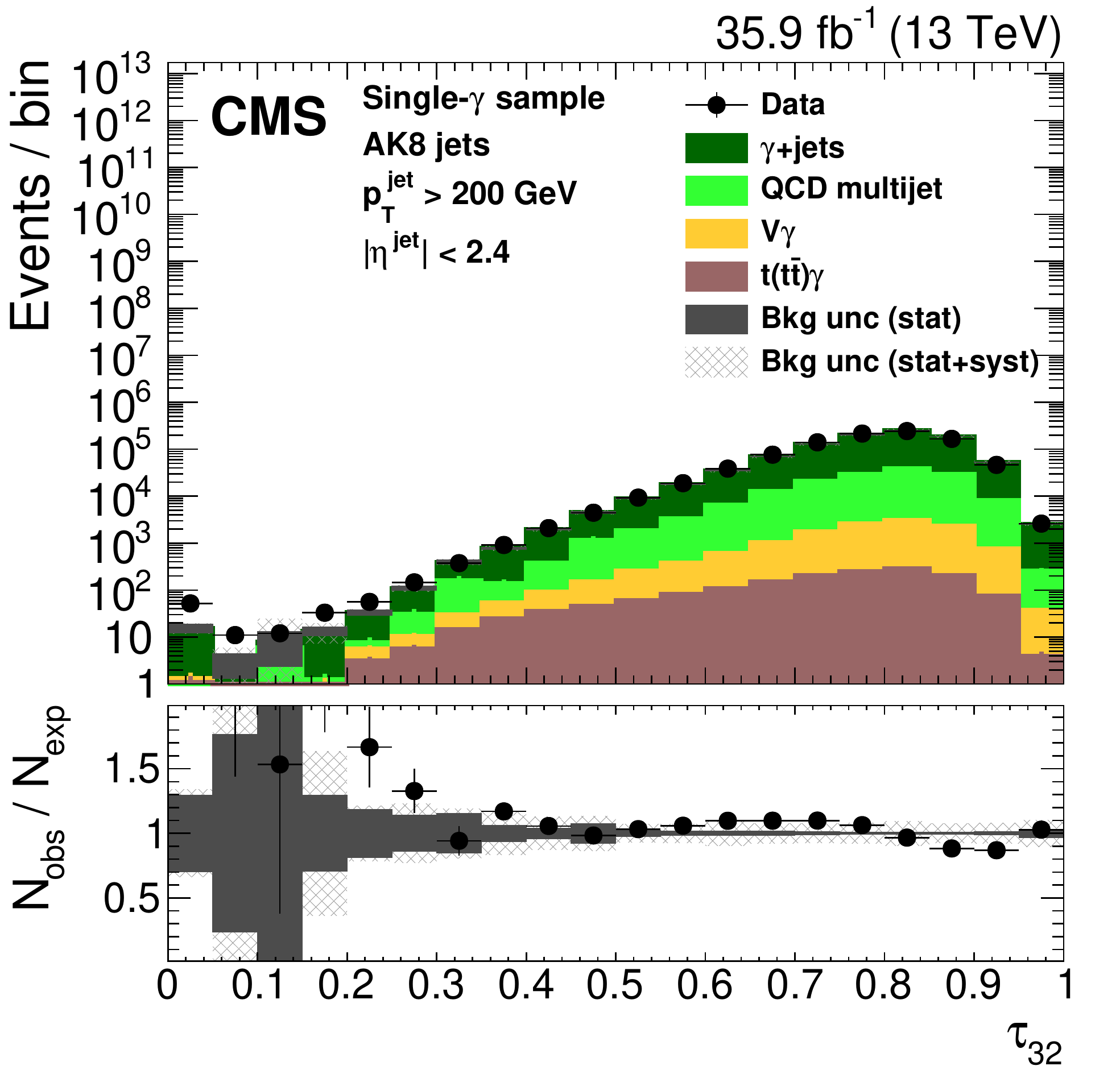}
\includegraphics[width=0.38\textwidth]{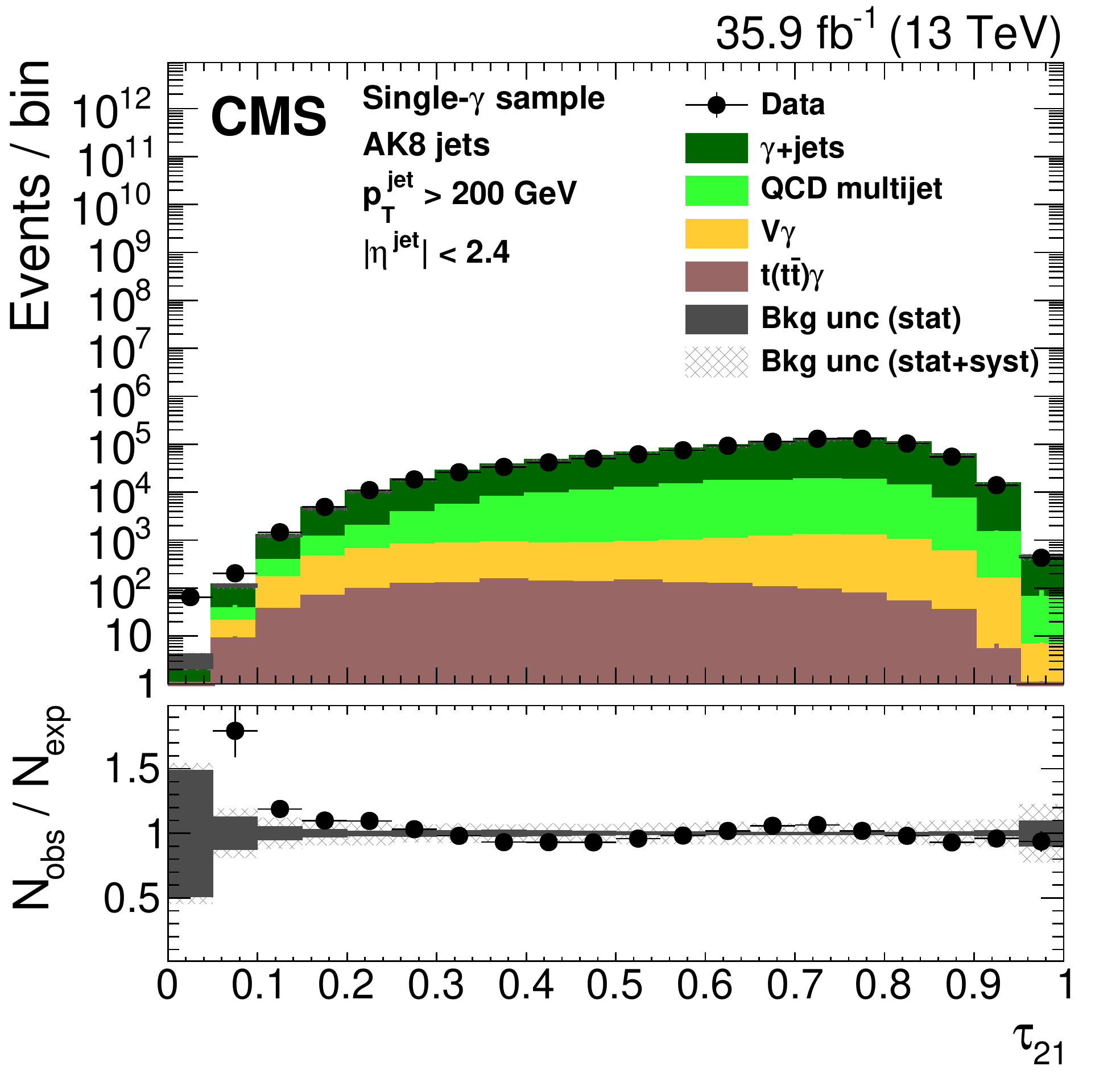}
\includegraphics[width=0.38\textwidth]{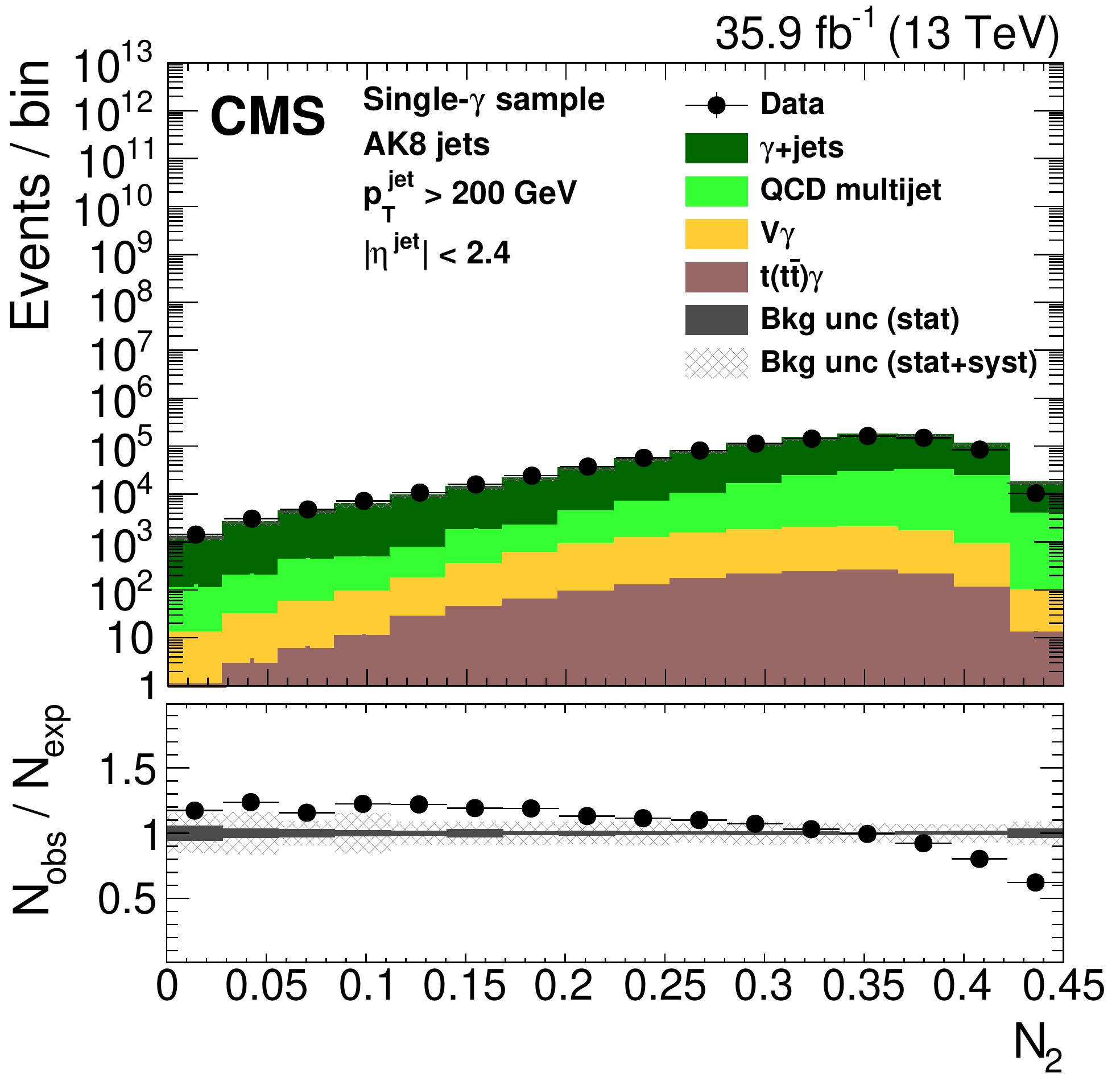}
\includegraphics[width=0.38\textwidth]{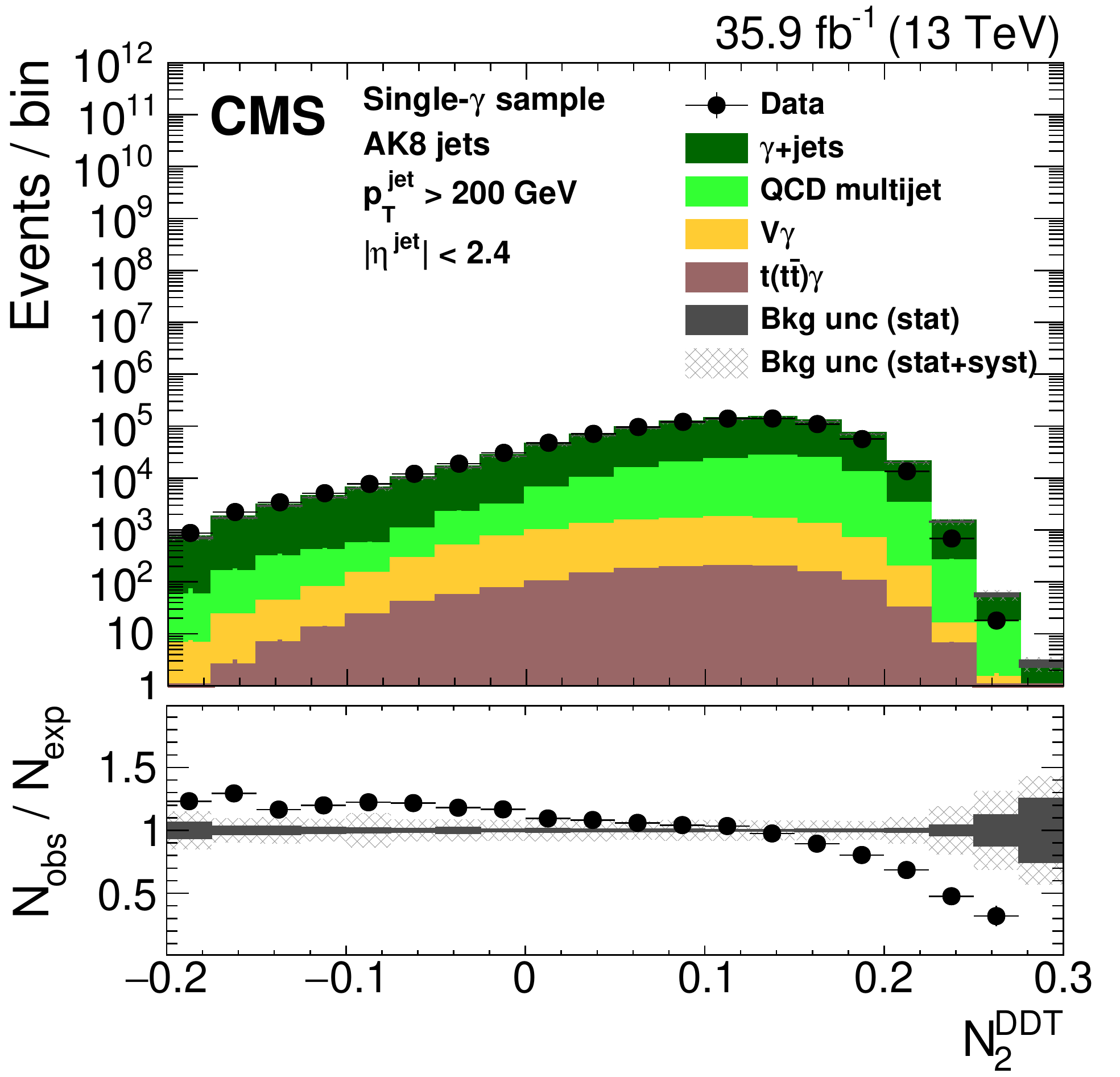}
\caption{\label{fig:pho_presel}Distribution of the jet \pt\
  (upper left), the jet mass, \msd (upper right), the $N$-subjettiness ratios
  $\tauthreetwo$ (middle left) and $\tautwoone$ (middle right), and
  the $N_2$ (lower left) and $N_{2}^{\text{DDT}}$ (lower right) in data and
  simulation in the single-$\gamma$ sample. The background event yield is normalized to the total observed data yield.
  The lower panel shows the data to simulation ratio. The solid dark-gray (shaded light-gray)
  band corresponds to the total uncertainty (statistical uncertainty of the simulated samples),
  and the vertical lines correspond to the statistical
  uncertainty of the data. The distributions are weighted so that the jet \pt
  distribution of the simulation matches the data.}
\end{figure}

The same set of variables is
presented in Fig.~\ref{fig:pho_presel} for the
single-$\gamma$ sample.
From previous measurements~\cite{Sirunyan:2018xdh},
the \msd agrees very well with simulation except at low
masses.
The modeling of the $N$-subjettiness and $N_2$ ratios is poorer in
the single-$\gamma$ sample.

\begin{figure}[hp!]
\centering
\includegraphics[width=0.45\textwidth]{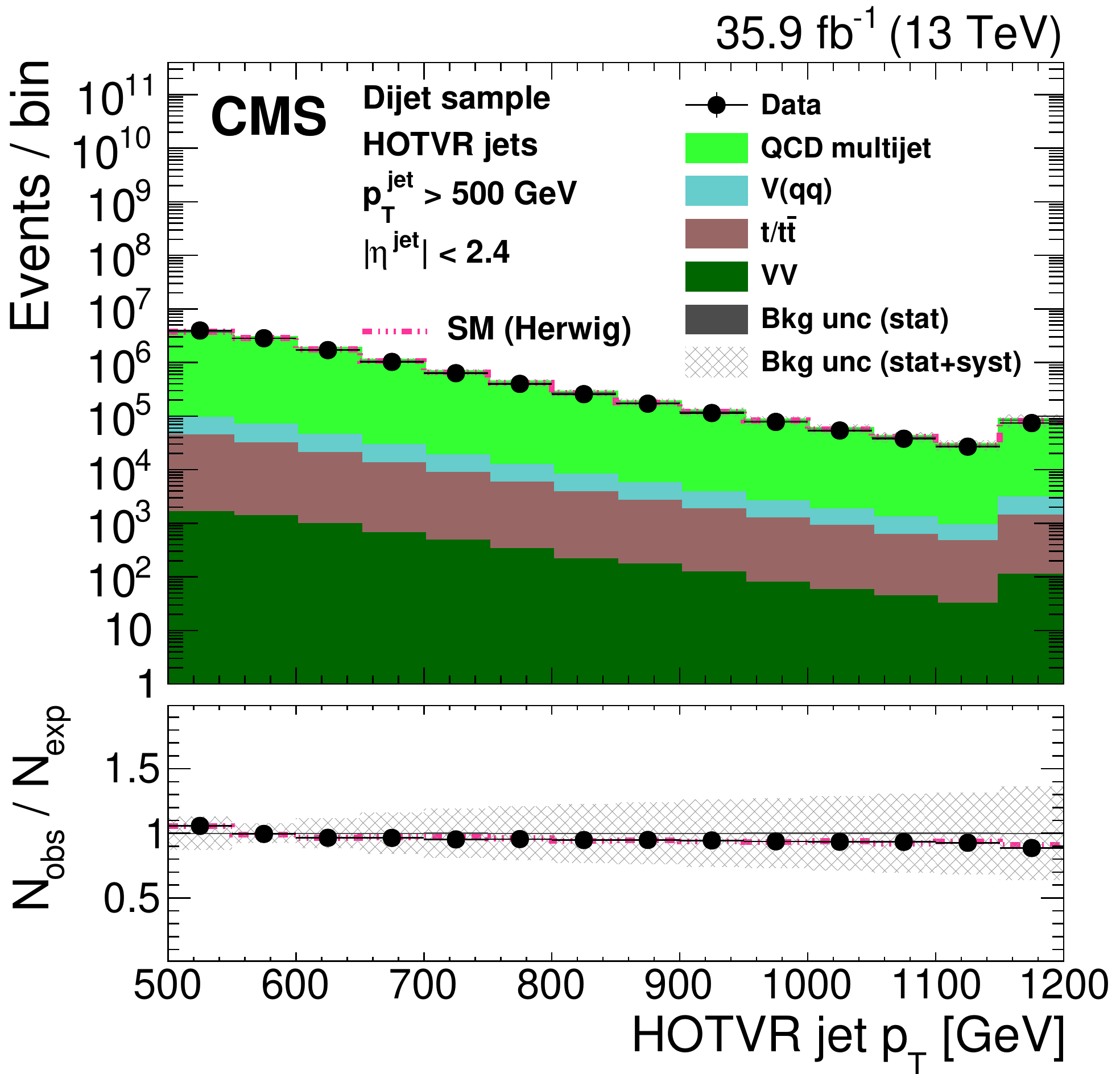}
\includegraphics[width=0.45\textwidth]{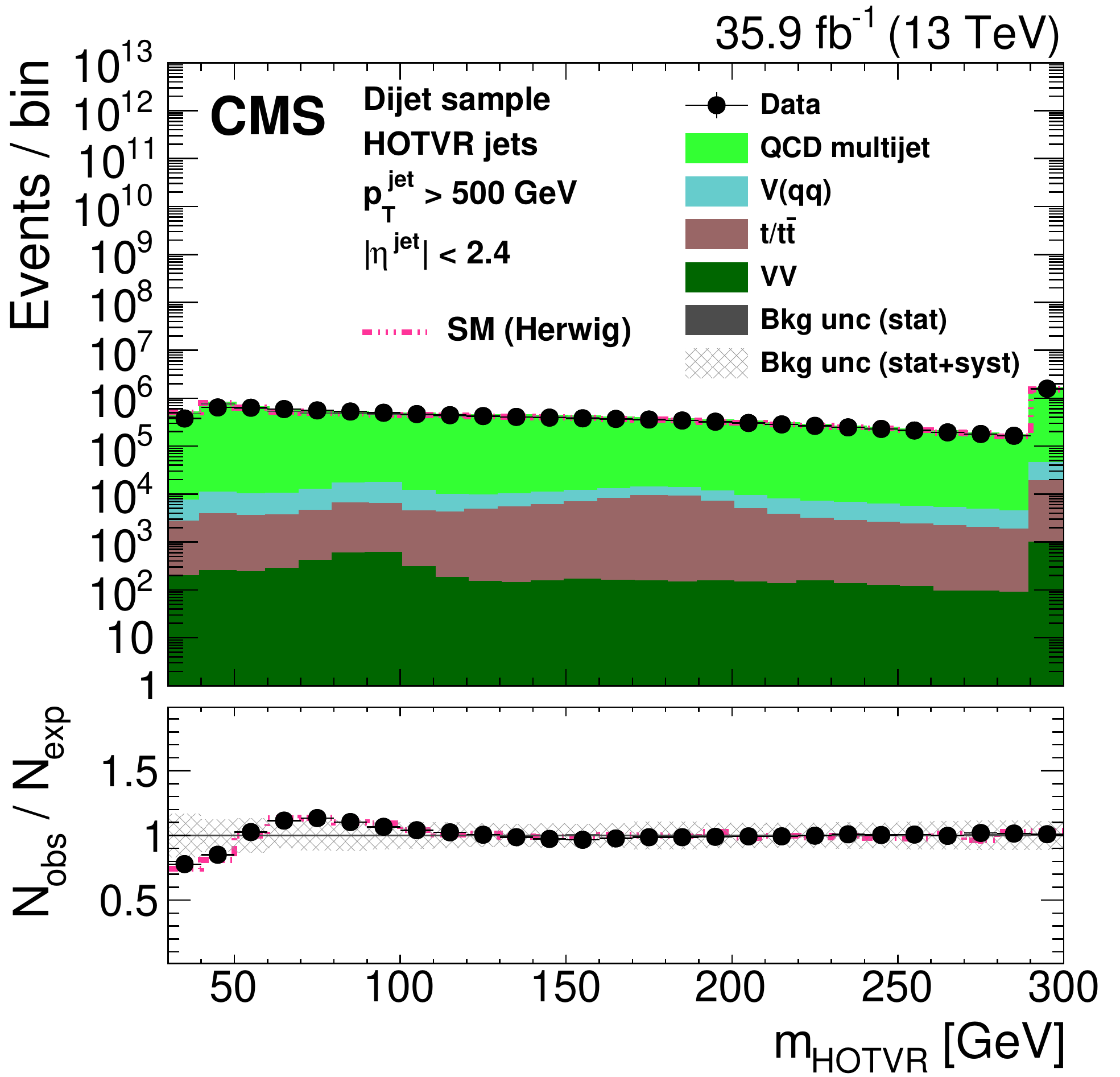}\\
\includegraphics[width=0.45\textwidth]{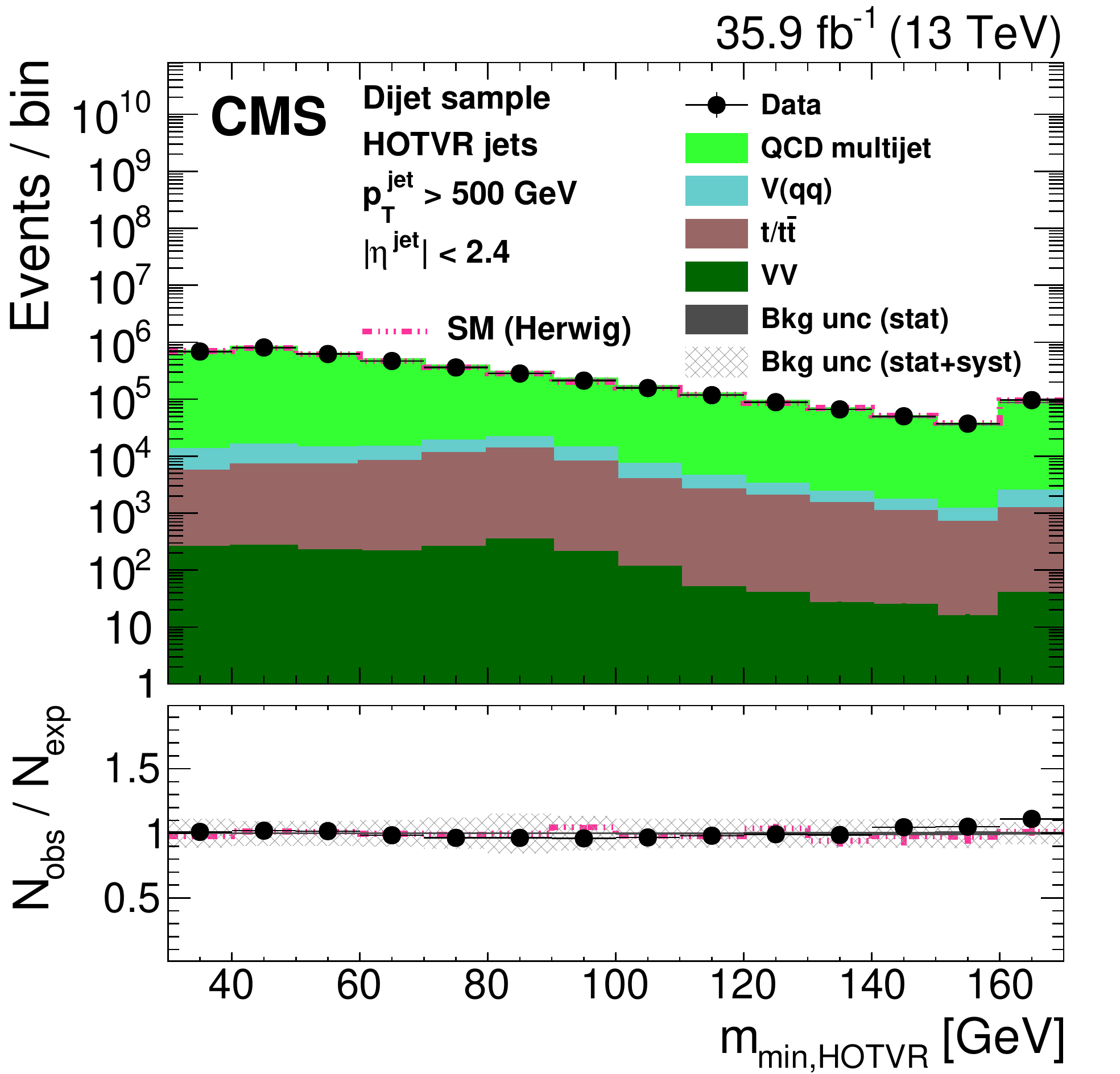}
\includegraphics[width=0.45\textwidth]{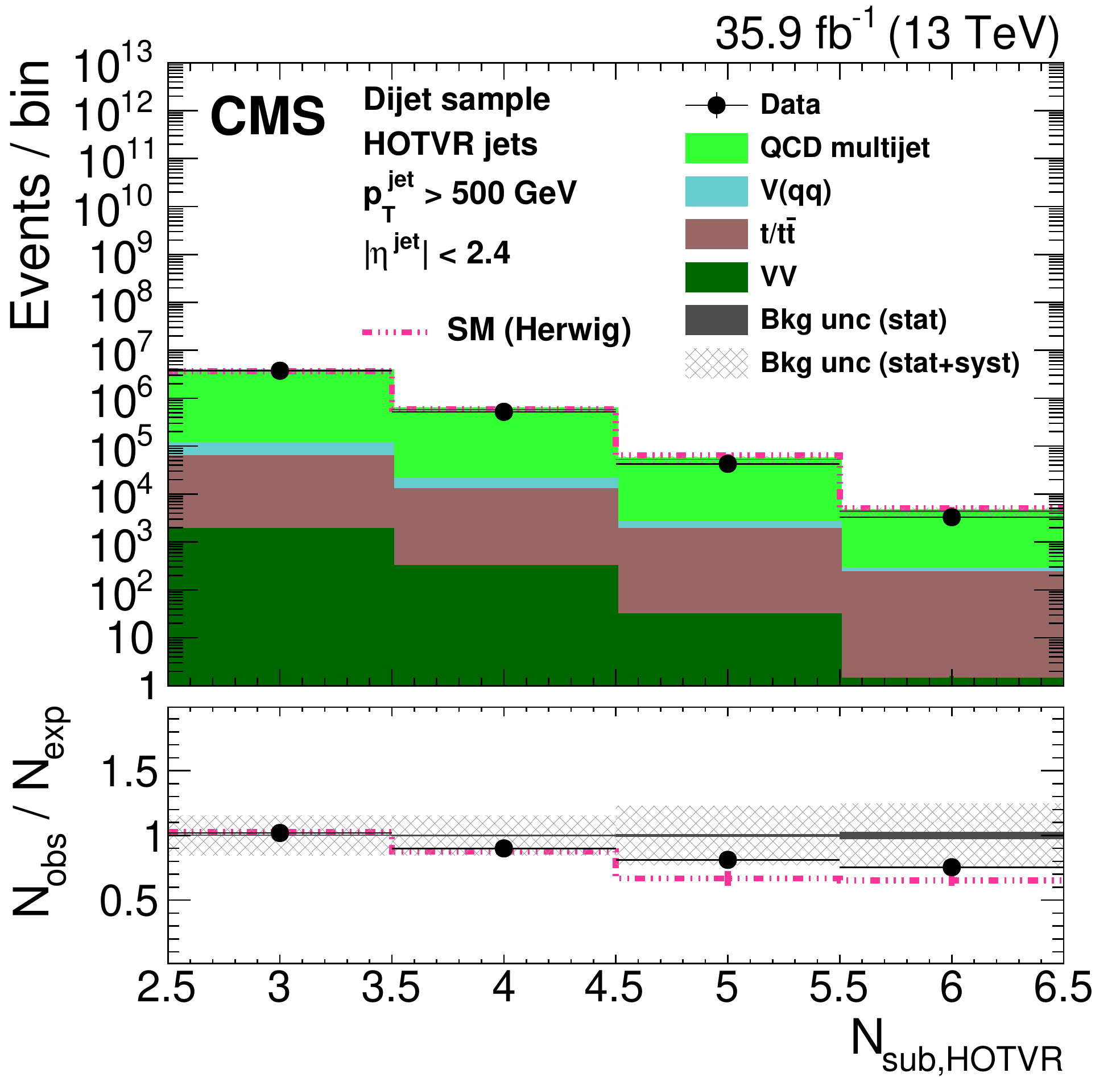}
\caption{\label{fig:qcd_hotvr}Distribution of the main observables of
  the HOTVR algorithm, HOTVR jet \pt (upper left), $m_{\text{HOTVR}}$ (upper right),
  $m_{\text{min,HOTVR}}$ (lower left) and $N_{\text{sub,HOTVR}}$
  (lower right) in data and simulation in the dijet sample. The pink
  line corresponds to the simulation distribution obtained using the
  alternative QCD multijet sample. The background event yield is normalized to the total observed data yield.
  The lower panel shows the data to simulation ratio. The solid dark-gray (shaded light-gray)
  band corresponds to the total uncertainty (statistical uncertainty of the simulated samples),
  the pink line to the data to simulation ratio using the alternative QCD multijet sample,
  and the vertical black lines correspond to the statistical
  uncertainty of the data. The vertical pink lines correspond to the
  statistical uncertainty of the alternative QCD multijet sample.
  The distributions are weighted so that the jet \pt
  distribution of the simulation matches the data. }
\end{figure}

\begin{figure}[hp!]
\centering
\includegraphics[width=0.45\textwidth]{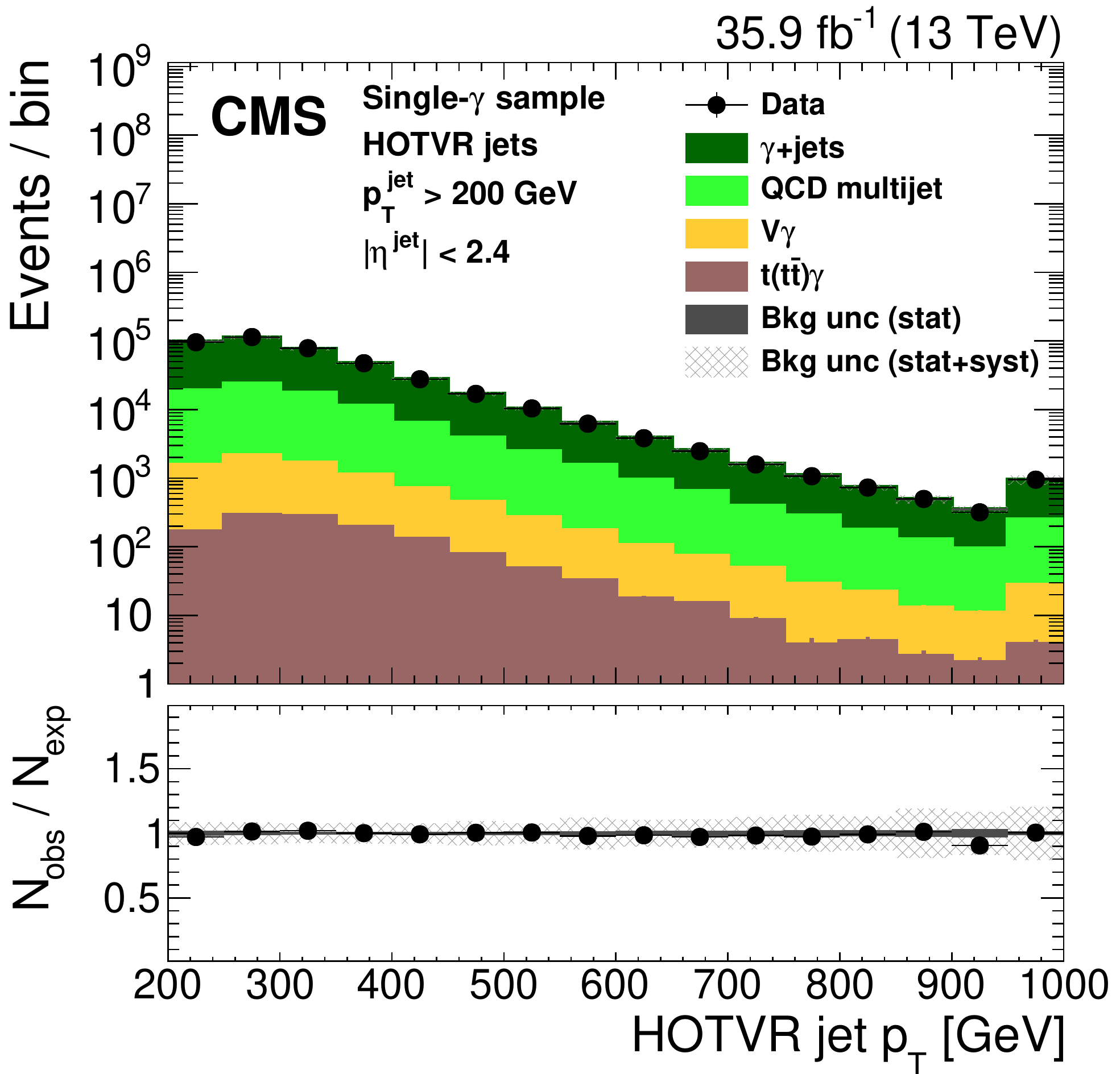}
\includegraphics[width=0.45\textwidth]{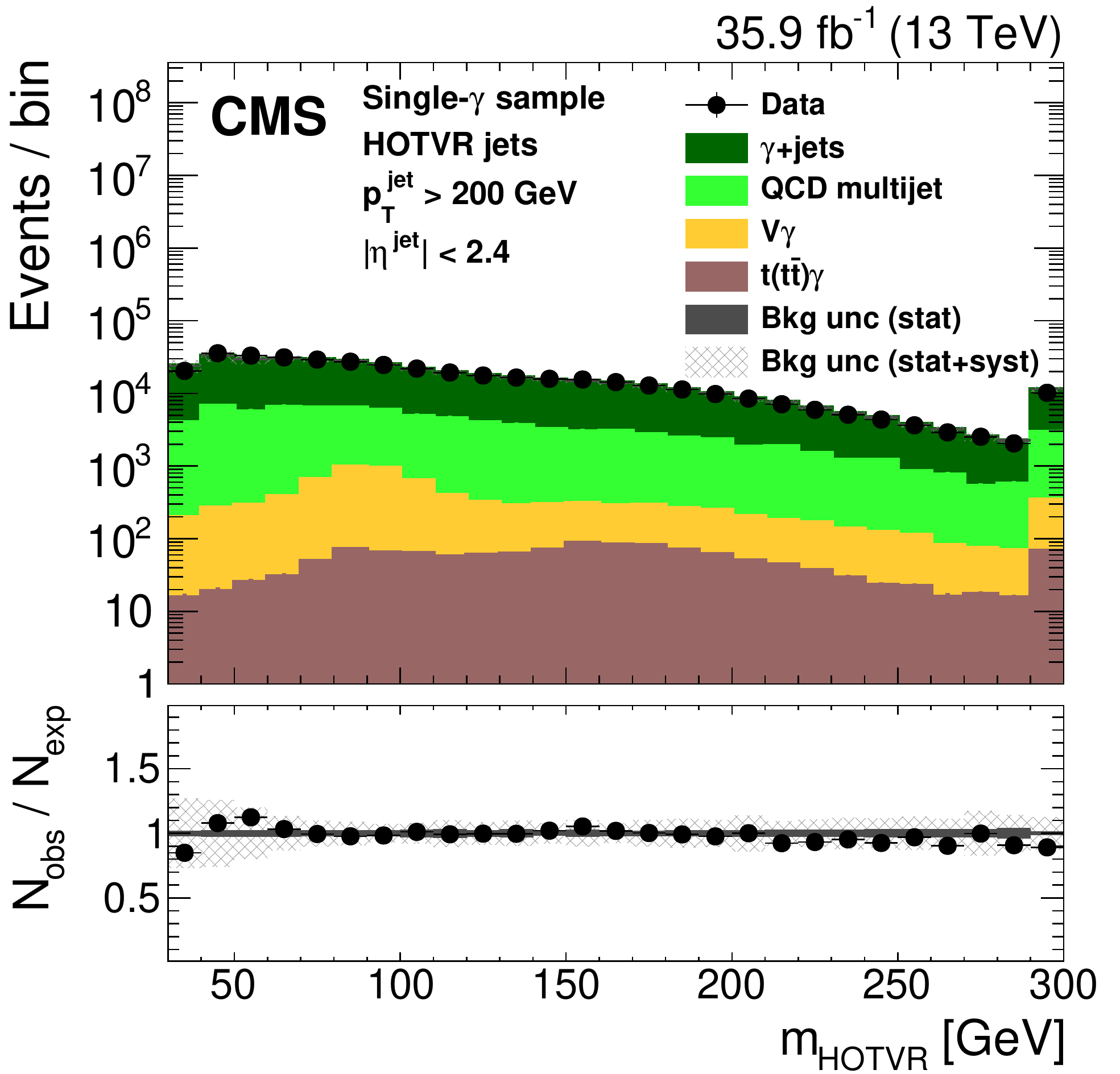}\\
\includegraphics[width=0.45\textwidth]{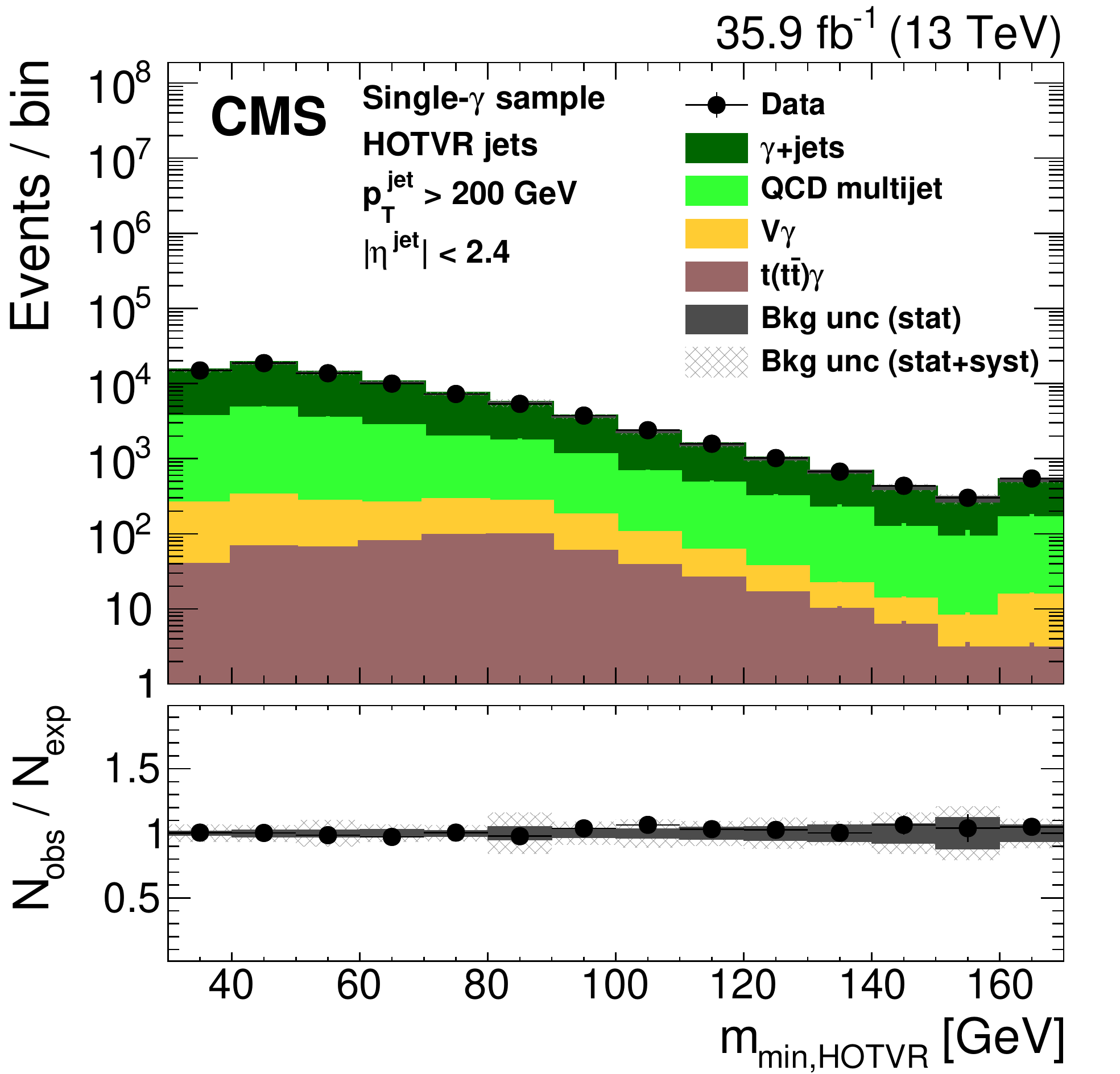}
\includegraphics[width=0.45\textwidth]{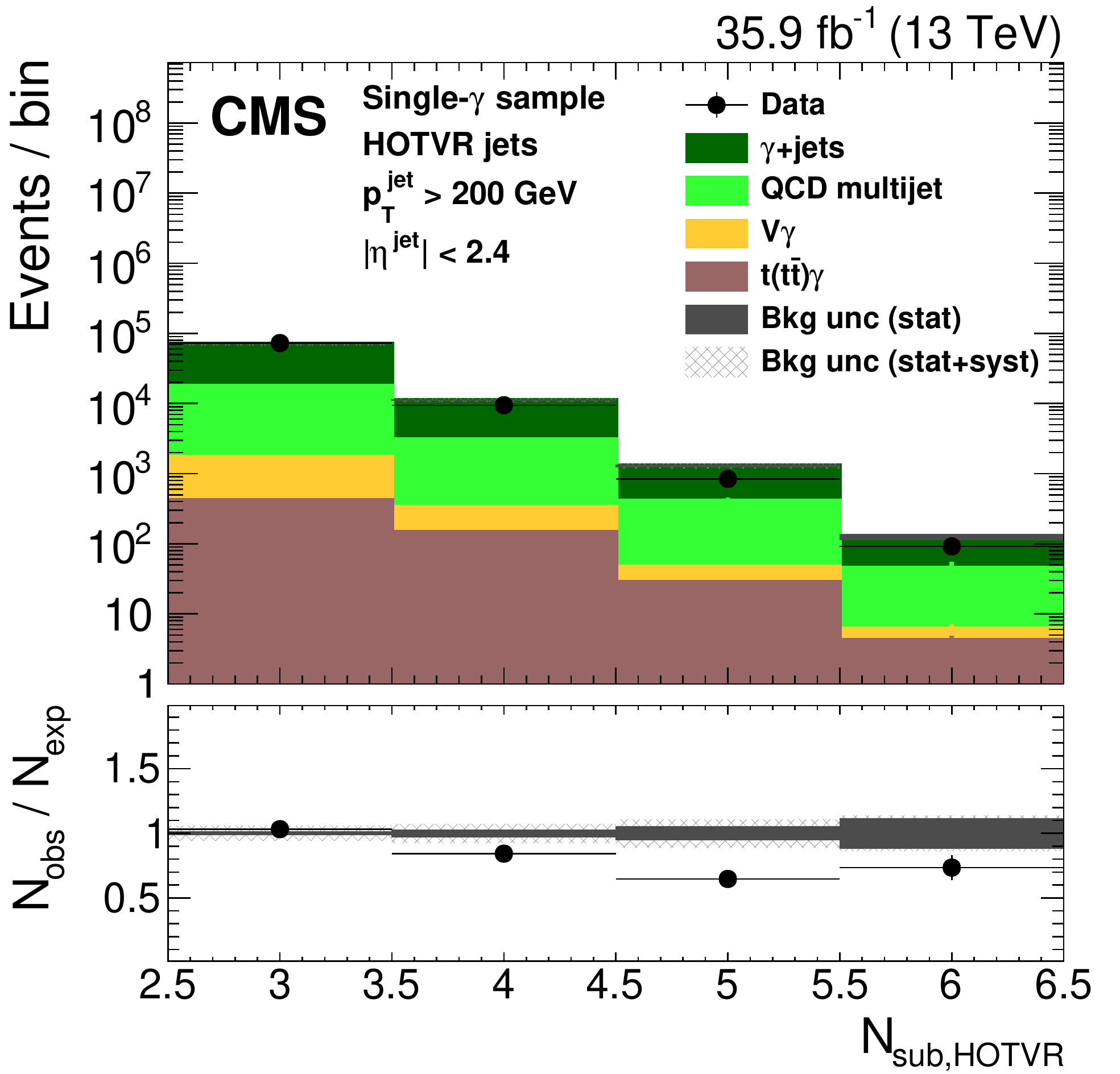}
\caption{\label{fig:pho_hotvr}Distribution of the main observables of
  the HOTVR algorithm, HOTVR jet \pt (upper left), $m_{\text{HOTVR}}$ (upper right),
  $m_{\text{min,HOTVR}}$ (lower left) and $N_{\text{sub,HOTVR}}$
  (lower right) in data and simulation in the single-$\gamma$ sample. The background event yield is normalized to the total observed data yield.
  The lower panel shows the data to simulation ratio. The solid dark-gray (shaded light-gray)
  band corresponds to the total uncertainty (statistical uncertainty of the simulated samples),
   and the vertical lines correspond to the statistical
  uncertainty of the data. The distributions are weighted so that the jet \pt
  distribution of the simulation matches the data.}
\end{figure}

Figures~\ref{fig:qcd_hotvr} and~\ref{fig:pho_hotvr} show the
distribution of the main observables of the HOTVR algorithm, namely
$m_{\text{HOTVR}}$, $m_{\text{min,HOTVR}}$, and $N_{\text{sub,HOTVR}}$,
in data and simulation, in the dijet and single-$\gamma$ samples,
respectively. In both samples, $m_{\text{HOTVR}}$ and
$m_{\text{min,HOTVR}}$ show good agreement between data and simulation. The
$N_{\text{sub,HOTVR}}$ distribution in data is softer than in
simulation. Similar conclusions hold using \HERWIGpp to simulate the QCD multijet events.
The difference is more pronounced in the single-$\gamma$ sample. The
$N_{\text{sub,HOTVR}}$ is particularly sensitive to the precise
modeling of the parton showering.

\begin{figure}[hp!]
\centering
\includegraphics[width=0.45\textwidth]{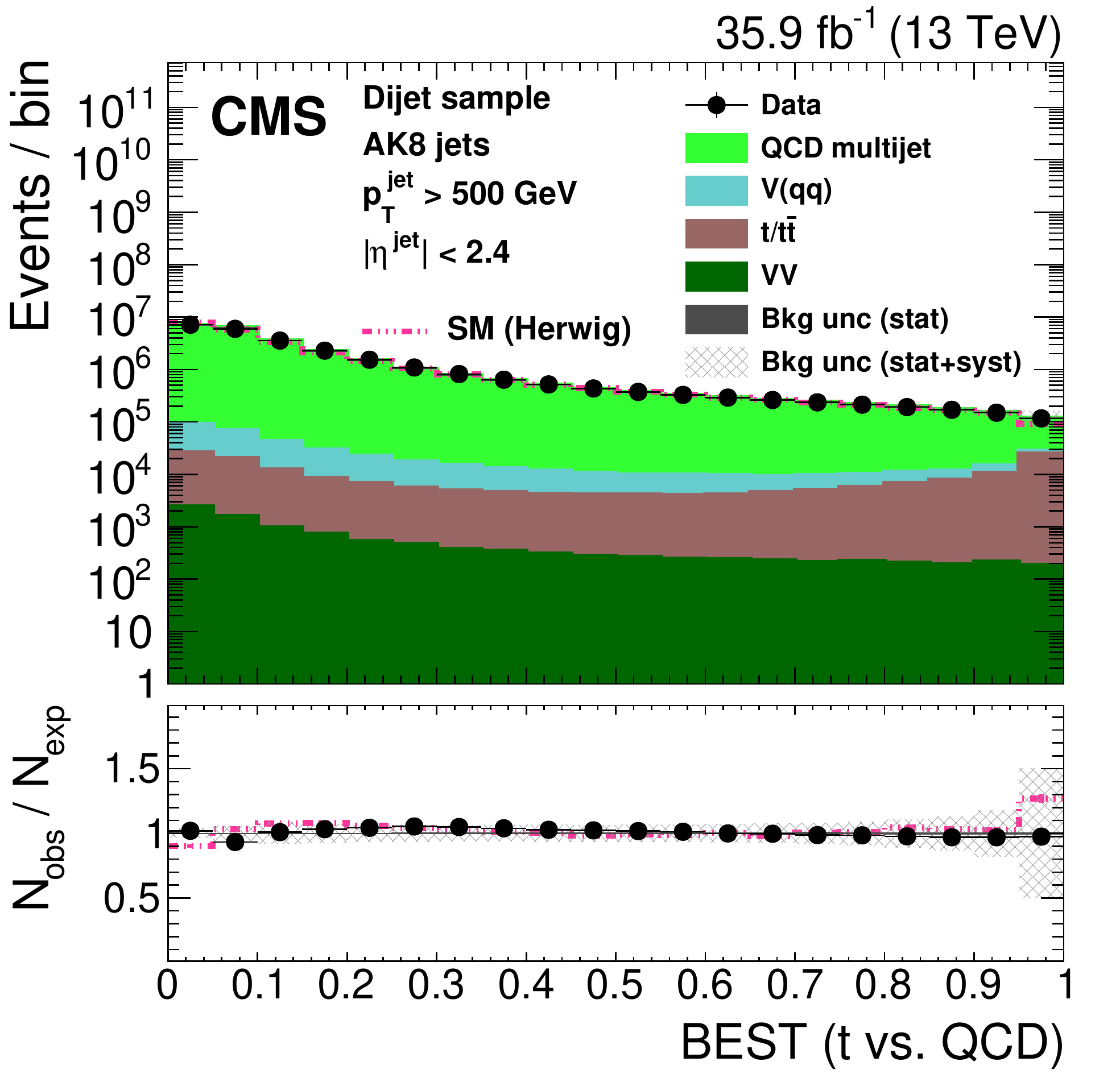}
\includegraphics[width=0.45\textwidth]{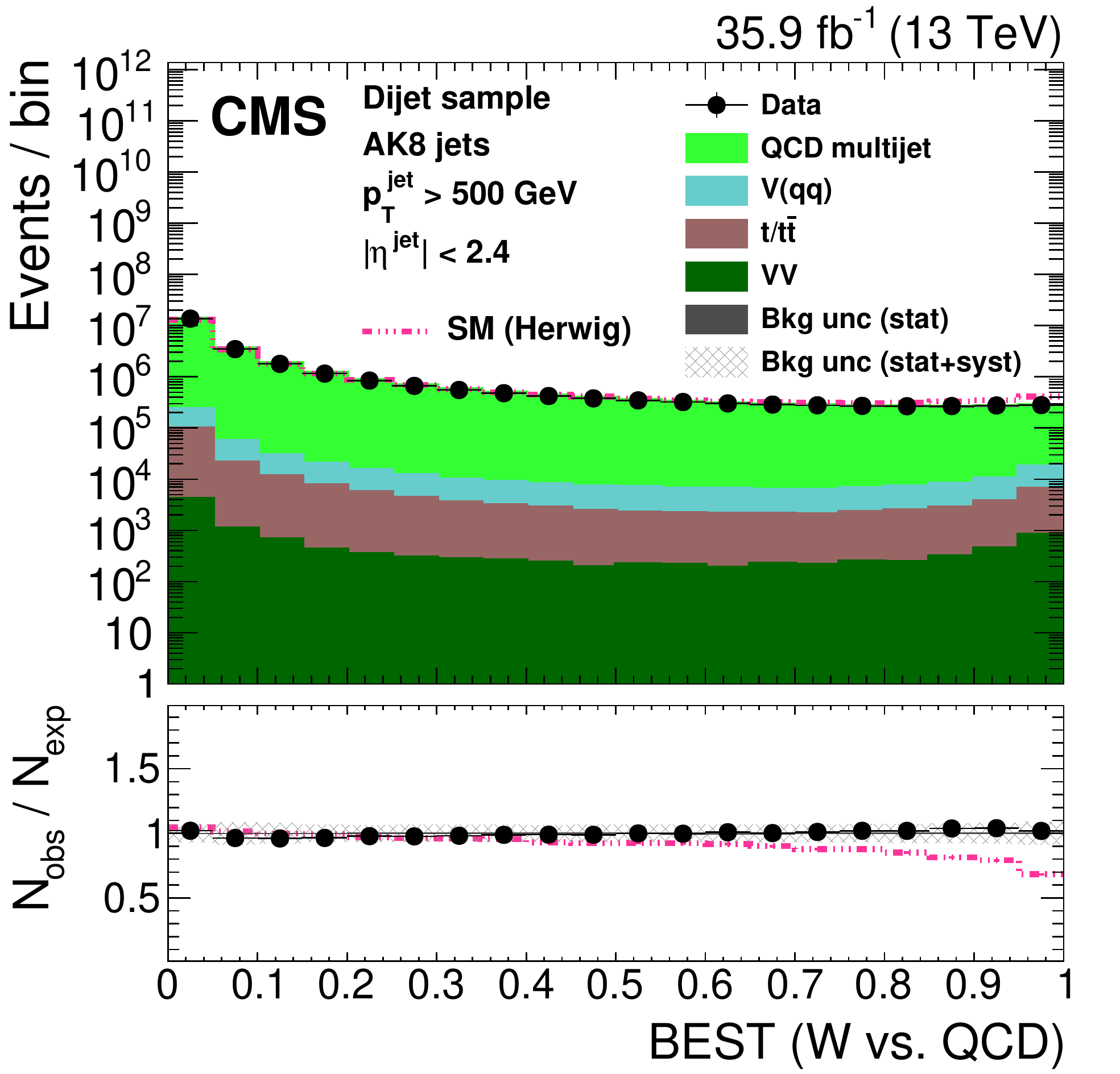}\\
\includegraphics[width=0.45\textwidth]{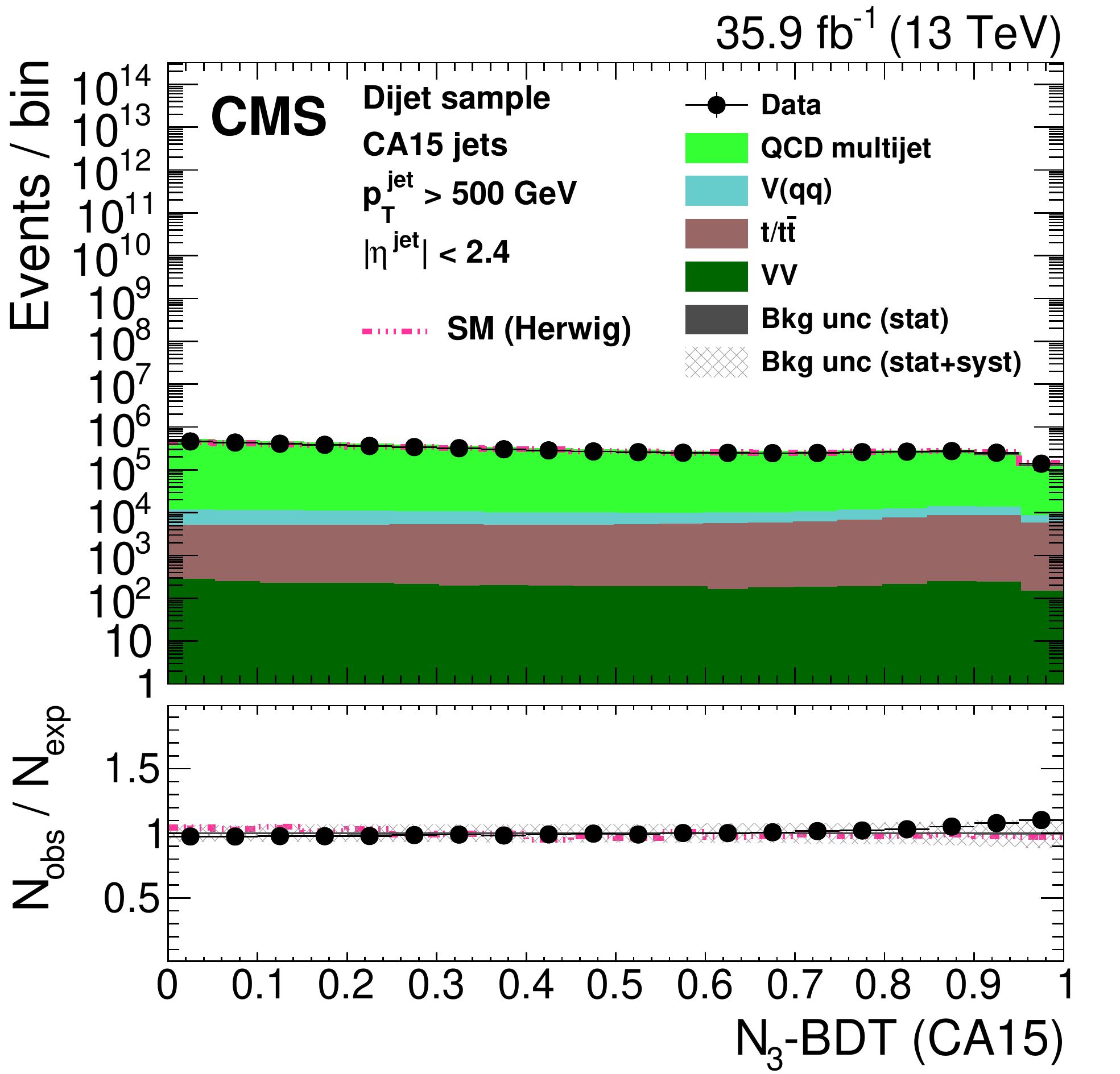}
\caption{\label{fig:qcd_hlv}Distribution of the \PQt~quark (upper
  left) and \PW~boson (upper right) identification probabilities for
  the BEST algorithm, and the \ecftop discriminant in data and simulation
  in the dijet sample.  The background event yield is
  normalized to the total observed data yield. The pink line
  corresponds to the simulation distribution obtained using the alternative
  QCD multijet sample. The background event yield is normalized to the total observed data yield.
  The lower panel shows the data to simulation ratio. The solid dark-gray (shaded light-gray)
  band corresponds to the total uncertainty (statistical uncertainty of the simulated samples),
  the pink line to the data to simulation ratio using the alternative QCD multijet sample,
  and the vertical black lines correspond to the statistical
  uncertainty of the data. The vertical pink lines correspond to the
  statistical uncertainty of the alternative QCD multijet sample.
  The distributions are weighted so that the jet \pt
  distribution of the simulation matches the data. }
\end{figure}

\begin{figure}[hp!]
\centering
\includegraphics[width=0.45\textwidth]{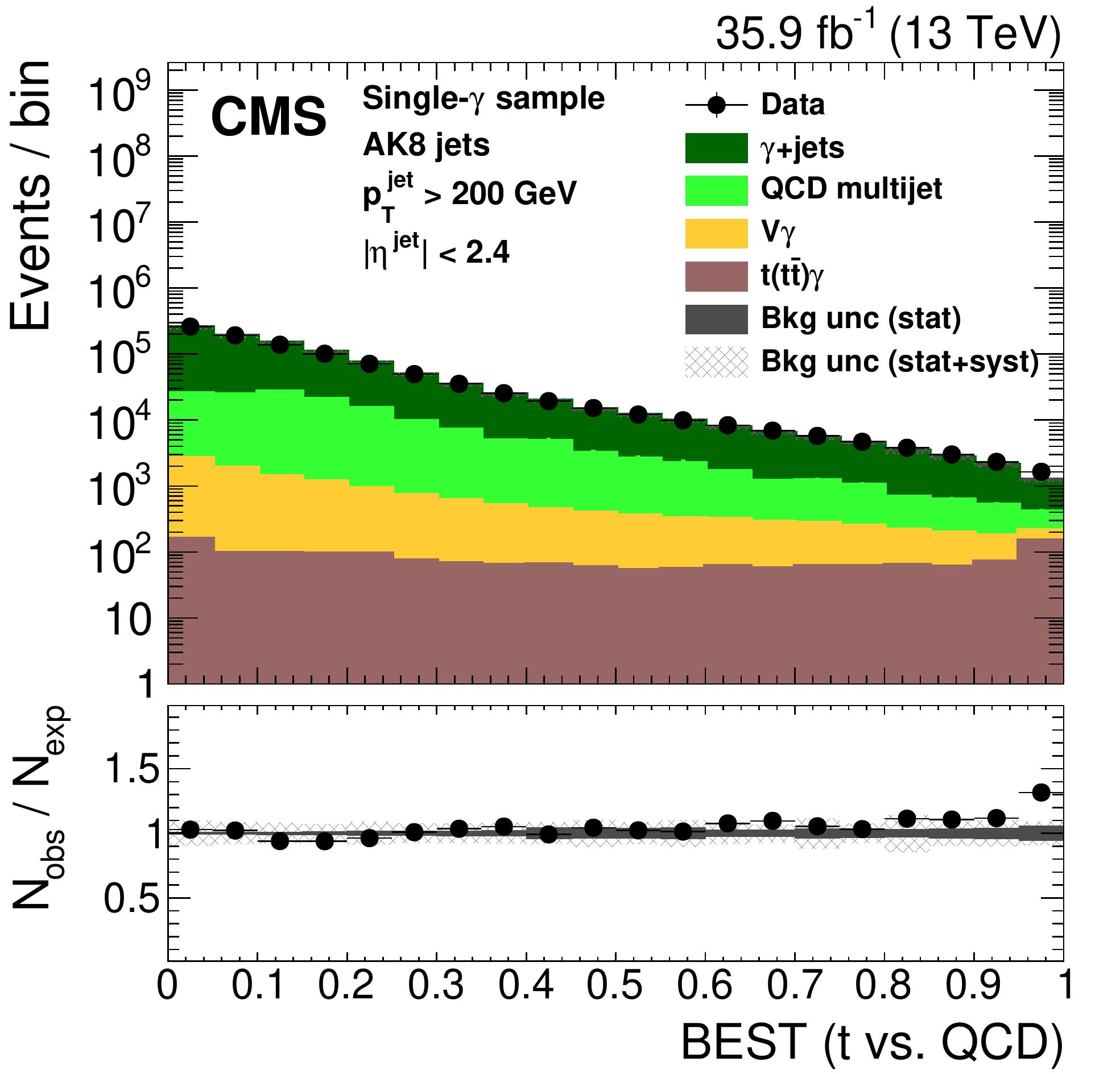}
\includegraphics[width=0.45\textwidth]{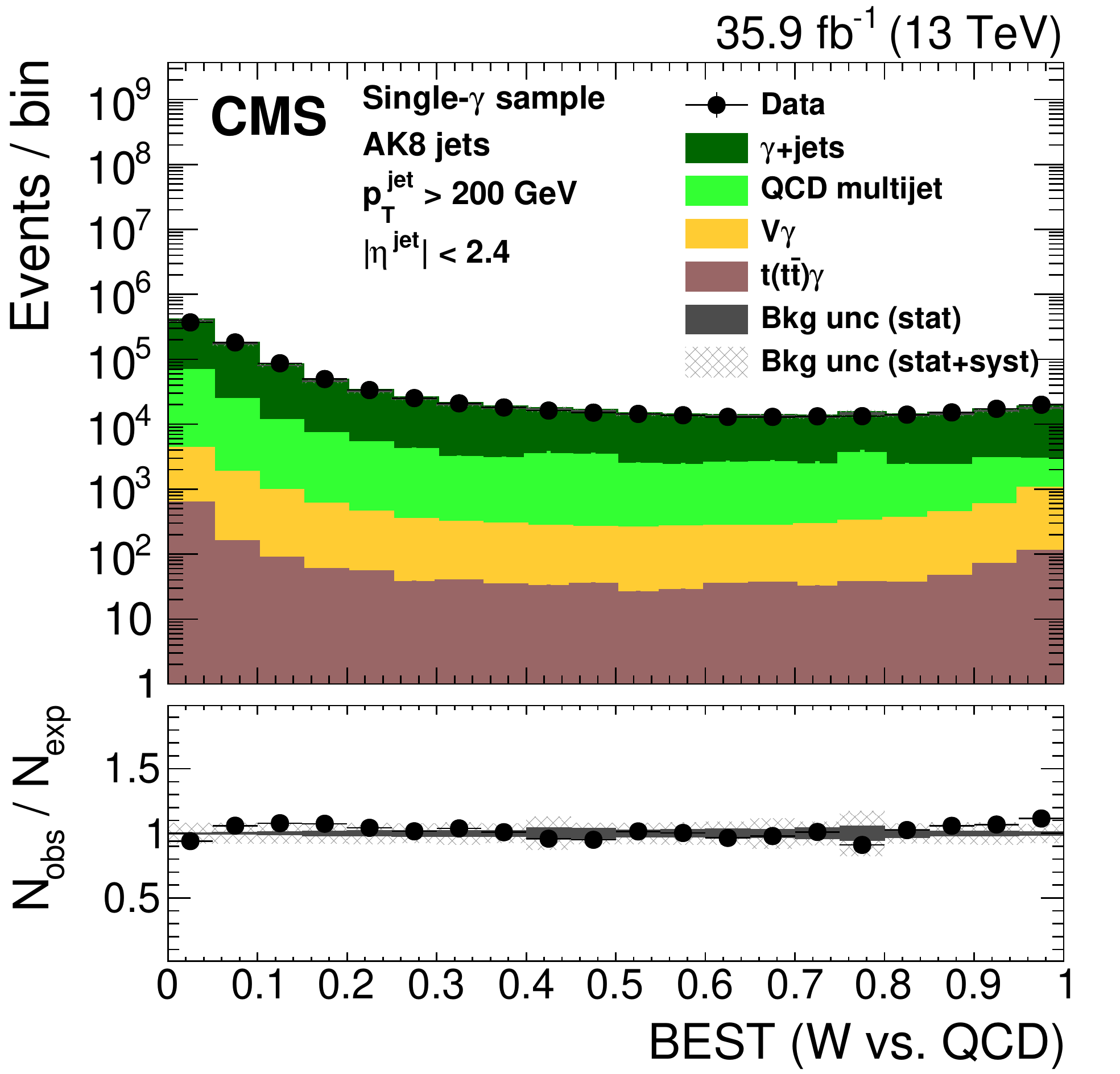}\\
\includegraphics[width=0.45\textwidth]{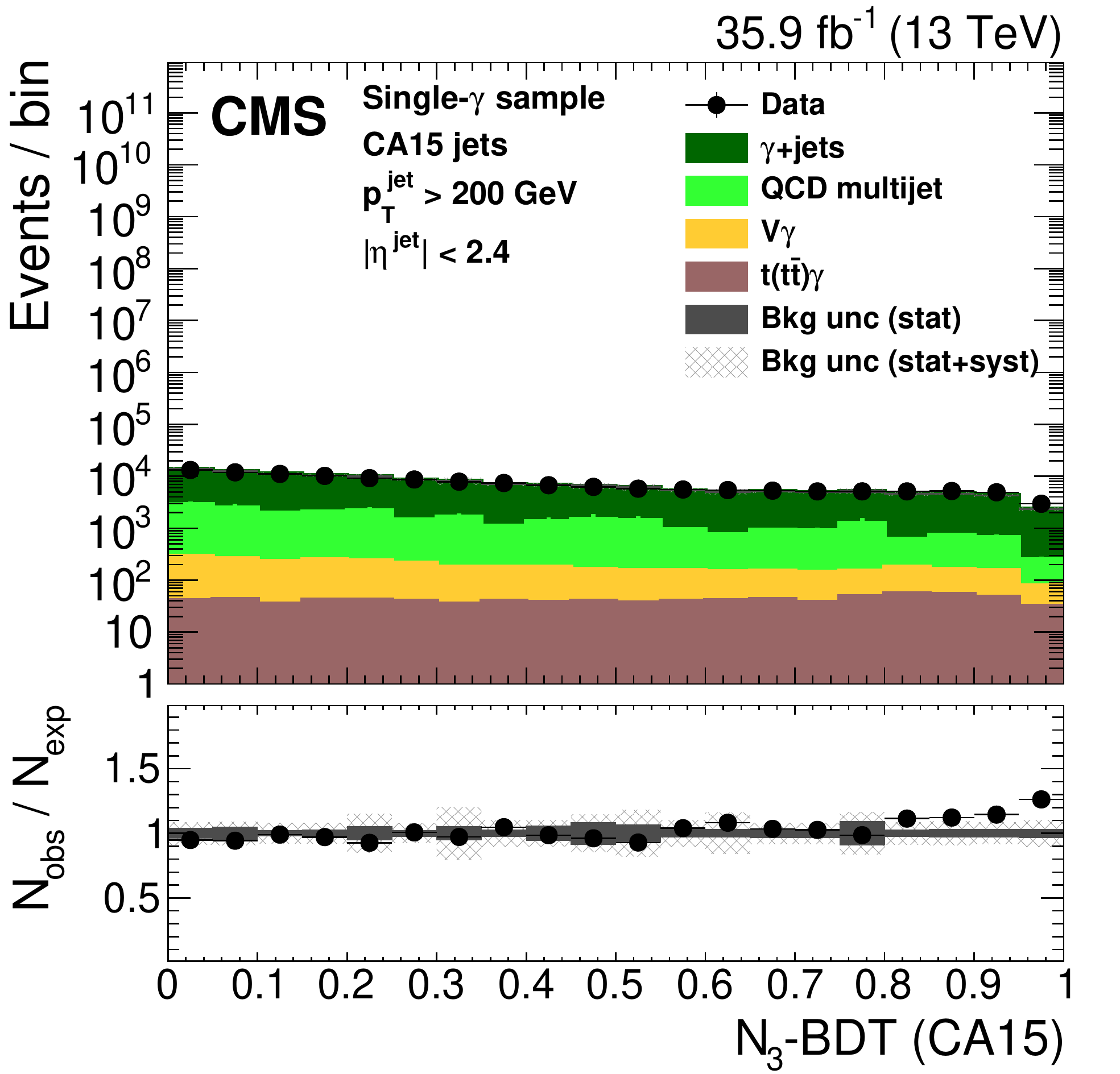}
\caption{\label{fig:pho_hlv}Distribution of the \PQt~quark (upper
  left) and \PW~boson (upper right) identification probabilities for
  the BEST algorithm, and the \ecftop discriminant in data and simulation
  in the single-$\gamma$ sample.  The background event yield is
  normalized to the total observed data yield. The background event yield is normalized to the total observed data yield.
  The lower panel shows the data to simulation ratio. The solid dark-gray (shaded light-gray)
  band corresponds to the total uncertainty (statistical uncertainty of the simulated samples),
  and the vertical lines correspond to the statistical
  uncertainty of the data. The distributions are weighted so that the jet \pt
  distribution of the simulation matches the data.}
\end{figure}

The distribution of the \PQt quark and \PW boson identification
probabilities for BEST and the \PQt quark tagging discriminant for the
\ecftop algorithm in the dijet sample are presented in
Fig.~\ref{fig:qcd_hlv}, and the equivalent plots for the
single-$\gamma$ selection are shown in Fig.~\ref{fig:pho_hlv}. In both
samples the agreement between data and simulation is
reasonable. Some disagreement is observed in the very high values ($\gtrsim$0.95)
for the \PQt quark identification probability of the BEST
algorithm in the single-$\gamma$ sample. The disagreement is observed in
the region of the \PQt quark probability greater than 0.95, which is
significantly higher than the recommended operating points.
Some disagreement is observed between the nominal QCD multijet simulated sample
and the alternative sample for large values of the \PW boson probability of the
 BEST algorithm, with the nominal sample showing better agreement with the data.

\begin{figure}[hp!]
\centering
\includegraphics[width=0.38\textwidth]{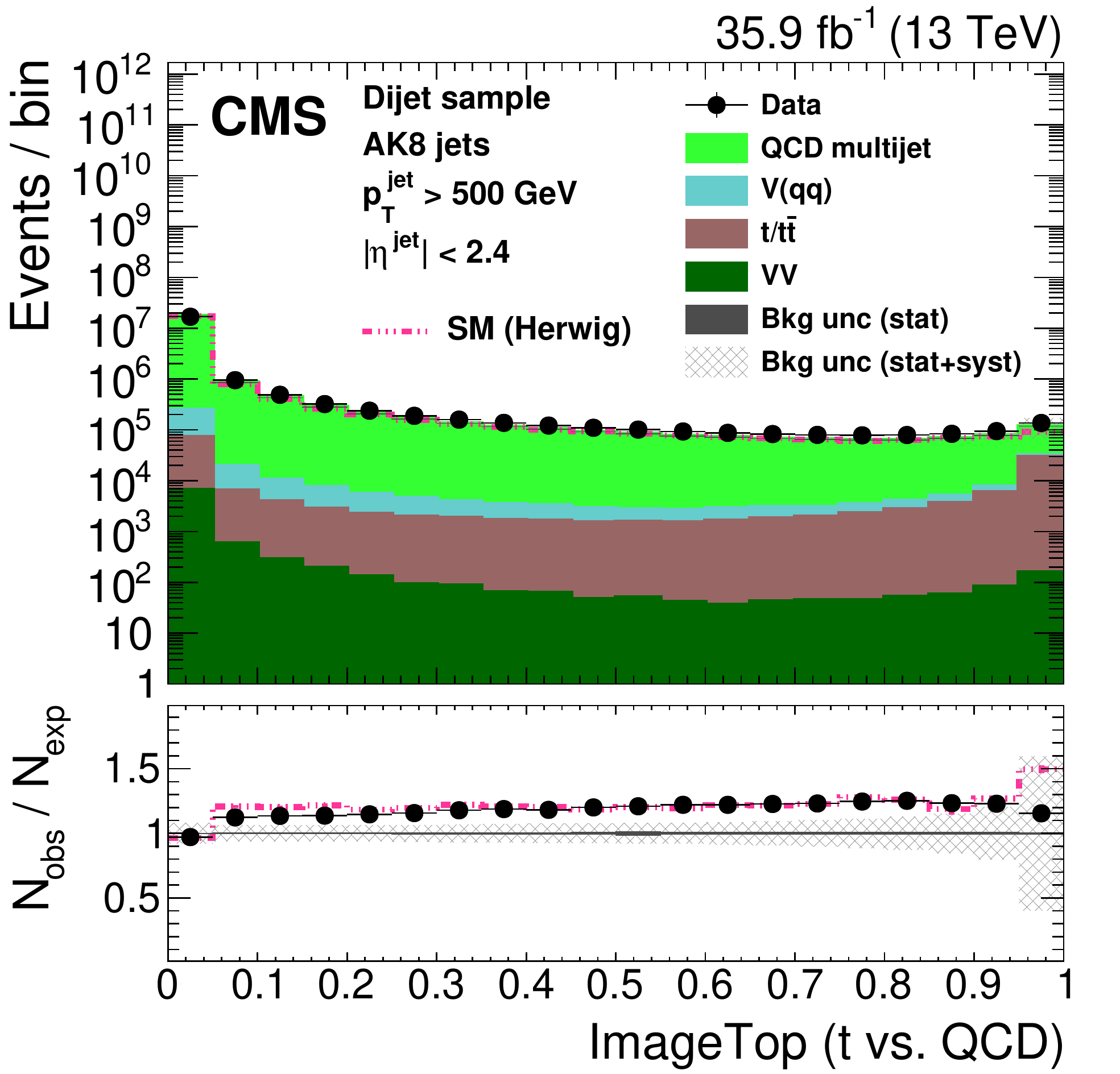}
\includegraphics[width=0.38\textwidth]{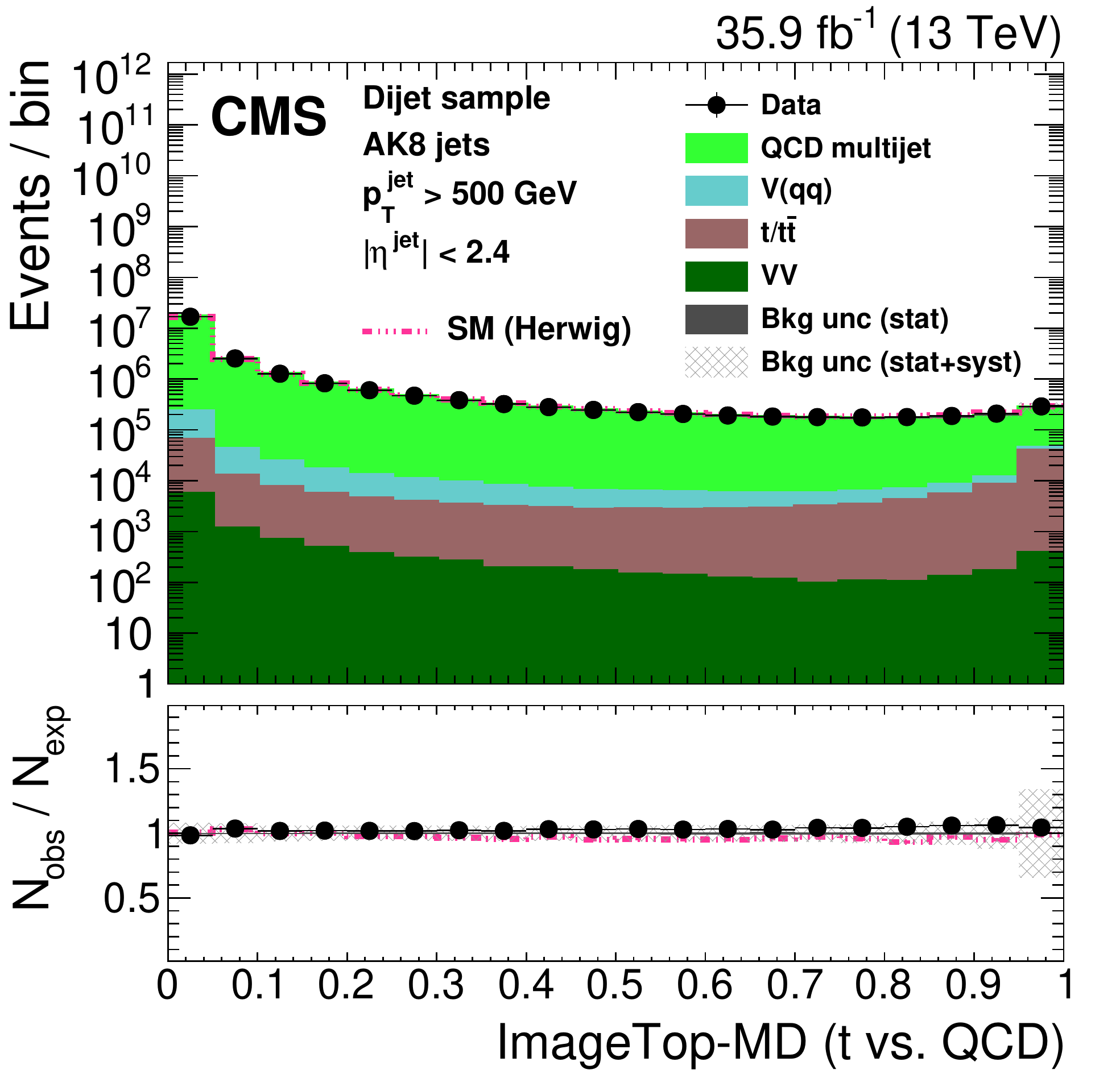}\\
\includegraphics[width=0.38\textwidth]{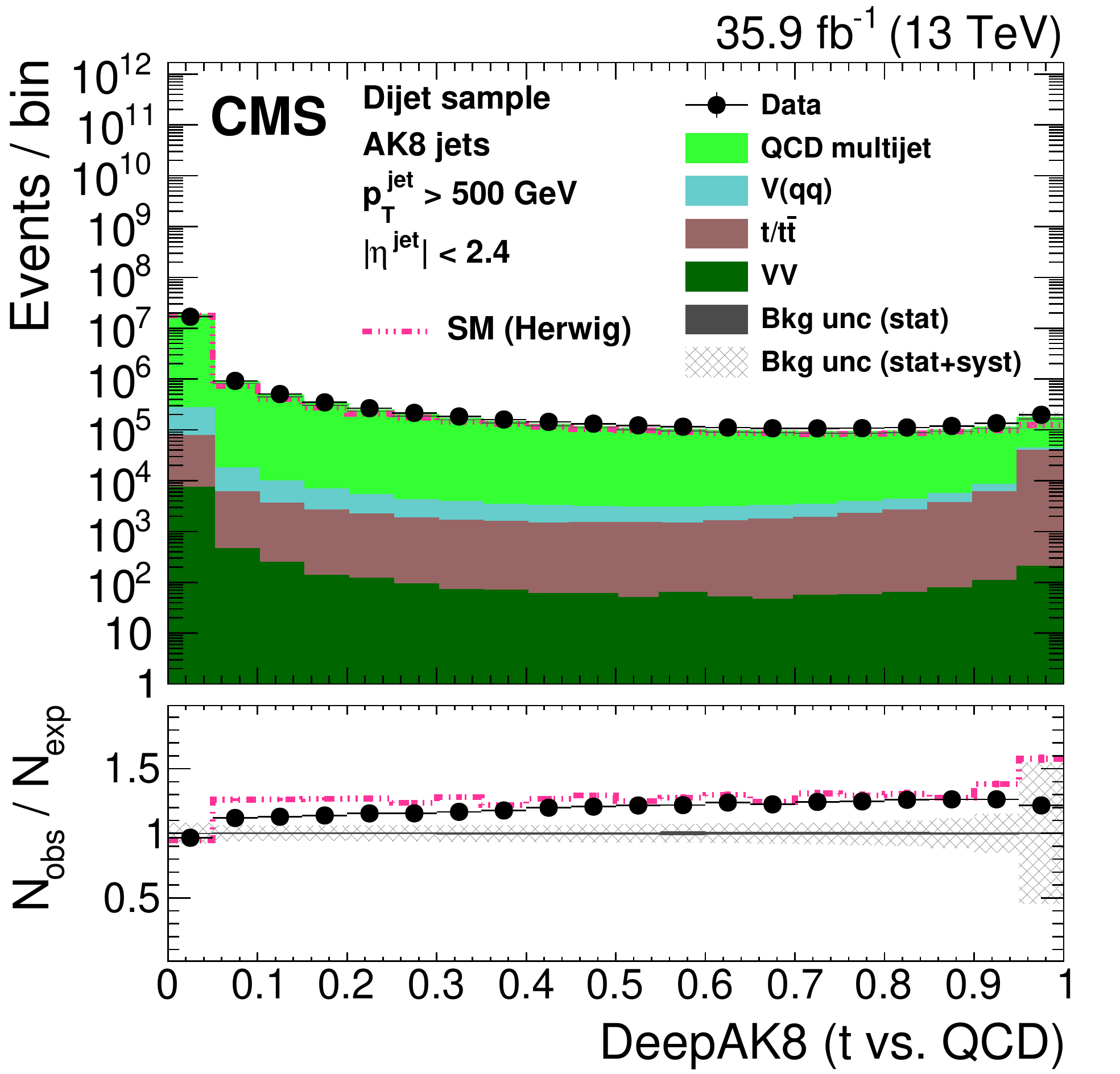}
\includegraphics[width=0.38\textwidth]{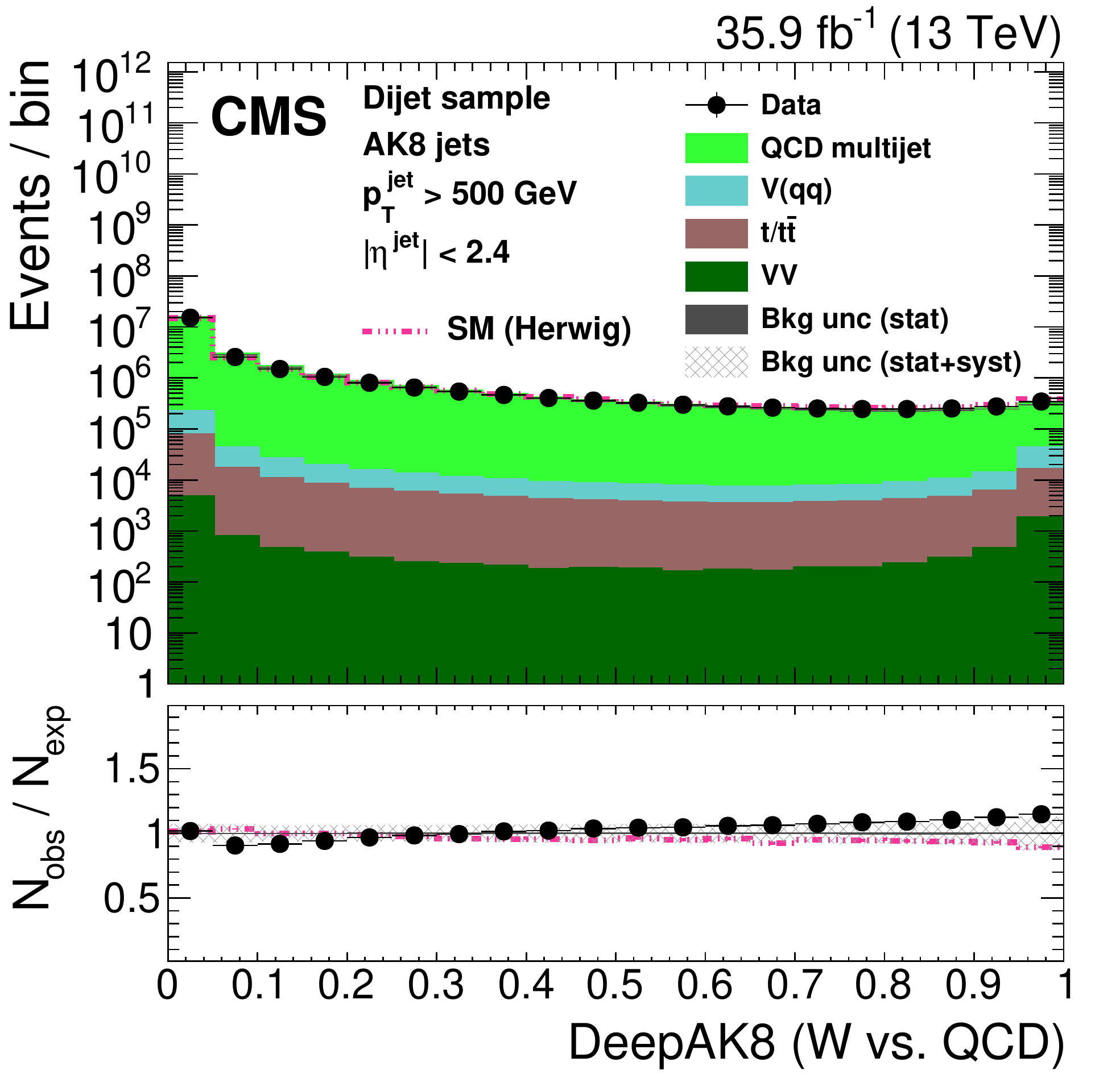} \\
\includegraphics[width=0.38\textwidth]{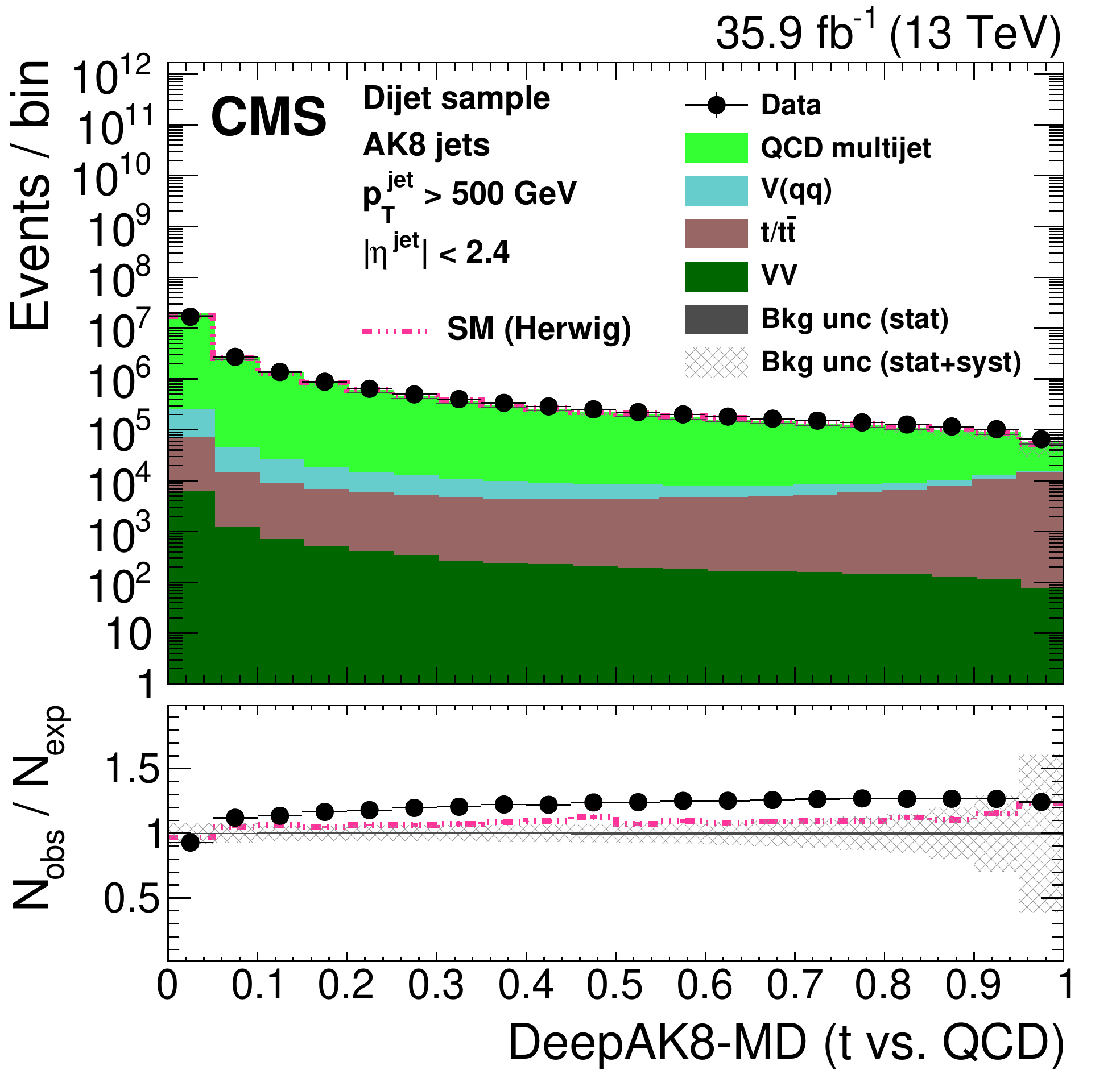}
\includegraphics[width=0.38\textwidth]{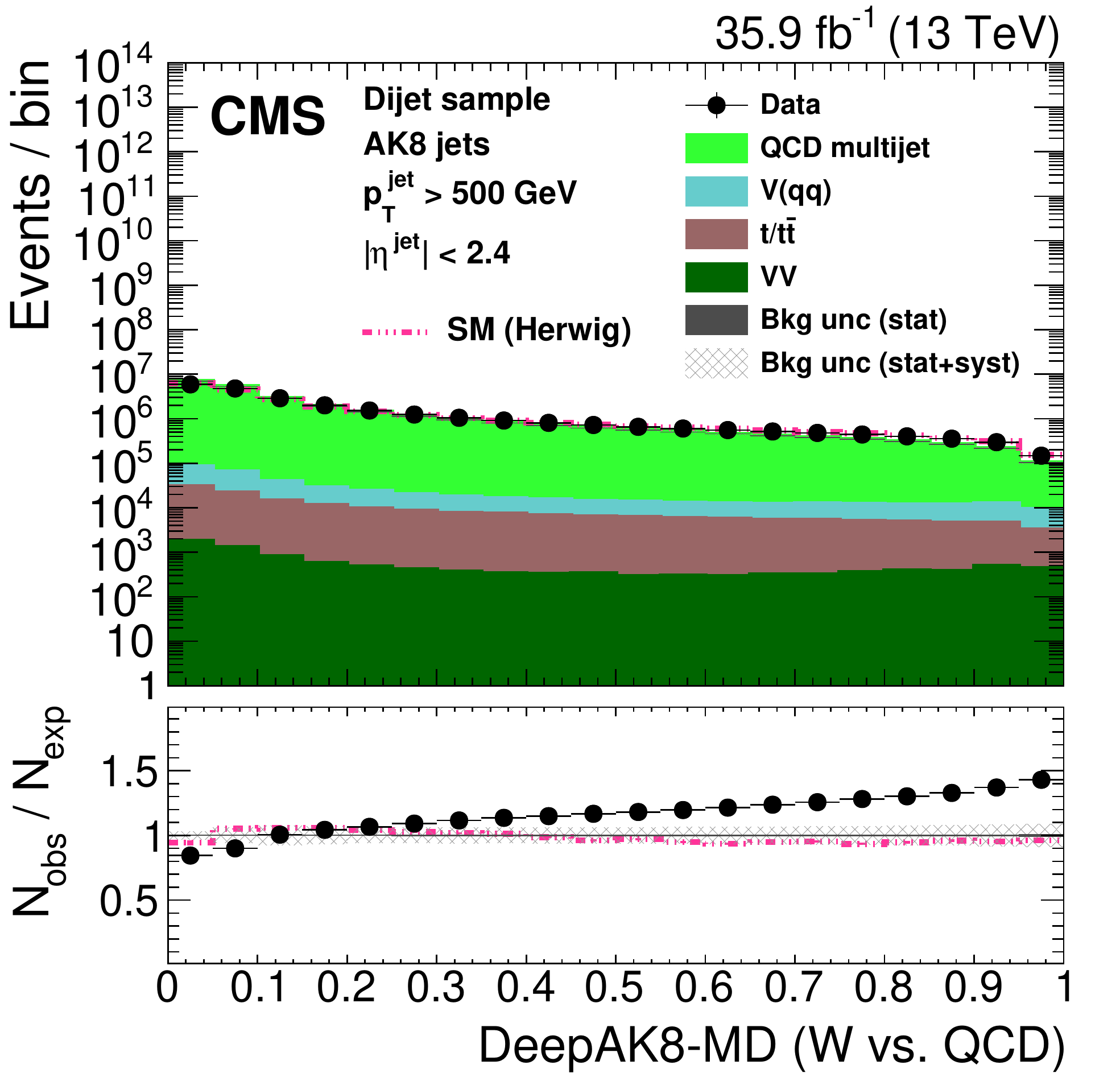} \\
\caption{\label{fig:qcd_low}Distribution of the ImageTop (upper left)
  and ImageTop-MD (upper right) discriminant in data and simulation in the
  dijet sample. The plots in the middle row show the \PQt~quark
  (left) and \PW~boson (right)  identification probabilities in data
  and simulation for the DeepAK8 algorithm. The corresponding plots for
  DeepAK8-MD are displayed in the lower row. The pink line
  corresponds to the simulation distribution obtained using the alternative
  QCD multijet sample. The background event yield is normalized to the total observed data yield.
  The lower panel shows the data to simulation ratio. The solid dark-gray (shaded light-gray)
  band corresponds to the total uncertainty (statistical uncertainty of the simulated samples),
  the pink line to the data to simulation ratio using the alternative QCD multijet sample,
  and the vertical black lines correspond to the statistical
  uncertainty of the data. The vertical pink lines correspond to the
  statistical uncertainty of the alternative QCD multijet sample.
  The distributions are weighted so that the jet \pt
  distribution of the simulation matches the data.}
\end{figure}

\begin{figure}[hp!]
\centering
\includegraphics[width=0.38\textwidth]{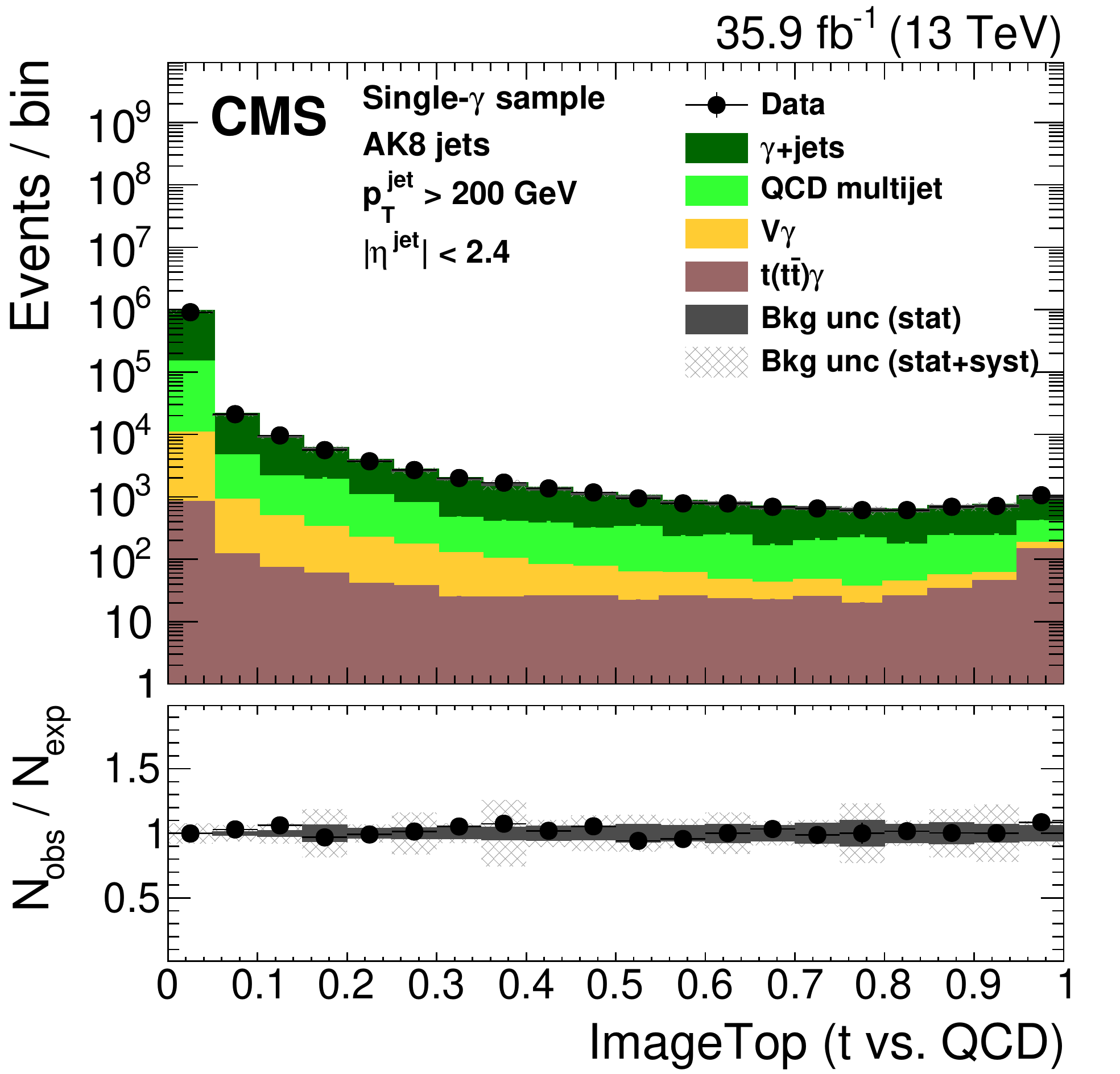}
\includegraphics[width=0.38\textwidth]{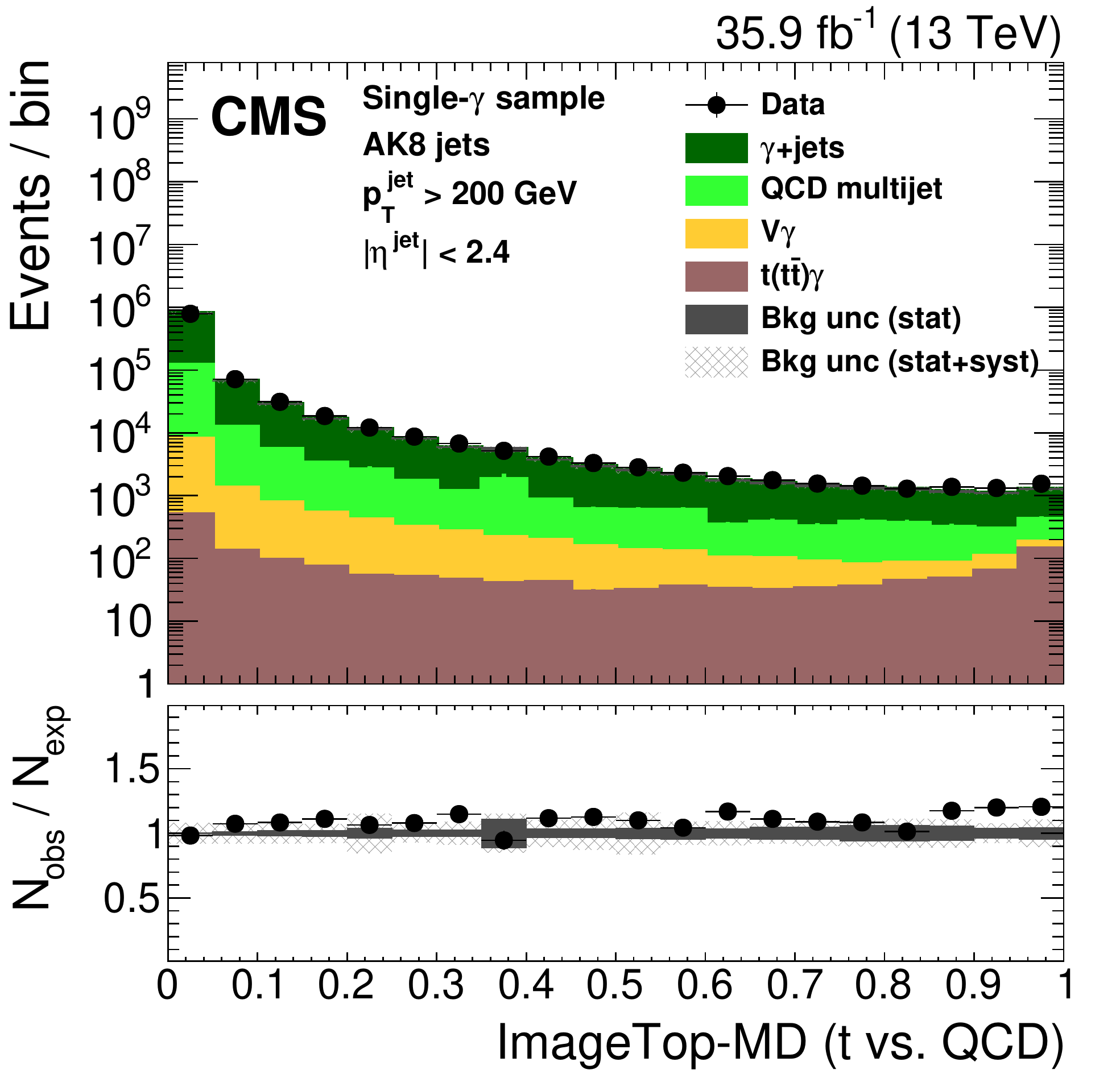}\\
\includegraphics[width=0.38\textwidth]{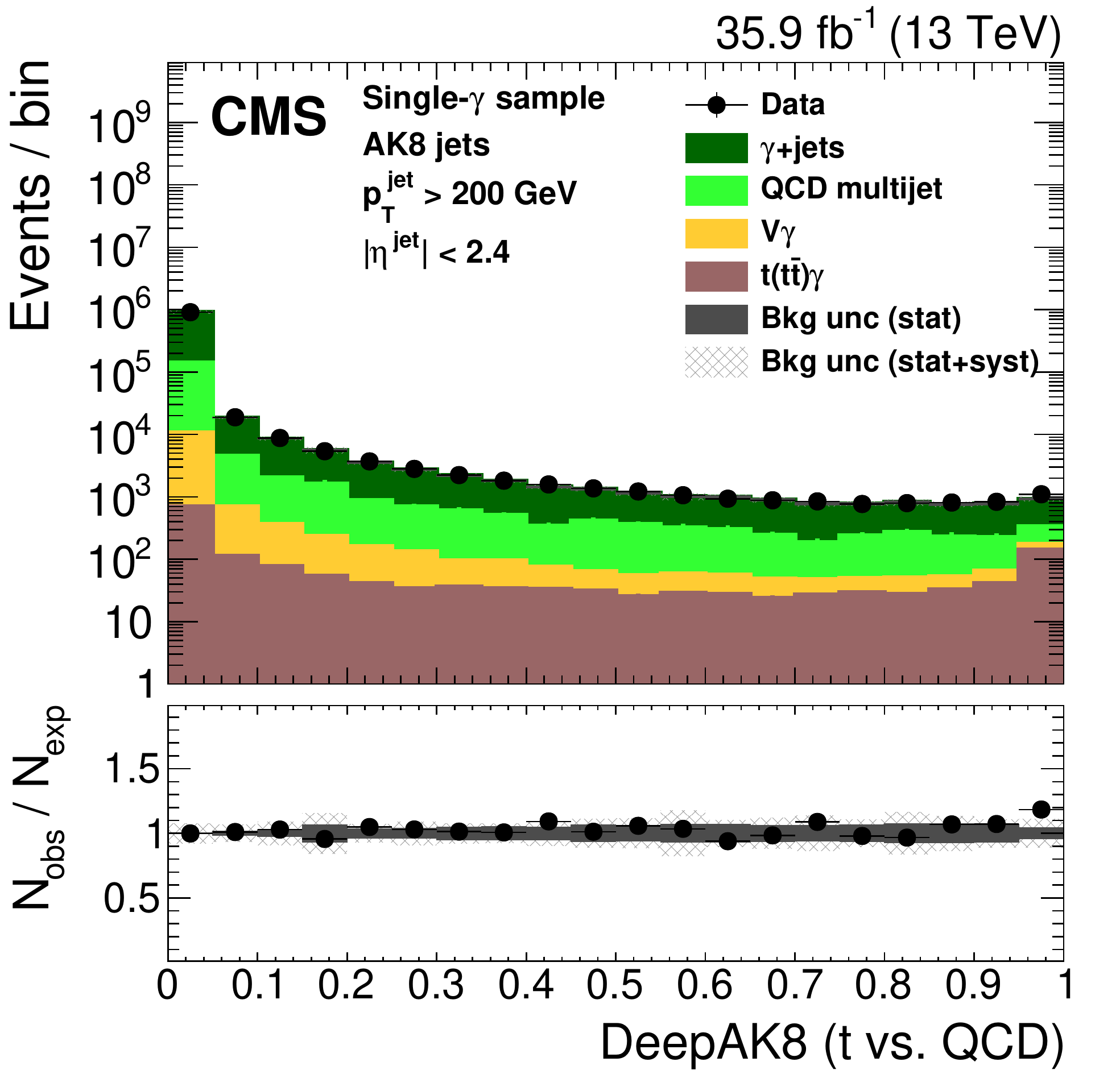}
\includegraphics[width=0.38\textwidth]{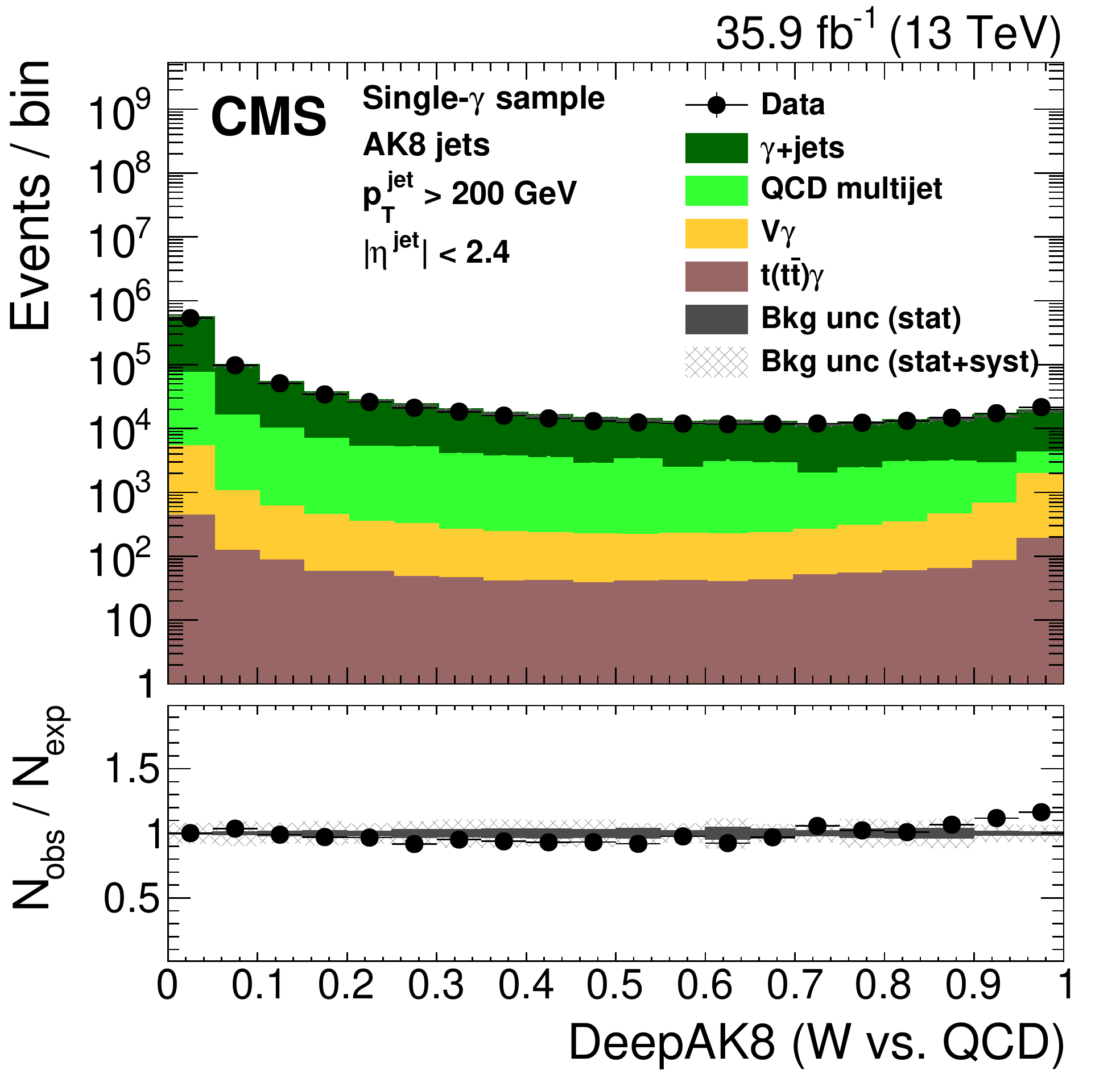} \\
\includegraphics[width=0.38\textwidth]{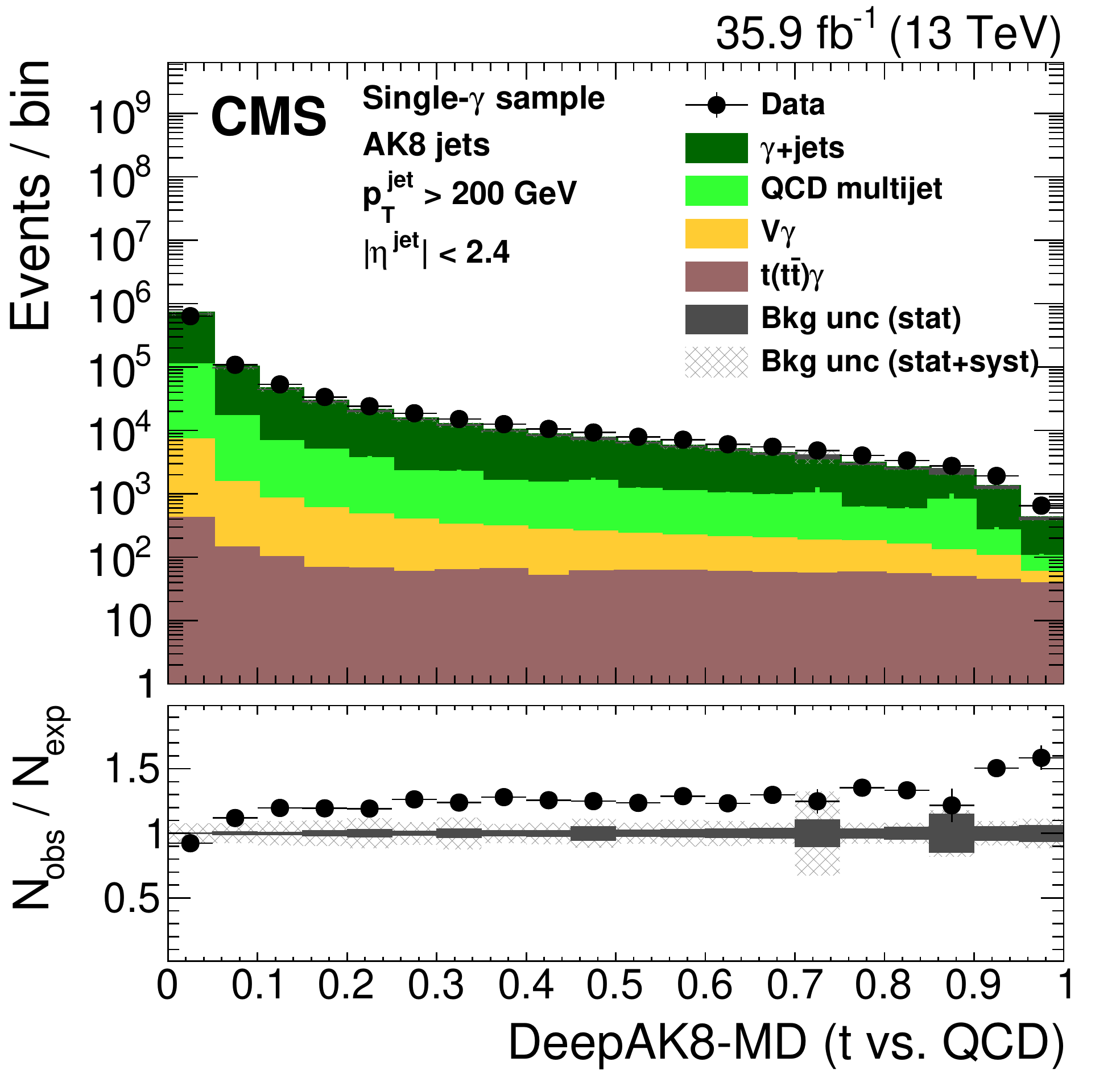}
\includegraphics[width=0.38\textwidth]{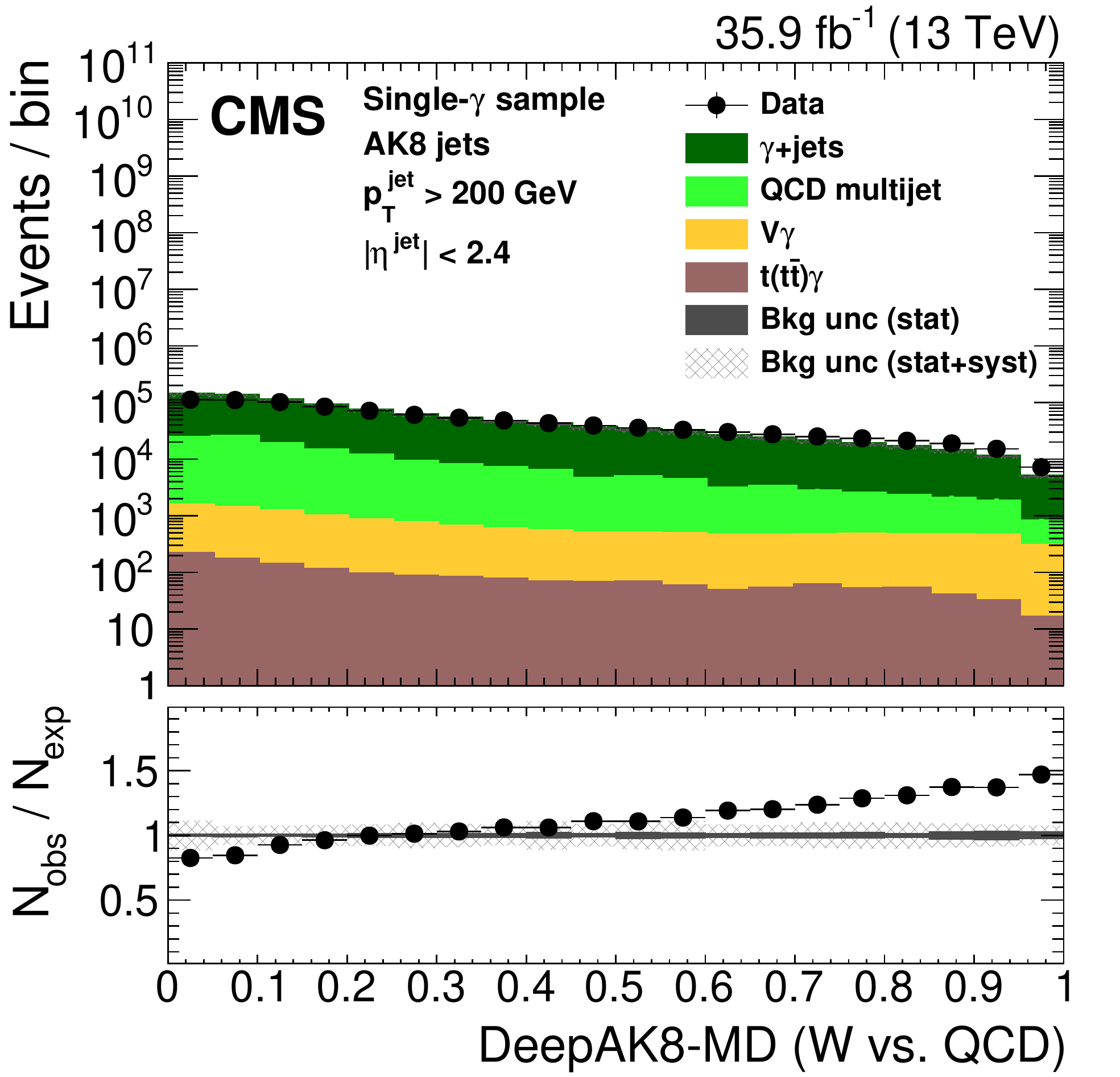} \\
\caption{\label{fig:pho_low}Distribution of the ImageTop (upper left)
  and ImageTop-MD (upper right) discriminant in data and simulation in the
  single-$\gamma$ sample. The plots in the middle row show the
  \PQt~quark (left) and \PW~boson (right)  identification
  probabilities in data and simulation for the DeepAK8 algorithm. The
  corresponding plots for DeepAK8-MD are displayed in the lower
  row. The background event yield is normalized to the total observed data yield.
  The lower panel shows the data to simulation ratio. The solid dark-gray (shaded light-gray)
  band corresponds to the total uncertainty (statistical uncertainty of the simulated samples),
  and the vertical lines correspond to the statistical
  uncertainty of the data. The distributions are weighted so that the jet \pt
  distribution of the simulation matches the data. }
\end{figure}

The distributions of the ImageTop and DeepAK8 discriminants are shown
in Figs.~\ref{fig:qcd_low} and~\ref{fig:pho_low} for jets in the dijet
 and single-$\gamma$ samples, respectively. The overall agreement
between data and simulation in the single-$\gamma$ is better than in the
dijet sample. Moreover, the discrepancy in the shape is mainly
observed at the very low values of the discriminant and is more enhanced
in the \PQt tagging case. The dijet sample is dominated by jets
initiated by gluons, especially at low values of the
discriminant.  In addition, ImageTop and DeepAK8 are very sensitive to mismodeling of quarks or
gluons in the simulation, and so exhibit more sample dependence. QCD multijet
events simulated using \HERWIGpp generally show better agreement with the data.

\subsection{Corrections to simulation}
\label{sec:sfmeasurement}
The measurement of the \PQt~quark and \PW~boson tagging efficiencies in
data are performed in the single-$\mu$ sample using a ``tag-and-probe''
method~\cite{Khachatryan:2010xn}. The muon, in combination
with the \PQb-tagged jet, is used as
the ``tag''. In the opposite hemisphere of the event, the jet is
considered as the ``probe jet''.

 The total SM  sample is decomposed into three
categories based on the spatial separation of the partons from the
\PQt~quark decay with respect to the AK8 jet, following the
discussion in Section~\ref{sec:evtreco}. The ``Merged \PQt~quark''
category includes cases where the three partons and the jet have
$\Delta R < 0.6$. The ``Merged \PW~boson'' category includes cases where
only the two partons from the \PW~boson decay are within $\Delta R <0.6$
of the jet and the \PQb quark from the top quark decay is outside the jet
cone. Any
other topology falls in the ``Nonmerged'' category. In the cases of the
HOTVR and \ecftop algorithms, the matching requirement is adjusted from
0.6 to 1.2.

 The jet mass distributions in
simulation of each one of the three categories are used to derive templates to
fit the jet mass distribution in data. For a given working point, the fit
is done for all three categories simultaneously for both the ``passing'' and ``failing''
events. The fit is performed in the range from
50 to 250\GeV with a bin width of 10\GeV. The sources of
systematic uncertainties discussed in Section~\ref{sec:systematics}
are considered and are treated as nuisance parameters in the
fit. After calculating the efficiencies in data ($\epsilon_{\text{Data}}$)
and simulation ($\epsilon_{\text{Simulation}}$), the SF is
determined as the ratio of $\epsilon_{\text{Data}}$ over $\epsilon_{\text{Simulation}}$.

\begin{figure}[htb!]
  \centering
    \includegraphics[width=.48\textwidth]{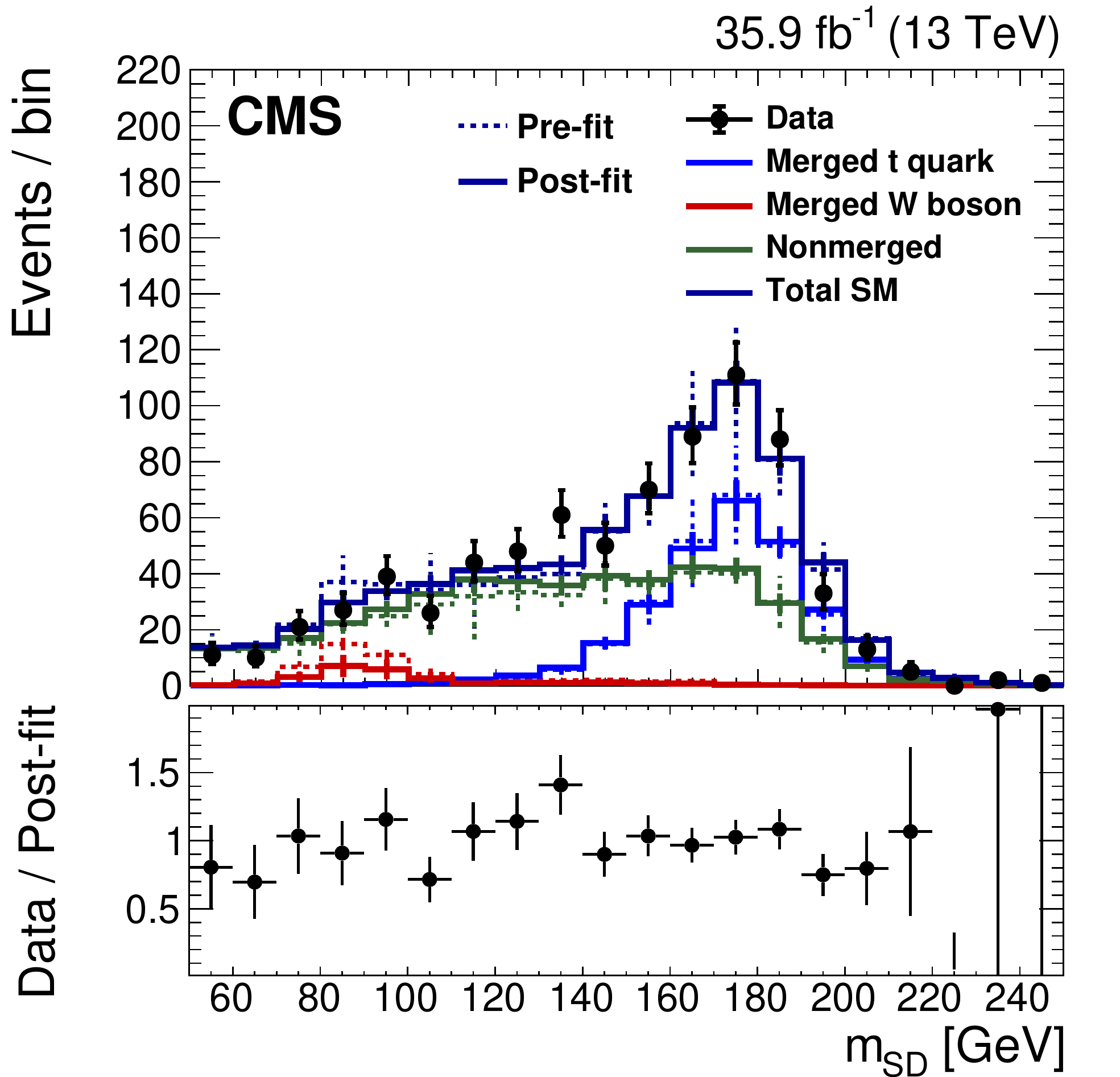}
   \includegraphics[width=.48\textwidth]{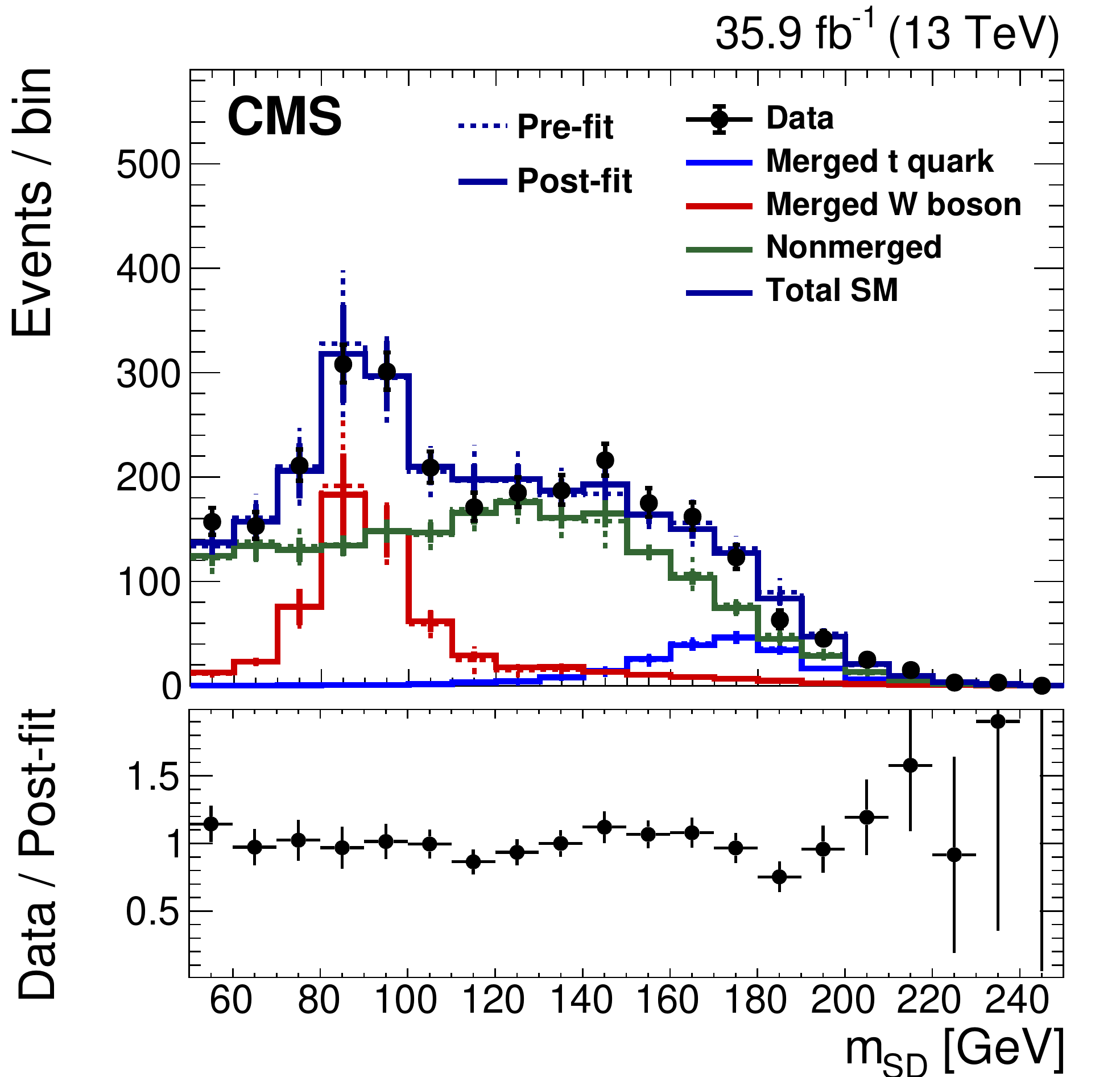}
\caption{\label{fig:fit_example}The \msd distribution in data and
  simulation in the passing (left) and failing (right) categories for DeepAK8-MD for
  the jet \pt in the 400--800\GeV range. The solid lines correspond to the contribution of
  each category after performing the maximum likelihood fit as described in the text.
  The dashed lines are the expectation from simulation before the fit.
  The lower panel shows the data to simulation ratio.  
  The vertical black lines correspond to the total uncertainty, including the 
  statistical uncertainty of the data, after the fit.
}
\end{figure}

The SFs are extracted differentially in jet \pt for the \PQt quark and \PW boson tagging working points discussed in Section~\ref{sec:robustnessstudies}. For the case of
\PQt~quark identification the following exclusive jet \pt regions
are considered: 300--400, 400--480, 480--600, and 600--1200\GeV.
To increase the purity of ``Merged \PW~boson''
candidates, we consider regions with lower jet \pt : 200--300,
300--400, 400--550, and 550--800\GeV. The effects of the systematic
sources discussed in Section~\ref{sec:systematics} are propagated to
uncertainties in the SF. The \msd distributions after
performing the maximum likelihood fit for data and
simulation in the passing and failing categories for 
DeepAK8-MD for $400<\pt<480\GeV$
 are displayed in
Fig.~\ref{fig:fit_example}.

\begin{figure}[htb!]
\centering
\includegraphics[width=0.60\textwidth]{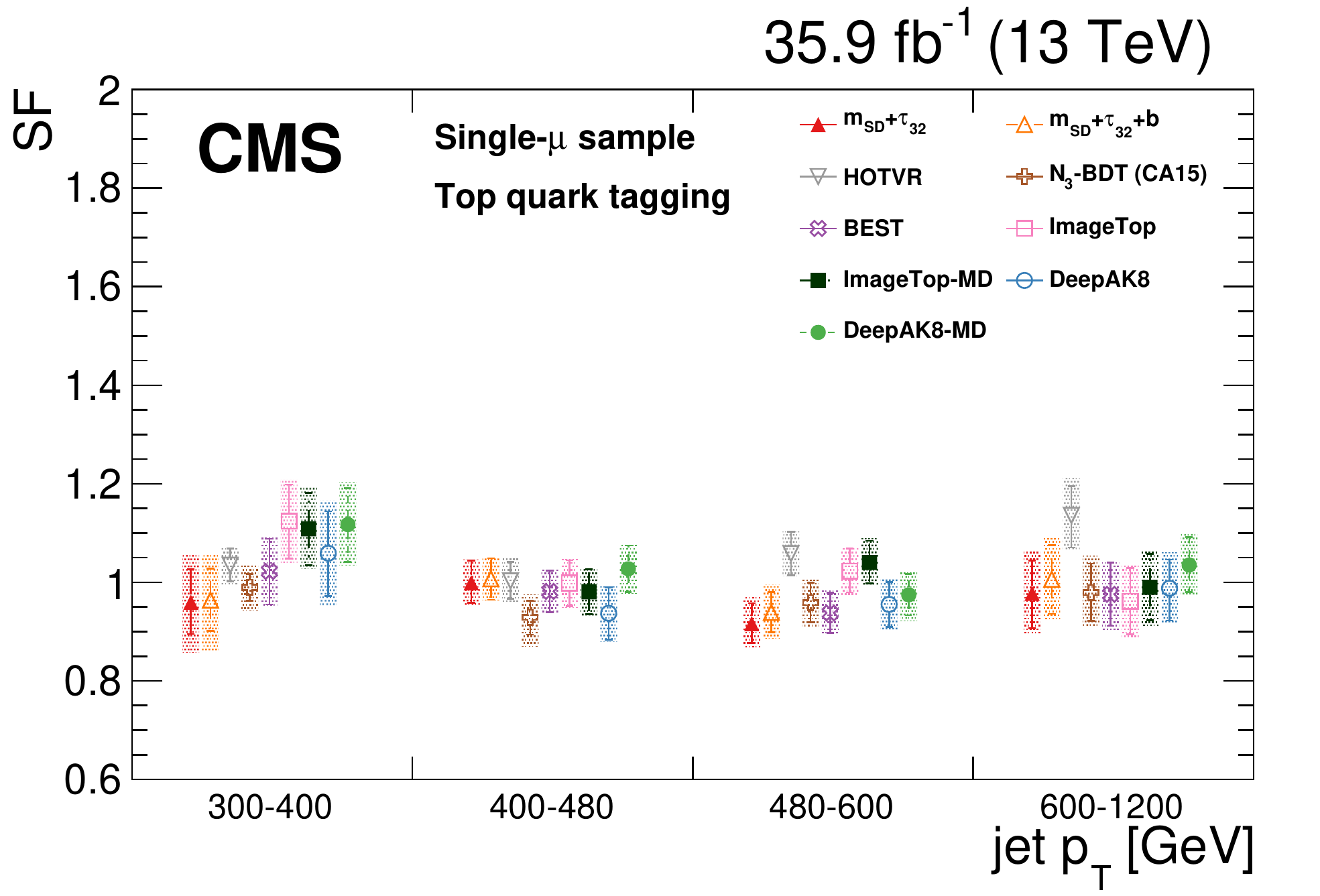}
        \includegraphics[width=0.60\textwidth]{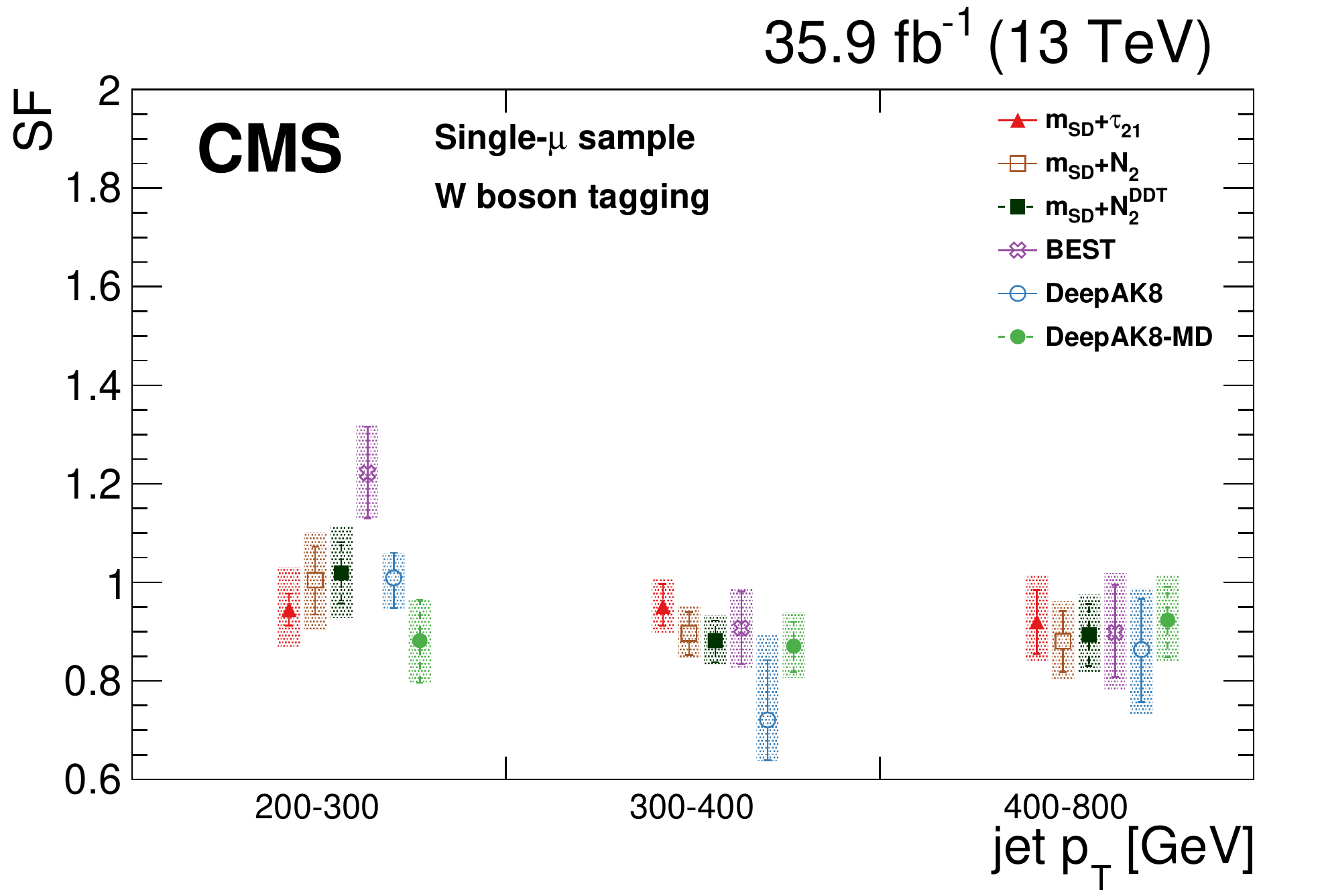}
\caption{\label{fig:top_sf} Summary of the scale factors (SF) measured for each of
  the \PQt~quark (upper) and \PW~boson (lower) identification algorithms.
  The markers correspond
  to the SF value, the error bars to the
  statistical uncertainty on the SF measurement, and the band is the
  total uncertainty, including the systematic component. }
\end{figure}

The SFs measured for each of the \PQt~quark and \PW~boson
identification algorithms are summarized in
Fig.~\ref{fig:top_sf}. The SFs are
typically consistent with unity, within the uncertainties. The largest SF is
measured for the identification of \PQt~quarks using DeepAK8-MD. The
statistical and parton shower uncertainties dominate the SF
measurement. The algorithms
designed to avoid strong dependence on the mass, such as the DeepAK8-MD,
have typically smaller uncertainties than the other
algorithms. The effect of systematic
uncertainties is more pronounced in algorithms that
utilize a larger set of observables to increase discrimination
power. These algorithms (i.e., BEST, ImageTop, and DeepAK8) are more
sensitive to the simulation details. The features are more evident in
the \PW boson case, due to the larger sample size of the ``Merged
\PW~boson'' category compared to the ``Merged \PQt~quark'' category, which
allows for more precise comparisons due to increased number of events. 

\begin{figure}[hp!]
\centering
\includegraphics[width=0.60\textwidth]{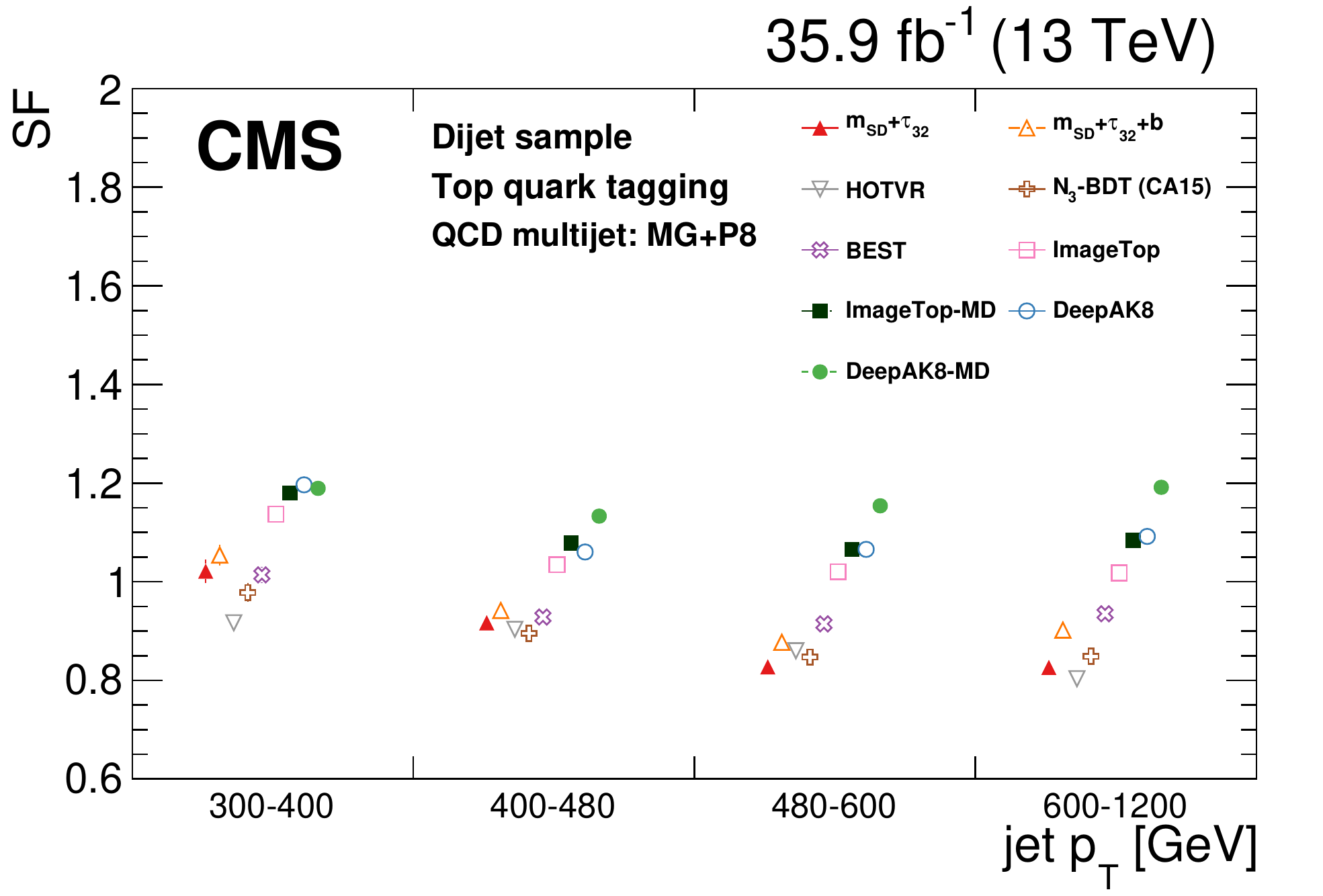}\\
\includegraphics[width=0.60\textwidth]{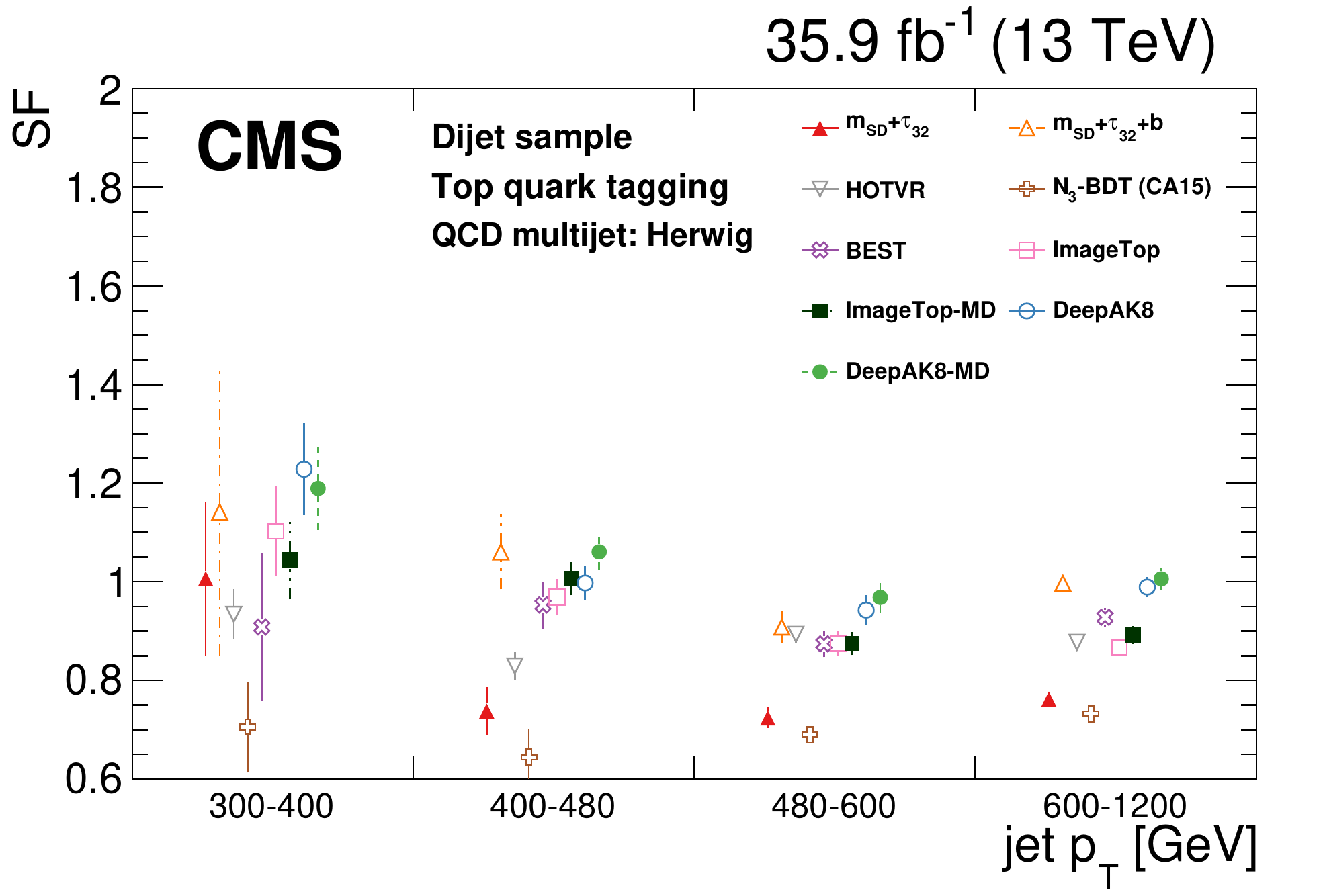}\\
\includegraphics[width=0.60\textwidth]{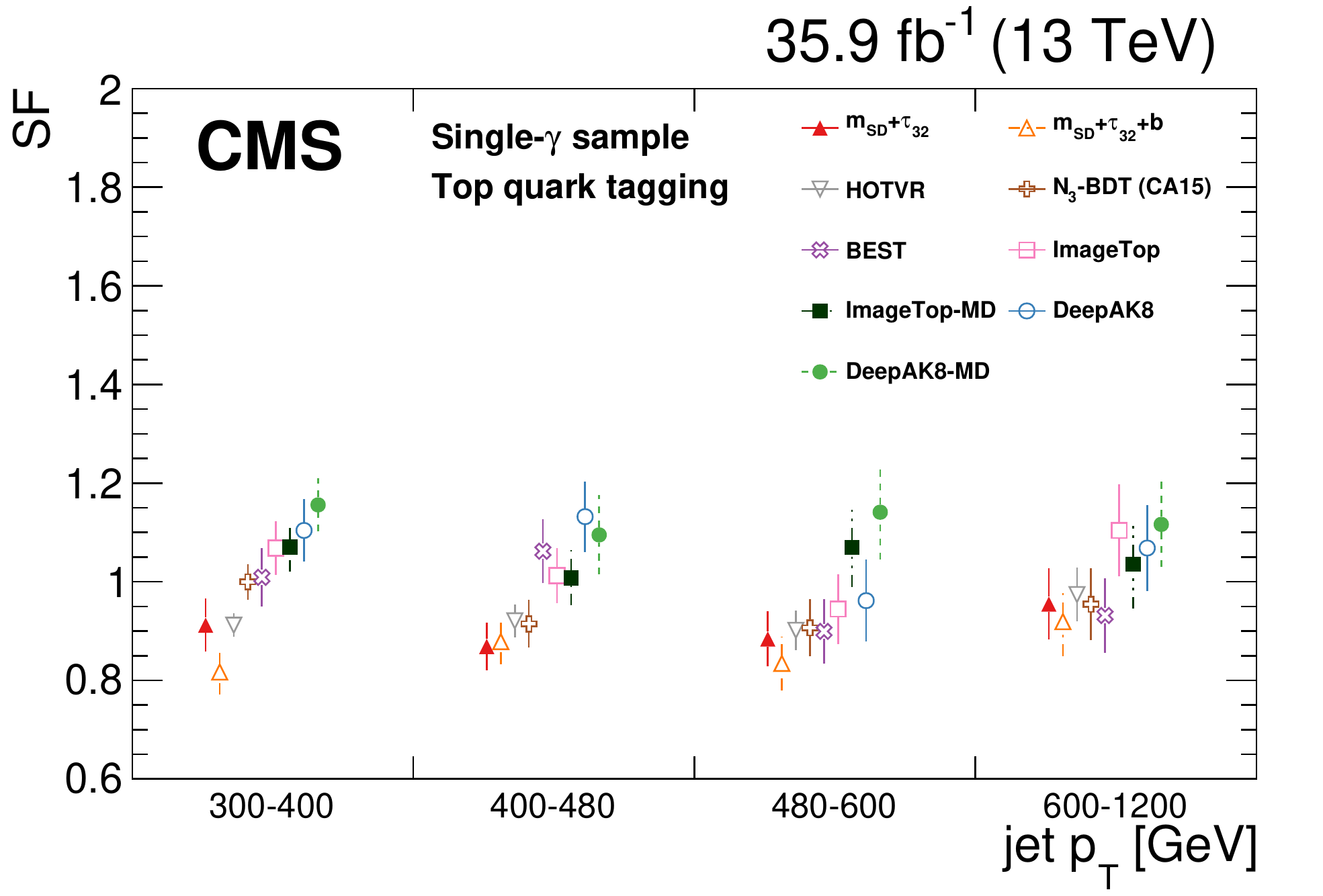}
\caption{\label{fig:misid_top} The ratio of the misidentification rate of
  \PQt~quarks in data and simulation in the dijet (upper and middle rows) and the
  single-$\gamma$ (lower row) samples. The QCD multijet process is
  simulated  using \MADGRAPH for the hard process and \PYTHIA for parton
  showering (upper) and \HERWIGpp for  both
  (middle). The vertical lines correspond to the statistical uncertainty of the data
and the simulated samples.}
\end{figure}

\begin{figure}[hp!]
\centering
\includegraphics[width=0.60\textwidth]{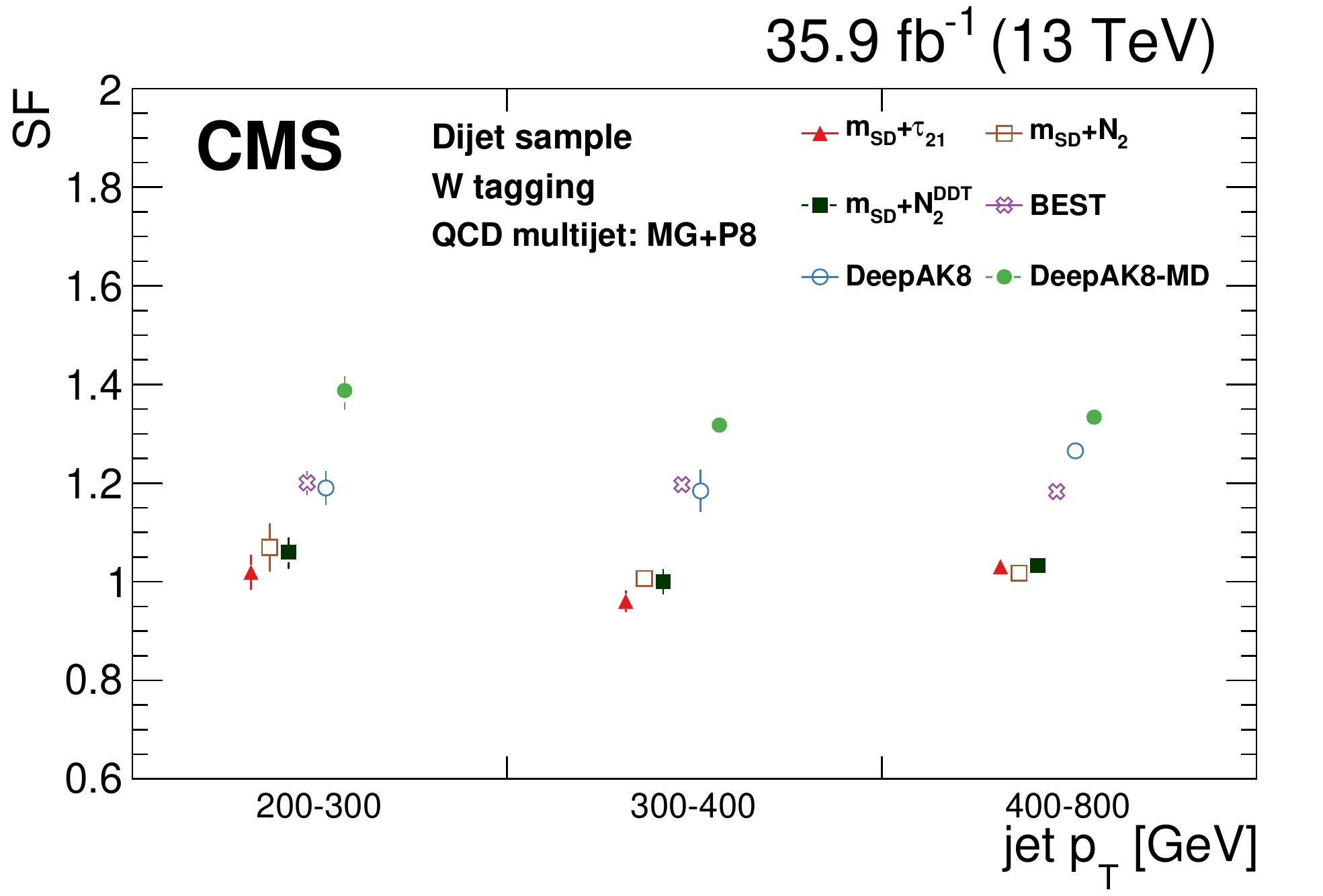}
\includegraphics[width=0.60\textwidth]{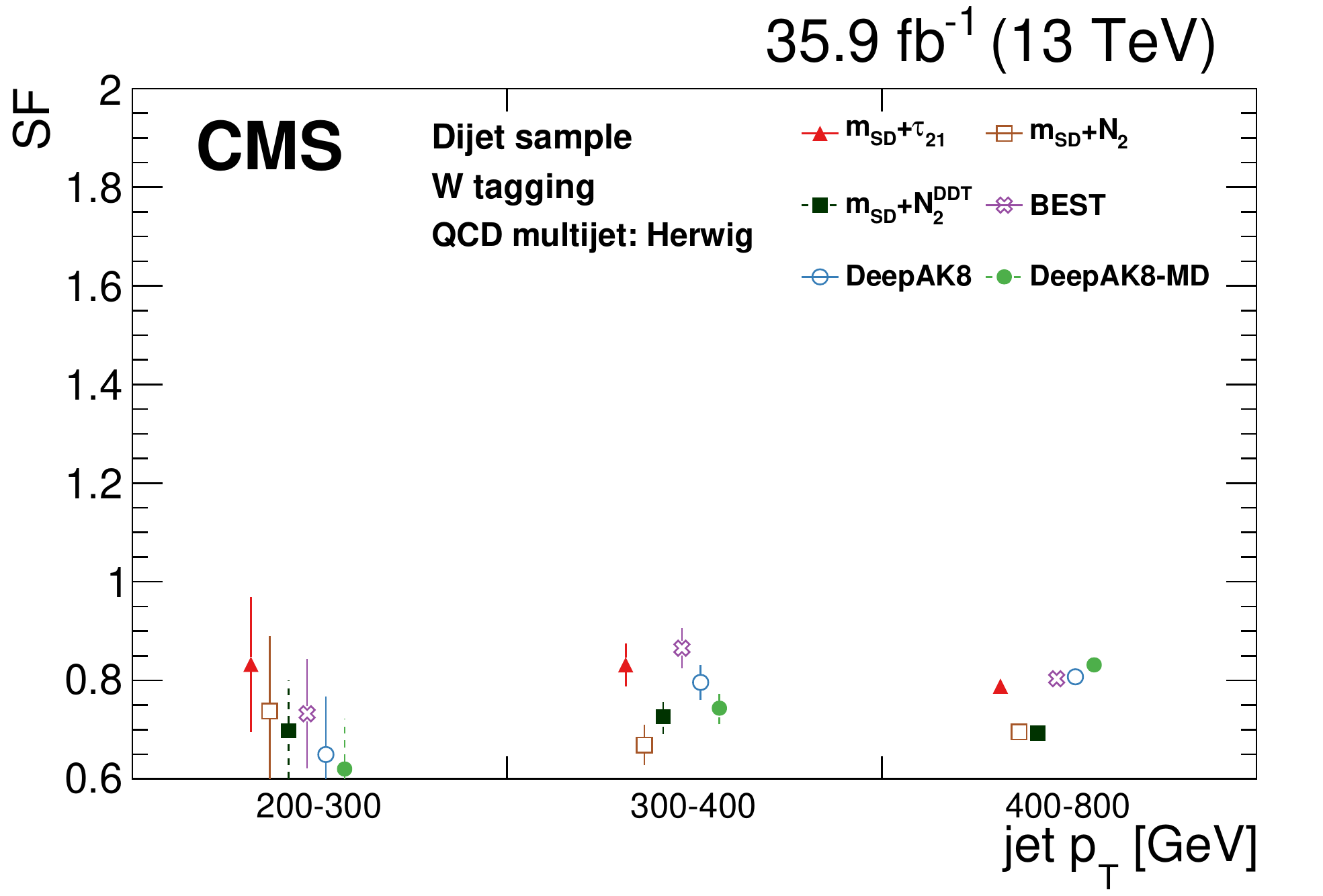}
\includegraphics[width=0.60\textwidth]{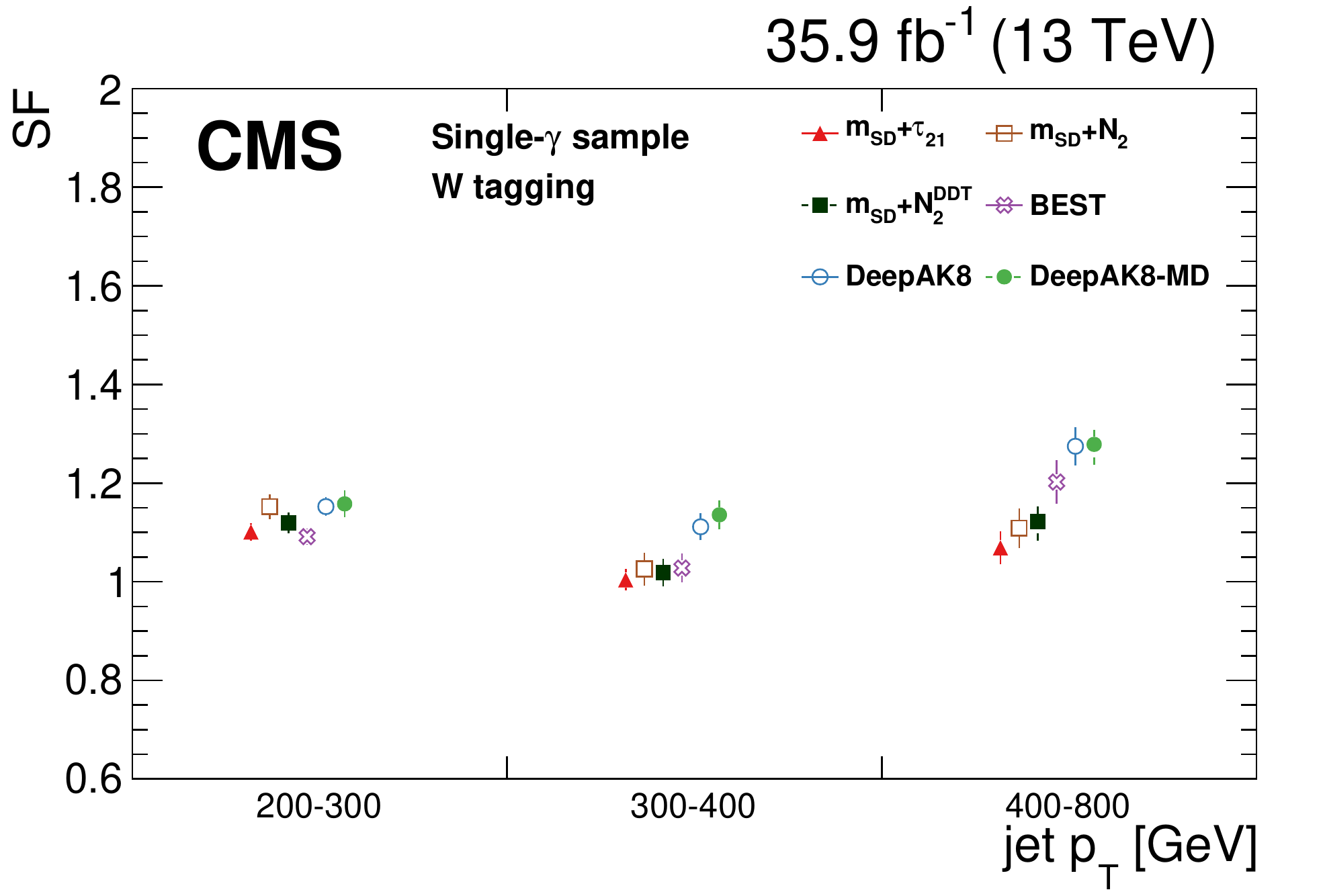}
\caption{\label{fig:misid_w} The ratio of the misidentification rate of
  \PW~bosons in data and simulation in the dijet (upper and middle rows) and the
  single-$\gamma$ (lower row) samples. The QCD multijet process is
  simulated using \MADGRAPH for the hard process and \PYTHIA for parton
  showering (upper) and \HERWIGpp for  both
  (middle). The vertical lines correspond to the statistical uncertainty of the data
and the simulated samples.} 
\end{figure}

The misidentification rate as a function of the jet \pt, is displayed in
Figs.~\ref{fig:misid_top} and \ref{fig:misid_w} for the \PQt and \PW
tagging algorithms. To study the dependence of the
misidentification probability on the matrix element  generator, and on
the parton showering, we use an  additional simulation sample
for the QCD multijet background, which uses
\HERWIGpp for both the hard scattering generation and parton
showering. In some cases, the misidentification probabilities show a significant
 dependence (up to $\sim$25\%) on the
simulation details, particularly for the ImageTop and DeepAK8
algorithms. The main source of this dependence is the description of
 gluon content; these are the only algorithms that have access to
quark-gluon separation to improve the performance. Differences in the
quark/gluon content can have large effects on the uncertainties.

The misidentification probability is also studied in the
single-$\gamma$ sample. Overall the performance in data and simulation
agrees better in this sample than in the dijet
sample. This can be attributed to the fact that the single-$\gamma$
sample has a larger fraction of light-flavor quarks, which are better modeled
in simulation~\cite{CMS-PAS-JME-16-003}.

\section{Summary}
\label{sec:discussion}
\label{sec:summary}

A review of the heavy-object tagging methods recently developed in CMS has been presented.
The variety of tagging strategies is diverse,
 including algorithms
based on more traditional theory-inspired high-level per-jet observables
with and without multivariate techniques,
as well as methods based
on lower-level information from individual particles.
New tagging approaches, such as the
Energy Correlation Functions (ECF) tagger and the Boosted Event Shape Tagger (BEST),
utilize multivariate methods (i.e.,  boosted decision trees and
deep neural networks) on physically motivated high-level observables and
attain enhanced performance. Two novel tagging algorithms, ImageTop and
DeepAK8, are developed based on candidate-level information, allowing the
exploitation of more information, where lower-level information is
processed using advanced machine-learning methods. Moreover, the BEST and
DeepAK8 algorithms are developed to provide multi-class tagging capabilities. Finally,
dedicated versions of the algorithms that are only weakly correlated with
the jet mass are developed. Such tools are particularly important for analyses
that rely on the jet mass sidebands to estimate the background contribution
under the heavy resonance mass. The mass-decorrelated algorithms 
(\ecfvddt, ImageTop-MD, and DeepAK8-MD) typically show
weaker discriminating power than their counterparts. However, they can yield better
sensitivity in some physics analyses because of smaller uncertainties in background
estimations. 

The performances of the various tagging algorithms are directly compared using
simulation in a jet \pt range from 200 to 2000\GeV. Overall, the
application of machine-learning techniques for jet tagging shows strong improvement
compared to cutoff-based methods. The approaches based on low-level
information yield the best performance, with as much as an order of magnitude
gain in background rejection for the same signal efficiency. Another important aspect
essential for the application of the new techniques in physics analysis is the systematic
uncertainties associated to each algorithm. Those based on low-level features and
advanced machine-learning techniques are typically prone to larger systematic uncertainties.
However, these uncertainties are usually small enough to
preserve the significant improvements observed. The techniques have also been validated in
collision data, with scale factors extracted, including systematic uncertainties.
The performances of these tagging algorithms are in good agreement between data and simulation.

\begin{acknowledgments}
  We congratulate our colleagues in the CERN accelerator departments for the excellent performance of the LHC and thank the technical and administrative staffs at CERN and at other CMS institutes for their contributions to the success of the CMS effort. In addition, we gratefully acknowledge the computing centers and personnel of the Worldwide LHC Computing Grid for delivering so effectively the computing infrastructure essential to our analyses. Finally, we acknowledge the enduring support for the construction and operation of the LHC and the CMS detector provided by the following funding agencies: BMBWF and FWF (Austria); FNRS and FWO (Belgium); CNPq, CAPES, FAPERJ, FAPERGS, and FAPESP (Brazil); MES (Bulgaria); CERN; CAS, MoST, and NSFC (China); COLCIENCIAS (Colombia); MSES and CSF (Croatia); RPF (Cyprus); SENESCYT (Ecuador); MoER, ERC IUT, PUT and ERDF (Estonia); Academy of Finland, MEC, and HIP (Finland); CEA and CNRS/IN2P3 (France); BMBF, DFG, and HGF (Germany); GSRT (Greece); NKFIA (Hungary); DAE and DST (India); IPM (Iran); SFI (Ireland); INFN (Italy); MSIP and NRF (Republic of Korea); MES (Latvia); LAS (Lithuania); MOE and UM (Malaysia); BUAP, CINVESTAV, CONACYT, LNS, SEP, and UASLP-FAI (Mexico); MOS (Montenegro); MBIE (New Zealand); PAEC (Pakistan); MSHE and NSC (Poland); FCT (Portugal); JINR (Dubna); MON, RosAtom, RAS, RFBR, and NRC KI (Russia); MESTD (Serbia); SEIDI, CPAN, PCTI, and FEDER (Spain); MOSTR (Sri Lanka); Swiss Funding Agencies (Switzerland); MST (Taipei); ThEPCenter, IPST, STAR, and NSTDA (Thailand); TUBITAK and TAEK (Turkey); NASU (Ukraine); STFC (United Kingdom); DOE and NSF (USA). 

 \hyphenation{Rachada-pisek} Individuals have received support from the Marie-Curie program and the European Research Council and Horizon 2020 Grant, contract Nos. 675440, 752730, and 765710 (European Union); the Leventis Foundation; the A.P. Sloan Foundation; the Alexander von Humboldt Foundation; the Belgian Federal Science Policy Office; the Fonds pour la Formation \`a la Recherche dans l'Industrie et dans l'Agriculture (FRIA-Belgium); the Agentschap voor Innovatie door Wetenschap en Technologie (IWT-Belgium); the F.R.S.-FNRS and FWO (Belgium) under the ``Excellence of Science -- EOS" -- be.h project n. 30820817; the Beijing Municipal Science \& Technology Commission, No. Z191100007219010; the Ministry of Education, Youth and Sports (MEYS) of the Czech Republic; the Deutsche Forschungsgemeinschaft (DFG) under Germany's Excellence Strategy -- EXC 2121 ``Quantum Universe" -- 390833306; the Lend\"ulet (``Momentum") Program and the J\'anos Bolyai Research Scholarship of the Hungarian Academy of Sciences, the New National Excellence Program \'UNKP, the NKFIA research grants 123842, 123959, 124845, 124850, 125105, 128713, 128786, and 129058 (Hungary); the Council of Science and Industrial Research, India; the HOMING PLUS program of the Foundation for Polish Science, cofinanced from European Union, Regional Development Fund, the Mobility Plus program of the Ministry of Science and Higher Education, the National Science Center (Poland), contracts Harmonia 2014/14/M/ST2/00428, Opus 2014/13/B/ST2/02543, 2014/15/B/ST2/03998, and 2015/19/B/ST2/02861, Sonata-bis 2012/07/E/ST2/01406; the National Priorities Research Program by Qatar National Research Fund; the Ministry of Science and Education, grant no. 14.W03.31.0026 (Russia); the Tomsk Polytechnic University Competitiveness Enhancement Program and ``Nauka" Project FSWW-2020-0008 (Russia); the Programa Estatal de Fomento de la Investigaci{\'o}n Cient{\'i}fica y T{\'e}cnica de Excelencia Mar\'{\i}a de Maeztu, grant MDM-2015-0509 and the Programa Severo Ochoa del Principado de Asturias; the Thalis and Aristeia programs cofinanced by EU-ESF and the Greek NSRF; the Rachadapisek Sompot Fund for Postdoctoral Fellowship, Chulalongkorn University and the Chulalongkorn Academic into Its 2nd Century Project Advancement Project (Thailand); the Kavli Foundation; the Nvidia Corporation; the SuperMicro Corporation; the Welch Foundation, contract C-1845; and the Weston Havens Foundation (USA).
\end{acknowledgments}
\clearpage
\bibliography{auto_generated} 

\providecommand{\href}[2]{#2}\begingroup\raggedright\begin{thebibliography}{100}%
\makeatletter
\providecommand{\hrefCMSnoop }[0]{\@secondoftwo}%
\makeatother
\providecommand{\doi}{\texttt{doi:}\begingroup \urlstyle{tt}\Url}

\bibitem{LHC}
\hrefCMSnoop {}{{L. Evans and P. Bryant (editors)}, ``{LHC machine}'',}
  \textit{ JINST} \textbf{ 3} (2008) S08001,
\href{http://dx.doi.org/10.1088/1748-0221/3/08/S08001}{\doi{10.1088/1748-0221/3/08/S08001}}.

\bibitem{Asquith:2018igt}
\hrefCMSnoop {}{R.~Kogler {et~al.}, ``{Jet substructure at the Large Hadron
  Collider : Experimental review}'',} \textit{ Rev. Mod. Phys.} (2018)
  \href{http://dx.doi.org/10.1103/RevModPhys.91.045003}{\doi{10.1103/RevModPhys.91.045003}},
\href{http://www.arXiv.org/abs/1803.06991}{\texttt{arXiv:1803.06991}}.

\bibitem{Larkoski:2017jix}
\hrefCMSnoop {}{A.~J. Larkoski, I.~Moult, and B.~Nachman, ``{Jet substructure
  at the Large Hadron Collider: a review of recent advances in theory and
  machine learning}'',} \textit{ Phys. Rept.} (2017)
  \href{http://dx.doi.org/10.1016/j.physrep.2019.11.001}{\doi{10.1016/j.physrep.2019.11.001}},
\href{http://www.arXiv.org/abs/1709.04464}{\texttt{arXiv:1709.04464}}.

\bibitem{JINST}
\hrefCMSnoop {}{{CMS Collaboration}, ``The {CMS} experiment at the {CERN}
  {LHC}'',} \textit{ JINST} \textbf{ 3} (2008) S08004,
\href{http://dx.doi.org/10.1088/1748-0221/3/08/S08004}{\doi{10.1088/1748-0221/3/08/S08004}}.

\bibitem{Khachatryan:2016gxp}
\hrefCMSnoop {}{{CMS Collaboration}, ``{Measurement of the integrated and
  differential $\text{t}\bar{\text{t}}$ production cross sections for
  high-$p_t$ top quarks in pp collisions at $\sqrt s =$ 8 TeV}'',} \textit{
  Phys. Rev. D} \textbf{ 94} (2016) 072002,
  \href{http://dx.doi.org/10.1103/PhysRevD.94.072002}{\doi{10.1103/PhysRevD.94.072002}},
\href{http://www.arXiv.org/abs/1605.00116}{\texttt{arXiv:1605.00116}}.

\bibitem{Sirunyan:2017yar}
\hrefCMSnoop {}{{CMS Collaboration}, ``{Measurement of the jet mass in highly
  boosted ${\mathrm{t}}\overline{\mathrm{t}}$ events from pp collisions at
  $\sqrt{s}=8$ $\,\text {TeV}$}'',} \textit{ Eur. Phys. J. C} \textbf{ 77}
  (2017) 467,
  \href{http://dx.doi.org/10.1140/epjc/s10052-017-5030-3}{\doi{10.1140/epjc/s10052-017-5030-3}},
\href{http://www.arXiv.org/abs/1703.06330}{\texttt{arXiv:1703.06330}}.

\bibitem{Chatrchyan:2013vbb}
\hrefCMSnoop {}{{CMS Collaboration}, ``{Studies of jet mass in dijet and W/Z +
  jet events}'',} \textit{ JHEP} \textbf{ 05} (2013) 090,
  \href{http://dx.doi.org/10.1007/JHEP05(2013)090}{\doi{10.1007/JHEP05(2013)090}},
\href{http://www.arXiv.org/abs/1303.4811}{\texttt{arXiv:1303.4811}}.

\bibitem{Sirunyan:2018xdh}
\hrefCMSnoop {}{{CMS Collaboration}, ``{Measurements of the differential jet
  cross section as a function of the jet mass in dijet events from
  proton-proton collisions at $ \sqrt{s}=13 $ TeV}'',} \textit{ JHEP} \textbf{
  11} (2018) 113,
  \href{http://dx.doi.org/10.1007/JHEP11(2018)113}{\doi{10.1007/JHEP11(2018)113}},
\href{http://www.arXiv.org/abs/1807.05974}{\texttt{arXiv:1807.05974}}.

\bibitem{Sirunyan:2018asm}
\hrefCMSnoop {}{{CMS Collaboration}, ``{Measurement of jet substructure
  observables in $\mathrm{t\overline{t}}$ events from proton-proton collisions
  at $\sqrt{s}=$ 13 TeV}'',} \textit{ Phys. Rev. D} \textbf{ 98} (2018) 092014,
  \href{http://dx.doi.org/10.1103/PhysRevD.98.092014}{\doi{10.1103/PhysRevD.98.092014}},
\href{http://www.arXiv.org/abs/1808.07340}{\texttt{arXiv:1808.07340}}.

\bibitem{ATLAS:2012am}
\hrefCMSnoop {}{{ATLAS Collaboration}, ``{Jet mass and substructure of
  inclusive jets in $\sqrt{s}=7$ TeV pp collisions with the ATLAS
  experiment}'',} \textit{ JHEP} \textbf{ 05} (2012) 128,
  \href{http://dx.doi.org/10.1007/JHEP05(2012)128}{\doi{10.1007/JHEP05(2012)128}},
\href{http://www.arXiv.org/abs/1203.4606}{\texttt{arXiv:1203.4606}}.

\bibitem{Aaboud:2017qwh}
\hrefCMSnoop {}{{ATLAS Collaboration}, ``{Measurement of the soft-drop jet mass
  in pp collisions at $\sqrt{s} = 13$ TeV with the ATLAS detector}'',} \textit{
  Phys. Rev. Lett.} \textbf{ 121} (2018) 092001,
  \href{http://dx.doi.org/10.1103/PhysRevLett.121.092001}{\doi{10.1103/PhysRevLett.121.092001}},
\href{http://www.arXiv.org/abs/1711.08341}{\texttt{arXiv:1711.08341}}.

\bibitem{Aad:2014haa}
\hrefCMSnoop {}{{ATLAS Collaboration}, ``{Measurement of the cross-section of
  high transverse momentum vector bosons reconstructed as single jets and
  studies of jet substructure in pp collisions at ${\sqrt{s}}$ = 7 TeV with the
  ATLAS detector}'',} \textit{ New J. Phys.} \textbf{ 16} (2014) 113013,
  \href{http://dx.doi.org/10.1088/1367-2630/16/11/113013}{\doi{10.1088/1367-2630/16/11/113013}},
\href{http://www.arXiv.org/abs/1407.0800}{\texttt{arXiv:1407.0800}}.

\bibitem{Aaboud:2018eqg}
\hrefCMSnoop {}{{ATLAS Collaboration}, ``{Measurements of $t\bar{t}$
  differential cross-sections of highly boosted top quarks decaying to
  all-hadronic final states in pp collisions at $\sqrt{s}=13\,$ TeV using the
  ATLAS detector}'',} \textit{ Phys. Rev. D} \textbf{ 98} (2018) 012003,
  \href{http://dx.doi.org/10.1103/PhysRevD.98.012003}{\doi{10.1103/PhysRevD.98.012003}},
\href{http://www.arXiv.org/abs/1801.02052}{\texttt{arXiv:1801.02052}}.

\bibitem{Aaboud:2019aii}
\hrefCMSnoop {}{{ATLAS Collaboration}, ``{Measurement of jet-substructure
  observables in top quark, $W$ boson and light jet production in proton-proton
  collisions at $\sqrt{s}=13$ TeV with the ATLAS detector}'',} \textit{ JHEP}
  \textbf{ 08} (2019) 033,
  \href{http://dx.doi.org/10.1007/JHEP08(2019)033}{\doi{10.1007/JHEP08(2019)033}},
\href{http://www.arXiv.org/abs/1903.02942}{\texttt{arXiv:1903.02942}}.

\bibitem{Dolen:2016kst}
J.~Dolen\hrefCMSnoop {}{ {et~al.}, ``{Thinking outside the ROCs: designing
  decorrelated taggers (DDT) for jet substructure}'',} \textit{ JHEP} \textbf{
  05} (2016) 156,
  \href{http://dx.doi.org/10.1007/JHEP05(2016)156}{\doi{10.1007/JHEP05(2016)156}},
\href{http://www.arXiv.org/abs/1603.00027}{\texttt{arXiv:1603.00027}}.

\bibitem{CMS-PAS-JME-13-007}
\href {http://cds.cern.ch/record/1647419}{{CMS Collaboration}, ``Boosted top
  jet tagging at {CMS}'',} CMS Physics Analysis Summary CMS-PAS-JME-13-007,
  2014.

\bibitem{CMS-PAS-JME-15-002}
\href {http://cds.cern.ch/record/2126325}{{CMS Collaboration}, ``Top tagging
  with new approaches'',} CMS Physics Analysis Summary CMS-PAS-JME-15-002,
  2016.

\bibitem{CMS-PAS-JME-16-003}
\href {http://cds.cern.ch/record/2256875}{{CMS Collaboration}, ``Jet algorithms
  performance in 13 {TeV} data'',} CMS Physics Analysis Summary
  CMS-PAS-JME-16-003, 2017.

\bibitem{Khachatryan:2014vla}
\hrefCMSnoop {}{{CMS Collaboration}, ``{Identification techniques for highly
  boosted W bosons that decay into hadrons}'',} \textit{ JHEP} \textbf{ 12}
  (2014) 017,
  \href{http://dx.doi.org/10.1007/JHEP12(2014)017}{\doi{10.1007/JHEP12(2014)017}},
\href{http://www.arXiv.org/abs/1410.4227}{\texttt{arXiv:1410.4227}}.

\bibitem{Sirunyan:2017ezt}
\hrefCMSnoop {}{{CMS Collaboration}, ``{Identification of heavy-flavour jets
  with the CMS detector in pp collisions at 13 TeV}'',} \textit{ JINST}
  \textbf{ 13} (2018) P05011,
  \href{http://dx.doi.org/10.1088/1748-0221/13/05/P05011}{\doi{10.1088/1748-0221/13/05/P05011}},
\href{http://www.arXiv.org/abs/1712.07158}{\texttt{arXiv:1712.07158}}.

\bibitem{CMS:EGM-14-001}
\hrefCMSnoop {}{{CMS Collaboration}, ``{Performance of photon reconstruction
  and identification with the CMS detector in proton-proton collisions at
  $\sqrt{s} = 8$\TeV}'',} \textit{ JINST} \textbf{ 10} (2015) P08010,
  \href{http://dx.doi.org/10.1088/1748-0221/10/08/P08010}{\doi{10.1088/1748-0221/10/08/P08010}},
\href{http://www.arXiv.org/abs/1502.02702}{\texttt{arXiv:1502.02702}}.

\bibitem{Chatrchyan:2012xi}
\hrefCMSnoop {}{{CMS Collaboration}, ``{Performance of CMS muon reconstruction
  in pp collision events at $\sqrt{s} = 7$\TeV}'',} \textit{ JINST} \textbf{ 7}
  (2012) P10002,
  \href{http://dx.doi.org/10.1088/1748-0221/7/10/P10002}{\doi{10.1088/1748-0221/7/10/P10002}},
\href{http://www.arXiv.org/abs/1206.4071}{\texttt{arXiv:1206.4071}}.

\bibitem{TRK-11-001}
\hrefCMSnoop {}{{CMS Collaboration}, ``{Description and performance of track
  and primary-vertex reconstruction with the CMS tracker}'',} \textit{ JINST}
  \textbf{ 9} (2014) P10009,
  \href{http://dx.doi.org/10.1088/1748-0221/9/10/P10009}{\doi{10.1088/1748-0221/9/10/P10009}},
\href{http://www.arXiv.org/abs/1405.6569}{\texttt{arXiv:1405.6569}}.

\bibitem{Khachatryan:2016bia}
\hrefCMSnoop {}{{CMS Collaboration}, ``{The CMS trigger system}'',} \textit{
  JINST} \textbf{ 12} (2017) P01020,
  \href{http://dx.doi.org/10.1088/1748-0221/12/01/P01020}{\doi{10.1088/1748-0221/12/01/P01020}},
\href{http://www.arXiv.org/abs/1609.02366}{\texttt{arXiv:1609.02366}}.

\bibitem{Randall:1999ee}
\hrefCMSnoop {}{L.~Randall and R.~Sundrum, ``{A large mass hierarchy from a
  small extra dimension}'',} \textit{ Phys. Rev. Lett.} \textbf{ 83} (1999)
  3370,
  \href{http://dx.doi.org/10.1103/PhysRevLett.83.3370}{\doi{10.1103/PhysRevLett.83.3370}},
\href{http://www.arXiv.org/abs/hep-ph/9905221}{\texttt{arXiv:hep-ph/9905221}}.

\bibitem{Randall:1999vf}
\hrefCMSnoop {}{L.~Randall and R.~Sundrum, ``{An alternative to
  compactification}'',} \textit{ Phys. Rev. Lett.} \textbf{ 83} (1999) 4690,
  \href{http://dx.doi.org/10.1103/PhysRevLett.83.4690}{\doi{10.1103/PhysRevLett.83.4690}},
\href{http://www.arXiv.org/abs/hep-th/9906064}{\texttt{arXiv:hep-th/9906064}}.

\bibitem{Alwall:2014hca}
J.~Alwall\hrefCMSnoop {}{ {et~al.}, ``The automated computation of tree-level
  and next-to-leading order differential cross sections, and their matching to
  parton shower simulations'',} \textit{ JHEP} \textbf{ 07} (2014) 079,
  \href{http://dx.doi.org/10.1007/JHEP07(2014)079}{\doi{10.1007/JHEP07(2014)079}},
\href{http://www.arXiv.org/abs/1405.0301}{\texttt{arXiv:1405.0301}}.

\bibitem{Sjostrand:2007gs}
\hrefCMSnoop {}{T.~Sj{\"o}strand, S.~Mrenna, and P.~Z. Skands, ``{A brief
  introduction to PYTHIA 8.1}'',} \textit{ Comput. Phys. Commun.} \textbf{ 178}
  (2008) 852,
  \href{http://dx.doi.org/10.1016/j.cpc.2008.01.036}{\doi{10.1016/j.cpc.2008.01.036}},
\href{http://www.arXiv.org/abs/0710.3820}{\texttt{arXiv:0710.3820}}.

\bibitem{Skands:2014pea}
\hrefCMSnoop {}{P.~Skands, S.~Carrazza, and J.~Rojo, ``{Tuning PYTHIA 8.1: the
  Monash 2013 tune}'',} \textit{ Eur. Phys. J. C} \textbf{ 74} (2014) 3024,
  \href{http://dx.doi.org/10.1140/epjc/s10052-014-3024-y}{\doi{10.1140/epjc/s10052-014-3024-y}},
\href{http://www.arXiv.org/abs/1404.5630}{\texttt{arXiv:1404.5630}}.

\bibitem{Khachatryan:2015pea}
\hrefCMSnoop {}{{CMS Collaboration}, ``{Event generator tunes obtained from
  underlying event and multiparton scattering measurements}'',} \textit{ Eur.
  Phys. J. C} \textbf{ 76} (2016) 155,
  \href{http://dx.doi.org/10.1140/epjc/s10052-016-3988-x}{\doi{10.1140/epjc/s10052-016-3988-x}},
\href{http://www.arXiv.org/abs/1512.00815}{\texttt{arXiv:1512.00815}}.

\bibitem{Ball:2012cx}
\hrefCMSnoop {}{{NNPDF} Collaboration, ``{Parton distributions with LHC
  data}'',} \textit{ Nucl. Phys. B} \textbf{ 867} (2013) 244,
  \href{http://dx.doi.org/10.1016/j.nuclphysb.2012.10.003}{\doi{10.1016/j.nuclphysb.2012.10.003}},
\href{http://www.arXiv.org/abs/1207.1303}{\texttt{arXiv:1207.1303}}.

\bibitem{Nason:2004rx}
\hrefCMSnoop {}{P.~Nason, ``{A new method for combining NLO QCD with shower
  Monte Carlo algorithms}'',} \textit{ JHEP} \textbf{ 11} (2004) 040,
  \href{http://dx.doi.org/10.1088/1126-6708/2004/11/040}{\doi{10.1088/1126-6708/2004/11/040}},
\href{http://www.arXiv.org/abs/hep-ph/0409146}{\texttt{arXiv:hep-ph/0409146}}.

\bibitem{Frixione:2007nw}
\hrefCMSnoop {}{S.~Frixione, P.~Nason, and G.~Ridolfi, ``{A positive-weight
  next-to-leading-order Monte Carlo for heavy flavour hadroproduction}'',}
  \textit{ JHEP} \textbf{ 09} (2007) 126,
  \href{http://dx.doi.org/10.1088/1126-6708/2007/09/126}{\doi{10.1088/1126-6708/2007/09/126}},
\href{http://www.arXiv.org/abs/0707.3088}{\texttt{arXiv:0707.3088}}.

\bibitem{Alioli:2010xd}
\hrefCMSnoop {}{S.~Alioli, P.~Nason, C.~Oleari, and E.~Re, ``{A general
  framework for implementing NLO calculations in shower Monte Carlo programs:
  the POWHEG BOX}'',} \textit{ JHEP} \textbf{ 06} (2010) 043,
  \href{http://dx.doi.org/10.1007/JHEP06(2010)043}{\doi{10.1007/JHEP06(2010)043}},
\href{http://www.arXiv.org/abs/1002.2581}{\texttt{arXiv:1002.2581}}.

\bibitem{Alwall:2007fs}
J.~Alwall\hrefCMSnoop {}{ {et~al.}, ``Comparative study of various algorithms
  for the merging of parton showers and matrix elements in hadronic
  collisions'',} \textit{ Eur. Phys. J. C} \textbf{ 53} (2008) 473,
  \href{http://dx.doi.org/10.1140/epjc/s10052-007-0490-5}{\doi{10.1140/epjc/s10052-007-0490-5}},
\href{http://www.arXiv.org/abs/0706.2569}{\texttt{arXiv:0706.2569}}.

\bibitem{Frederix:2012ps}
\hrefCMSnoop {}{R.~Frederix and S.~Frixione, ``{Merging meets matching in
  MC@NLO}'',} \textit{ JHEP} \textbf{ 12} (2012) 061,
  \href{http://dx.doi.org/10.1007/JHEP12(2012)061}{\doi{10.1007/JHEP12(2012)061}},
\href{http://www.arXiv.org/abs/1209.6215}{\texttt{arXiv:1209.6215}}.

\bibitem{Bahr:2008pv}
M.~B{\"a}hr\hrefCMSnoop {}{ {et~al.}, ``Herwig++ physics and manual'',}
  \textit{ Eur. Phys. J. C} \textbf{ 58} (2008) 639,
  \href{http://dx.doi.org/10.1140/epjc/s10052-008-0798-9}{\doi{10.1140/epjc/s10052-008-0798-9}},
\href{http://www.arXiv.org/abs/0803.0883}{\texttt{arXiv:0803.0883}}.

\bibitem{Bellm:2013hwb}
J.~Bellm\hrefCMSnoop {}{ {et~al.}, ``{Herwig++ 2.7} release note'',} (2013).
\href{http://www.arXiv.org/abs/1310.6877}{\texttt{arXiv:1310.6877}}.

\bibitem{Seymour:2013qka}
\hrefCMSnoop {}{M.~H. Seymour and A.~Siodmok, ``{Constraining MPI models using
  $\sigma_{eff}$ and recent Tevatron and LHC underlying event data}'',}
  \textit{ JHEP} \textbf{ 10} (2013) 113,
  \href{http://dx.doi.org/10.1007/JHEP10(2013)113}{\doi{10.1007/JHEP10(2013)113}},
\href{http://www.arXiv.org/abs/1307.5015}{\texttt{arXiv:1307.5015}}.

\bibitem{PhysRevD.95.092001}
\hrefCMSnoop {}{{CMS Collaboration}, ``Measurement of differential cross
  sections for top quark pair production using the $\text{lepton}+\text{jets}$
  final state in proton-proton collisions at 13 {TeV}'',} \textit{ Phys. Rev.
  D} \textbf{ 95} (2017) 092001,
  \href{http://dx.doi.org/10.1103/PhysRevD.95.092001}{\doi{10.1103/PhysRevD.95.092001}},
  \href{http://www.arXiv.org/abs/1610.04191}{\texttt{arXiv:1610.04191}}.

\bibitem{Agostinelli:2002hh}
\hrefCMSnoop {}{{GEANT4} Collaboration, ``{\GEANTfour}---a simulation
  toolkit'',} \textit{ Nucl. Instrum. Meth. A} \textbf{ 506} (2003) 250,
\href{http://dx.doi.org/10.1016/S0168-9002(03)01368-8}{\doi{10.1016/S0168-9002(03)01368-8}}.

\bibitem{CMS-PRF-14-001}
\hrefCMSnoop {}{{CMS Collaboration}, ``{Particle-flow reconstruction and global
  event description with the CMS detector}'',} \textit{ JINST} \textbf{ 12}
  (2017) P10003,
  \href{http://dx.doi.org/10.1088/1748-0221/12/10/P10003}{\doi{10.1088/1748-0221/12/10/P10003}},
\href{http://www.arXiv.org/abs/1706.04965}{\texttt{arXiv:1706.04965}}.

\bibitem{Cacciari:2008gp}
\hrefCMSnoop {}{M.~Cacciari, G.~P. Salam, and G.~Soyez, ``{The
  anti-$k_\text{T}$ jet clustering algorithm}'',} \textit{ JHEP} \textbf{ 04}
  (2008) 063,
  \href{http://dx.doi.org/10.1088/1126-6708/2008/04/063}{\doi{10.1088/1126-6708/2008/04/063}},
  \href{http://www.arXiv.org/abs/0802.1189}{\texttt{arXiv:0802.1189}}.

\bibitem{Cacciari:2011ma}
\hrefCMSnoop {}{M.~Cacciari, G.~P. Salam, and G.~Soyez, ``{FastJet} user
  manual'',} \textit{ Eur. Phys. J. C} \textbf{ 72} (2012) 1896,
  \href{http://dx.doi.org/10.1140/epjc/s10052-012-1896-2}{\doi{10.1140/epjc/s10052-012-1896-2}},
\href{http://www.arXiv.org/abs/1111.6097}{\texttt{arXiv:1111.6097}}.

\bibitem{Khachatryan:2015hwa}
\hrefCMSnoop {}{{CMS Collaboration}, ``{Performance of electron reconstruction
  and selection with the CMS detector in proton-proton collisions at $\sqrt{s}$
  = 8 TeV}'',} \textit{ JINST} \textbf{ 10} (2015) P06005,
  \href{http://dx.doi.org/10.1088/1748-0221/10/06/P06005}{\doi{10.1088/1748-0221/10/06/P06005}},
\href{http://www.arXiv.org/abs/1502.02701}{\texttt{arXiv:1502.02701}}.

\bibitem{CMS-PAS-JME-14-001}
\href {http://cds.cern.ch/record/1751454}{{CMS Collaboration}, ``Pileup removal
  algorithms'',} CMS Physics Analysis Summary CMS-PAS-JME-14-001, 2014.

\bibitem{Dokshitzer:1997in}
\hrefCMSnoop {}{Y.~L. Dokshitzer, G.~D. Leder, S.~Moretti, and B.~R. Webber,
  ``Better jet clustering algorithms'',} \textit{ JHEP} \textbf{ 08} (1997)
  001,
  \href{http://dx.doi.org/10.1088/1126-6708/1997/08/001}{\doi{10.1088/1126-6708/1997/08/001}},
\href{http://www.arXiv.org/abs/hep-ph/9707323}{\texttt{arXiv:hep-ph/9707323}}.

\bibitem{Wobisch:1998wt}
\href {https://inspirehep.net/record/484872}{M.~Wobisch and T.~Wengler,
  ``Hadronization corrections to jet cross-sections in deep inelastic
  scattering'',} in \textit{ {Proceedings of the Workshop on Monte Carlo
  Generators for HERA Physics, Hamburg, Germany}}, p.~270.
\newblock 1998.
\newblock
\href{http://www.arXiv.org/abs/hep-ph/9907280}{\texttt{arXiv:hep-ph/9907280}}.
\newblock

\bibitem{Bertolini:2014bba}
\hrefCMSnoop {}{D.~Bertolini, P.~Harris, M.~Low, and N.~Tran, ``{Pileup per
  particle identification}'',} \textit{ JHEP} \textbf{ 10} (2014) 059,
  \href{http://dx.doi.org/10.1007/JHEP10(2014)059}{\doi{10.1007/JHEP10(2014)059}},
\href{http://www.arXiv.org/abs/1407.6013}{\texttt{arXiv:1407.6013}}.

\bibitem{puppicms}
\href {http://cdsweb.cern.ch/record/PENDING}{{CMS Collaboration}, ``Pileup
  mitigation at {CMS} in 13 {TeV} data'',} CMS Physics Analysis Summary
  CMS-PAS-JME-18-001, 2019.
\newblock
  \href{http://www.arXiv.org/abs/2003.00503}{\texttt{arXiv:2003.00503}},
  Submitted to JINST.

\bibitem{Khachatryan:2016kdb}
\hrefCMSnoop {}{{CMS Collaboration}, ``{Jet energy scale and resolution in the
  CMS experiment in pp collisions at 8 TeV}'',} \textit{ JINST} \textbf{ 12}
  (2017) P02014,
  \href{http://dx.doi.org/10.1088/1748-0221/12/02/P02014}{\doi{10.1088/1748-0221/12/02/P02014}},
\href{http://www.arXiv.org/abs/1607.03663}{\texttt{arXiv:1607.03663}}.

\bibitem{Sirunyan:2019kia}
\hrefCMSnoop {}{{CMS Collaboration}, ``{Performance of missing transverse
  momentum reconstruction in proton-proton collisions at $\sqrt{s} =$ 13 TeV
  using the CMS detector}'',} \textit{ JINST} \textbf{ 14} (2019), no.~07,
  P07004,
  \href{http://dx.doi.org/10.1088/1748-0221/14/07/P07004}{\doi{10.1088/1748-0221/14/07/P07004}},
\href{http://www.arXiv.org/abs/1903.06078}{\texttt{arXiv:1903.06078}}.

\bibitem{Chatrchyan:2013lca}
\hrefCMSnoop {}{{CMS Collaboration}, ``{Searches for new physics using the
  t$\bar{\text{t}}$ invariant mass distribution in pp collisions at
  $\sqrt{s}$=8 TeV}'',} \textit{ Phys. Rev. Lett.} \textbf{ 111} (2013) 211804,
  \href{http://dx.doi.org/10.1103/PhysRevLett.111.211804}{\doi{10.1103/PhysRevLett.111.211804}},
  \href{http://www.arXiv.org/abs/1309.2030}{\texttt{arXiv:1309.2030}}.
[Erratum: \DOI{10.1103/PhysRevLett.112.119903}].

\bibitem{Khachatryan:2015sma}
\hrefCMSnoop {}{{CMS Collaboration}, ``{Search for resonant t$\bar{\text{t}}$
  production in proton-proton collisions at $\sqrt s=$ 8 TeV}'',} \textit{
  Phys. Rev. D} \textbf{ 93} (2016) 012001,
  \href{http://dx.doi.org/10.1103/PhysRevD.93.012001}{\doi{10.1103/PhysRevD.93.012001}},
\href{http://www.arXiv.org/abs/1506.03062}{\texttt{arXiv:1506.03062}}.

\bibitem{Sirunyan:2017uhk}
\hrefCMSnoop {}{{CMS Collaboration}, ``{Search for $
  \mathrm{t}\overline{\mathrm{t}} $ resonances in highly boosted lepton+jets
  and fully hadronic final states in proton-proton collisions at $ \sqrt{s}=13
  $ TeV}'',} \textit{ JHEP} \textbf{ 07} (2017) 001,
  \href{http://dx.doi.org/10.1007/JHEP07(2017)001}{\doi{10.1007/JHEP07(2017)001}},
\href{http://www.arXiv.org/abs/1704.03366}{\texttt{arXiv:1704.03366}}.

\bibitem{Sirunyan:2018ryr}
\hrefCMSnoop {}{{CMS Collaboration}, ``{Search for resonant $
  \mathrm{t}\overline{\mathrm{t}} $ production in proton-proton collisions at $
  \sqrt{s}=13 $ TeV}'',} \textit{ JHEP} \textbf{ 04} (2019) 031,
  \href{http://dx.doi.org/10.1007/JHEP04(2019)031}{\doi{10.1007/JHEP04(2019)031}},
\href{http://www.arXiv.org/abs/1810.05905}{\texttt{arXiv:1810.05905}}.

\bibitem{Chen:2016wkt}
\hrefCMSnoop {}{J.~Chen, T.~Han, and B.~Tweedie, ``{Electroweak splitting
  functions and high energy showering}'',} \textit{ JHEP} \textbf{ 11} (2017)
  093,
  \href{http://dx.doi.org/10.1007/JHEP11(2017)093}{\doi{10.1007/JHEP11(2017)093}},
  \href{http://www.arXiv.org/abs/1611.00788}{\texttt{arXiv:1611.00788}}.

\bibitem{Dasgupta:2013ihk}
\hrefCMSnoop {}{M.~Dasgupta, A.~Fregoso, S.~Marzani, and G.~P. Salam, ``Towards
  an understanding of jet substructure'',} \textit{ JHEP} \textbf{ 09} (2013)
  029,
  \href{http://dx.doi.org/10.1007/JHEP09(2013)029}{\doi{10.1007/JHEP09(2013)029}},
\href{http://www.arXiv.org/abs/1307.0007}{\texttt{arXiv:1307.0007}}.

\bibitem{Larkoski:2014wba}
\hrefCMSnoop {}{A.~J. Larkoski, S.~Marzani, G.~Soyez, and J.~Thaler, ``Soft
  drop'',} \textit{ JHEP} \textbf{ 05} (2014) 146,
  \href{http://dx.doi.org/10.1007/JHEP05(2014)146}{\doi{10.1007/JHEP05(2014)146}},
\href{http://www.arXiv.org/abs/1402.2657}{\texttt{arXiv:1402.2657}}.

\bibitem{Frye:2016aiz}
\hrefCMSnoop {}{C.~Frye, A.~J. Larkoski, M.~D. Schwartz, and K.~Yan,
  ``{Factorization for groomed jet substructure beyond the next-to-leading
  logarithm}'',} \textit{ JHEP} \textbf{ 07} (2016) 064,
  \href{http://dx.doi.org/10.1007/JHEP07(2016)064}{\doi{10.1007/JHEP07(2016)064}},
\href{http://www.arXiv.org/abs/1603.09338}{\texttt{arXiv:1603.09338}}.

\bibitem{Marzani:2017mva}
\hrefCMSnoop {}{S.~Marzani, L.~Schunk, and G.~Soyez, ``{A study of jet mass
  distributions with grooming}'',} \textit{ JHEP} \textbf{ 07} (2017) 132,
  \href{http://dx.doi.org/10.1007/JHEP07(2017)132}{\doi{10.1007/JHEP07(2017)132}},
\href{http://www.arXiv.org/abs/1704.02210}{\texttt{arXiv:1704.02210}}.

\bibitem{Thaler:2010tr}
\hrefCMSnoop {}{J.~Thaler and K.~Van~Tilburg, ``{Identifying boosted objects
  with $N$-subjettiness}'',} \textit{ JHEP} \textbf{ 03} (2011) 015,
  \href{http://dx.doi.org/10.1007/JHEP03(2011)015}{\doi{10.1007/JHEP03(2011)015}},
\href{http://www.arXiv.org/abs/1011.2268}{\texttt{arXiv:1011.2268}}.

\bibitem{Thaler:2011gf}
\hrefCMSnoop {}{J.~Thaler and K.~Van~Tilburg, ``{Maximizing boosted top
  identification by minimizing $N$-subjettiness}'',} \textit{ JHEP} \textbf{
  02} (2012) 093,
  \href{http://dx.doi.org/10.1007/JHEP02(2012)093}{\doi{10.1007/JHEP02(2012)093}},
\href{http://www.arXiv.org/abs/1108.2701}{\texttt{arXiv:1108.2701}}.

\bibitem{Catani:1993hr}
\hrefCMSnoop {}{S.~Catani, Y.~L. Dokshitzer, M.~H. Seymour, and B.~R. Webber,
  ``{Longitudinally invariant $k_\perp$ clustering algorithms for hadron hadron
  collisions}'',} \textit{ Nucl. Phys. B} \textbf{ 406} (1993) 187,
\href{http://dx.doi.org/10.1016/0550-3213(93)90166-M}{\doi{10.1016/0550-3213(93)90166-M}}.

\bibitem{Ellis:1993tq}
\hrefCMSnoop {}{S.~D. Ellis and D.~E. Soper, ``{Successive combination jet
  algorithm for hadron collisions}'',} \textit{ Phys. Rev. D} \textbf{ 48}
  (1993) 3160,
  \href{http://dx.doi.org/10.1103/PhysRevD.48.3160}{\doi{10.1103/PhysRevD.48.3160}},
\href{http://www.arXiv.org/abs/hep-ph/9305266}{\texttt{arXiv:hep-ph/9305266}}.

\bibitem{Sirunyan:2019jbg}
\hrefCMSnoop {}{{CMS Collaboration}, ``{A multi-dimensional search for new
  heavy resonances decaying to boosted WW, WZ, or ZZ boson pairs in the dijet
  final state at 13 TeV}'',} (2019).
  \href{http://www.arXiv.org/abs/1906.05977}{\texttt{arXiv:1906.05977}}.
Submitted to Eur. Phys. J. C.

\bibitem{Sirunyan:2017wto}
\hrefCMSnoop {}{{CMS Collaboration}, ``{Search for heavy resonances that decay
  into a vector boson and a Higgs boson in hadronic final states at $\sqrt{s} =
  13$ $\,\text {TeV}$}'',} \textit{ Eur. Phys. J. C} \textbf{ 77} (2017) 636,
  \href{http://dx.doi.org/10.1140/epjc/s10052-017-5192-z}{\doi{10.1140/epjc/s10052-017-5192-z}},
\href{http://www.arXiv.org/abs/1707.01303}{\texttt{arXiv:1707.01303}}.

\bibitem{Sirunyan:2018omb}
\hrefCMSnoop {}{{CMS Collaboration}, ``{Search for vector-like T and B quark
  pairs in final states with leptons at $\sqrt{s} =$ 13 TeV}'',} \textit{ JHEP}
  \textbf{ 08} (2018) 177,
  \href{http://dx.doi.org/10.1007/JHEP08(2018)177}{\doi{10.1007/JHEP08(2018)177}},
\href{http://www.arXiv.org/abs/1805.04758}{\texttt{arXiv:1805.04758}}.

\bibitem{Sirunyan:2019sza}
\hrefCMSnoop {}{{CMS Collaboration}, ``{Search for pair production of
  vector-like quarks in the fully hadronic final state}'',} (2019).
  \href{http://www.arXiv.org/abs/1906.11903}{\texttt{arXiv:1906.11903}}.
Submitted to Phys. Rev. D.

\bibitem{Sirunyan:2018fki}
\hrefCMSnoop {}{{CMS Collaboration}, ``{Search for a W$^\prime$ boson decaying
  to a vector-like quark and a top or bottom quark in the all-jets final
  state}'',} \textit{ JHEP} \textbf{ 03} (2019) 127,
  \href{http://dx.doi.org/10.1007/JHEP03(2019)127}{\doi{10.1007/JHEP03(2019)127}},
\href{http://www.arXiv.org/abs/1811.07010}{\texttt{arXiv:1811.07010}}.

\bibitem{Sirunyan:2018rfo}
\hrefCMSnoop {}{{CMS Collaboration}, ``{Search for a heavy resonance decaying
  to a top quark and a vector-like top quark in the lepton+jets final state in
  pp collisions at $\sqrt{s} =$ 13 TeV}'',} \textit{ Eur. Phys. J. C} \textbf{
  79} (2019) 208,
  \href{http://dx.doi.org/10.1140/epjc/s10052-019-6688-5}{\doi{10.1140/epjc/s10052-019-6688-5}},
\href{http://www.arXiv.org/abs/1812.06489}{\texttt{arXiv:1812.06489}}.

\bibitem{Sirunyan:2017ukk}
\hrefCMSnoop {}{{CMS Collaboration}, ``{Searches for W' bosons decaying to a
  top quark and a bottom quark in proton-proton collisions at 13 TeV}'',}
  \textit{ JHEP} \textbf{ 08} (2017) 029,
  \href{http://dx.doi.org/10.1007/JHEP08(2017)029}{\doi{10.1007/JHEP08(2017)029}},
\href{http://www.arXiv.org/abs/1706.04260}{\texttt{arXiv:1706.04260}}.

\bibitem{Lapsien2016}
\hrefCMSnoop {}{T.~Lapsien, R.~Kogler, and J.~Haller, ``A new tagger for
  hadronically decaying heavy particles at the {LHC}'',} \textit{ Eur. Phys. J.
  C} \textbf{ 76} (2016) 600,
  \href{http://dx.doi.org/10.1140/epjc/s10052-016-4443-8}{\doi{10.1140/epjc/s10052-016-4443-8}},
  \href{http://www.arXiv.org/abs/1606.04961}{\texttt{arXiv:1606.04961}}.

\bibitem{Moult2016}
\hrefCMSnoop {}{I.~Moult, L.~Necib, and J.~Thaler, ``{New angles on energy
  correlation functions}'',} \textit{ JHEP} \textbf{ 12} (2016) 153,
  \href{http://dx.doi.org/10.1007/JHEP12(2016)153}{\doi{10.1007/JHEP12(2016)153}},
\href{http://www.arXiv.org/abs/1609.07483}{\texttt{arXiv:1609.07483}}.

\bibitem{Plehn:2009rk}
\hrefCMSnoop {}{T.~Plehn, G.~P. Salam, and M.~Spannowsky, ``{Fat jets for a
  light Higgs}'',} \textit{ Phys. Rev. Lett.} \textbf{ 104} (2010) 111801,
  \href{http://dx.doi.org/10.1103/PhysRevLett.104.111801}{\doi{10.1103/PhysRevLett.104.111801}},
\href{http://www.arXiv.org/abs/0910.5472}{\texttt{arXiv:0910.5472}}.

\bibitem{Plehn:2010st}
\hrefCMSnoop {}{T.~Plehn, M.~Spannowsky, M.~Takeuchi, and D.~Zerwas, ``{Stop
  reconstruction with tagged tops}'',} \textit{ JHEP} \textbf{ 10} (2010) 078,
  \href{http://dx.doi.org/10.1007/JHEP10(2010)078}{\doi{10.1007/JHEP10(2010)078}},
\href{http://www.arXiv.org/abs/1006.2833}{\texttt{arXiv:1006.2833}}.

\bibitem{Kasieczka:2015jma}
G.~Kasieczka\hrefCMSnoop {}{ {et~al.}, ``{Resonance searches with an updated
  top tagger}'',} \textit{ JHEP} \textbf{ 06} (2015) 203,
  \href{http://dx.doi.org/10.1007/JHEP06(2015)203}{\doi{10.1007/JHEP06(2015)203}},
\href{http://www.arXiv.org/abs/1503.05921}{\texttt{arXiv:1503.05921}}.

\bibitem{tmva}
\href {http://pos.sissa.it/archive/conferences/050/040/ACAT_040.pdf}{H.~Voss,
  A.~H{\"o}cker, J.~Stelzer, and F.~Tegenfeldt, ``{TMVA}, the toolkit for
  multivariate data analysis with {ROOT}'',} in \textit{ XIth International
  Workshop on Advanced Computing and Analysis Techniques in Physics Research
  (ACAT)}, p.~40.
\newblock 2007.
\newblock
\href{http://www.arXiv.org/abs/physics/0703039}{\texttt{arXiv:physics/0703039}}.
\newblock

\bibitem{Sirunyan:2018gka}
\hrefCMSnoop {}{{CMS Collaboration}, ``{Search for dark matter in events with
  energetic, hadronically decaying top quarks and missing transverse momentum
  at $ \sqrt{s}=13 $ TeV}'',} \textit{ JHEP} \textbf{ 06} (2018) 027,
  \href{http://dx.doi.org/10.1007/JHEP06(2018)027}{\doi{10.1007/JHEP06(2018)027}},
\href{http://www.arXiv.org/abs/1801.08427}{\texttt{arXiv:1801.08427}}.

\bibitem{Sirunyan:2017nvi}
\hrefCMSnoop {}{{CMS Collaboration}, ``{Search for low mass vector resonances
  decaying into quark-antiquark pairs in proton-proton collisions at $
  \sqrt{s}=13 $ TeV}'',} \textit{ JHEP} \textbf{ 01} (2018) 097,
  \href{http://dx.doi.org/10.1007/JHEP01(2018)097}{\doi{10.1007/JHEP01(2018)097}},
\href{http://www.arXiv.org/abs/1710.00159}{\texttt{arXiv:1710.00159}}.

\bibitem{Sirunyan:2017dgc}
\hrefCMSnoop {}{{CMS Collaboration}, ``{Inclusive search for a highly boosted
  Higgs boson decaying to a bottom quark-antiquark pair}'',} \textit{ Phys.
  Rev. Lett.} \textbf{ 120} (2018) 071802,
  \href{http://dx.doi.org/10.1103/PhysRevLett.120.071802}{\doi{10.1103/PhysRevLett.120.071802}},
\href{http://www.arXiv.org/abs/1709.05543}{\texttt{arXiv:1709.05543}}.

\bibitem{BEST}
\hrefCMSnoop {}{J.~S. Conway, R.~Bhaskar, R.~D. Erbacher, and J.~Pilot,
  ``{Identification of high-momentum top quarks, Higgs bosons, and W and Z
  bosons using boosted event shapes}'',} \textit{ Phys. Rev. D} \textbf{ 94}
  (2016) 094027,
  \href{http://dx.doi.org/10.1103/PhysRevD.94.094027}{\doi{10.1103/PhysRevD.94.094027}},
  \href{http://www.arXiv.org/abs/1606.06859}{\texttt{arXiv:1606.06859}}.

\bibitem{fwm}
\hrefCMSnoop {}{G.~C. Fox and S.~Wolfram, ``Observables for the analysis of
  event shapes in ${\text{e}}^{+}{\text{e}}^{\ensuremath{-}}$ annihilation and
  other processes'',} \textit{ Phys. Rev. Lett.} \textbf{ 41} (1978) 1581,
  \href{http://dx.doi.org/10.1103/PhysRevLett.41.1581}{\doi{10.1103/PhysRevLett.41.1581}}.

\bibitem{sphericity}
\hrefCMSnoop {}{J.~D. Bjorken and S.~J. Brodsky, ``Statistical model for
  electron-positron annihilation into hadrons'',} \textit{ Phys. Rev. D}
  \textbf{ 1} (1970) 1416,
  \href{http://dx.doi.org/10.1103/PhysRevD.1.1416}{\doi{10.1103/PhysRevD.1.1416}}.

\bibitem{thrust}
\hrefCMSnoop {}{E.~Farhi, ``Quantum chromodynamics test for jets'',} \textit{
  Phys. Rev. Lett.} \textbf{ 39} (1977) 1587,
  \href{http://dx.doi.org/10.1103/PhysRevLett.39.1587}{\doi{10.1103/PhysRevLett.39.1587}}.

\bibitem{scikitlearn}
\hrefCMSnoop {}{F.~Pedregosa {et~al.}, ``Scikit-learn: Machine learning in
  {P}ython'',} \textit{ J. Mach. Learn. Res.} \textbf{ 12} (2011) 2825,
  \href{http://www.arXiv.org/abs/1201.0490}{\texttt{arXiv:1201.0490}}.

\bibitem{Nair:2010:RLU:3104322.3104425}
\href {http://dl.acm.org/citation.cfm?id=3104322.3104425}{V.~Nair and G.~E.
  Hinton, ``{Rectified linear units improve restricted Boltzmann machines}'',}
  in \textit{ Proceedings of the 27th International Conference on International
  Conference on Machine Learning}, ICML'10, p.~807.
\newblock Omnipress, USA, 2010.

\bibitem{kingma2014method}
\hrefCMSnoop {}{D.~P. Kingma and J.~Ba, ``Adam: {A} method for stochastic
  optimization'',} in \textit{ Proceedings of the 3rd International Conference
  on Learning Representations}, ICLR'15.
\newblock ICLR, USA, 2015.
\newblock \href{http://www.arXiv.org/abs/1412.6980}{\texttt{arXiv:1412.6980}}.

\bibitem{Macaluso:2018tck}
\hrefCMSnoop {}{S.~Macaluso and D.~Shih, ``{Pulling out all the tops with
  computer vision and deep learning}'',} \textit{ JHEP} \textbf{ 10} (2018)
  121,
  \href{http://dx.doi.org/10.1007/JHEP10(2018)121}{\doi{10.1007/JHEP10(2018)121}},
\href{http://www.arXiv.org/abs/1803.00107}{\texttt{arXiv:1803.00107}}.

\bibitem{Kasieczka:2017nvn}
\hrefCMSnoop {}{G.~Kasieczka, T.~Plehn, M.~Russell, and T.~Schell,
  ``{Deep-learning top taggers or the end of QCD?}'',} \textit{ JHEP} \textbf{
  05} (2017) 006,
  \href{http://dx.doi.org/10.1007/JHEP05(2017)006}{\doi{10.1007/JHEP05(2017)006}},
\href{http://www.arXiv.org/abs/1701.08784}{\texttt{arXiv:1701.08784}}.

\bibitem{tensorflow2015-whitepaper}
\href {http://tensorflow.org/}{M.~Abadi {et~al.}, ``{TensorFlow}: Large-scale
  machine learning on heterogeneous systems'',} 2015.
\newblock Software available from tensorflow.org. \url
  {http://tensorflow.org/}.

\bibitem{DBLP:journals/corr/abs-1212-5701}
\href {http://arxiv.org/abs/1212.5701}{M.~D. Zeiler, ``{ADADELTA:} an adaptive
  learning rate method'',} \textit{ CoRR} (2012)
  \href{http://www.arXiv.org/abs/1212.5701}{\texttt{arXiv:1212.5701}}.

\bibitem{CMS-DP-2018-033}
\href {http://cds.cern.ch/record/2627468}{{CMS Collaboration}, ``Performance of
  b tagging algorithms in proton-proton collisions at 13 {TeV} with phase 1
  {CMS} detector'',} CMS Detector Performance Note CMS-DP-2018-033, 2018.

\bibitem{DBLP:journals/corr/HeZRS15}
\hrefCMSnoop {}{K.~He, X.~Zhang, S.~Ren, and J.~Sun, ``Deep residual learning
  for image recognition'',} (2015).
  \href{http://www.arXiv.org/abs/1512.03385}{\texttt{arXiv:1512.03385}}.

\bibitem{JMLR:v15:srivastava14a}
N.~Srivastava\href {http://jmlr.org/papers/v15/srivastava14a.html}{ {et~al.},
  ``Dropout: A simple way to prevent neural networks from overfitting'',}
  \textit{ J. Mach. Learn. Res} \textbf{ 15} (2014) 1929.

\bibitem{DBLP:journals/corr/ChenLLLWWXXZZ15}
T.~Chen\hrefCMSnoop {}{ {et~al.}, ``Mxnet: A flexible and efficient machine
  learning library for heterogeneous distributed systems'',} in \textit{
  Proceedings of the 30th Conference In Neural Information Processing Systems,
  Workshop on Machine Learning Systems, 2016}, NIPS'16.
\newblock NIPS, Spain, 2016.

\bibitem{NIPS2017_6699}
\href
  {http://papers.nips.cc/paper/6699-learning-to-pivot-with-adversarial-networks.pdf}{G.~Louppe,
  M.~Kagan, and K.~Cranmer, ``Learning to pivot with adversarial networks'',}
  in \textit{ Advances in Neural Information Processing Systems 30}, I.~Guyon
  {et~al.}, eds., p.~981.
\newblock Curran Associates, Inc., 2017.

\bibitem{Klambauer:2017:SNN:3294771.3294864}
\href {http://dl.acm.org/citation.cfm?id=3294771.3294864}{G.~Klambauer,
  T.~Unterthiner, A.~Mayr, and S.~Hochreiter, ``Self-normalizing neural
  networks'',} in \textit{ Proceedings of the 31st International Conference on
  Neural Information Processing Systems}, NIPS'17, p.~972.
\newblock Curran Associates Inc., USA, 2017.

\bibitem{Sirunyan:2018qca}
\hrefCMSnoop {}{{CMS Collaboration}, ``{Search for production of Higgs boson
  pairs in the four b quark final state using large-area jets in proton-proton
  collisions at $\sqrt{s}=$ 13 TeV}'',} \textit{ JHEP} \textbf{ 01} (2019) 040,
  \href{http://dx.doi.org/10.1007/JHEP01(2019)040}{\doi{10.1007/JHEP01(2019)040}},
\href{http://www.arXiv.org/abs/1808.01473}{\texttt{arXiv:1808.01473}}.

\bibitem{Sirunyan:2018ikr}
\hrefCMSnoop {}{{CMS Collaboration}, ``{Search for low-mass resonances decaying
  into bottom quark-antiquark pairs in proton-proton collisions at $\sqrt{s} =$
  13 TeV}'',} \textit{ Phys. Rev. D} \textbf{ 99} (2019) 012005,
  \href{http://dx.doi.org/10.1103/PhysRevD.99.012005}{\doi{10.1103/PhysRevD.99.012005}},
\href{http://www.arXiv.org/abs/1810.11822}{\texttt{arXiv:1810.11822}}.

\bibitem{Sirunyan:2017wif}
\hrefCMSnoop {}{{CMS Collaboration}, ``{Search for direct production of
  supersymmetric partners of the top quark in the all-jets final state in
  proton-proton collisions at $ \sqrt{s}=13 $ TeV}'',} \textit{ JHEP} \textbf{
  10} (2017) 005,
  \href{http://dx.doi.org/10.1007/JHEP10(2017)005}{\doi{10.1007/JHEP10(2017)005}},
\href{http://www.arXiv.org/abs/1707.03316}{\texttt{arXiv:1707.03316}}.

\bibitem{jsdref}
\hrefCMSnoop {}{J.~Lin, ``{Divergence measures based on the Shannon
  entropy}'',} \textit{ IEEE Trans. on Inf. Th.} \textbf{ 37} (1991) 145,
  \href{http://dx.doi.org/10.1109/18.61115}{\doi{10.1109/18.61115}}.

\bibitem{Kullback51klDivergence}
\hrefCMSnoop {}{S.~Kullback and R.~A. Leibler, ``On information and
  sufficiency'',} \textit{ Ann. Math. Statist.} \textbf{ 22} (1951) 79,
  \href{http://dx.doi.org/10.1214/aoms/1177729694}{\doi{10.1214/aoms/1177729694}}.

\bibitem{CMS:2019tbh}
\hrefCMSnoop {}{{CMS Collaboration}, ``{Search for the standard model Higgs
  boson decaying to charm quarks}'',} \textit{ JHEP} \textbf{ 03} (2019) 131,
  \href{http://dx.doi.org/10.1007/JHEP03(2020)131}{\doi{10.1007/JHEP03(2020)131}},
\href{http://www.arXiv.org/abs/1912.01662}{\texttt{arXiv:1912.01662}}.

\bibitem{Sirunyan:2018nqx}
\hrefCMSnoop {}{{CMS Collaboration}, ``{Measurement of the inelastic
  proton-proton cross section at $ \sqrt{s}=13 $ TeV}'',} \textit{ JHEP}
  \textbf{ 07} (2018) 161,
  \href{http://dx.doi.org/10.1007/JHEP07(2018)161}{\doi{10.1007/JHEP07(2018)161}},
\href{http://www.arXiv.org/abs/1802.02613}{\texttt{arXiv:1802.02613}}.

\bibitem{Aaboud:2016mmw}
\hrefCMSnoop {}{{ATLAS Collaboration}, ``{Measurement of the inelastic
  proton-proton cross section at $\sqrt{s} = 13$ TeV with the ATLAS detector at
  the LHC}'',} \textit{ Phys. Rev. Lett.} \textbf{ 117} (2016) 182002,
  \href{http://dx.doi.org/10.1103/PhysRevLett.117.182002}{\doi{10.1103/PhysRevLett.117.182002}},
\href{http://www.arXiv.org/abs/1606.02625}{\texttt{arXiv:1606.02625}}.

\bibitem{Khachatryan:2010xn}
\hrefCMSnoop {}{{CMS Collaboration}, ``{Measurements of inclusive $W$ and $Z$
  cross sections in pp collisions at $\sqrt{s}=7$ TeV}'',} \textit{ JHEP}
  \textbf{ 01} (2011) 080,
  \href{http://dx.doi.org/10.1007/JHEP01(2011)080}{\doi{10.1007/JHEP01(2011)080}},
\href{http://www.arXiv.org/abs/1012.2466}{\texttt{arXiv:1012.2466}}.

\end{thebibliography}\endgroup
\cleardoublepage \appendix\section{The CMS Collaboration \label{app:collab}}\begin{sloppypar}\hyphenpenalty=5000\widowpenalty=500\clubpenalty=5000\vskip\cmsinstskip
\textbf{Yerevan Physics Institute, Yerevan, Armenia}\\*[0pt]
A.M.~Sirunyan$^{\textrm{\dag}}$, A.~Tumasyan
\vskip\cmsinstskip
\textbf{Institut f\"{u}r Hochenergiephysik, Wien, Austria}\\*[0pt]
W.~Adam, F.~Ambrogi, T.~Bergauer, M.~Dragicevic, J.~Er\"{o}, A.~Escalante~Del~Valle, M.~Flechl, R.~Fr\"{u}hwirth\cmsAuthorMark{1}, M.~Jeitler\cmsAuthorMark{1}, N.~Krammer, I.~Kr\"{a}tschmer, D.~Liko, T.~Madlener, I.~Mikulec, N.~Rad, J.~Schieck\cmsAuthorMark{1}, R.~Sch\"{o}fbeck, M.~Spanring, W.~Waltenberger, C.-E.~Wulz\cmsAuthorMark{1}, M.~Zarucki
% ...
\vskip\cmsinstskip
\dag: Deceased\\
1:  Also at Vienna University of Technology, Vienna, Austria\\
% ...
\end{sloppypar}
\end{document}